%% file: splines.tex
\documentclass[%
final,         
a4paper,       
UKenglish,     
headsepline,   
DIV10          
]{scrartcl}
\usepackage[utf8]{inputenc}
\usepackage[UKenglish]{babel}
\usepackage[T1]{fontenc}
\usepackage{textcomp}

\usepackage[verbose,
bookmarksopen=true, 
breaklinks=false,
pdftitle={Mixtures of g-Priors for Generalised Additive Model Selection with
  Penalised Splines},
pdfauthor={Daniel Saban\'es Bov\'e, Leonhard Held, G\"oran Kauermann},
pdfsubject={Statistics},
pdfkeywords={penalised splines, Bayesian variable selection, $g$-prior,
  shrinkage, objective Bayes},
colorlinks=true,
linkcolor=blue,
anchorcolor=black,
citecolor=teal,
urlcolor=purple]{hyperref}  
\usepackage{url}                
\usepackage[sumlimits, intlimits, namelimits]{amsmath} 
\usepackage{amssymb}            
\usepackage[caption = true]{subfig} 
\usepackage{array}              
\usepackage{booktabs}           
\usepackage{ae}                 
\usepackage[longnamesfirst,round]{natbib} 
\usepackage{color}              
\usepackage{tikz}               
\newcommand{\tikzexternaldisable}{}
\usepackage[nogin]{Sweave}      
\usepackage[sc]{mathpazo}       
\usepackage{helvet}             

\input{newCommands}

\begin{document}

\title{{\sffamily Mixtures of $g$-Priors for Generalised Additive Model Selection with
  Penalised Splines}} 
\author{%
  {\sffamily Daniel Saban\'es Bov\'e}%
  \thanks{Division of Biostatistics, Institute of Social
    and Preventive Medicine, University of Zurich, Switzerland. E-mail:
    {\tt \{\mailto[daniel.sabanesbove]{daniel.sabanesbove@ifspm.uzh.ch},%
    \mailto[leonhard.held]{leonhard.held@ifspm.uzh.ch}\}@ifspm.uzh.ch}}
  \and
  {\sffamily Leonhard Held}%
  \footnotemark[1] 
  \and 
  {\sffamily Göran Kauermann}%
  \thanks{Department of Statistics, Ludwig-Maximilians-Universität München,
    Germany. E-mail: \mailto{goeran.kauermann@stat.uni-muenchen.de}}
}
\date{}
\maketitle

\input{abstract}

\input{contents}
\input{appendix}

\input{literatur}
\end{document}

%% file: newCommands.tex
\usepackage{xifthen}            
\usepackage{array}
\usepackage{amssymb}

\newcommand{\blanco}[1]{  } 

\newcommand{\latin}[1]{\textit{#1}}

\newcommand{\abk}[1]{\mbox{#1}\xdot}
\DeclareRobustCommand\xdot{\futurelet\token\Xdot}
\def\Xdot{\ifx\token\bgroup.\else\ifx\token\egroup.\else
  \ifx\token\/.\else\ifx\token\ .\else\ifx\token!.\else
  \ifx\token,.\else\ifx\token:.\else\ifx\token;.\else
  \ifx\token?.\else\ifx\token/.\else\ifx\token'.\else
  \ifx\token).\else\ifx\token-.\else\ifx\token+.\else
  \ifx\token~.\else
  \ifx\token.\else.\ \fi\fi\fi\fi\fi\fi\fi\fi\fi\fi\fi\fi\fi\fi\fi\fi}

\newcommand{\etc}{\abk{\latin{etc}}}

\newcommand{\eg}{\abk{\latin{e.\,g}}}   
\newcommand{\ie}{\abk{\latin{i.\,e}}}

\newcommand{\cp}{\abk{\latin{cp}}}

\newcolumntype{L}[1]{>{$}p{#1}<{$}} 
\newcolumntype{M}{>{$}l<{$}}    
\newcolumntype{N}{>{$}c<{$}}    

{\section*{Acknowledgments}}%
{}

\DeclareMathOperator{\Bin}{Bin} 
\DeclareMathOperator{\Nor}{N} 
  
\DeclareMathOperator{\Unif}{U} 
\DeclareMathOperator{\Var}{Var} 
\DeclareMathOperator{\E}{\mathbb{E}} 
\DeclareMathOperator{\diag}{diag} 
\DeclareMathOperator{\trace}{tr} 
\renewcommand{\P}{\operatorname{\mathbb{P}}} 

\newcommand{\R}{\mathbb{R}}

\newcommand{\ml}[2][]{
  \ifthenelse{\isempty{#1}}%
  {\widehat{#2}_{\scriptscriptstyle{ML}}}%
  {\widehat{#2}^{#1}_{\scriptscriptstyle{ML}}}
}
\newcommand{\sima}{\mathrel{\overset{a}{\thicksim}}} 

\newcommand{\given}{\,\vert\,} 

\newcommand{\abs}[1]{\left\lvert#1\right\rvert} 
\newcommand{\norm}[1]{\left\lVert#1\right\rVert} 
\newcommand{\normsmall}[1]{\lVert#1\rVert} 

\newcommand{\partialv}[3][]{%
  \ifthenelse{\isempty{#1}}
  {\frac{\partial\,#2}{\partial\,#3}}
  {\frac{\partial^{#1} #2}{\partial\,#3^{#1}}} 
} 

\newcommand{\partials}[3][]{%
  \ifthenelse{\isempty{#1}}
  {\frac{d\,#2}{d\,#3}}
  {\frac{d^{#1} #2}{d\,#3^{#1}}}
} 

\newcommand{\dseps}[2][]{%
  \ifthenelse{\isempty{#1}}
  {\frac{d}{d#2}}
  {\frac{d^{#1}}{d#2^{#1}}}
}

\newcommand{\dsepv}[2][]{%
  \ifthenelse{\isempty{#1}}
  {\frac{\partial\,}{\partial\,#2}}
  {\frac{\partial^{#1}}{\partial\,#2^{#1}}}
}

\newcommand{\mailto}[2][]{%
  \ifthenelse{\isempty{#1}}%
  {\href{mailto:#2}{\nolinkurl{#2}}}%
  {\href{mailto:#2}{\nolinkurl{#1}}}%
}



%% file: abstract.tex

\begin{abstract}
  \noindent 

  We propose an objective Bayesian approach to the selection of covariates and
  their penalised splines transformations in generalised additive models.
  Specification of a reasonable default prior for the model parameters and
  combination with a multiplicity-correction prior for the models themselves is
  crucial for this task. Here we use well-studied and well-behaved continuous
  mixtures of $g$-priors as default priors. We introduce the methodology in the
  normal model and extend it to non-normal exponential families. A simulation
  study and an application from the literature illustrate the proposed approach.
  An efficient implementation is available in the \texttt{R}-package
  ``\texttt{hypergsplines}''.

  \noindent \emph{Keywords}: 
  penalised splines, Bayesian variable selection, $g$-prior, shrinkage,
  objective Bayes.
\end{abstract}

%% file: contents.tex
\section{Introduction}

Semiparametric regression has achieved an impressive dissemination over the last
years. Its central idea is to replace parametric regression functions by smooth,
semiparametric components. Following \citet{HastieTibshirani90}, suppose we have
$p$ continuous covariates $x_{1}, \dotsc, x_{p}$ and use the additive model
\begin{equation}
  \label{eq:additive-model}
  y = \beta_{0} + \sum_{j=1}^{p} m_{j} (x_{j}) + \epsilon,
\end{equation}
where $m_{j}$ are smooth but otherwise unspecified functions and $\epsilon \sim
\Nor(0, \sigma^{2})$. For identifiability purposes we further assume that 
$\E\{m_{j}(X_{j})\} = 0$ with respect to the marginal distribution of covariate
$X_{j}$. Estimation of the smooth terms in \eqref{eq:additive-model} can be
carried out in different ways, where we here make use of penalised splines, see
\eg \citet{eilers.marx2010} or \citet{Wood2006b}. 
%
%
A general introduction to penalised spline smoothing has been provided by
\citet{Ruppert:Wand:2003} and the approach has become a popular smoothing
technique since then, see \citet{ruppert.etal2009}. The general idea is to
decompose the function $m_{j}$ into a linear and a nonlinear part, where the
latter is represented through a spline basis, that is
\begin{equation}
  \label{eq:spline-function}
  m_{j}(x_{j}) = x_{j}\beta_{j} +
  \boldsymbol{Z}_{j}(x_{j})^{T}\boldsymbol{u}_{j}.
\end{equation}  
Here $\boldsymbol{Z}_{j}(x_{j})$ is a $K \times 1$ spline basis vector at
position $x_{j}$ and $\boldsymbol{u}_{j}$ is the corresponding coefficient
vector. Conveniently one may choose a truncated polynomial basis for
$\boldsymbol{Z}_{j}(\cdot)$ but representation \eqref{eq:spline-function} holds
in general as well, see \citet{wand.ormerod2008}. To achieve a smooth fit one
imposes a (quadratic) penalty on the spline coefficient vector
$\boldsymbol{u}_{j}$ which is formulated as the normal prior
\begin{equation}
  \label{eq:penalty-prior}
  \boldsymbol{u}_{j} \sim 
  \Nor_{K}(\boldsymbol{0}_{K}, \sigma^{2}\rho_{j}\boldsymbol{I}_{K}),
\end{equation}
where $\boldsymbol{0}_{K}$ is the all-zeros vector and $\boldsymbol{I}_{K}$ is
the identity matrix of dimension $K$. 
Here the variance factor $\rho_{j}$ steers the amount of penalisation (relative
to the regression variance $\sigma^{2}$). A larger $\rho_{j}$ leads to a higher
prior variance of the spline coefficients and hence a more wiggly function
$m_{j}$, while a smaller $\rho_{j}$ leads to a stronger penalty on
$\norm{\boldsymbol{u}_{j}}$ and thus a smoother function $m_{j}$. Setting
$\rho_{j}$ to zero imposes $\boldsymbol{u}_{j}\equiv \boldsymbol{0}_{K}$ so that
$m_{j}(x_{j})$ collapses to a linear term $m_{j}(x_{j})=x_{j} \beta_{j}$. Hence
the role of $\rho_{j}$ ($j=1, \dotsc, p$) extends to the selection of
(generalised) additive models, which will be the focus of this paper. Variable
selection will be treated by allowing the alternative $m_{j}(x_{j}) \equiv 0$.

Variable selection in generalised additive models is important to reduce the
variance of effect estimates due to uninformative covariates. The field is wide
and many different approaches have been proposed in the last years.
\citet{friedman2001} and \citet{tutz.binder2006} describe boosting algorithms,
which are extended by \citet{kneib.etal2009} to geoadditive regression models
\citep{fahrmeir.etal2004}. For the same model class, \citet{belitz.lang2008}
propose to use information-criteria or cross-validation, while
\citet{fahrmeir.etal2010} and \citet{panagiotelis.smith2008} use spike-and-slab
priors for variable and function selection. \citet{brezger.lang2008} adopt the
concept of Bayesian contour probabilities \citep{held2004} to decide on the
inclusion and form of covariate effects. \citet{CottetKohnNott2008}
generalise earlier work by \citet{YauKohnWood2003} to Bayesian
double-exponential regression models, which comprise generalised additive models
as a special case. Shrinkage approaches are proposed by \citet{wood2011} and
\citet{marra.wood2011}. \citet{zhang.lin2006} use a lasso-type penalised
likelihood approach, and \citet{ravikumar.etal2008} and \citet{meier.etal2009}
use penalties favouring both sparsity and smoothness of high-dimensional models.
Likelihood-ratio testing methods are described by \citet{kauermann.tutz2001} and
\citet{cantoni.hastie2002}. This list mirrors the multitude as well as the
variety of the different approaches and the enumeration is, of course, in no way
exhaustive.

In this paper we propose a novel Bayesian variable and function selection
approach based on mixtures of (generalised) $g$-priors. This type of prior for
the parameters in the generalised additive model traces back to the $g$-prior in
the linear model \citep{Zellner1986}. Its hyper-parameter $g$ acts as an inverse
relative prior sample size, and assigning it a hyper-prior solves the
information paradox \citep[section~4.1]{LiangPauloMolinaClydeBerger2008} of the
fixed-$g$ case \citep[p.~148]{BergerPericchi2001} in the linear model. We will
subsequently refer to such mixtures of $g$-priors generically as hyper-$g$
priors. One specific example are the hyper-$g$ priors of
\citet[section~3.2]{LiangPauloMolinaClydeBerger2008}, which enjoy a closed form
for the marginal likelihood and lead to consistent model selection and
model-averaged prediction. We will proceed to use these, because they have been
well studied and have shown good frequentist properties in the Gaussian linear
model, and have already been extended to generalised linear models by
\citet{sabanesbove.held2011}. We follow the conventional prior approach
\citep[section~2.1]{BergerPericchi2001} by using non-informative improper priors
for parameters which are common to all models, and default proper hyper-$g$
priors for model-specific parameters.

While hyper-$g$ priors have been discussed extensively in the Bayesian variable
selection literature, \eg by \citet{CuiGeorge2008},
\citet{LiangPauloMolinaClydeBerger2008}, \citet{forte2011} and
\citet{celeux.etal2012}, this is the first paper to our knowledge that applies
hyper-$g$ priors to generalised additive models. The general idea of applying
default priors (as hyper-$g$ priors) which have originally been developed for
linear models to generalised additive models is new. The methodology presented
here is straightforward to use with other default priors. The rationale is that
default priors have carefully and exhaustively been constructed for the linear
model, so their advantages should be used when inferring about generalised
additive models. Moreover, this paper is one of the few Bayesian papers
considering automatic and simultaneous variable selection and transformation.

The current work generalises the hyper-$g$ priors for generalised linear models
\citep{sabanesbove.held2011}. In the same paper, we showed how fractional
polynomials \citep[FPs, ][]{sabanesbove.held2011a}, which extend ordinary polynomials by
square roots, reciprocals and the logarithm, can be used to model nonlinear
covariate effects. However, FPs have the disadvantage of
being not invariant to linear transformations of the covariates.
For variable and function selection, \citet{fahrmeir.etal2010} and
\citet{scheipl.etal2011} use a mixture of two inverse-gamma distributions with a very
small (``spike'') and a larger mean (``slab'') as a hyper-prior for the
variances of the regression coefficients' independent normal priors. The
posterior probability for inclusion of a coefficient is then estimated from the
proportion of Markov chain Monte Carlo (MCMC) variance samples in the ``slab''.
While this prior structure eases the MCMC algorithm, it does not take into
account the correlation structure of the covariates, and depends on the
specification of the prior means in the two mixture components.
\citet{CottetKohnNott2008} also use independent normal inverse-gamma priors for
the regression coefficients, but they explicitly exclude coefficients from the
model. For nonlinear effects they utilise low-rank approximations of smoothing
splines, which require the choice of a threshold on the eigenvalue scale.

The paper is organised as follows. We first apply the hyper-$g$ priors of
\citet{LiangPauloMolinaClydeBerger2008} to additive models in
Section~\ref{sec:normal-models}. The methodology is extended to generalised
additive models in Section~\ref{sec:non-normal-models}. A
multiplicity-correction prior on the model space and a stochastic search
procedure are described in Section~\ref{sec:model-prior-and-search}. We apply our
approach to simulated and real data in Section~\ref{sec:applications} and suggest
postprocessing techniques in Section~\ref{sec:postprocessing}.
Section~\ref{sec:discussion} closes the paper with a discussion.

\section{Hyper-$g$ Priors for Additive Models}
\label{sec:normal-models}

Assume we have observed independent responses $y_{i}$ at covariate values
$x_{i1}, \dotsc, x_{ip}$, $i=1, \dotsc, n$, from the additive normal
model~\eqref{eq:additive-model}. For each covariate $j=1, \dotsc, p$, we stack
the covariate values into the $n \times 1$ vector $\tilde{\boldsymbol{x}}_{j} =
(x_{1j}, \dotsc, x_{nj})^{T}$ and the spline basis vectors into the $n \times K$
matrix $\tilde{\boldsymbol{Z}}_{j} = (\boldsymbol{Z}_{j}(x_{1j}), \dotsc,
\boldsymbol{Z}_{j}(x_{nj}))^{T}$. The subsequent Gram-Schmidt process
\citep[see][]{bjoerck1967} 
\begin{align}
  \boldsymbol{x}_{j} &= 
  \tilde{\boldsymbol{x}}_{j}
  - \boldsymbol{1}_{n}
  \frac{\boldsymbol{1}_{n}^{T}\tilde{\boldsymbol{x}}_{j}}
  {\boldsymbol{1}_{n}^{T}\boldsymbol{1}_{n}} 
  = \tilde{\boldsymbol{x}}_{j}
  - \boldsymbol{1}_{n}\bar{x}_{j},
  \label{eq:normal-models:center-vector}
  \\
  \boldsymbol{Z}_{j} &=  
  \tilde{\boldsymbol{Z}}_{j}
  - \boldsymbol{1}_{n}
  \frac{\boldsymbol{1}_{n}^{T}\tilde{\boldsymbol{Z}}_{j}}
  {\boldsymbol{1}_{n}^{T}\boldsymbol{1}_{n}} 
  - \boldsymbol{x}_{j}
  \frac{\boldsymbol{x}_{j}^{T}\tilde{\boldsymbol{Z}}_{j}}
  {\boldsymbol{x}_{j}^{T}\boldsymbol{x}_{j}},
  \label{eq:normal-models:center-matrix}
\end{align}
where $\boldsymbol{1}_{n}$ denotes the all-ones vector of dimension $n$, 
ensures that $\boldsymbol{1}_{n}$, $\boldsymbol{x}_{j}$ and the columns of
$\boldsymbol{Z}_{j}$ are orthogonal to each other, \ie
$\boldsymbol{1}_{n}^{T}\boldsymbol{x}_{j} = 0$ and
$\boldsymbol{1}_{n}^{T}\boldsymbol{Z}_{j} =
\boldsymbol{x}_{j}^{T}\boldsymbol{Z}_{j} = \boldsymbol{0}_{K}$. 

A central measure of model complexity is the degrees of freedom. While in
parametric models this is just the number of parameters, for smoothing and
mixed models \citet[section~2.2]{aerts.etal2002} translate the variance factor
$\rho_{j}$ into the corresponding degrees of freedom
\begin{equation}
  \label{eq:normal-models:degrees-of-freedom}
  d_{j}(\rho_{j}) = 
  \trace \{ 
  (\boldsymbol{Z}_{j}^{T}
  \boldsymbol{Z}_{j} + \rho_{j}^{-1} \boldsymbol{I})^{-1} \boldsymbol{Z}_{j}^{T}
  \boldsymbol{Z}_{j} 
  \} 
  + 1
  \in (1, K+1)
\end{equation}
for a smoothly modelled covariate effect $m_{j}$. Note that $d_{j}(\rho_{j}) =
\sum_{k=1}^{K} \lambda_{jk} / (\lambda_{jk} + \rho_{j}^{-1})$ is easy to
calculate via the (positive) eigenvalues $\lambda_{jk}$ of
$\boldsymbol{Z}_{j}^{T}\boldsymbol{Z}_{j}$. This also shows that
$d_{j}(\rho_{j})$ is strictly increasing with derivative
$\sum_{k=1}^{K}\lambda_{jk}/(\rho_{j}\lambda_{jk} + 1)^{2} > 0$, which implies
that we may (numerically) invert the function to $\rho_{j}(d_{j})$. In fact, by
fixing the degrees of freedom $d_{j}$ for function $m_{j}(x_{j})$ we define the
variance factor $\rho_{j}$. Subsequently we will restrict the degrees of freedom
to take values in a finite set $\mathcal{D} \subset \{0\} \cup [1, K+1)$, say
$\mathcal{D} = \{0, 1, 2, 3, \dotsc, K\}$. For $d_{j}=0$ we set
$m_{j}(x_{j})\equiv 0$ while for $d_{j}=1$ we have the linear model
$m_{j}(x_{j})=x_{j}\beta_{j}$. In general, model \eqref{eq:additive-model} is
indexed by $\boldsymbol{d}=(d_{1}, \dotsc, d_{p})$ giving the degrees of freedom
for each functional component and hence the structure of the model.

After combining the $I = \sum_{j=1}^{p}\mathbb{I}(d_{j} \geq 1)$
vectors $\boldsymbol{x}_{j}$ to the $n \times I$ linear design
matrix $\boldsymbol{X}_{\boldsymbol{d}} = (\boldsymbol{x}_{j} : d_{j} \geq 1)$
and the $J = \sum_{j=1}^{p}\mathbb{I}(d_{j} > 1)$ matrices
$\boldsymbol{Z}_{j}$ to the $n \times JK$ spline design
matrix $\boldsymbol{Z}_{\boldsymbol{d}} = (\boldsymbol{Z}_{j} : d_{j} > 1)$, and
analogously constructing the respective coefficient vectors
$\boldsymbol{\beta}_{\boldsymbol{d}}$ and $\boldsymbol{u}_{\boldsymbol{d}}$, the
conditional additive model for the response vector $\boldsymbol{y} = (y_{1},
\dotsc, y_{n})^{T}$ is
\begin{equation}
  \boldsymbol{y} \given 
  \beta_{0}, \boldsymbol{\beta}_{\boldsymbol{d}},
  \boldsymbol{u}_{\boldsymbol{d}}, \sigma^{2}
  \sim
  \Nor_{n} \left(
    \boldsymbol{1}_{n} \beta_{0} + 
    \boldsymbol{X}_{\boldsymbol{d}}\boldsymbol{\beta}_{\boldsymbol{d}} +
    \boldsymbol{Z}_{\boldsymbol{d}}\boldsymbol{u}_{\boldsymbol{d}},
    \,
    \sigma^{2}\boldsymbol{I}_{n}
  \right).
  \label{eq:normal-models:conditional}
\end{equation}
Integrating out the the spline coefficient vector
$\boldsymbol{u}_{\boldsymbol{d}} \sim \Nor_{JK} (\boldsymbol{0}_{JK},
\sigma^{2}\boldsymbol{D}_{\boldsymbol{d}})$, where
$\boldsymbol{D}_{\boldsymbol{d}}$ is block-diagonal
with $J$ blocks $\rho_{j}\boldsymbol{I}_{K}$ ($d_{j} > 1$), yields the marginal
model
\begin{equation}
  \boldsymbol{y} \given
  \beta_{0}, \boldsymbol{\beta}_{\boldsymbol{d}}, \sigma^{2}
  \sim
  \Nor_{n} \left(
    \boldsymbol{1}_{n} \beta_{0} + 
    \boldsymbol{X}_{\boldsymbol{d}}\boldsymbol{\beta}_{\boldsymbol{d}},
    \,
    \sigma^{2}\boldsymbol{V}_{\boldsymbol{d}}
  \right)
  \label{eq:normal-models:marginal}  
\end{equation}
with $\boldsymbol{V}_{\boldsymbol{d}} = \boldsymbol{I}_{n} +
\boldsymbol{Z}_{\boldsymbol{d}}\boldsymbol{D}_{\boldsymbol{d}}
\boldsymbol{Z}_{\boldsymbol{d}}^{T}$. This general linear model can be
decorrelated into a standard linear model by using the Cholesky decomposition
$\boldsymbol{V}_{\boldsymbol{d}} = \boldsymbol{V}_{\boldsymbol{d}}^{T/2}
\boldsymbol{V}_{\boldsymbol{d}}^{1/2}$: For the transformed response vector
$\tilde{\boldsymbol{y}} = \boldsymbol{V}_{\boldsymbol{d}}^{-T/2}\boldsymbol{y}$
we have
\begin{equation}
  \tilde{\boldsymbol{y}} \given
  \beta_{0}, \boldsymbol{\beta}_{\boldsymbol{d}}, \sigma^{2}
  \sim 
  \Nor_{n}
  \left(
    \tilde{\boldsymbol{1}}_{n}\beta_{0} + 
    \tilde{\boldsymbol{X}}_{\boldsymbol{d}}\boldsymbol{\beta}_{\boldsymbol{d}},
    \,
    \sigma^{2} \boldsymbol{I}_{n}
  \right)
  \label{eq:normal-models:decorrelated}
\end{equation}
with analogously transformed all-ones vector $\tilde{\boldsymbol{1}}_{n}=
\boldsymbol{V}_{\boldsymbol{d}}^{-T/2}\boldsymbol{1}_{n}$ and design matrix
$\tilde{\boldsymbol{X}}_{\boldsymbol{d}} =
\boldsymbol{V}_{\boldsymbol{d}}^{-T/2}\boldsymbol{X}_{\boldsymbol{d}}$.
Note that now also $\tilde{\boldsymbol{y}}$ and $\tilde{\boldsymbol{1}}_{n}$
depend on the model $\boldsymbol{d}$, but we suppress this dependence for ease
of notation.

We will now show how to use the hyper-$g$ priors of
\citet{LiangPauloMolinaClydeBerger2008} for the parameters
$\beta_{0}$, $\boldsymbol{\beta}_{\boldsymbol{d}}$ and $\sigma^{2}$ in the
decorrelated model~\eqref{eq:normal-models:decorrelated}. The hyper-$g$ priors
comprise a locally uniform prior $f(\beta_{0}) \propto 1$ on the intercept,
Jeffreys' prior $f(\sigma^{2}) \propto (\sigma^{2})^{-1}$ on the regression
variance and the $g$-prior \citep{Zellner1986}
\begin{equation}
  \boldsymbol{\beta}_{\boldsymbol{d}} \given g, \sigma^{2} 
  \sim 
  \Nor_{I}\left(
    \boldsymbol{0}_{I},
    \,
    g\sigma^{2} 
    (\tilde{\boldsymbol{X}}_{\boldsymbol{d}}^{T}
    \tilde{\boldsymbol{X}}_{\boldsymbol{d}})^{-1}
  \right)
  \label{eq:normal-models:g-prior} 
\end{equation}
on the linear coefficient vector. Note that the prior precision matrix
in~\eqref{eq:normal-models:g-prior} is proportional to
$\sigma^{-2}\tilde{\boldsymbol{X}}_{\boldsymbol{d}}^{T}
\tilde{\boldsymbol{X}}_{\boldsymbol{d}} =
\sigma^{-2}\boldsymbol{X}_{\boldsymbol{d}}^{T}
\boldsymbol{V}_{\boldsymbol{d}}^{-1}\boldsymbol{X}_{\boldsymbol{d}}$, which is
the Fisher information matrix of $\boldsymbol{\beta}_{\boldsymbol{d}}$ in
model~\eqref{eq:normal-models:marginal}. The prior construction is completed
with either a uniform hyper-prior on the shrinkage coefficient $g/(1 + g)$,
\[
\frac{g}{1 + g} \sim \Unif(0, 1),
\]
leading to the hyper-$g$ prior, or with
\[
\frac{g/n}{1 + g/n} \sim \Unif(0, 1),
\]
leading to the hyper-$g/n$ prior. We recommend to use the latter, because it
also leads to consistent posterior model probabilities if the true model is the
null model (see Table~\ref{tab:med-post-prob-true-model} in
Section~\ref{sec:applications:sim-study} for an illustration of 
this). 
 
Basically all formulae given by \citet{LiangPauloMolinaClydeBerger2008} carry
over to our setting, since inner products of the response vector
$\boldsymbol{y}$, the all-ones vector $\boldsymbol{1}_{n}$ and the design matrix
$\boldsymbol{X}_{\boldsymbol{d}}$ in model~\eqref{eq:normal-models:marginal}
carry over to their transformed counterparts $\tilde{\boldsymbol{y}}$,
$\tilde{\boldsymbol{1}}_{n}$ and $\tilde{\boldsymbol{X}}_{\boldsymbol{d}}$ in
model~\eqref{eq:normal-models:decorrelated}. This is due to
\begin{equation}
  \label{eq:normal-models:marginal-precision}
  \boldsymbol{V}_{\boldsymbol{d}}^{-1} 
  = 
  (\boldsymbol{I}_{n} +
  \boldsymbol{Z}_{\boldsymbol{d}} 
  \boldsymbol{D}_{\boldsymbol{d}}
  \boldsymbol{Z}_{\boldsymbol{d}}^{T})^{-1}  
  =
  \boldsymbol{I}_{n} -
  \boldsymbol{Z}_{\boldsymbol{d}}
  (\boldsymbol{Z}_{\boldsymbol{d}}^{T}\boldsymbol{Z}_{\boldsymbol{d}}
  + \boldsymbol{D}_{\boldsymbol{d}}^{-1})^{-1} 
  \boldsymbol{Z}_{\boldsymbol{d}}^{T},
\end{equation}
which follows from the matrix inversion lemma
\citep[see][]{henderson.searle1981} and leads to
$\tilde{\boldsymbol{1}}_{n}^{T}\tilde{\boldsymbol{1}}_{n} =
\boldsymbol{1}_{n}^{T}\boldsymbol{1}_{n} = n$,
$\tilde{\boldsymbol{1}}_{n}^{T}\tilde{\boldsymbol{X}}_{\boldsymbol{d}} =
\boldsymbol{1}_{n}^{T}\boldsymbol{X}_{\boldsymbol{d}} = \boldsymbol{0}_{I}$ and
$\tilde{\boldsymbol{1}}_{n}^{T}\tilde{\boldsymbol{y}} =
\boldsymbol{1}_{n}^{T}\boldsymbol{y}$ by straightforward calculations. A most
convenient property of the hyper-$g$ priors is that they yield closed form
marginal likelihoods, which need to be computed on the original response scale
via the change of variables formula:
\begin{equation}
  f(\boldsymbol{y} \given \boldsymbol{d})
  \propto
  f(\tilde{\boldsymbol{y}} \given \boldsymbol{d})
  \abs{\boldsymbol{V}_{\boldsymbol{d}}^{1/2}}^{-1},
  \label{eq:normal-models:marginal-likelihood}  
\end{equation}
where $f(\tilde{\boldsymbol{y}} \given \boldsymbol{d})$ is the marginal
likelihood of the transformed response vector $\tilde{\boldsymbol{y}}$ in the
standard linear model~\eqref{eq:normal-models:decorrelated}. The closed forms
for $f(\tilde{\boldsymbol{y}} \given \boldsymbol{d})$ under the hyper-$g$ priors
are given in Appendix~\ref{sec:impl-deta:marg-lik-comp}, along with other
implementation details.

Other hyper-priors could be assigned to $g$, but will typically not lead to a
closed form of the marginal likelihood. Examples are the incomplete
inverse-gamma prior on $1 + g$ \citep[p.~891]{CuiGeorge2008}, which generalises
the above uniform prior on $g/(1 + g)$, and an inverse-gamma prior on $g$, which
corresponds to the Cauchy prior of \citet{ZellnerSiow1980}. The hyper-$g/n$
prior is a special case of the conventional robust prior proposed by
\citet{forte2011}, for which a closed form of the marginal likelihood exists.
An overview of mixtures of $g$-priors is given by \citet{ley.steel2011}. 

It is not clear that the good properties of hyper-$g$ priors (or other default
priors in the Gaussian linear model) would be retained if we based them on the
conditional model~\eqref{eq:normal-models:conditional} without integrating out
the random effects. We followed the natural idea of transforming the mixed model
into a standard model, where default priors have already been studied
extensively. Moreover, computation would be harder if we proceeded otherwise.
Hence we prefer to keep the good properties of the default priors by integrating
out the random effects.

Posterior inference in a given model $\boldsymbol{d}$ is based on Monte Carlo
estimation of the parameters in model~\eqref{eq:normal-models:conditional},
using the factorisation 
\begin{equation}
  \label{eq:normal-models:sampling}
  f(\beta_{0}, \boldsymbol{\beta}_{\boldsymbol{d}},
  \boldsymbol{u}_{\boldsymbol{d}}, \sigma^{2}, g \given \boldsymbol{y})
  =
  f(\boldsymbol{u}_{\boldsymbol{d}} \given \beta_{0},
  \boldsymbol{\beta}_{\boldsymbol{d}}, \sigma^{2}, \boldsymbol{y})
  f(\beta_{0}, \boldsymbol{\beta}_{\boldsymbol{d}} \given \sigma^{2}, g,
  \boldsymbol{y}) 
  f(\sigma^{2} \given \boldsymbol{y})
  f(g \given \boldsymbol{y}).
\end{equation}
Sampling of $g$, $\sigma^{2}$ and subsequently $\beta_{0},
\boldsymbol{\beta}_{\boldsymbol{d}}$ can be done along the lines of
\citet[section~2.3]{sabanesbove.held2011a}, by adapting their algorithm to the
transformations in model~\eqref{eq:normal-models:decorrelated}. Finally, the
spline coefficient vector $\boldsymbol{u}_{\boldsymbol{d}}$ is sampled from
\begin{align}
  f(\boldsymbol{u}_{\boldsymbol{d}} \given \beta_{0},
  \boldsymbol{\beta}_{\boldsymbol{d}}, \sigma^{2}, \boldsymbol{y})
  &\propto 
  f(\boldsymbol{u}_{\boldsymbol{d}} \given \sigma^{2})
  f(\boldsymbol{y} \given \beta_{0}, \boldsymbol{\beta}_{\boldsymbol{d}},
  \boldsymbol{u}_{\boldsymbol{d}}, \sigma^{2})
  \notag \\
  &\propto
  \exp
  \left\{
    - \frac{1}{2\sigma^{2}}
    \left[
      \boldsymbol{u}_{\boldsymbol{d}}^{T}
      \boldsymbol{D}_{\boldsymbol{d}}^{-1}
    \boldsymbol{u}_{\boldsymbol{d}}
    + \norm{\boldsymbol{y} - 
      \boldsymbol{1}_{n}\beta_{0} -
      \boldsymbol{X}_{\boldsymbol{d}}\boldsymbol{\beta}_{\boldsymbol{d}} - 
      \boldsymbol{Z}_{\boldsymbol{d}}\boldsymbol{u}_{\boldsymbol{d}}}^{2}
    \right]
  \right\}
  \notag \\
  &\propto
  \Nor_{JK}
  \left(
    \boldsymbol{u}_{\boldsymbol{d}} \given 
    \boldsymbol{\Sigma}_{\boldsymbol{d}}
    \boldsymbol{Z}_{\boldsymbol{d}}^{T}
    (\boldsymbol{y} - 
    \boldsymbol{X}_{\boldsymbol{d}}\boldsymbol{\beta}_{\boldsymbol{d}}),
    \,
    \sigma^{2}\boldsymbol{\Sigma}_{\boldsymbol{d}}
  \right),
  \label{eq:normal-models:spline-coefs-conditional}
\end{align}
where $\boldsymbol{\Sigma}_{\boldsymbol{d}} =
(\boldsymbol{Z}_{\boldsymbol{d}}^{T}\boldsymbol{Z}_{\boldsymbol{d}} +
\boldsymbol{D}_{\boldsymbol{d}}^{-1})^{-1}$ and $\beta_{0}$ disappears because
$\boldsymbol{Z}_{\boldsymbol{d}}^{T}\boldsymbol{1}_{n} =
\boldsymbol{0}_{JK}$. 

Given posterior samples for the linear coefficient $\beta_{j}$ and the spline
coefficient vector $\boldsymbol{u}_{j}$ for covariate $j$ ($d_{j} > 1$), we would
like to transform these into samples for the function $m_{j}(x_{j})$, along a
grid vector $\tilde{\boldsymbol{x}}_{j}^{*}$ of $n^{*}$ points (on the same
scale as the original $\tilde{\boldsymbol{x}}_{j}$ used for the model fitting).
This is in principle straightforward, but one has to carefully apply
analogous transformations as in~\eqref{eq:normal-models:center-vector} 
and~\eqref{eq:normal-models:center-matrix} to $\tilde{\boldsymbol{x}}_{j}^{*}$
and the corresponding spline basis matrix $\tilde{\boldsymbol{Z}}_{j}^{*}$:  
\begin{align}
  \boldsymbol{x}_{j}^{*} &= 
  \tilde{\boldsymbol{x}}_{j}^{*}
  - \boldsymbol{1}_{n^{*}}
  \frac{\boldsymbol{1}_{n}^{T}\tilde{\boldsymbol{x}}_{j}}
  {\boldsymbol{1}_{n}^{T}\boldsymbol{1}_{n}},
  \label{eq:normal-models:center-grid-vector}
  \\
  \boldsymbol{Z}_{j}^{*} &=  
  \tilde{\boldsymbol{Z}}_{j}^{*}
  - \boldsymbol{1}_{n^{*}}
  \frac{\boldsymbol{1}_{n}^{T}\tilde{\boldsymbol{Z}}_{j}}
  {\boldsymbol{1}_{n}^{T}\boldsymbol{1}_{n}} 
  - \boldsymbol{x}_{j}^{*}
  \frac{\boldsymbol{x}_{j}^{T}\tilde{\boldsymbol{Z}}_{j}}
  {\boldsymbol{x}_{j}^{T}\boldsymbol{x}_{j}}.
  \label{eq:normal-models:center-grid-matrix}
\end{align}
Afterwards, for each coefficient sample one can compute the corresponding
vector of function values $m_{j}(\tilde{\boldsymbol{x}}_{j}^{*}) = 
\boldsymbol{x}_{j}^{*}\beta_{j} + \boldsymbol{Z}_{j}^{*}\boldsymbol{u}_{j}$.
Similarly, prediction samples for the corresponding response vector
$\boldsymbol{y}^{*}$ can be extracted from the sampling output.

\section{Hyper-$g$ Priors for Generalised Additive Models}
\label{sec:non-normal-models}

Now we extend the above setting and assume that the covariate effects
$m_{j}(x_{j})$ enter additively into the linear predictor
\begin{equation}
  \label{eq:generalised-additive-model}
  \eta = \beta_{0} + \sum_{j=1}^{p} m_{j}(x_{j})
\end{equation}
of an exponential family distribution with canonical parameter $\theta$, mean
$\E(y) = h(\eta) = db(\theta)/d\theta$ and variance $\Var(y) = \phi/w\cdot
d^{2}b(\theta)/d\theta^{2}$ \citep[see][]{McCullaghNelder1989}. We restrict our
attention to non-normal distributions with fixed dispersion~$\phi$ (as $\phi =
1$ for the Bernoulli and Poisson distribution) and known weight~$w$. For $n$
observations, the linear predictor vector $\boldsymbol{\eta} = (\eta_{1},
\dotsc, \eta_{n})^{T}$ is
\begin{equation}
  \label{eq:lin-pred-vector}
  \boldsymbol{\eta} = \boldsymbol{1}_{n} \beta_{0} + 
  \boldsymbol{X}_{\boldsymbol{d}}\boldsymbol{\beta}_{\boldsymbol{d}} +
  \boldsymbol{Z}_{\boldsymbol{d}}\boldsymbol{u}_{\boldsymbol{d}}
\end{equation}
and the likelihood is
\begin{equation}
  \label{eq:non-normal-models:conditional}
  f(\boldsymbol{y} \given \beta_{0}, \boldsymbol{\beta}_{\boldsymbol{d}},
  \boldsymbol{u}_{\boldsymbol{d}}) 
  \propto
  \exp\left\{
    \sum_{i=1}^{n} 
    \frac{y_{i}\theta_{i} - b(\theta_{i})}{\phi/w_{i}} 
  \right\}.
\end{equation}
The main challenge for the derivation of a generalised $g$-prior is that the
marginal density $f(\boldsymbol{y} \given \beta_{0},
\boldsymbol{\beta}_{\boldsymbol{d}})$, which results from integrating out the 
spline coefficient vector
\begin{equation}
  \label{eq:non-normal-models:spline-coef-prior}
  \boldsymbol{u}_{\boldsymbol{d}} \sim \Nor_{JK}
  (\boldsymbol{0}_{JK}, \boldsymbol{D}_{\boldsymbol{d}})
\end{equation}
from~\eqref{eq:non-normal-models:conditional}, has no closed form. In
particular, it is not Gaussian, in contrast
to~\eqref{eq:normal-models:marginal}.

Before addressing this problem we first consider appropriate
construction of the design matrices $\boldsymbol{X}_{\boldsymbol{d}}$ and
$\boldsymbol{Z}_{\boldsymbol{d}}$ and calculation of the degrees of freedom
$d_{j}(\rho_{j})$ for a smoothly modelled term $m_{j}$. Starting with the
latter, a reasonable generalisation
of~\eqref{eq:normal-models:degrees-of-freedom} is
\citep[see][section~11.4]{Ruppert:Wand:2003}
\begin{equation}
  \label{eq:non-normal-models:degrees-of-freedom}
  d_{j}(\rho_{j}) = 
  \trace \{ 
  (\boldsymbol{Z}_{j}^{T} \widehat{\boldsymbol{W}} \boldsymbol{Z}_{j} +
  \rho_{j}^{-1} \boldsymbol{I})^{-1} 
  \boldsymbol{Z}_{j}^{T} \widehat{\boldsymbol{W}} \boldsymbol{Z}_{j} 
  \} 
  + 1,
\end{equation}
which uses a fixed weight matrix $\widehat{\boldsymbol{W}} =
\boldsymbol{W}(\boldsymbol{1}_{n}\widehat{\beta}_{0})$, where
$\boldsymbol{W}(\boldsymbol{\eta}) = \diag\{ (dh(\eta_{i})/d\eta)^{2} /
\Var(y_{i})\}_{i=1}^{n}$ is the usual generalised linear model weight matrix and
$\widehat{\beta}_{0}$ is the intercept estimate from the null model
$\boldsymbol{d}=\boldsymbol{0}_{p}$. This definition avoids dependence of
$\rho_{j}(d_{j})$ on the model $\boldsymbol{d}$ under consideration.
As a consequence, we need to generalise the orthogonalisation of the
original covariate vector $\tilde{\boldsymbol{x}}_{j}$ and spline basis matrix
$\tilde{\boldsymbol{Z}}_{j}$ from \eqref{eq:normal-models:center-vector} and
\eqref{eq:normal-models:center-matrix} to
\begin{align}
  \boldsymbol{x}_{j} &= 
  \tilde{\boldsymbol{x}}_{j}
  - \boldsymbol{1}_{n}
  \frac{\boldsymbol{1}_{n}^{T}\widehat{\boldsymbol{W}}\tilde{\boldsymbol{x}}_{j}}
  {\boldsymbol{1}_{n}^{T}\widehat{\boldsymbol{W}}\boldsymbol{1}_{n}}
  \label{eq:non-normal-models:center-vector}
  \\
  \text{and}\quad
  \boldsymbol{Z}_{j} &=  
  \tilde{\boldsymbol{Z}}_{j}
  - \boldsymbol{1}_{n}
  \frac{\boldsymbol{1}_{n}^{T}\widehat{\boldsymbol{W}}\tilde{\boldsymbol{Z}}_{j}}
  {\boldsymbol{1}_{n}^{T}\widehat{\boldsymbol{W}}\boldsymbol{1}_{n}} 
  - \boldsymbol{x}_{j}
  \frac{\boldsymbol{x}_{j}^{T}\widehat{\boldsymbol{W}}\tilde{\boldsymbol{Z}}_{j}}
  {\boldsymbol{x}_{j}^{T}\widehat{\boldsymbol{W}}\boldsymbol{x}_{j}}
  \label{eq:non-normal-models:center-matrix},
\end{align}
implying that $\boldsymbol{1}_{n}$, $\boldsymbol{x}_{j}$ and the columns of
$\boldsymbol{Z}_{j}$ are orthogonal to each other with respect to the inner
product in terms of $\widehat{\boldsymbol{W}}$. This ensures
that~\eqref{eq:non-normal-models:degrees-of-freedom} correctly captures only the
degrees of freedom associated with the nonlinear part of $m_{j}$. Note that
\eqref{eq:normal-models:center-grid-vector} and
\eqref{eq:normal-models:center-grid-matrix} are adapted analogously.

We will now derive a generalised $g$-prior analogous
to~\eqref{eq:normal-models:g-prior} for the linear coefficient vector
$\boldsymbol{\beta}_{\boldsymbol{d}}$ in the generalised additive model. The
idea is to use the iterative weighted least squares (IWLS) approximation
to~\eqref{eq:non-normal-models:conditional} to obtain an approximate normal
model of the form \eqref{eq:normal-models:conditional} and then derive the
resulting $g$-prior~\eqref{eq:normal-models:g-prior}. With a slight abuse of
notation, \eg $h(\boldsymbol{\eta}) = (h(\eta_{1}), \dotsc, h(\eta_{n}))^{T}$,
let 
\begin{equation}
  \label{eq:adjusted-response-variable}
  \boldsymbol{z}_{0} = 
  \boldsymbol{\eta}_{0} + 
  \diag\{dh(\boldsymbol{\eta}_{0})/d\boldsymbol{\eta}\}^{-1}
  (\boldsymbol{y} - h(\boldsymbol{\eta}_{0}))
\end{equation}
be the adjusted response vector resulting from a first-order approximation to
$h^{-1}(\boldsymbol{y})$ around $\boldsymbol{y} = h(\boldsymbol{\eta}_{0})$. Then
\begin{equation}
  \label{eq:iwls-model}
  \boldsymbol{z}_{0} \given
  \beta_{0}, \boldsymbol{\beta}_{\boldsymbol{d}},
  \boldsymbol{u}_{\boldsymbol{d}} 
  \sima
  \Nor\bigl(\boldsymbol{1}_{n}\beta_{0} + 
  \boldsymbol{X}_{\boldsymbol{d}}\boldsymbol{\beta}_{\boldsymbol{d}} +
  \boldsymbol{Z}_{\boldsymbol{d}}\boldsymbol{u}_{\boldsymbol{d}},
  \,
  \boldsymbol{W}_{0}^{-1}\bigr)  
\end{equation}
with $\boldsymbol{W}_{0} = \boldsymbol{W}(\boldsymbol{\eta}_{0})$ is the working
normal model \citep[see {\eg}][p.~40]{McCullaghNelder1989}. Remember that the
IWLS algorithm iteratively updates $\boldsymbol{\eta}_{0}$ by weighted least
squares estimation of the coefficients in~\eqref{eq:iwls-model}. Here, we fix
$\boldsymbol{\eta}_{0} = \boldsymbol{0}_{n}$, which is the value expected
\latin{a priori}. Then we rewrite~\eqref{eq:iwls-model} using
$\tilde{\boldsymbol{z}}_{0} = \boldsymbol{W}_{0}^{1/2}\boldsymbol{z}_{0}$ \etc
as
\begin{equation}
  \label{eq:iwls-model-rewritten}
  \tilde{\boldsymbol{z}}_{0} \given
  \beta_{0}, \boldsymbol{\beta}_{\boldsymbol{d}},
  \boldsymbol{u}_{\boldsymbol{d}} 
  \sima 
  \Nor(\tilde{\boldsymbol{1}}_{n}\beta_{0} + 
  \tilde{\boldsymbol{X}}_{\boldsymbol{d}}\boldsymbol{\beta}_{\boldsymbol{d}} +
  \tilde{\boldsymbol{Z}}_{\boldsymbol{d}}\boldsymbol{u}_{\boldsymbol{d}},
  \,
  \boldsymbol{I}_{n}),
\end{equation}
which brings us back to a normal model of the form
in~\eqref{eq:normal-models:conditional}. By
computing the corresponding $g$-prior~\eqref{eq:normal-models:g-prior}, we
arrive at the generalised $g$-prior
\begin{equation}
  \label{eq:generalised-g-prior}
  \boldsymbol{\beta}_{\boldsymbol{d}} \given g
  \sim
  \Nor_{I}(\boldsymbol{0}_{I},
  g \boldsymbol{J}_{0}^{-1})
\end{equation}
with prior precision matrix proportional to
\begin{align}
  \boldsymbol{J}_{0} 
  &= 
  \tilde{\boldsymbol{X}}_{\boldsymbol{d}}^{T}
  (\boldsymbol{I}_{n} + 
  \tilde{\boldsymbol{Z}}_{\boldsymbol{d}} 
  \boldsymbol{D}_{\boldsymbol{d}}
  \tilde{\boldsymbol{Z}}_{\boldsymbol{d}}^{T})^{-1}
  \tilde{\boldsymbol{X}}_{\boldsymbol{d}}
  \notag \\
  &=
  \boldsymbol{X}_{\boldsymbol{d}}^{T}
  \boldsymbol{W}_{0}^{1/2}
  (\boldsymbol{I}_{n} + 
  \boldsymbol{W}_{0}^{1/2}
  \boldsymbol{Z}_{\boldsymbol{d}} 
  \boldsymbol{D}_{\boldsymbol{d}}
  \boldsymbol{Z}_{\boldsymbol{d}}^{T}
  \boldsymbol{W}_{0}^{1/2})^{-1}
  \boldsymbol{W}_{0}^{1/2}
  \boldsymbol{X}_{\boldsymbol{d}}.
  \label{eq:approx-fisher-info}
\end{align}
An appealing feature of this prior is that it directly generalises the $g$-prior
proposed by \citet{sabanesbove.held2011} for generalised linear models, to which
it reduces when there are no spline effects in the model, \ie
$\boldsymbol{J}_{0} = \boldsymbol{X}_{\boldsymbol{d}}^{T} \boldsymbol{W}_{0}
\boldsymbol{X}_{\boldsymbol{d}}$. An alternative and more rigorous derivation of
\eqref{eq:approx-fisher-info} as the Fisher information obtained from a Laplace
approximation to the marginal model $f(\boldsymbol{y} \given \beta_{0},
\boldsymbol{\beta}_{\boldsymbol{d}})$ is presented in the supplementary material
available at Biometrika online.

The generalised hyper-$g$ prior 
\begin{equation}
  \label{eq:generalised-hyper-g-prior}
  f(\beta_{0}, \boldsymbol{\beta}_{\boldsymbol{d}},
  \boldsymbol{u}_{\boldsymbol{d}}, g) =
  f(\beta_{0}) f(\boldsymbol{\beta}_{\boldsymbol{d}} \given g) f(g)
  f(\boldsymbol{u}_{\boldsymbol{d}}) 
\end{equation}
is defined to comprise the locally uniform prior $f(\beta_{0}) \propto 1$ on the
intercept $\beta_{0}$, the generalised $g$-prior \eqref{eq:generalised-g-prior}
on the linear coefficient vector $\boldsymbol{\beta}_{\boldsymbol{d}}$, the
penalty prior \eqref{eq:non-normal-models:spline-coef-prior} on the spline
coefficient vector $\boldsymbol{u}_{\boldsymbol{d}}$, and some proper
hyper-prior $f(g)$ on the hyper-parameter $g$. Posterior inference under this
prior can be implemented by a straightforward extension of the approach of
\citet[section~3]{sabanesbove.held2011}, which is outlined in the following.
The efficient \texttt{R}-package ``\texttt{hypergsplines}'' for this and all
other computations in this paper is available from
\texttt{R}-Forge.\footnote{The website is
  \url{http://hypergsplines.r-forge.r-project.org/}. To install the
  \texttt{R}-package, just type
  \texttt{install.packages("hypergsplines",repos="http://r-forge.r-project.org")}
  into \texttt{R}.} 

Let $\boldsymbol{X}_{a} = (\boldsymbol{1}_{n}, \boldsymbol{X}_{\boldsymbol{d}},
\boldsymbol{Z}_{\boldsymbol{d}})$ and $\boldsymbol{\beta}_{a} = (\beta_{0},
\boldsymbol{\beta}_{\boldsymbol{d}}^{T},
\boldsymbol{u}_{\boldsymbol{d}}^{T})^{T}$ denote the grand design matrix and
regression coefficient vector, respectively, such that $\boldsymbol{\eta} =
\boldsymbol{X}_{a}\boldsymbol{\beta}_{a}$. The prior for
$\boldsymbol{\beta}_{a}$ conditional on $g$ has a Gaussian form with mean zero and
singular precision matrix $\diag\{0, g^{-1} \boldsymbol{J}_{0},
\boldsymbol{D}_{\boldsymbol{d}}^{-1}\}$. Thus, the Gaussian approximation of
$f(\boldsymbol{\beta}_{a} \given \boldsymbol{y}, g, \boldsymbol{d})$, which is
necessary for the Laplace approximation of $f(\boldsymbol{y} \given g,
\boldsymbol{d})$, can be obtained by the Bayesian IWLS algorithm
\citep{West1985}. Afterwards, an approximation of the marginal likelihood of
model $\boldsymbol{d}$,
\begin{equation}
  \label{eq:marginal-likelihood}
  f(\boldsymbol{y} \given \boldsymbol{d}) =
  \int_{0}^{\infty} f(\boldsymbol{y} \given g, \boldsymbol{d})
  f(g)\, dg,
\end{equation}
is obtained by numerical integration of the Laplace approximation
$\tilde{f}(\boldsymbol{y} \given g, \boldsymbol{d})$. Note that recently
integrated Laplace approximations have successfully been applied in a more
general context \citep{RueMartinoChopin2009}. Finally, we can use a tuning-free
Metropolis-Hastings algorithm to sample from the joint posterior of
$\boldsymbol{\beta}_{a}$ and $g$ in a specific model $\boldsymbol{d}$.

\section{Model Prior and Stochastic Search}
\label{sec:model-prior-and-search}

We propose a prior $f(\boldsymbol{d})$ on the model space $\mathcal{D}^{p}$
which explicitly corrects for the multiplicity of testing inherent in the
simultaneous analysis of the $p$ covariates \citep[see][]{scott.berger2010}:
\latin{A priori}, the number of covariates included in the model
($I$) is uniformly distributed on $\{0, 1, \dotsc, p\}$. 
The choice of the $I$ covariates is then uniformly distributed on all possible
configurations, and their degrees of freedom are independent and uniformly
distributed on $\mathcal{D} \setminus \{0\} = \{1, 2, 3, \dotsc, K\}$.
Altogether, this gives
\begin{equation}
  1 / f(\boldsymbol{d}) = 
  (p+1)
  \binom{p}{I} 
  K^I.
  \label{eq:model-prior}
\end{equation}
A nice property of this prior is that it leads to marginal prior probabilities
$\P(d_{j} = 0) = \P(d_{j} > 0) = 1/2$. Elsewhere this is often achieved by
assigning independent priors to the $p$ covariates, which implies that averaged
over all models, $I \sim \Bin(p, 1/2)$.
It is clear that our uniform prior on $I$ allows the data $\boldsymbol{y}$ to
have a maximum effect on the posterior of $I$ because it is the reference prior
\citep{bernardo1979}. Note that this prior actually favours models with high or
low numbers of covariates, as there are fewer such models. This or similar model
priors have been used in a number of previous papers, including \eg
\citet{GeorgeMcCulloch1993}, \citet{panagiotelis.smith2008} and
\citet{LeySteel2009}.

Alternatively, one might also use a fixed (independent of $K$) prior probability
for a linear effect ($d_{j} = 1$). This is appropriate for the
situation where one explicitly wants to test linearity versus nonlinearity of
each effect. Furthermore, a multiplicity correction for these tests can be
implemented by assuming that the number of smoothly included covariates ($J$) is
uniformly distributed on $\{0, 1, \dotsc, I\}$ and their choice is uniform on
all possible choices. This would add one level to the prior hierarchy.

As the model space $\mathcal{D}^{p}$ grows exponentially in the number of
covariates $p$, only for small values of $p$ all possible models can be
evaluated. Otherwise the marginal likelihoods $f(\boldsymbol{y} \given
\boldsymbol{d})$ and posterior model probabilities $f(\boldsymbol{d}
\given \boldsymbol{y}) \propto f(\boldsymbol{y} \given \boldsymbol{d})
f(\boldsymbol{d})$ 
can be computed only for a subset of the model space. Usually
this subset is determined by stochastic search procedures. Here we propose to
use a simple Metropolis-Hastings algorithm with two possible move types in the
proposal kernel:
\begin{description}
\item[Move] Sample a covariate index $j \sim \Unif\{1, 2, \dotsc, p\}$ and
  decrease or increase $d_{j}$ to the next adjacent value in $\mathcal{D}$ (with
  probability $1/2$ each, or deterministically if $d_{j} = 0$ or
  $d_{j} = K$, respectively).
\item[Swap] Sample a pair $(i, j) \sim \Unif\{(1, 1), (1, 2), \dotsc, (p,
  p)\}$ of covariate indices ($i \leq j$) and swap $d_{i}$ and $d_{j}$. 
\end{description}
The `Swap' move is designed to efficiently trace models with high posterior
probability even in situations where covariates are almost collinear. For each
Metropolis-Hastings iteration, a `Move' is chosen with some fixed probability
(we use $3/4$), and otherwise a `Swap'. Denote the current model by
$\boldsymbol{d}$, then the proposed model $\boldsymbol{d}'$ is accepted with
probability
\begin{equation*}
  \alpha(\boldsymbol{d}'\given \boldsymbol{d}) = 
  1
  \wedge
  \frac
  {f(\boldsymbol{y}\given \boldsymbol{d}') 
    f(\boldsymbol{d}')
    q(\boldsymbol{d}'\given  \boldsymbol{d})}
  {f(\boldsymbol{y}\given \boldsymbol{d})
    f(\boldsymbol{d})
    q(\boldsymbol{d}\given  \boldsymbol{d}')} 
\end{equation*}
where the calculation of the proposal probability ratio $q(\boldsymbol{d}'\given
\boldsymbol{d}) / q(\boldsymbol{d}\given \boldsymbol{d}')$ is
straightforward, see Appendix~\ref{sec:impl-deta:proposal-probs}.

The advantage of such an MCMC based model exploration compared to more elaborate
stochastic search algorithms \citep[\eg][]{HansDobraWest2007,clyde.etal2011} is
that it does not preclude estimation of posterior model probabilities via
sampling frequencies, as it was originally proposed for MCMC model composition
by \citet{MadiganYork1995}. Recently reported problems with renormalized
probability estimates \citep{clyde.ghosh2010,garcia-donato.martinez-beneito2011}
can be avoided by using the model sampling frequencies instead.
Nevertheless, other search procedures might be beneficial when only the
\latin{maximum a posteriori} (MAP) model and not \eg the marginal posterior
inclusion probabilities for the covariates are of interest.

\section{Applications}
\label{sec:applications}

We examine the performance of the proposed additive model selection methodology
with a simulation study in Section~\ref{sec:applications:sim-study}, and
illustrate logistic regression using the Pima Indian data set in
Section~\ref{sec:applications:pima}.

\subsection{Simulation Study in Additive Models}
\label{sec:applications:sim-study}

In order to study the frequentist properties of our approach, we performed a
simulation study. The full details are provided as supplementary material which
is available at Biometrika online. Here we summarise the main results.

Three different true models were simulated: The first model (``null'') was the
null model with $p=20$ nuisance covariates. The second model (``small'') also
had $p=20$ covariates of which 3 had a linear effect and 3 had a nonlinear
(quadratic, sine, and skew-normal density) effect. Correlations of different
strength were generated between some the covariates. The third model (``large'')
was identical to the second model, but included additional 80 nuisance
covariates, which were independent of the first 20 covariates. For the ``small''
and ``large'' model, one covariate was chosen to be a surrogate for the true,
quadratic, effect of another covariate. It masks the quadratic effect if only
linear effects can be fitted by a variable selection algorithm. For three
different sample sizes $n=50, 100, 1000$, and for the three different true
models, we simulated $n$ observations from the Gaussian additive
model~\eqref{eq:additive-model} with $\beta_{0} = 0$ and $\sigma^{2} = 0.2^{2}$.
This was repeated 50 times for each combination of model and sample size, in
order to assess the sampling variability.

We applied the proposed additive model selection approaches to each data set,
using the hyper-$g$ and hyper-$g/n$ priors. As the computational complexity of
the marginal likelihood \eqref{eq:normal-models:marginal-likelihood} is cubic in
the spline basis dimension $K$ (see Appendix~\ref{sec:impl-deta:marg-lik-comp}),
we want to use splines with few, quantile-based knots. Therefore, we choose
cubic O'Sullivan splines \citep{wand.ormerod2008}. Here, we got basis matrices
$\boldsymbol{Z}_{j}$ with $K=8$ columns from 6~inner knots at the septiles. We
applied the stochastic search algorithm described in
Section~\ref{sec:model-prior-and-search} with $10^{6}$~iterations.

We compared the results with those from pure variable selection including only
linear functions, Bayesian FPs \citep{sabanesbove.held2011a}, spike-and-slab
function selection \citep{scheipl.etal2011} and splines knot selection
(\citeauthor{denison.etal1998}, \citeyear{denison.etal1998}, using code from
chapters 3 and 4 in \citeauthor{DenisonHolmes02}, \citeyear{DenisonHolmes02}).

Concerning discovery of the true set of influential covariates, the additive
model selection procedures introduced in this paper were very competitive with
the considered alternative methods, as is illustrated in
Table~\ref{tab:med-post-prob-true-model}. In particular, they showed clear
advantages in the case of small and moderate sample sizes. Using splines instead
of only linear functions proved essential for the discovery of the masked
quadratic effect and hence convergence to the true model.

\begin{table}
  \centering
  {\footnotesize
%
\begin{tabular}{lrrrcrrrcrrr}
\toprule
\multicolumn{1}{l}{\small }&
\multicolumn{3}{c}{\small null}&
\multicolumn{1}{c}{\small }&
\multicolumn{3}{c}{\small small}&
\multicolumn{1}{c}{\small }&
\multicolumn{3}{c}{\small large}
\tabularnewline
\cline{2-4} \cline{6-8} \cline{10-12}
\multicolumn{1}{l}{}&\multicolumn{1}{c}{$n=50,$}&\multicolumn{1}{c}{$100,$}&\multicolumn{1}{c}{$1000$}&\multicolumn{1}{c}{}&\multicolumn{1}{c}{$n=50,$}&\multicolumn{1}{c}{$100,$}&\multicolumn{1}{c}{$1000$}&\multicolumn{1}{c}{}&\multicolumn{1}{c}{$n=50,$}&\multicolumn{1}{c}{$100,$}&\multicolumn{1}{c}{$1000$}\tabularnewline
\midrule
Hyper-$g$ splines&$83$&$84$&$84$&&$49$&$65$&$86$&&$2$&$74$&$87$\tabularnewline
Hyper-$g/n$ splines&$86$&$91$&$97$&&$47$&$68$&$87$&&$0$&$75$&$89$\tabularnewline
Hyper-$g$ linear&$20$&$21$&$23$&&$ 0$&$ 0$&$ 0$&&$0$&$ 0$&$ 0$\tabularnewline
Hyper-$g/n$ linear&$50$&$64$&$90$&&$ 0$&$ 0$&$ 0$&&$0$&$ 0$&$ 0$\tabularnewline
Bayesian FPs&$37$&$37$&$37$&&$ 2$&$35$&$ 3$&&$0$&$47$&$37$\tabularnewline
Spike-and-slab&$89$&$93$&$98$&&$ 3$&$45$&$79$&&$0$&$10$&$71$\tabularnewline
Knot selection&$92$&$94$&$98$&&$ 0$&$34$&$95$&&$0$&$ 0$&$89$\tabularnewline
\bottomrule
\end{tabular}}
\caption{Median posterior probability of the true model in percentage, when the
  true model is defined by correct variable inclusion.}
\label{tab:med-post-prob-true-model}
\end{table}

Variable inclusion performance did not differ substantively with respect to
sensitivity, specificity and area under the ROC curve between the considered
methods, with the exception of a slightly worse performance of the two linear
methods. However, as shown in Table~\ref{tab:diff-inc-probs}, the hyper-$g$ and
hyper-$g/n$ spline methods were clearly better in distinguishing truly effective
covariates from highly correlated nuisance covariates. Moreover, for small
sample sizes, they outperformed the other nonlinear methodologies concerning
discovery of the masked quadratic effect. In this task the merely linear methods
obviously failed.

\begin{table}
  \centering
  {\footnotesize
%
\begin{tabular}{lrrrcrrr}
\toprule
\multicolumn{1}{l}{\small }&
\multicolumn{3}{c}{\small small}&
\multicolumn{1}{c}{\small }&
\multicolumn{3}{c}{\small large}
\tabularnewline
\cline{2-4} \cline{6-8}
\multicolumn{1}{l}{}&\multicolumn{1}{c}{$n=50,$}&\multicolumn{1}{c}{$100,$}&\multicolumn{1}{c}{$1000$}&\multicolumn{1}{c}{}&\multicolumn{1}{c}{$n=50,$}&\multicolumn{1}{c}{$100,$}&\multicolumn{1}{c}{$1000$}\tabularnewline
\midrule
Hyper-$g$ splines&$75$&$97$&$98$&&$26$&$100$&$100$\tabularnewline
Hyper-$g/n$ splines&$79$&$97$&$98$&&$20$&$100$&$100$\tabularnewline
Hyper-$g$ linear&$18$&$44$&$87$&&$ 6$&$ 26$&$ 98$\tabularnewline
Hyper-$g/n$ linear&$22$&$48$&$90$&&$17$&$ 26$&$ 98$\tabularnewline
Bayesian FPs&$41$&$89$&$68$&&$ 9$&$ 92$&$ 81$\tabularnewline
Spike-and-slab&$30$&$88$&$97$&&$ 1$&$ 60$&$ 97$\tabularnewline
Knot selection&$ 9$&$78$&$99$&&$ 4$&$ 13$&$ 99$\tabularnewline
\bottomrule
\end{tabular}}
\caption{Average difference $\frac{1}{2}(P_{16} + P_{17}) - \frac{1}{3}(P_{18} +
  P_{19} + P_{20})$ of inclusion probabilities $P_{j} = \P\{m_{j}(x_{j}) \neq 0
  \given \boldsymbol{y}\}$ (in percentage points) between the truly effective
  covariates $x_{16}$ and $x_{17}$ and the nuisance covariates $x_{18}, x_{19},
  x_{20}$, which had correlation $0.8$ with $x_{16}$ and $x_{17}$. (The optimal
  value is $100$, the worst value is $-100$.)}
\label{tab:diff-inc-probs}
\end{table}

Concerning the average mean squared errors of the model-averaged posterior mean
function estimates $\hat{m}_{j}(x_{j})$, the proposed additive model selection
procedures were very competitive. They performed well or better than the best
compared method each, as is shown in Table~\ref{tab:errors-bma-overall}. It is
interesting that the hyper-$g$ splines were slightly but consistently better
than the hyper-$g/n$ splines. We also investigated the coverage rates of
pointwise 95\% credible intervals for the functions, and found that the two
proposed methods were slightly conservative.

\begin{table}
  \centering
  {\footnotesize
%
\begin{tabular}{lrrrcrrrcrrr}
\toprule
\multicolumn{1}{l}{\small }&
\multicolumn{3}{c}{\small null}&
\multicolumn{1}{c}{\small }&
\multicolumn{3}{c}{\small small}&
\multicolumn{1}{c}{\small }&
\multicolumn{3}{c}{\small large}
\tabularnewline
\cline{2-4} \cline{6-8} \cline{10-12}
\multicolumn{1}{l}{}&\multicolumn{1}{c}{$n=50,$}&\multicolumn{1}{c}{$100,$}&\multicolumn{1}{c}{$1000$}&\multicolumn{1}{c}{}&\multicolumn{1}{c}{$n=50,$}&\multicolumn{1}{c}{$100,$}&\multicolumn{1}{c}{$1000$}&\multicolumn{1}{c}{}&\multicolumn{1}{c}{$n=50,$}&\multicolumn{1}{c}{$100,$}&\multicolumn{1}{c}{$1000$}\tabularnewline
\midrule
Hyper-$g$ splines&$0.03$&$0.01$&$0.00$&&$   39.15$&$  10.32$&$  1.68$&&$ 30.42$&$  1.88$&$ 0.33$\tabularnewline
Hyper-$g/n$ splines&$0.05$&$0.01$&$0.00$&&$   47.82$&$  18.33$&$  3.20$&&$784.44$&$  2.78$&$ 0.61$\tabularnewline
Hyper-$g$ linear&$0.76$&$0.14$&$0.01$&&$  158.10$&$ 133.55$&$121.97$&&$ 45.11$&$ 32.26$&$24.36$\tabularnewline
Hyper-$g/n$ linear&$0.22$&$0.02$&$0.00$&&$  189.57$&$ 169.00$&$120.96$&&$378.07$&$ 36.23$&$26.09$\tabularnewline
Bayesian FPs&$0.14$&$0.03$&$0.00$&&$16837.92$&$3026.61$&$ 29.51$&&$ 76.78$&$356.30$&$ 5.80$\tabularnewline
Spike-and-slab&$1.90$&$1.82$&$0.57$&&$   80.94$&$  14.00$&$  2.09$&&$ 45.45$&$  8.71$&$ 0.81$\tabularnewline
Knot selection&$0.03$&$0.00$&$0.00$&&$  180.03$&$  35.29$&$  2.07$&&$ 47.23$&$ 29.33$&$ 0.78$\tabularnewline
\bottomrule
\end{tabular}}
\caption{Average mean squared errors (in $10^{-4}$ units) of function estimates.
 Numbers are averaged over all covariates and the 50~replications.}
\label{tab:errors-bma-overall}
\end{table}

Finally, the average computational effort of the two proposed additive model
selection procedures ranged between one minute for $n=100$ in a ``null'' data
set to about 50~minutes for $n=50$ in a ``large'' data set.

\subsection{Pima Indian Diabetes Data}
\label{sec:applications:pima}

We now apply the generalised additive model selection approach to the logistic
regression of $p=7$ potential risk factors on the presence of diabetes in
$n=532$ women of Pima Indian heritage
\citep{FrankAsuncion2010,Ripley1996}, see Table~\ref{tab:pima-description} for
details. We use cubic O'Sullivan splines with 4~inner knots at the quintiles and
the generalised hyper-$g/n$ prior, and explore the model space of dimension
$7^{7} = 823\,543$ with $10^{6}$
iterations of the stochastic search algorithm. The computational complexity is
higher than for the normal response case, with
95~minutes required for the evaluation
of 39\,081~models. We validated the
results with an exhaustive evaluation of all models, requiring
33~hours. Indeed, the stochastic
search found 99\% of the posterior
probability mass and the 733~top models.

\begin{table}[htbp]
  \centering
  \begin{tabular}{ll}
    \toprule
    Variable & Description \\ 
    \midrule
    $y$ & Signs of diabetes according to WHO criteria (Yes = $1$, No = $0$) \\
    $x_{1}$ & Number of pregnancies \\
    $x_{2}$ & Plasma glucose concentration in an oral glucose tolerance
    test [mg/dl]\\
    $x_{3}$ & Diastolic blood pressure [mm Hg] \\
    $x_{4}$ & Triceps skin fold thickness [mm] \\
    $x_{5}$ & Body mass index (BMI) [kg/m\textsuperscript{2}]\\
    $x_{6}$ & Diabetes pedigree function \\
    $x_{7}$ & Age [years] \\    
    \bottomrule
  \end{tabular}
  \caption{Description of the variables in the Pima Indian diabetes data set.} 
  \label{tab:pima-description}
\end{table}

In Table~\ref{tab:pima-inc-probs} the marginal posterior probabilities for
linear and smooth inclusion of the covariates are shown.
There is clear evidence for inclusion of the covariates $x_{2}$, $x_{5}$,
$x_{6}$ and $x_{7}$, which have posterior inclusion probabilities over 96\%. For
the other three covariates, the inclusion probability is below 30\%. Smooth
modelling of the effects of $x_{5}$, $x_{6}$ and $x_{7}$ seems to be necessary,
while this is not so clear for $x_{2}$.

\begin{table}
%
\centering
\begin{tabular}{llllllll}
\toprule
\multicolumn{1}{l}{}&\multicolumn{1}{c}{$x_{1}$}&\multicolumn{1}{c}{$x_{2}$}&\multicolumn{1}{c}{$x_{3}$}&\multicolumn{1}{c}{$x_{4}$}&\multicolumn{1}{c}{$x_{5}$}&\multicolumn{1}{c}{$x_{6}$}&\multicolumn{1}{c}{$x_{7}$}\tabularnewline
\midrule
not included ($d_{j} = 0$)&0.74&0.00&0.88&0.91&0.00&0.04&0.01\tabularnewline
linear ($d_{j} = 1$)&0.07&0.48&0.06&0.04&0.11&0.26&0.00\tabularnewline
smooth ($d_{j} \textgreater  1$)&0.19&0.52&0.06&0.05&0.89&0.70&0.99\tabularnewline
\bottomrule
\end{tabular}\caption{Marginal posterior inclusion probabilities in the Pima Indian diabetes
  data set.}
\label{tab:pima-inc-probs}
\end{table}

In order to examine the mixing properties of the stochastic search algorithm
proposed in Section~\ref{sec:model-prior-and-search}, we compared the results
based on starting the MCMC chain from the full model with $d_{j} = 4$ instead of
the previously used null model with $d_{j} = 0$ ($j=1, \dotsc, p$). The results
are very close: for example, the entries in Table~\ref{tab:pima-inc-probs}
differ by at most $2.28\cdot 10^{-4}$, and the top 500
models which were visited by the chains are identical. These results are an
indication that slow mixing is not a problem for the presented stochastic search
algorithm.

\begin{figure}
  \centering
  \begin{tabular}{cc}
{\tikzexternaldisable
\input{out-pima-plot-x2.tikz}
}
&
{\tikzexternaldisable
\input{out-pima-plot-x5.tikz}
}
\\
{\tikzexternaldisable
\input{out-pima-plot-x6.tikz}
}
&
{\tikzexternaldisable
\input{out-pima-plot-x7.tikz}
}
\end{tabular}
\caption{Estimated covariate effects in the MAP model for the Pima Indian
  diabetes data set, based on
  10\,000~MCMC samples: Posterior means
  (solid lines), pointwise (dashed lines) and simultaneous (dotted lines)
  95\%-credible intervals are shown.}
  \label{fig:pima-hpm}
\end{figure}
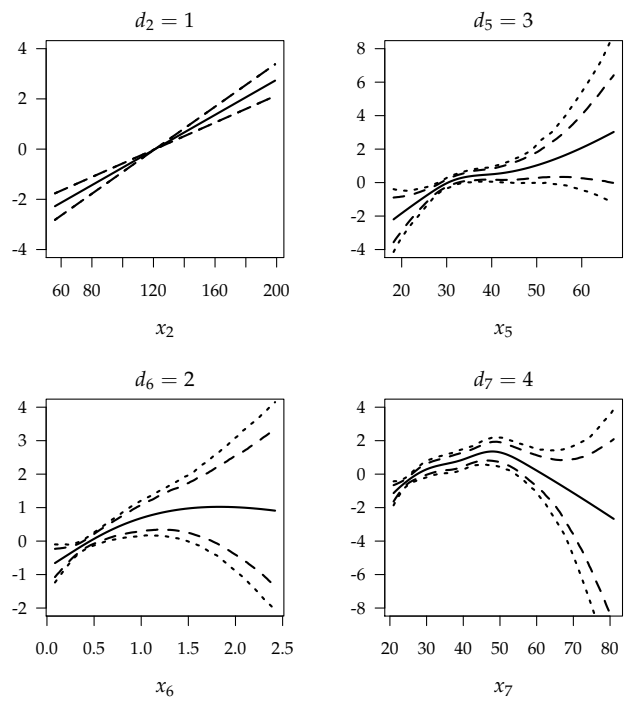

Figure~\ref{fig:pima-hpm} shows the estimated covariate effects in the MAP model
which features a linear term for $x_{2}$ and smooth terms for $x_{5}$, $x_{6}$
and $x_{7}$. The estimates are obtained from
10\,000~MCMC samples.\footnote{Every
  2nd sample was saved after burning the first
  1000 iterations, with acceptance
  rate 67\% using two
  IWLS steps per proposal.} 
Note that for linear functions $m_{j}$, the pointwise credible intervals
coincide with the simultaneous credible intervals
\citep[p.~30]{BesagGreenHigdonMengersen1995}. This is because all straight lines
samples intersect in one point, which is due to the centring of the covariates
in \eqref{eq:non-normal-models:center-vector}.
Furthermore, we observe that the \citet{ChibJeliazkov2001} estimate
($-240.924$, MCMC standard
error $0.008$) of the log
marginal likelihood of the MAP model, which was also computed, is quite close to
the integrated Laplace approximation
($-241.01$). This
indicates that the integrated Laplace approximation is fairly accurate.

The results are qualitatively similar to those obtained with a FP modelling
approach by \citet[section~5]{sabanesbove.held2011} and with a cubic smoothing
spline approach by \citet[section~3.2]{CottetKohnNott2008}. It is interesting
that in the earlier work by \citet[section~5.2]{YauKohnWood2003}, a very low
posterior inclusion probability ($0.07$) for $x_{6}$ was reported for a
different subset of the original Pima Indian diabetes data set. If pure variable
selection without covariate transformation is considered, as in
\citet[section~2.6]{HolmesHeld2006} and \citet[section~4]{sabanesbove.held2011},
the strong nonlinear effect of $x_{7}$ is missed completely, and instead $x_{1}$
gets a higher posterior inclusion probability. This may be a case of a masked
nonlinear effect, as was simulated in Section~\ref{sec:applications:sim-study},
and highlights the importance of allowing for nonlinear covariate effects.

\section{Postprocessing}
\label{sec:postprocessing}

Given the list of all possible models $\boldsymbol{d} \in \mathcal{D}^{p}$, or a
subset found by the stochastic search procedure described in
Section~\ref{sec:model-prior-and-search}, one may consider postprocessing the
results.

\begin{figure}
  \centering
  \begin{tabular}{cc}
{\tikzexternaldisable
\input{out-pima-topInc-plot-x2.tikz}
}
&
{\tikzexternaldisable
\input{out-pima-topInc-plot-x5.tikz}
}
\\
{\tikzexternaldisable
\input{out-pima-topInc-plot-x6.tikz}
}
&
{\tikzexternaldisable
\input{out-pima-topInc-plot-x7.tikz}
}
\end{tabular}
  \caption{Estimated covariate effects in the best meta-model (and median
    probability meta-model) for the Pima Indian diabetes data,
    based on
    20\,000
    samples: 
    Posterior means (solid lines), pointwise (dashed lines) and simultaneous
    (dotted lines) 95\%-credible intervals are shown.}
  \label{fig:pima-best-meta-model}
\end{figure}
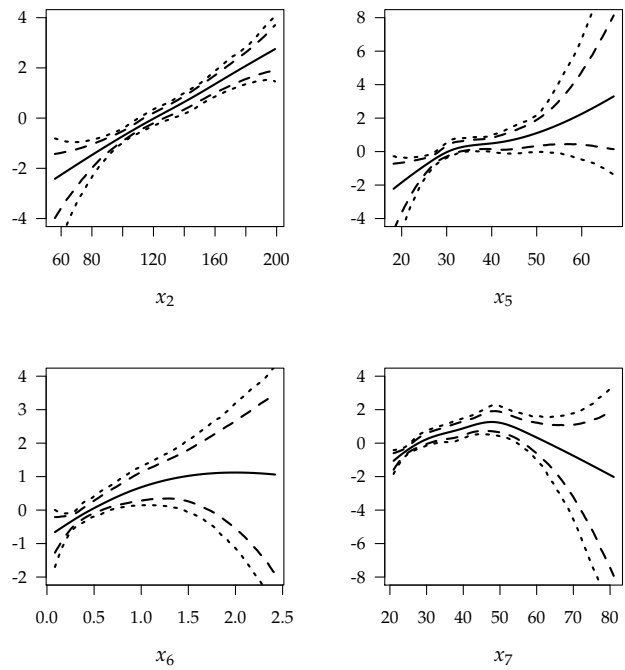

First, when the main interest lies in variable selection, the models which
feature the same covariates can be summarised into a meta-model as follows: The
posterior probabilities of the sub-models are summed up to give the posterior
probability of the meta-model, and estimates in the meta-model are obtained by
averaging the sub-models with weights proportional to their posterior
probabilities \citep[see \eg][for model averaging]{hoeting.etal1999}.
For example, the best meta-model for the Pima Indian diabetes data includes
$x_{2}$, $x_{5}$, $x_{6}$ and $x_{7}$ and has posterior probability
$0.598$. The
corresponding estimates of the covariate effects are shown in
Figure~\ref{fig:pima-best-meta-model}. This best meta-model happens to be
identical with the median probability meta-model, which features all covariates
having marginal posterior inclusion probability greater than 50\%
\citep{BarbieriBerger2004}, \cp Table~\ref{tab:pima-inc-probs}.
Similarly, it could be interesting to summarise models which only differ in the
degrees of freedom for smooth terms. This would correspond to the situation of
testing linearity versus nonlinearity of covariate effects (\cp
Section~\ref{sec:model-prior-and-search}). 

Second, in order to allow for \emph{continuous} degrees of freedom, one can optimise
the marginal likelihood of the MAP model with respect to the degrees of freedom
of the covariates included. That is, an optimisation of $f(\boldsymbol{y} \given
\boldsymbol{d})$ over the continuous range $1 < d_{j} < K + 1$ is
performed for all covariates included in the MAP model.
For example, the MAP configuration for the Pima Indian diabetes data is
$(0,1,0,0,3,2,4)$ and the resulting optimised
configuration is $(0,1,0,0,3.42,2.1,3.74)$,
which increases the log marginal likelihood from
$-241.01$ to
$-240.86$. Although $d_{5}$ and $d_{7}$
changed considerably in the optimisation, the resulting function estimates are
very similar to those from the MAP model in Figure~\ref{fig:pima-hpm} and are
hence omitted.

\section{Discussion}
\label{sec:discussion}

Our Bayesian approach to simultaneous variable and function selection in
generalised regression is based on fixed-dimensional spline bases and
penalty-parameter smoothness control. In this respect, it differs from
knot-selection approaches such as \citet{smith.kohn1996} and
\citet{denison.etal1998}. We found that fixed-dimensional spline bases are
flexible enough to capture the functional forms we expect \citep[see
\eg][]{abrahamowicz.etal1996}. Moreover, by using fixed-dimensional smooth
components we can constrain a covariate effect to be exactly linear. This
enables us to look at posterior probabilities of linear \latin{versus} smooth
inclusion of covariates. Approaches which use variable-dimensional smooth
components and select knots, as \citet{denison.etal1998}, cannot fit linear
functions.

We are only considering roughness penalties on a fixed grid of values, which
scales automatically for each covariate via the degrees of freedom
transformation. We found that it is a very useful approximation of a continuous
scale, and postprocessing is possible to remove the restriction to the grid
values. In the examples we have looked at, the resulting optimised models
yielded very similar results compared to the MAP model.
In this regard, our approach is close to many popular Lasso-type proposals,
which optimise the tuning-parameters on a fixed grid via cross-validation
\citep[\eg][]{zou.hastie2005}. \citet{cantoni.hastie2002} propose a
likelihood-ratio-type test statistic to compare additive models with different
degrees of freedom.
\citet{FongRueWakefield2010} use a similar scaling to examine the prior on the
degrees of freedom implied by the prior on the variance component in a
generalised linear mixed model. They also use O'Sullivan spline bases as we did
in our applications, but they do not consider variable selection.

In a frequentist setting, \citet[section~2.1]{marra.wood2011} propose to use an
additional penalty on the linear part of the spline function in order to shrink
it adaptively to zero. To include variable selection, a lower threshold for the
effective degrees of freedom must be chosen. Our generalised $g$-prior
\eqref{eq:generalised-g-prior} also shrinks the linear parts of the spline
functions to zero, where the prior covariance matrix takes the correlations
between the covariates into account. Incorporating the covariates correlation in
the coefficients prior allows for better discrimination between influential and
correlated nuisance covariates. Empirical results from our simulation study in
Section~\ref{sec:applications:sim-study} support this. Furthermore, we
explicitly ex- or include covariates and then compare the resulting models based
on their posterior probabilities.

We propose a conventional prior for the intercept and the linear coefficients,
which directly generalises the hyper-$g$ priors in the linear model
\citep{LiangPauloMolinaClydeBerger2008} and in the generalised linear model
\citep{sabanesbove.held2011}. \citet{pauler1998} proposes a related
unit-information prior for the fixed effects in linear mixed models, but fixes
$g = n$ in \eqref{eq:normal-models:g-prior}. \citet{OverstallForster2010}
propose a unit-information prior for the fixed effects in generalised linear
mixed models, but the information matrix is based on the first-stage likelihood
and not on the integrated likelihood as in our approach. Also, no hyper-prior on
the parameter $g$ is considered, because it is fixed at $g = n$. As they use an
inverse-Wishart prior on the covariance matrix of the random effects, their
approach is perhaps better suited to generic random effects models.
\citet{forster.etal2012} propose a novel reversible-jump MCMC algorithm to infer
the corresponding posterior model probabilities.

In future work, we would like to combine the semiparametric splines with the
parametric FPs \citep{sabanesbove.held2011a}. The idea is that a smooth term
$m_{j}(x_{j})$ could also be modelled by a FP instead of a spline. This
extension could be implemented coherently, because the prior formulations are
compatible. With such a general framework, the important question whether a
parsimonious FP (\eg $m_{7} = x_{7}\beta_{71} + x_{7}^{2}\beta_{72}$ in the Pima
Indian diabetes data example) is sufficient could be answered via posterior
probabilities (see \citet{strasak.etal2011} for a simulation study comparing the
stepwise FP approach by \citet{Royston:Sauerbrei:2008-Buch} with penalised
spline approaches). Moreover we will apply mixtures of $g$-priors to additive
Cox-type survival models.



%% file: out-pima-plot-x2.tikz
\begin{tikzpicture}[x=1pt,y=1pt]
\definecolor[named]{drawColor}{rgb}{0.00,0.00,0.00}
\definecolor[named]{fillColor}{rgb}{1.00,1.00,1.00}
\fill[color=fillColor,fill opacity=0.00,] (0,0) rectangle (108.41,130.09);
\begin{scope}
\path[clip] (  0.00,  0.00) rectangle (108.41,130.09);
\definecolor[named]{drawColor}{rgb}{0.00,0.00,0.00}

\draw[color=drawColor,line cap=round,line join=round,fill opacity=0.00,] ( 20.72, 33.12) -- (101.37, 33.12);

\draw[color=drawColor,line cap=round,line join=round,fill opacity=0.00,] ( 20.72, 33.12) -- ( 20.72, 29.52);

\draw[color=drawColor,line cap=round,line join=round,fill opacity=0.00,] ( 32.24, 33.12) -- ( 32.24, 29.52);

\draw[color=drawColor,line cap=round,line join=round,fill opacity=0.00,] ( 43.76, 33.12) -- ( 43.76, 29.52);

\draw[color=drawColor,line cap=round,line join=round,fill opacity=0.00,] ( 55.28, 33.12) -- ( 55.28, 29.52);

\draw[color=drawColor,line cap=round,line join=round,fill opacity=0.00,] ( 66.80, 33.12) -- ( 66.80, 29.52);

\draw[color=drawColor,line cap=round,line join=round,fill opacity=0.00,] ( 78.32, 33.12) -- ( 78.32, 29.52);

\draw[color=drawColor,line cap=round,line join=round,fill opacity=0.00,] ( 89.85, 33.12) -- ( 89.85, 29.52);

\draw[color=drawColor,line cap=round,line join=round,fill opacity=0.00,] (101.37, 33.12) -- (101.37, 29.52);

\node[color=drawColor,anchor=base,inner sep=0pt, outer sep=0pt, scale=  0.60] at ( 20.72, 18.72) {60};

\node[color=drawColor,anchor=base,inner sep=0pt, outer sep=0pt, scale=  0.60] at ( 32.24, 18.72) {80};

\node[color=drawColor,anchor=base,inner sep=0pt, outer sep=0pt, scale=  0.60] at ( 55.28, 18.72) {120};

\node[color=drawColor,anchor=base,inner sep=0pt, outer sep=0pt, scale=  0.60] at ( 78.32, 18.72) {160};

\node[color=drawColor,anchor=base,inner sep=0pt, outer sep=0pt, scale=  0.60] at (101.37, 18.72) {200};

\draw[color=drawColor,line cap=round,line join=round,fill opacity=0.00,] ( 15.12, 36.15) -- ( 15.12,111.93);

\draw[color=drawColor,line cap=round,line join=round,fill opacity=0.00,] ( 15.12, 36.15) -- ( 11.52, 36.15);

\draw[color=drawColor,line cap=round,line join=round,fill opacity=0.00,] ( 15.12, 55.10) -- ( 11.52, 55.10);

\draw[color=drawColor,line cap=round,line join=round,fill opacity=0.00,] ( 15.12, 74.04) -- ( 11.52, 74.04);

\draw[color=drawColor,line cap=round,line join=round,fill opacity=0.00,] ( 15.12, 92.99) -- ( 11.52, 92.99);

\draw[color=drawColor,line cap=round,line join=round,fill opacity=0.00,] ( 15.12,111.93) -- ( 11.52,111.93);

\node[color=drawColor,anchor=base east,inner sep=0pt, outer sep=0pt, scale=  0.60] at (  7.92, 34.09) {-4};

\node[color=drawColor,anchor=base east,inner sep=0pt, outer sep=0pt, scale=  0.60] at (  7.92, 53.03) {-2};

\node[color=drawColor,anchor=base east,inner sep=0pt, outer sep=0pt, scale=  0.60] at (  7.92, 71.98) {0};

\node[color=drawColor,anchor=base east,inner sep=0pt, outer sep=0pt, scale=  0.60] at (  7.92, 90.92) {2};

\node[color=drawColor,anchor=base east,inner sep=0pt, outer sep=0pt, scale=  0.60] at (  7.92,109.87) {4};

\draw[color=drawColor,line cap=round,line join=round,fill opacity=0.00,] ( 15.12, 33.12) --
	(104.08, 33.12) --
	(104.08,114.97) --
	( 15.12,114.97) --
	( 15.12, 33.12);
\end{scope}
\begin{scope}
\path[clip] (  0.00,  0.00) rectangle (108.41,130.09);
\definecolor[named]{drawColor}{rgb}{0.00,0.00,0.00}

\node[color=drawColor,anchor=base,inner sep=0pt, outer sep=0pt, scale=  0.72] at ( 59.60,120.04) {\bfseries $d_{2}= 1$};

\node[color=drawColor,anchor=base,inner sep=0pt, outer sep=0pt, scale=  0.72] at ( 59.60,  4.32) {$x_{2}$};
\end{scope}
\begin{scope}
\path[clip] ( 15.12, 33.12) rectangle (104.08,114.97);
\definecolor[named]{drawColor}{rgb}{0.00,0.00,0.00}

\draw[color=drawColor,line width= 0.8pt,line cap=round,line join=round,fill opacity=0.00,] ( 18.42, 52.49) --
	( 18.83, 52.73) --
	( 19.24, 52.97) --
	( 19.66, 53.21) --
	( 20.07, 53.44) --
	( 20.48, 53.68) --
	( 20.90, 53.92) --
	( 21.31, 54.16) --
	( 21.73, 54.40) --
	( 22.14, 54.63) --
	( 22.55, 54.87) --
	( 22.97, 55.11) --
	( 23.38, 55.35) --
	( 23.80, 55.59) --
	( 24.21, 55.83) --
	( 24.62, 56.06) --
	( 25.04, 56.30) --
	( 25.45, 56.54) --
	( 25.87, 56.78) --
	( 26.28, 57.02) --
	( 26.69, 57.25) --
	( 27.11, 57.49) --
	( 27.52, 57.73) --
	( 27.94, 57.97) --
	( 28.35, 58.21) --
	( 28.76, 58.45) --
	( 29.18, 58.68) --
	( 29.59, 58.92) --
	( 30.01, 59.16) --
	( 30.42, 59.40) --
	( 30.83, 59.64) --
	( 31.25, 59.87) --
	( 31.66, 60.11) --
	( 32.08, 60.35) --
	( 32.49, 60.59) --
	( 32.90, 60.83) --
	( 33.32, 61.06) --
	( 33.73, 61.30) --
	( 34.14, 61.54) --
	( 34.56, 61.78) --
	( 34.97, 62.02) --
	( 35.39, 62.26) --
	( 35.80, 62.49) --
	( 36.21, 62.73) --
	( 36.63, 62.97) --
	( 37.04, 63.21) --
	( 37.46, 63.45) --
	( 37.87, 63.68) --
	( 38.28, 63.92) --
	( 38.70, 64.16) --
	( 39.11, 64.40) --
	( 39.53, 64.64) --
	( 39.94, 64.88) --
	( 40.35, 65.11) --
	( 40.77, 65.35) --
	( 41.18, 65.59) --
	( 41.60, 65.83) --
	( 42.01, 66.07) --
	( 42.42, 66.30) --
	( 42.84, 66.54) --
	( 43.25, 66.78) --
	( 43.67, 67.02) --
	( 44.08, 67.26) --
	( 44.49, 67.49) --
	( 44.91, 67.73) --
	( 45.32, 67.97) --
	( 45.74, 68.21) --
	( 46.15, 68.45) --
	( 46.56, 68.69) --
	( 46.98, 68.92) --
	( 47.39, 69.16) --
	( 47.81, 69.40) --
	( 48.22, 69.64) --
	( 48.63, 69.88) --
	( 49.05, 70.11) --
	( 49.46, 70.35) --
	( 49.87, 70.59) --
	( 50.29, 70.83) --
	( 50.70, 71.07) --
	( 51.12, 71.31) --
	( 51.53, 71.54) --
	( 51.94, 71.78) --
	( 52.36, 72.02) --
	( 52.77, 72.26) --
	( 53.19, 72.50) --
	( 53.60, 72.73) --
	( 54.01, 72.97) --
	( 54.43, 73.21) --
	( 54.84, 73.45) --
	( 55.26, 73.69) --
	( 55.67, 73.92) --
	( 56.08, 74.16) --
	( 56.50, 74.40) --
	( 56.91, 74.64) --
	( 57.33, 74.88) --
	( 57.74, 75.12) --
	( 58.15, 75.35) --
	( 58.57, 75.59) --
	( 58.98, 75.83) --
	( 59.40, 76.07) --
	( 59.81, 76.31) --
	( 60.22, 76.54) --
	( 60.64, 76.78) --
	( 61.05, 77.02) --
	( 61.47, 77.26) --
	( 61.88, 77.50) --
	( 62.29, 77.74) --
	( 62.71, 77.97) --
	( 63.12, 78.21) --
	( 63.53, 78.45) --
	( 63.95, 78.69) --
	( 64.36, 78.93) --
	( 64.78, 79.16) --
	( 65.19, 79.40) --
	( 65.60, 79.64) --
	( 66.02, 79.88) --
	( 66.43, 80.12) --
	( 66.85, 80.35) --
	( 67.26, 80.59) --
	( 67.67, 80.83) --
	( 68.09, 81.07) --
	( 68.50, 81.31) --
	( 68.92, 81.55) --
	( 69.33, 81.78) --
	( 69.74, 82.02) --
	( 70.16, 82.26) --
	( 70.57, 82.50) --
	( 70.99, 82.74) --
	( 71.40, 82.97) --
	( 71.81, 83.21) --
	( 72.23, 83.45) --
	( 72.64, 83.69) --
	( 73.06, 83.93) --
	( 73.47, 84.17) --
	( 73.88, 84.40) --
	( 74.30, 84.64) --
	( 74.71, 84.88) --
	( 75.13, 85.12) --
	( 75.54, 85.36) --
	( 75.95, 85.59) --
	( 76.37, 85.83) --
	( 76.78, 86.07) --
	( 77.20, 86.31) --
	( 77.61, 86.55) --
	( 78.02, 86.78) --
	( 78.44, 87.02) --
	( 78.85, 87.26) --
	( 79.26, 87.50) --
	( 79.68, 87.74) --
	( 80.09, 87.98) --
	( 80.51, 88.21) --
	( 80.92, 88.45) --
	( 81.33, 88.69) --
	( 81.75, 88.93) --
	( 82.16, 89.17) --
	( 82.58, 89.40) --
	( 82.99, 89.64) --
	( 83.40, 89.88) --
	( 83.82, 90.12) --
	( 84.23, 90.36) --
	( 84.65, 90.59) --
	( 85.06, 90.83) --
	( 85.47, 91.07) --
	( 85.89, 91.31) --
	( 86.30, 91.55) --
	( 86.72, 91.79) --
	( 87.13, 92.02) --
	( 87.54, 92.26) --
	( 87.96, 92.50) --
	( 88.37, 92.74) --
	( 88.79, 92.98) --
	( 89.20, 93.21) --
	( 89.61, 93.45) --
	( 90.03, 93.69) --
	( 90.44, 93.93) --
	( 90.86, 94.17) --
	( 91.27, 94.41) --
	( 91.68, 94.64) --
	( 92.10, 94.88) --
	( 92.51, 95.12) --
	( 92.93, 95.36) --
	( 93.34, 95.60) --
	( 93.75, 95.83) --
	( 94.17, 96.07) --
	( 94.58, 96.31) --
	( 94.99, 96.55) --
	( 95.41, 96.79) --
	( 95.82, 97.02) --
	( 96.24, 97.26) --
	( 96.65, 97.50) --
	( 97.06, 97.74) --
	( 97.48, 97.98) --
	( 97.89, 98.22) --
	( 98.31, 98.45) --
	( 98.72, 98.69) --
	( 99.13, 98.93) --
	( 99.55, 99.17) --
	( 99.96, 99.41) --
	(100.38, 99.64) --
	(100.79, 99.88);

\draw[color=drawColor,line width= 0.8pt,dash pattern=on 4pt off 4pt ,line cap=round,line join=round,fill opacity=0.00,] ( 18.42, 47.35) --
	( 18.83, 47.64) --
	( 19.24, 47.94) --
	( 19.66, 48.23) --
	( 20.07, 48.53) --
	( 20.48, 48.82) --
	( 20.90, 49.12) --
	( 21.31, 49.41) --
	( 21.73, 49.71) --
	( 22.14, 50.00) --
	( 22.55, 50.30) --
	( 22.97, 50.59) --
	( 23.38, 50.89) --
	( 23.80, 51.18) --
	( 24.21, 51.48) --
	( 24.62, 51.77) --
	( 25.04, 52.07) --
	( 25.45, 52.36) --
	( 25.87, 52.66) --
	( 26.28, 52.95) --
	( 26.69, 53.25) --
	( 27.11, 53.54) --
	( 27.52, 53.84) --
	( 27.94, 54.13) --
	( 28.35, 54.43) --
	( 28.76, 54.72) --
	( 29.18, 55.02) --
	( 29.59, 55.31) --
	( 30.01, 55.61) --
	( 30.42, 55.90) --
	( 30.83, 56.20) --
	( 31.25, 56.49) --
	( 31.66, 56.79) --
	( 32.08, 57.08) --
	( 32.49, 57.38) --
	( 32.90, 57.67) --
	( 33.32, 57.97) --
	( 33.73, 58.26) --
	( 34.14, 58.56) --
	( 34.56, 58.85) --
	( 34.97, 59.15) --
	( 35.39, 59.44) --
	( 35.80, 59.74) --
	( 36.21, 60.03) --
	( 36.63, 60.33) --
	( 37.04, 60.62) --
	( 37.46, 60.92) --
	( 37.87, 61.21) --
	( 38.28, 61.51) --
	( 38.70, 61.80) --
	( 39.11, 62.10) --
	( 39.53, 62.39) --
	( 39.94, 62.69) --
	( 40.35, 62.98) --
	( 40.77, 63.28) --
	( 41.18, 63.57) --
	( 41.60, 63.87) --
	( 42.01, 64.16) --
	( 42.42, 64.46) --
	( 42.84, 64.75) --
	( 43.25, 65.05) --
	( 43.67, 65.34) --
	( 44.08, 65.64) --
	( 44.49, 65.93) --
	( 44.91, 66.23) --
	( 45.32, 66.52) --
	( 45.74, 66.82) --
	( 46.15, 67.11) --
	( 46.56, 67.41) --
	( 46.98, 67.70) --
	( 47.39, 68.00) --
	( 47.81, 68.29) --
	( 48.22, 68.59) --
	( 48.63, 68.88) --
	( 49.05, 69.18) --
	( 49.46, 69.47) --
	( 49.87, 69.77) --
	( 50.29, 70.06) --
	( 50.70, 70.36) --
	( 51.12, 70.65) --
	( 51.53, 70.95) --
	( 51.94, 71.24) --
	( 52.36, 71.54) --
	( 52.77, 71.83) --
	( 53.19, 72.13) --
	( 53.60, 72.42) --
	( 54.01, 72.72) --
	( 54.43, 73.01) --
	( 54.84, 73.31) --
	( 55.26, 73.60) --
	( 55.67, 73.90) --
	( 56.08, 74.14) --
	( 56.50, 74.32) --
	( 56.91, 74.50) --
	( 57.33, 74.69) --
	( 57.74, 74.87) --
	( 58.15, 75.06) --
	( 58.57, 75.24) --
	( 58.98, 75.43) --
	( 59.40, 75.61) --
	( 59.81, 75.80) --
	( 60.22, 75.98) --
	( 60.64, 76.17) --
	( 61.05, 76.35) --
	( 61.47, 76.53) --
	( 61.88, 76.72) --
	( 62.29, 76.90) --
	( 62.71, 77.09) --
	( 63.12, 77.27) --
	( 63.53, 77.46) --
	( 63.95, 77.64) --
	( 64.36, 77.83) --
	( 64.78, 78.01) --
	( 65.19, 78.19) --
	( 65.60, 78.38) --
	( 66.02, 78.56) --
	( 66.43, 78.75) --
	( 66.85, 78.93) --
	( 67.26, 79.12) --
	( 67.67, 79.30) --
	( 68.09, 79.49) --
	( 68.50, 79.67) --
	( 68.92, 79.86) --
	( 69.33, 80.04) --
	( 69.74, 80.22) --
	( 70.16, 80.41) --
	( 70.57, 80.59) --
	( 70.99, 80.78) --
	( 71.40, 80.96) --
	( 71.81, 81.15) --
	( 72.23, 81.33) --
	( 72.64, 81.52) --
	( 73.06, 81.70) --
	( 73.47, 81.88) --
	( 73.88, 82.07) --
	( 74.30, 82.25) --
	( 74.71, 82.44) --
	( 75.13, 82.62) --
	( 75.54, 82.81) --
	( 75.95, 82.99) --
	( 76.37, 83.18) --
	( 76.78, 83.36) --
	( 77.20, 83.55) --
	( 77.61, 83.73) --
	( 78.02, 83.91) --
	( 78.44, 84.10) --
	( 78.85, 84.28) --
	( 79.26, 84.47) --
	( 79.68, 84.65) --
	( 80.09, 84.84) --
	( 80.51, 85.02) --
	( 80.92, 85.21) --
	( 81.33, 85.39) --
	( 81.75, 85.57) --
	( 82.16, 85.76) --
	( 82.58, 85.94) --
	( 82.99, 86.13) --
	( 83.40, 86.31) --
	( 83.82, 86.50) --
	( 84.23, 86.68) --
	( 84.65, 86.87) --
	( 85.06, 87.05) --
	( 85.47, 87.24) --
	( 85.89, 87.42) --
	( 86.30, 87.60) --
	( 86.72, 87.79) --
	( 87.13, 87.97) --
	( 87.54, 88.16) --
	( 87.96, 88.34) --
	( 88.37, 88.53) --
	( 88.79, 88.71) --
	( 89.20, 88.90) --
	( 89.61, 89.08) --
	( 90.03, 89.26) --
	( 90.44, 89.45) --
	( 90.86, 89.63) --
	( 91.27, 89.82) --
	( 91.68, 90.00) --
	( 92.10, 90.19) --
	( 92.51, 90.37) --
	( 92.93, 90.56) --
	( 93.34, 90.74) --
	( 93.75, 90.93) --
	( 94.17, 91.11) --
	( 94.58, 91.29) --
	( 94.99, 91.48) --
	( 95.41, 91.66) --
	( 95.82, 91.85) --
	( 96.24, 92.03) --
	( 96.65, 92.22) --
	( 97.06, 92.40) --
	( 97.48, 92.59) --
	( 97.89, 92.77) --
	( 98.31, 92.95) --
	( 98.72, 93.14) --
	( 99.13, 93.32) --
	( 99.55, 93.51) --
	( 99.96, 93.69) --
	(100.38, 93.88) --
	(100.79, 94.06);

\draw[color=drawColor,line width= 0.8pt,dash pattern=on 4pt off 4pt ,line cap=round,line join=round,fill opacity=0.00,] ( 18.42, 57.35) --
	( 18.83, 57.53) --
	( 19.24, 57.72) --
	( 19.66, 57.90) --
	( 20.07, 58.08) --
	( 20.48, 58.27) --
	( 20.90, 58.45) --
	( 21.31, 58.64) --
	( 21.73, 58.82) --
	( 22.14, 59.01) --
	( 22.55, 59.19) --
	( 22.97, 59.38) --
	( 23.38, 59.56) --
	( 23.80, 59.75) --
	( 24.21, 59.93) --
	( 24.62, 60.11) --
	( 25.04, 60.30) --
	( 25.45, 60.48) --
	( 25.87, 60.67) --
	( 26.28, 60.85) --
	( 26.69, 61.04) --
	( 27.11, 61.22) --
	( 27.52, 61.41) --
	( 27.94, 61.59) --
	( 28.35, 61.77) --
	( 28.76, 61.96) --
	( 29.18, 62.14) --
	( 29.59, 62.33) --
	( 30.01, 62.51) --
	( 30.42, 62.70) --
	( 30.83, 62.88) --
	( 31.25, 63.07) --
	( 31.66, 63.25) --
	( 32.08, 63.44) --
	( 32.49, 63.62) --
	( 32.90, 63.80) --
	( 33.32, 63.99) --
	( 33.73, 64.17) --
	( 34.14, 64.36) --
	( 34.56, 64.54) --
	( 34.97, 64.73) --
	( 35.39, 64.91) --
	( 35.80, 65.10) --
	( 36.21, 65.28) --
	( 36.63, 65.46) --
	( 37.04, 65.65) --
	( 37.46, 65.83) --
	( 37.87, 66.02) --
	( 38.28, 66.20) --
	( 38.70, 66.39) --
	( 39.11, 66.57) --
	( 39.53, 66.76) --
	( 39.94, 66.94) --
	( 40.35, 67.12) --
	( 40.77, 67.31) --
	( 41.18, 67.49) --
	( 41.60, 67.68) --
	( 42.01, 67.86) --
	( 42.42, 68.05) --
	( 42.84, 68.23) --
	( 43.25, 68.42) --
	( 43.67, 68.60) --
	( 44.08, 68.79) --
	( 44.49, 68.97) --
	( 44.91, 69.15) --
	( 45.32, 69.34) --
	( 45.74, 69.52) --
	( 46.15, 69.71) --
	( 46.56, 69.89) --
	( 46.98, 70.08) --
	( 47.39, 70.26) --
	( 47.81, 70.45) --
	( 48.22, 70.63) --
	( 48.63, 70.81) --
	( 49.05, 71.00) --
	( 49.46, 71.18) --
	( 49.87, 71.37) --
	( 50.29, 71.55) --
	( 50.70, 71.74) --
	( 51.12, 71.92) --
	( 51.53, 72.11) --
	( 51.94, 72.29) --
	( 52.36, 72.48) --
	( 52.77, 72.66) --
	( 53.19, 72.84) --
	( 53.60, 73.03) --
	( 54.01, 73.21) --
	( 54.43, 73.40) --
	( 54.84, 73.58) --
	( 55.26, 73.77) --
	( 55.67, 73.95) --
	( 56.08, 74.19) --
	( 56.50, 74.49) --
	( 56.91, 74.78) --
	( 57.33, 75.08) --
	( 57.74, 75.37) --
	( 58.15, 75.67) --
	( 58.57, 75.96) --
	( 58.98, 76.26) --
	( 59.40, 76.55) --
	( 59.81, 76.85) --
	( 60.22, 77.14) --
	( 60.64, 77.44) --
	( 61.05, 77.73) --
	( 61.47, 78.03) --
	( 61.88, 78.32) --
	( 62.29, 78.62) --
	( 62.71, 78.91) --
	( 63.12, 79.21) --
	( 63.53, 79.50) --
	( 63.95, 79.80) --
	( 64.36, 80.09) --
	( 64.78, 80.39) --
	( 65.19, 80.68) --
	( 65.60, 80.98) --
	( 66.02, 81.27) --
	( 66.43, 81.57) --
	( 66.85, 81.86) --
	( 67.26, 82.16) --
	( 67.67, 82.45) --
	( 68.09, 82.75) --
	( 68.50, 83.04) --
	( 68.92, 83.34) --
	( 69.33, 83.63) --
	( 69.74, 83.93) --
	( 70.16, 84.22) --
	( 70.57, 84.52) --
	( 70.99, 84.81) --
	( 71.40, 85.11) --
	( 71.81, 85.40) --
	( 72.23, 85.70) --
	( 72.64, 85.99) --
	( 73.06, 86.29) --
	( 73.47, 86.58) --
	( 73.88, 86.88) --
	( 74.30, 87.17) --
	( 74.71, 87.47) --
	( 75.13, 87.76) --
	( 75.54, 88.06) --
	( 75.95, 88.35) --
	( 76.37, 88.65) --
	( 76.78, 88.94) --
	( 77.20, 89.24) --
	( 77.61, 89.53) --
	( 78.02, 89.83) --
	( 78.44, 90.12) --
	( 78.85, 90.42) --
	( 79.26, 90.71) --
	( 79.68, 91.01) --
	( 80.09, 91.30) --
	( 80.51, 91.60) --
	( 80.92, 91.89) --
	( 81.33, 92.19) --
	( 81.75, 92.48) --
	( 82.16, 92.78) --
	( 82.58, 93.07) --
	( 82.99, 93.37) --
	( 83.40, 93.66) --
	( 83.82, 93.96) --
	( 84.23, 94.25) --
	( 84.65, 94.55) --
	( 85.06, 94.84) --
	( 85.47, 95.14) --
	( 85.89, 95.43) --
	( 86.30, 95.73) --
	( 86.72, 96.02) --
	( 87.13, 96.32) --
	( 87.54, 96.61) --
	( 87.96, 96.91) --
	( 88.37, 97.20) --
	( 88.79, 97.50) --
	( 89.20, 97.79) --
	( 89.61, 98.09) --
	( 90.03, 98.38) --
	( 90.44, 98.68) --
	( 90.86, 98.97) --
	( 91.27, 99.27) --
	( 91.68, 99.56) --
	( 92.10, 99.86) --
	( 92.51,100.15) --
	( 92.93,100.45) --
	( 93.34,100.74) --
	( 93.75,101.04) --
	( 94.17,101.33) --
	( 94.58,101.63) --
	( 94.99,101.92) --
	( 95.41,102.22) --
	( 95.82,102.51) --
	( 96.24,102.81) --
	( 96.65,103.10) --
	( 97.06,103.40) --
	( 97.48,103.69) --
	( 97.89,103.99) --
	( 98.31,104.28) --
	( 98.72,104.58) --
	( 99.13,104.87) --
	( 99.55,105.17) --
	( 99.96,105.46) --
	(100.38,105.76) --
	(100.79,106.05);

\draw[color=drawColor,line width= 0.8pt,dash pattern=on 1pt off 3pt ,line cap=round,line join=round,fill opacity=0.00,] ( 18.42, 47.35) --
	( 18.83, 47.64) --
	( 19.24, 47.94) --
	( 19.66, 48.23) --
	( 20.07, 48.53) --
	( 20.48, 48.82) --
	( 20.90, 49.12) --
	( 21.31, 49.41) --
	( 21.73, 49.71) --
	( 22.14, 50.00) --
	( 22.55, 50.30) --
	( 22.97, 50.59) --
	( 23.38, 50.89) --
	( 23.80, 51.18) --
	( 24.21, 51.48) --
	( 24.62, 51.77) --
	( 25.04, 52.07) --
	( 25.45, 52.36) --
	( 25.87, 52.66) --
	( 26.28, 52.95) --
	( 26.69, 53.25) --
	( 27.11, 53.54) --
	( 27.52, 53.84) --
	( 27.94, 54.13) --
	( 28.35, 54.43) --
	( 28.76, 54.72) --
	( 29.18, 55.02) --
	( 29.59, 55.31) --
	( 30.01, 55.61) --
	( 30.42, 55.90) --
	( 30.83, 56.20) --
	( 31.25, 56.49) --
	( 31.66, 56.79) --
	( 32.08, 57.08) --
	( 32.49, 57.38) --
	( 32.90, 57.67) --
	( 33.32, 57.97) --
	( 33.73, 58.26) --
	( 34.14, 58.56) --
	( 34.56, 58.85) --
	( 34.97, 59.15) --
	( 35.39, 59.44) --
	( 35.80, 59.74) --
	( 36.21, 60.03) --
	( 36.63, 60.33) --
	( 37.04, 60.62) --
	( 37.46, 60.92) --
	( 37.87, 61.21) --
	( 38.28, 61.51) --
	( 38.70, 61.80) --
	( 39.11, 62.10) --
	( 39.53, 62.39) --
	( 39.94, 62.69) --
	( 40.35, 62.98) --
	( 40.77, 63.28) --
	( 41.18, 63.57) --
	( 41.60, 63.87) --
	( 42.01, 64.16) --
	( 42.42, 64.46) --
	( 42.84, 64.75) --
	( 43.25, 65.05) --
	( 43.67, 65.34) --
	( 44.08, 65.64) --
	( 44.49, 65.93) --
	( 44.91, 66.23) --
	( 45.32, 66.52) --
	( 45.74, 66.82) --
	( 46.15, 67.11) --
	( 46.56, 67.41) --
	( 46.98, 67.70) --
	( 47.39, 68.00) --
	( 47.81, 68.29) --
	( 48.22, 68.59) --
	( 48.63, 68.88) --
	( 49.05, 69.18) --
	( 49.46, 69.47) --
	( 49.87, 69.77) --
	( 50.29, 70.06) --
	( 50.70, 70.36) --
	( 51.12, 70.65) --
	( 51.53, 70.95) --
	( 51.94, 71.24) --
	( 52.36, 71.54) --
	( 52.77, 71.83) --
	( 53.19, 72.13) --
	( 53.60, 72.42) --
	( 54.01, 72.72) --
	( 54.43, 73.01) --
	( 54.84, 73.31) --
	( 55.26, 73.60) --
	( 55.67, 73.90) --
	( 56.08, 74.14) --
	( 56.50, 74.32) --
	( 56.91, 74.50) --
	( 57.33, 74.69) --
	( 57.74, 74.87) --
	( 58.15, 75.06) --
	( 58.57, 75.24) --
	( 58.98, 75.43) --
	( 59.40, 75.61) --
	( 59.81, 75.80) --
	( 60.22, 75.98) --
	( 60.64, 76.16) --
	( 61.05, 76.35) --
	( 61.47, 76.53) --
	( 61.88, 76.72) --
	( 62.29, 76.90) --
	( 62.71, 77.09) --
	( 63.12, 77.27) --
	( 63.53, 77.46) --
	( 63.95, 77.64) --
	( 64.36, 77.83) --
	( 64.78, 78.01) --
	( 65.19, 78.19) --
	( 65.60, 78.38) --
	( 66.02, 78.56) --
	( 66.43, 78.75) --
	( 66.85, 78.93) --
	( 67.26, 79.12) --
	( 67.67, 79.30) --
	( 68.09, 79.49) --
	( 68.50, 79.67) --
	( 68.92, 79.85) --
	( 69.33, 80.04) --
	( 69.74, 80.22) --
	( 70.16, 80.41) --
	( 70.57, 80.59) --
	( 70.99, 80.78) --
	( 71.40, 80.96) --
	( 71.81, 81.15) --
	( 72.23, 81.33) --
	( 72.64, 81.51) --
	( 73.06, 81.70) --
	( 73.47, 81.88) --
	( 73.88, 82.07) --
	( 74.30, 82.25) --
	( 74.71, 82.44) --
	( 75.13, 82.62) --
	( 75.54, 82.81) --
	( 75.95, 82.99) --
	( 76.37, 83.17) --
	( 76.78, 83.36) --
	( 77.20, 83.54) --
	( 77.61, 83.73) --
	( 78.02, 83.91) --
	( 78.44, 84.10) --
	( 78.85, 84.28) --
	( 79.26, 84.47) --
	( 79.68, 84.65) --
	( 80.09, 84.83) --
	( 80.51, 85.02) --
	( 80.92, 85.20) --
	( 81.33, 85.39) --
	( 81.75, 85.57) --
	( 82.16, 85.76) --
	( 82.58, 85.94) --
	( 82.99, 86.13) --
	( 83.40, 86.31) --
	( 83.82, 86.49) --
	( 84.23, 86.68) --
	( 84.65, 86.86) --
	( 85.06, 87.05) --
	( 85.47, 87.23) --
	( 85.89, 87.42) --
	( 86.30, 87.60) --
	( 86.72, 87.79) --
	( 87.13, 87.97) --
	( 87.54, 88.16) --
	( 87.96, 88.34) --
	( 88.37, 88.52) --
	( 88.79, 88.71) --
	( 89.20, 88.89) --
	( 89.61, 89.08) --
	( 90.03, 89.26) --
	( 90.44, 89.45) --
	( 90.86, 89.63) --
	( 91.27, 89.82) --
	( 91.68, 90.00) --
	( 92.10, 90.18) --
	( 92.51, 90.37) --
	( 92.93, 90.55) --
	( 93.34, 90.74) --
	( 93.75, 90.92) --
	( 94.17, 91.11) --
	( 94.58, 91.29) --
	( 94.99, 91.48) --
	( 95.41, 91.66) --
	( 95.82, 91.84) --
	( 96.24, 92.03) --
	( 96.65, 92.21) --
	( 97.06, 92.40) --
	( 97.48, 92.58) --
	( 97.89, 92.77) --
	( 98.31, 92.95) --
	( 98.72, 93.14) --
	( 99.13, 93.32) --
	( 99.55, 93.50) --
	( 99.96, 93.69) --
	(100.38, 93.87) --
	(100.79, 94.06);

\draw[color=drawColor,line width= 0.8pt,dash pattern=on 1pt off 3pt ,line cap=round,line join=round,fill opacity=0.00,] ( 18.42, 57.35) --
	( 18.83, 57.53) --
	( 19.24, 57.72) --
	( 19.66, 57.90) --
	( 20.07, 58.09) --
	( 20.48, 58.27) --
	( 20.90, 58.46) --
	( 21.31, 58.64) --
	( 21.73, 58.83) --
	( 22.14, 59.01) --
	( 22.55, 59.19) --
	( 22.97, 59.38) --
	( 23.38, 59.56) --
	( 23.80, 59.75) --
	( 24.21, 59.93) --
	( 24.62, 60.12) --
	( 25.04, 60.30) --
	( 25.45, 60.49) --
	( 25.87, 60.67) --
	( 26.28, 60.85) --
	( 26.69, 61.04) --
	( 27.11, 61.22) --
	( 27.52, 61.41) --
	( 27.94, 61.59) --
	( 28.35, 61.78) --
	( 28.76, 61.96) --
	( 29.18, 62.15) --
	( 29.59, 62.33) --
	( 30.01, 62.51) --
	( 30.42, 62.70) --
	( 30.83, 62.88) --
	( 31.25, 63.07) --
	( 31.66, 63.25) --
	( 32.08, 63.44) --
	( 32.49, 63.62) --
	( 32.90, 63.81) --
	( 33.32, 63.99) --
	( 33.73, 64.17) --
	( 34.14, 64.36) --
	( 34.56, 64.54) --
	( 34.97, 64.73) --
	( 35.39, 64.91) --
	( 35.80, 65.10) --
	( 36.21, 65.28) --
	( 36.63, 65.47) --
	( 37.04, 65.65) --
	( 37.46, 65.84) --
	( 37.87, 66.02) --
	( 38.28, 66.20) --
	( 38.70, 66.39) --
	( 39.11, 66.57) --
	( 39.53, 66.76) --
	( 39.94, 66.94) --
	( 40.35, 67.13) --
	( 40.77, 67.31) --
	( 41.18, 67.50) --
	( 41.60, 67.68) --
	( 42.01, 67.86) --
	( 42.42, 68.05) --
	( 42.84, 68.23) --
	( 43.25, 68.42) --
	( 43.67, 68.60) --
	( 44.08, 68.79) --
	( 44.49, 68.97) --
	( 44.91, 69.16) --
	( 45.32, 69.34) --
	( 45.74, 69.52) --
	( 46.15, 69.71) --
	( 46.56, 69.89) --
	( 46.98, 70.08) --
	( 47.39, 70.26) --
	( 47.81, 70.45) --
	( 48.22, 70.63) --
	( 48.63, 70.82) --
	( 49.05, 71.00) --
	( 49.46, 71.18) --
	( 49.87, 71.37) --
	( 50.29, 71.55) --
	( 50.70, 71.74) --
	( 51.12, 71.92) --
	( 51.53, 72.11) --
	( 51.94, 72.29) --
	( 52.36, 72.48) --
	( 52.77, 72.66) --
	( 53.19, 72.84) --
	( 53.60, 73.03) --
	( 54.01, 73.21) --
	( 54.43, 73.40) --
	( 54.84, 73.58) --
	( 55.26, 73.77) --
	( 55.67, 73.95) --
	( 56.08, 74.19) --
	( 56.50, 74.49) --
	( 56.91, 74.78) --
	( 57.33, 75.08) --
	( 57.74, 75.37) --
	( 58.15, 75.67) --
	( 58.57, 75.96) --
	( 58.98, 76.26) --
	( 59.40, 76.55) --
	( 59.81, 76.85) --
	( 60.22, 77.14) --
	( 60.64, 77.44) --
	( 61.05, 77.73) --
	( 61.47, 78.03) --
	( 61.88, 78.32) --
	( 62.29, 78.62) --
	( 62.71, 78.91) --
	( 63.12, 79.21) --
	( 63.53, 79.50) --
	( 63.95, 79.80) --
	( 64.36, 80.09) --
	( 64.78, 80.39) --
	( 65.19, 80.68) --
	( 65.60, 80.98) --
	( 66.02, 81.27) --
	( 66.43, 81.57) --
	( 66.85, 81.86) --
	( 67.26, 82.16) --
	( 67.67, 82.45) --
	( 68.09, 82.75) --
	( 68.50, 83.04) --
	( 68.92, 83.34) --
	( 69.33, 83.63) --
	( 69.74, 83.93) --
	( 70.16, 84.22) --
	( 70.57, 84.52) --
	( 70.99, 84.81) --
	( 71.40, 85.11) --
	( 71.81, 85.40) --
	( 72.23, 85.70) --
	( 72.64, 85.99) --
	( 73.06, 86.29) --
	( 73.47, 86.58) --
	( 73.88, 86.88) --
	( 74.30, 87.17) --
	( 74.71, 87.47) --
	( 75.13, 87.76) --
	( 75.54, 88.06) --
	( 75.95, 88.35) --
	( 76.37, 88.65) --
	( 76.78, 88.94) --
	( 77.20, 89.24) --
	( 77.61, 89.53) --
	( 78.02, 89.83) --
	( 78.44, 90.12) --
	( 78.85, 90.42) --
	( 79.26, 90.71) --
	( 79.68, 91.01) --
	( 80.09, 91.30) --
	( 80.51, 91.60) --
	( 80.92, 91.89) --
	( 81.33, 92.19) --
	( 81.75, 92.48) --
	( 82.16, 92.78) --
	( 82.58, 93.07) --
	( 82.99, 93.37) --
	( 83.40, 93.66) --
	( 83.82, 93.96) --
	( 84.23, 94.25) --
	( 84.65, 94.55) --
	( 85.06, 94.84) --
	( 85.47, 95.14) --
	( 85.89, 95.43) --
	( 86.30, 95.73) --
	( 86.72, 96.02) --
	( 87.13, 96.32) --
	( 87.54, 96.61) --
	( 87.96, 96.91) --
	( 88.37, 97.20) --
	( 88.79, 97.50) --
	( 89.20, 97.79) --
	( 89.61, 98.09) --
	( 90.03, 98.38) --
	( 90.44, 98.68) --
	( 90.86, 98.97) --
	( 91.27, 99.27) --
	( 91.68, 99.56) --
	( 92.10, 99.86) --
	( 92.51,100.15) --
	( 92.93,100.45) --
	( 93.34,100.74) --
	( 93.75,101.04) --
	( 94.17,101.33) --
	( 94.58,101.63) --
	( 94.99,101.92) --
	( 95.41,102.22) --
	( 95.82,102.51) --
	( 96.24,102.81) --
	( 96.65,103.10) --
	( 97.06,103.40) --
	( 97.48,103.69) --
	( 97.89,103.99) --
	( 98.31,104.28) --
	( 98.72,104.58) --
	( 99.13,104.87) --
	( 99.55,105.17) --
	( 99.96,105.46) --
	(100.38,105.76) --
	(100.79,106.05);
\end{scope}
\end{tikzpicture}

%% file: out-pima-plot-x5.tikz
\begin{tikzpicture}[x=1pt,y=1pt]
\definecolor[named]{drawColor}{rgb}{0.00,0.00,0.00}
\definecolor[named]{fillColor}{rgb}{1.00,1.00,1.00}
\fill[color=fillColor,fill opacity=0.00,] (0,0) rectangle (108.41,130.09);
\begin{scope}
\path[clip] (  0.00,  0.00) rectangle (108.41,130.09);
\definecolor[named]{drawColor}{rgb}{0.00,0.00,0.00}

\draw[color=drawColor,line cap=round,line join=round,fill opacity=0.00,] ( 21.45, 33.12) -- ( 88.83, 33.12);

\draw[color=drawColor,line cap=round,line join=round,fill opacity=0.00,] ( 21.45, 33.12) -- ( 21.45, 29.52);

\draw[color=drawColor,line cap=round,line join=round,fill opacity=0.00,] ( 38.29, 33.12) -- ( 38.29, 29.52);

\draw[color=drawColor,line cap=round,line join=round,fill opacity=0.00,] ( 55.14, 33.12) -- ( 55.14, 29.52);

\draw[color=drawColor,line cap=round,line join=round,fill opacity=0.00,] ( 71.98, 33.12) -- ( 71.98, 29.52);

\draw[color=drawColor,line cap=round,line join=round,fill opacity=0.00,] ( 88.83, 33.12) -- ( 88.83, 29.52);

\node[color=drawColor,anchor=base,inner sep=0pt, outer sep=0pt, scale=  0.60] at ( 21.45, 18.72) {20};

\node[color=drawColor,anchor=base,inner sep=0pt, outer sep=0pt, scale=  0.60] at ( 38.29, 18.72) {30};

\node[color=drawColor,anchor=base,inner sep=0pt, outer sep=0pt, scale=  0.60] at ( 55.14, 18.72) {40};

\node[color=drawColor,anchor=base,inner sep=0pt, outer sep=0pt, scale=  0.60] at ( 71.98, 18.72) {50};

\node[color=drawColor,anchor=base,inner sep=0pt, outer sep=0pt, scale=  0.60] at ( 88.83, 18.72) {60};

\draw[color=drawColor,line cap=round,line join=round,fill opacity=0.00,] ( 15.12, 36.15) -- ( 15.12,111.93);

\draw[color=drawColor,line cap=round,line join=round,fill opacity=0.00,] ( 15.12, 36.15) -- ( 11.52, 36.15);

\draw[color=drawColor,line cap=round,line join=round,fill opacity=0.00,] ( 15.12, 48.78) -- ( 11.52, 48.78);

\draw[color=drawColor,line cap=round,line join=round,fill opacity=0.00,] ( 15.12, 61.41) -- ( 11.52, 61.41);

\draw[color=drawColor,line cap=round,line join=round,fill opacity=0.00,] ( 15.12, 74.04) -- ( 11.52, 74.04);

\draw[color=drawColor,line cap=round,line join=round,fill opacity=0.00,] ( 15.12, 86.67) -- ( 11.52, 86.67);

\draw[color=drawColor,line cap=round,line join=round,fill opacity=0.00,] ( 15.12, 99.30) -- ( 11.52, 99.30);

\draw[color=drawColor,line cap=round,line join=round,fill opacity=0.00,] ( 15.12,111.93) -- ( 11.52,111.93);

\node[color=drawColor,anchor=base east,inner sep=0pt, outer sep=0pt, scale=  0.60] at (  7.92, 34.09) {-4};

\node[color=drawColor,anchor=base east,inner sep=0pt, outer sep=0pt, scale=  0.60] at (  7.92, 46.72) {-2};

\node[color=drawColor,anchor=base east,inner sep=0pt, outer sep=0pt, scale=  0.60] at (  7.92, 59.35) {0};

\node[color=drawColor,anchor=base east,inner sep=0pt, outer sep=0pt, scale=  0.60] at (  7.92, 71.98) {2};

\node[color=drawColor,anchor=base east,inner sep=0pt, outer sep=0pt, scale=  0.60] at (  7.92, 84.61) {4};

\node[color=drawColor,anchor=base east,inner sep=0pt, outer sep=0pt, scale=  0.60] at (  7.92, 97.24) {6};

\node[color=drawColor,anchor=base east,inner sep=0pt, outer sep=0pt, scale=  0.60] at (  7.92,109.87) {8};

\draw[color=drawColor,line cap=round,line join=round,fill opacity=0.00,] ( 15.12, 33.12) --
	(104.08, 33.12) --
	(104.08,114.97) --
	( 15.12,114.97) --
	( 15.12, 33.12);
\end{scope}
\begin{scope}
\path[clip] (  0.00,  0.00) rectangle (108.41,130.09);
\definecolor[named]{drawColor}{rgb}{0.00,0.00,0.00}

\node[color=drawColor,anchor=base,inner sep=0pt, outer sep=0pt, scale=  0.72] at ( 59.60,120.04) {\bfseries $d_{5}= 3$};

\node[color=drawColor,anchor=base,inner sep=0pt, outer sep=0pt, scale=  0.72] at ( 59.60,  4.32) {$x_{5}$};
\end{scope}
\begin{scope}
\path[clip] ( 15.12, 33.12) rectangle (104.08,114.97);
\definecolor[named]{drawColor}{rgb}{0.00,0.00,0.00}

\draw[color=drawColor,line width= 0.8pt,line cap=round,line join=round,fill opacity=0.00,] ( 18.42, 47.53) --
	( 18.83, 47.85) --
	( 19.24, 48.16) --
	( 19.66, 48.47) --
	( 20.07, 48.79) --
	( 20.48, 49.10) --
	( 20.90, 49.41) --
	( 21.31, 49.72) --
	( 21.73, 50.02) --
	( 22.14, 50.33) --
	( 22.55, 50.64) --
	( 22.97, 50.94) --
	( 23.38, 51.25) --
	( 23.80, 51.55) --
	( 24.21, 51.86) --
	( 24.62, 52.16) --
	( 25.04, 52.46) --
	( 25.45, 52.76) --
	( 25.87, 53.06) --
	( 26.28, 53.36) --
	( 26.69, 53.65) --
	( 27.11, 53.95) --
	( 27.52, 54.24) --
	( 27.94, 54.54) --
	( 28.35, 54.83) --
	( 28.76, 55.12) --
	( 29.18, 55.41) --
	( 29.59, 55.70) --
	( 30.01, 55.99) --
	( 30.42, 56.27) --
	( 30.83, 56.56) --
	( 31.25, 56.84) --
	( 31.66, 57.13) --
	( 32.08, 57.41) --
	( 32.49, 57.69) --
	( 32.90, 57.96) --
	( 33.32, 58.24) --
	( 33.73, 58.51) --
	( 34.14, 58.77) --
	( 34.56, 59.04) --
	( 34.97, 59.30) --
	( 35.39, 59.55) --
	( 35.80, 59.80) --
	( 36.21, 60.04) --
	( 36.63, 60.28) --
	( 37.04, 60.51) --
	( 37.46, 60.73) --
	( 37.87, 60.95) --
	( 38.28, 61.16) --
	( 38.70, 61.36) --
	( 39.11, 61.56) --
	( 39.53, 61.74) --
	( 39.94, 61.92) --
	( 40.35, 62.08) --
	( 40.77, 62.24) --
	( 41.18, 62.39) --
	( 41.60, 62.54) --
	( 42.01, 62.67) --
	( 42.42, 62.80) --
	( 42.84, 62.92) --
	( 43.25, 63.03) --
	( 43.67, 63.14) --
	( 44.08, 63.24) --
	( 44.49, 63.33) --
	( 44.91, 63.42) --
	( 45.32, 63.50) --
	( 45.74, 63.58) --
	( 46.15, 63.65) --
	( 46.56, 63.72) --
	( 46.98, 63.78) --
	( 47.39, 63.84) --
	( 47.81, 63.90) --
	( 48.22, 63.95) --
	( 48.63, 64.00) --
	( 49.05, 64.04) --
	( 49.46, 64.08) --
	( 49.87, 64.12) --
	( 50.29, 64.16) --
	( 50.70, 64.20) --
	( 51.12, 64.23) --
	( 51.53, 64.27) --
	( 51.94, 64.30) --
	( 52.36, 64.33) --
	( 52.77, 64.36) --
	( 53.19, 64.39) --
	( 53.60, 64.42) --
	( 54.01, 64.45) --
	( 54.43, 64.49) --
	( 54.84, 64.52) --
	( 55.26, 64.56) --
	( 55.67, 64.59) --
	( 56.08, 64.63) --
	( 56.50, 64.67) --
	( 56.91, 64.72) --
	( 57.33, 64.76) --
	( 57.74, 64.81) --
	( 58.15, 64.86) --
	( 58.57, 64.91) --
	( 58.98, 64.97) --
	( 59.40, 65.03) --
	( 59.81, 65.09) --
	( 60.22, 65.15) --
	( 60.64, 65.22) --
	( 61.05, 65.29) --
	( 61.47, 65.36) --
	( 61.88, 65.43) --
	( 62.29, 65.51) --
	( 62.71, 65.58) --
	( 63.12, 65.66) --
	( 63.53, 65.75) --
	( 63.95, 65.83) --
	( 64.36, 65.92) --
	( 64.78, 66.01) --
	( 65.19, 66.10) --
	( 65.60, 66.19) --
	( 66.02, 66.29) --
	( 66.43, 66.39) --
	( 66.85, 66.49) --
	( 67.26, 66.59) --
	( 67.67, 66.69) --
	( 68.09, 66.80) --
	( 68.50, 66.91) --
	( 68.92, 67.02) --
	( 69.33, 67.13) --
	( 69.74, 67.25) --
	( 70.16, 67.37) --
	( 70.57, 67.49) --
	( 70.99, 67.61) --
	( 71.40, 67.73) --
	( 71.81, 67.86) --
	( 72.23, 67.98) --
	( 72.64, 68.11) --
	( 73.06, 68.24) --
	( 73.47, 68.38) --
	( 73.88, 68.51) --
	( 74.30, 68.65) --
	( 74.71, 68.79) --
	( 75.13, 68.93) --
	( 75.54, 69.07) --
	( 75.95, 69.21) --
	( 76.37, 69.36) --
	( 76.78, 69.51) --
	( 77.20, 69.66) --
	( 77.61, 69.81) --
	( 78.02, 69.96) --
	( 78.44, 70.12) --
	( 78.85, 70.27) --
	( 79.26, 70.43) --
	( 79.68, 70.59) --
	( 80.09, 70.75) --
	( 80.51, 70.91) --
	( 80.92, 71.08) --
	( 81.33, 71.24) --
	( 81.75, 71.41) --
	( 82.16, 71.58) --
	( 82.58, 71.75) --
	( 82.99, 71.92) --
	( 83.40, 72.10) --
	( 83.82, 72.27) --
	( 84.23, 72.45) --
	( 84.65, 72.63) --
	( 85.06, 72.81) --
	( 85.47, 72.99) --
	( 85.89, 73.17) --
	( 86.30, 73.36) --
	( 86.72, 73.54) --
	( 87.13, 73.73) --
	( 87.54, 73.91) --
	( 87.96, 74.10) --
	( 88.37, 74.29) --
	( 88.79, 74.49) --
	( 89.20, 74.68) --
	( 89.61, 74.87) --
	( 90.03, 75.07) --
	( 90.44, 75.26) --
	( 90.86, 75.46) --
	( 91.27, 75.66) --
	( 91.68, 75.86) --
	( 92.10, 76.06) --
	( 92.51, 76.27) --
	( 92.93, 76.47) --
	( 93.34, 76.67) --
	( 93.75, 76.88) --
	( 94.17, 77.09) --
	( 94.58, 77.29) --
	( 94.99, 77.50) --
	( 95.41, 77.71) --
	( 95.82, 77.92) --
	( 96.24, 78.13) --
	( 96.65, 78.35) --
	( 97.06, 78.56) --
	( 97.48, 78.78) --
	( 97.89, 78.99) --
	( 98.31, 79.21) --
	( 98.72, 79.42) --
	( 99.13, 79.64) --
	( 99.55, 79.86) --
	( 99.96, 80.08) --
	(100.38, 80.30) --
	(100.79, 80.52);

\draw[color=drawColor,line width= 0.8pt,dash pattern=on 4pt off 4pt ,line cap=round,line join=round,fill opacity=0.00,] ( 18.42, 38.90) --
	( 18.83, 39.52) --
	( 19.24, 40.14) --
	( 19.66, 40.72) --
	( 20.07, 41.26) --
	( 20.48, 41.79) --
	( 20.90, 42.37) --
	( 21.31, 42.94) --
	( 21.73, 43.53) --
	( 22.14, 44.09) --
	( 22.55, 44.63) --
	( 22.97, 45.14) --
	( 23.38, 45.67) --
	( 23.80, 46.21) --
	( 24.21, 46.71) --
	( 24.62, 47.22) --
	( 25.04, 47.71) --
	( 25.45, 48.20) --
	( 25.87, 48.64) --
	( 26.28, 49.13) --
	( 26.69, 49.61) --
	( 27.11, 50.07) --
	( 27.52, 50.54) --
	( 27.94, 51.02) --
	( 28.35, 51.48) --
	( 28.76, 51.91) --
	( 29.18, 52.35) --
	( 29.59, 52.77) --
	( 30.01, 53.24) --
	( 30.42, 53.67) --
	( 30.83, 54.08) --
	( 31.25, 54.50) --
	( 31.66, 54.92) --
	( 32.08, 55.32) --
	( 32.49, 55.74) --
	( 32.90, 56.12) --
	( 33.32, 56.50) --
	( 33.73, 56.87) --
	( 34.14, 57.19) --
	( 34.56, 57.50) --
	( 34.97, 57.82) --
	( 35.39, 58.12) --
	( 35.80, 58.38) --
	( 36.21, 58.65) --
	( 36.63, 58.89) --
	( 37.04, 59.13) --
	( 37.46, 59.35) --
	( 37.87, 59.57) --
	( 38.28, 59.76) --
	( 38.70, 59.97) --
	( 39.11, 60.14) --
	( 39.53, 60.32) --
	( 39.94, 60.50) --
	( 40.35, 60.67) --
	( 40.77, 60.82) --
	( 41.18, 60.99) --
	( 41.60, 61.13) --
	( 42.01, 61.26) --
	( 42.42, 61.38) --
	( 42.84, 61.49) --
	( 43.25, 61.61) --
	( 43.67, 61.70) --
	( 44.08, 61.78) --
	( 44.49, 61.87) --
	( 44.91, 61.94) --
	( 45.32, 62.00) --
	( 45.74, 62.05) --
	( 46.15, 62.10) --
	( 46.56, 62.14) --
	( 46.98, 62.16) --
	( 47.39, 62.20) --
	( 47.81, 62.23) --
	( 48.22, 62.26) --
	( 48.63, 62.28) --
	( 49.05, 62.32) --
	( 49.46, 62.35) --
	( 49.87, 62.37) --
	( 50.29, 62.41) --
	( 50.70, 62.44) --
	( 51.12, 62.47) --
	( 51.53, 62.50) --
	( 51.94, 62.51) --
	( 52.36, 62.52) --
	( 52.77, 62.53) --
	( 53.19, 62.53) --
	( 53.60, 62.53) --
	( 54.01, 62.53) --
	( 54.43, 62.52) --
	( 54.84, 62.51) --
	( 55.26, 62.50) --
	( 55.67, 62.48) --
	( 56.08, 62.47) --
	( 56.50, 62.45) --
	( 56.91, 62.45) --
	( 57.33, 62.45) --
	( 57.74, 62.44) --
	( 58.15, 62.42) --
	( 58.57, 62.41) --
	( 58.98, 62.41) --
	( 59.40, 62.41) --
	( 59.81, 62.42) --
	( 60.22, 62.42) --
	( 60.64, 62.43) --
	( 61.05, 62.44) --
	( 61.47, 62.46) --
	( 61.88, 62.47) --
	( 62.29, 62.48) --
	( 62.71, 62.50) --
	( 63.12, 62.49) --
	( 63.53, 62.51) --
	( 63.95, 62.50) --
	( 64.36, 62.52) --
	( 64.78, 62.55) --
	( 65.19, 62.59) --
	( 65.60, 62.63) --
	( 66.02, 62.67) --
	( 66.43, 62.70) --
	( 66.85, 62.72) --
	( 67.26, 62.76) --
	( 67.67, 62.80) --
	( 68.09, 62.86) --
	( 68.50, 62.90) --
	( 68.92, 62.93) --
	( 69.33, 62.95) --
	( 69.74, 62.99) --
	( 70.16, 63.02) --
	( 70.57, 63.09) --
	( 70.99, 63.13) --
	( 71.40, 63.15) --
	( 71.81, 63.18) --
	( 72.23, 63.21) --
	( 72.64, 63.25) --
	( 73.06, 63.26) --
	( 73.47, 63.32) --
	( 73.88, 63.36) --
	( 74.30, 63.41) --
	( 74.71, 63.41) --
	( 75.13, 63.42) --
	( 75.54, 63.41) --
	( 75.95, 63.46) --
	( 76.37, 63.49) --
	( 76.78, 63.50) --
	( 77.20, 63.50) --
	( 77.61, 63.48) --
	( 78.02, 63.54) --
	( 78.44, 63.54) --
	( 78.85, 63.56) --
	( 79.26, 63.54) --
	( 79.68, 63.55) --
	( 80.09, 63.55) --
	( 80.51, 63.54) --
	( 80.92, 63.53) --
	( 81.33, 63.54) --
	( 81.75, 63.54) --
	( 82.16, 63.54) --
	( 82.58, 63.53) --
	( 82.99, 63.52) --
	( 83.40, 63.52) --
	( 83.82, 63.51) --
	( 84.23, 63.48) --
	( 84.65, 63.47) --
	( 85.06, 63.46) --
	( 85.47, 63.41) --
	( 85.89, 63.37) --
	( 86.30, 63.32) --
	( 86.72, 63.27) --
	( 87.13, 63.24) --
	( 87.54, 63.19) --
	( 87.96, 63.14) --
	( 88.37, 63.10) --
	( 88.79, 63.05) --
	( 89.20, 63.02) --
	( 89.61, 63.00) --
	( 90.03, 62.92) --
	( 90.44, 62.87) --
	( 90.86, 62.81) --
	( 91.27, 62.76) --
	( 91.68, 62.72) --
	( 92.10, 62.68) --
	( 92.51, 62.63) --
	( 92.93, 62.57) --
	( 93.34, 62.54) --
	( 93.75, 62.48) --
	( 94.17, 62.43) --
	( 94.58, 62.33) --
	( 94.99, 62.28) --
	( 95.41, 62.26) --
	( 95.82, 62.21) --
	( 96.24, 62.12) --
	( 96.65, 62.05) --
	( 97.06, 62.01) --
	( 97.48, 61.98) --
	( 97.89, 61.86) --
	( 98.31, 61.72) --
	( 98.72, 61.62) --
	( 99.13, 61.56) --
	( 99.55, 61.49) --
	( 99.96, 61.40) --
	(100.38, 61.34) --
	(100.79, 61.26);

\draw[color=drawColor,line width= 0.8pt,dash pattern=on 4pt off 4pt ,line cap=round,line join=round,fill opacity=0.00,] ( 18.42, 55.76) --
	( 18.83, 55.82) --
	( 19.24, 55.85) --
	( 19.66, 55.90) --
	( 20.07, 55.95) --
	( 20.48, 55.96) --
	( 20.90, 56.06) --
	( 21.31, 56.11) --
	( 21.73, 56.15) --
	( 22.14, 56.22) --
	( 22.55, 56.32) --
	( 22.97, 56.40) --
	( 23.38, 56.45) --
	( 23.80, 56.57) --
	( 24.21, 56.64) --
	( 24.62, 56.73) --
	( 25.04, 56.83) --
	( 25.45, 56.93) --
	( 25.87, 57.05) --
	( 26.28, 57.18) --
	( 26.69, 57.34) --
	( 27.11, 57.49) --
	( 27.52, 57.63) --
	( 27.94, 57.76) --
	( 28.35, 57.91) --
	( 28.76, 58.06) --
	( 29.18, 58.23) --
	( 29.59, 58.38) --
	( 30.01, 58.53) --
	( 30.42, 58.67) --
	( 30.83, 58.84) --
	( 31.25, 59.00) --
	( 31.66, 59.18) --
	( 32.08, 59.35) --
	( 32.49, 59.51) --
	( 32.90, 59.69) --
	( 33.32, 59.88) --
	( 33.73, 60.08) --
	( 34.14, 60.28) --
	( 34.56, 60.50) --
	( 34.97, 60.72) --
	( 35.39, 60.97) --
	( 35.80, 61.20) --
	( 36.21, 61.46) --
	( 36.63, 61.68) --
	( 37.04, 61.91) --
	( 37.46, 62.15) --
	( 37.87, 62.37) --
	( 38.28, 62.59) --
	( 38.70, 62.81) --
	( 39.11, 63.00) --
	( 39.53, 63.19) --
	( 39.94, 63.36) --
	( 40.35, 63.53) --
	( 40.77, 63.68) --
	( 41.18, 63.84) --
	( 41.60, 63.98) --
	( 42.01, 64.14) --
	( 42.42, 64.27) --
	( 42.84, 64.39) --
	( 43.25, 64.50) --
	( 43.67, 64.61) --
	( 44.08, 64.75) --
	( 44.49, 64.86) --
	( 44.91, 64.99) --
	( 45.32, 65.11) --
	( 45.74, 65.20) --
	( 46.15, 65.31) --
	( 46.56, 65.41) --
	( 46.98, 65.51) --
	( 47.39, 65.62) --
	( 47.81, 65.71) --
	( 48.22, 65.79) --
	( 48.63, 65.87) --
	( 49.05, 65.93) --
	( 49.46, 65.97) --
	( 49.87, 66.01) --
	( 50.29, 66.08) --
	( 50.70, 66.12) --
	( 51.12, 66.16) --
	( 51.53, 66.22) --
	( 51.94, 66.26) --
	( 52.36, 66.30) --
	( 52.77, 66.33) --
	( 53.19, 66.39) --
	( 53.60, 66.44) --
	( 54.01, 66.52) --
	( 54.43, 66.58) --
	( 54.84, 66.63) --
	( 55.26, 66.68) --
	( 55.67, 66.76) --
	( 56.08, 66.84) --
	( 56.50, 66.92) --
	( 56.91, 67.02) --
	( 57.33, 67.14) --
	( 57.74, 67.23) --
	( 58.15, 67.35) --
	( 58.57, 67.47) --
	( 58.98, 67.57) --
	( 59.40, 67.68) --
	( 59.81, 67.81) --
	( 60.22, 67.93) --
	( 60.64, 68.04) --
	( 61.05, 68.19) --
	( 61.47, 68.34) --
	( 61.88, 68.49) --
	( 62.29, 68.62) --
	( 62.71, 68.76) --
	( 63.12, 68.91) --
	( 63.53, 69.04) --
	( 63.95, 69.19) --
	( 64.36, 69.33) --
	( 64.78, 69.49) --
	( 65.19, 69.62) --
	( 65.60, 69.79) --
	( 66.02, 69.99) --
	( 66.43, 70.15) --
	( 66.85, 70.31) --
	( 67.26, 70.51) --
	( 67.67, 70.72) --
	( 68.09, 70.94) --
	( 68.50, 71.10) --
	( 68.92, 71.30) --
	( 69.33, 71.50) --
	( 69.74, 71.69) --
	( 70.16, 71.91) --
	( 70.57, 72.12) --
	( 70.99, 72.34) --
	( 71.40, 72.53) --
	( 71.81, 72.77) --
	( 72.23, 72.96) --
	( 72.64, 73.23) --
	( 73.06, 73.49) --
	( 73.47, 73.77) --
	( 73.88, 74.10) --
	( 74.30, 74.37) --
	( 74.71, 74.63) --
	( 75.13, 74.95) --
	( 75.54, 75.23) --
	( 75.95, 75.49) --
	( 76.37, 75.79) --
	( 76.78, 76.14) --
	( 77.20, 76.41) --
	( 77.61, 76.69) --
	( 78.02, 76.98) --
	( 78.44, 77.26) --
	( 78.85, 77.55) --
	( 79.26, 77.83) --
	( 79.68, 78.15) --
	( 80.09, 78.51) --
	( 80.51, 78.84) --
	( 80.92, 79.16) --
	( 81.33, 79.53) --
	( 81.75, 79.85) --
	( 82.16, 80.22) --
	( 82.58, 80.60) --
	( 82.99, 81.03) --
	( 83.40, 81.43) --
	( 83.82, 81.79) --
	( 84.23, 82.18) --
	( 84.65, 82.57) --
	( 85.06, 82.99) --
	( 85.47, 83.36) --
	( 85.89, 83.83) --
	( 86.30, 84.27) --
	( 86.72, 84.69) --
	( 87.13, 85.09) --
	( 87.54, 85.50) --
	( 87.96, 85.91) --
	( 88.37, 86.36) --
	( 88.79, 86.86) --
	( 89.20, 87.36) --
	( 89.61, 87.87) --
	( 90.03, 88.37) --
	( 90.44, 88.84) --
	( 90.86, 89.31) --
	( 91.27, 89.82) --
	( 91.68, 90.34) --
	( 92.10, 90.89) --
	( 92.51, 91.40) --
	( 92.93, 91.71) --
	( 93.34, 92.35) --
	( 93.75, 92.88) --
	( 94.17, 93.36) --
	( 94.58, 93.84) --
	( 94.99, 94.33) --
	( 95.41, 94.86) --
	( 95.82, 95.43) --
	( 96.24, 95.96) --
	( 96.65, 96.54) --
	( 97.06, 97.10) --
	( 97.48, 97.71) --
	( 97.89, 98.18) --
	( 98.31, 98.72) --
	( 98.72, 99.28) --
	( 99.13, 99.79) --
	( 99.55,100.34) --
	( 99.96,100.81) --
	(100.38,101.32) --
	(100.79,101.90);

\draw[color=drawColor,line width= 0.8pt,dash pattern=on 1pt off 3pt ,line cap=round,line join=round,fill opacity=0.00,] ( 18.42, 35.23) --
	( 18.83, 36.02) --
	( 19.24, 36.80) --
	( 19.66, 37.56) --
	( 20.07, 38.32) --
	( 20.48, 39.07) --
	( 20.90, 39.76) --
	( 21.31, 40.37) --
	( 21.73, 40.97) --
	( 22.14, 41.66) --
	( 22.55, 42.23) --
	( 22.97, 42.88) --
	( 23.38, 43.38) --
	( 23.80, 43.91) --
	( 24.21, 44.47) --
	( 24.62, 45.13) --
	( 25.04, 45.69) --
	( 25.45, 46.17) --
	( 25.87, 46.66) --
	( 26.28, 47.15) --
	( 26.69, 47.60) --
	( 27.11, 48.13) --
	( 27.52, 48.62) --
	( 27.94, 49.11) --
	( 28.35, 49.66) --
	( 28.76, 50.17) --
	( 29.18, 50.69) --
	( 29.59, 51.20) --
	( 30.01, 51.72) --
	( 30.42, 52.23) --
	( 30.83, 52.77) --
	( 31.25, 53.21) --
	( 31.66, 53.61) --
	( 32.08, 54.11) --
	( 32.49, 54.63) --
	( 32.90, 55.13) --
	( 33.32, 55.66) --
	( 33.73, 56.12) --
	( 34.14, 56.50) --
	( 34.56, 56.82) --
	( 34.97, 57.11) --
	( 35.39, 57.48) --
	( 35.80, 57.80) --
	( 36.21, 58.09) --
	( 36.63, 58.34) --
	( 37.04, 58.57) --
	( 37.46, 58.79) --
	( 37.87, 58.98) --
	( 38.28, 59.14) --
	( 38.70, 59.34) --
	( 39.11, 59.49) --
	( 39.53, 59.65) --
	( 39.94, 59.85) --
	( 40.35, 60.05) --
	( 40.77, 60.20) --
	( 41.18, 60.37) --
	( 41.60, 60.51) --
	( 42.01, 60.66) --
	( 42.42, 60.83) --
	( 42.84, 60.95) --
	( 43.25, 61.06) --
	( 43.67, 61.18) --
	( 44.08, 61.27) --
	( 44.49, 61.29) --
	( 44.91, 61.32) --
	( 45.32, 61.40) --
	( 45.74, 61.46) --
	( 46.15, 61.50) --
	( 46.56, 61.52) --
	( 46.98, 61.49) --
	( 47.39, 61.53) --
	( 47.81, 61.56) --
	( 48.22, 61.61) --
	( 48.63, 61.64) --
	( 49.05, 61.65) --
	( 49.46, 61.68) --
	( 49.87, 61.70) --
	( 50.29, 61.74) --
	( 50.70, 61.77) --
	( 51.12, 61.77) --
	( 51.53, 61.79) --
	( 51.94, 61.82) --
	( 52.36, 61.82) --
	( 52.77, 61.82) --
	( 53.19, 61.84) --
	( 53.60, 61.84) --
	( 54.01, 61.82) --
	( 54.43, 61.78) --
	( 54.84, 61.75) --
	( 55.26, 61.74) --
	( 55.67, 61.74) --
	( 56.08, 61.72) --
	( 56.50, 61.70) --
	( 56.91, 61.63) --
	( 57.33, 61.61) --
	( 57.74, 61.60) --
	( 58.15, 61.56) --
	( 58.57, 61.53) --
	( 58.98, 61.50) --
	( 59.40, 61.48) --
	( 59.81, 61.45) --
	( 60.22, 61.42) --
	( 60.64, 61.39) --
	( 61.05, 61.41) --
	( 61.47, 61.38) --
	( 61.88, 61.38) --
	( 62.29, 61.40) --
	( 62.71, 61.35) --
	( 63.12, 61.31) --
	( 63.53, 61.29) --
	( 63.95, 61.30) --
	( 64.36, 61.29) --
	( 64.78, 61.26) --
	( 65.19, 61.25) --
	( 65.60, 61.28) --
	( 66.02, 61.26) --
	( 66.43, 61.24) --
	( 66.85, 61.28) --
	( 67.26, 61.28) --
	( 67.67, 61.26) --
	( 68.09, 61.28) --
	( 68.50, 61.27) --
	( 68.92, 61.23) --
	( 69.33, 61.23) --
	( 69.74, 61.26) --
	( 70.16, 61.28) --
	( 70.57, 61.29) --
	( 70.99, 61.31) --
	( 71.40, 61.31) --
	( 71.81, 61.33) --
	( 72.23, 61.33) --
	( 72.64, 61.33) --
	( 73.06, 61.33) --
	( 73.47, 61.32) --
	( 73.88, 61.28) --
	( 74.30, 61.22) --
	( 74.71, 61.27) --
	( 75.13, 61.33) --
	( 75.54, 61.31) --
	( 75.95, 61.25) --
	( 76.37, 61.19) --
	( 76.78, 61.13) --
	( 77.20, 61.07) --
	( 77.61, 61.01) --
	( 78.02, 60.95) --
	( 78.44, 60.85) --
	( 78.85, 60.83) --
	( 79.26, 60.79) --
	( 79.68, 60.70) --
	( 80.09, 60.61) --
	( 80.51, 60.57) --
	( 80.92, 60.54) --
	( 81.33, 60.52) --
	( 81.75, 60.49) --
	( 82.16, 60.43) --
	( 82.58, 60.39) --
	( 82.99, 60.33) --
	( 83.40, 60.23) --
	( 83.82, 60.17) --
	( 84.23, 60.09) --
	( 84.65, 59.92) --
	( 85.06, 59.76) --
	( 85.47, 59.64) --
	( 85.89, 59.52) --
	( 86.30, 59.37) --
	( 86.72, 59.29) --
	( 87.13, 59.17) --
	( 87.54, 59.10) --
	( 87.96, 59.03) --
	( 88.37, 58.87) --
	( 88.79, 58.70) --
	( 89.20, 58.53) --
	( 89.61, 58.37) --
	( 90.03, 58.26) --
	( 90.44, 58.16) --
	( 90.86, 58.05) --
	( 91.27, 57.95) --
	( 91.68, 57.85) --
	( 92.10, 57.67) --
	( 92.51, 57.44) --
	( 92.93, 57.29) --
	( 93.34, 57.14) --
	( 93.75, 56.99) --
	( 94.17, 56.83) --
	( 94.58, 56.65) --
	( 94.99, 56.50) --
	( 95.41, 56.34) --
	( 95.82, 56.19) --
	( 96.24, 56.03) --
	( 96.65, 55.86) --
	( 97.06, 55.51) --
	( 97.48, 55.32) --
	( 97.89, 55.12) --
	( 98.31, 54.93) --
	( 98.72, 54.73) --
	( 99.13, 54.53) --
	( 99.55, 54.31) --
	( 99.96, 54.04) --
	(100.38, 53.78) --
	(100.79, 53.55);

\draw[color=drawColor,line width= 0.8pt,dash pattern=on 1pt off 3pt ,line cap=round,line join=round,fill opacity=0.00,] ( 18.42, 58.94) --
	( 18.83, 58.83) --
	( 19.24, 58.73) --
	( 19.66, 58.72) --
	( 20.07, 58.65) --
	( 20.48, 58.55) --
	( 20.90, 58.58) --
	( 21.31, 58.60) --
	( 21.73, 58.44) --
	( 22.14, 58.34) --
	( 22.55, 58.39) --
	( 22.97, 58.34) --
	( 23.38, 58.32) --
	( 23.80, 58.39) --
	( 24.21, 58.45) --
	( 24.62, 58.47) --
	( 25.04, 58.47) --
	( 25.45, 58.64) --
	( 25.87, 58.76) --
	( 26.28, 58.74) --
	( 26.69, 58.83) --
	( 27.11, 58.94) --
	( 27.52, 58.99) --
	( 27.94, 59.05) --
	( 28.35, 59.17) --
	( 28.76, 59.27) --
	( 29.18, 59.36) --
	( 29.59, 59.40) --
	( 30.01, 59.53) --
	( 30.42, 59.62) --
	( 30.83, 59.72) --
	( 31.25, 59.85) --
	( 31.66, 59.99) --
	( 32.08, 60.12) --
	( 32.49, 60.29) --
	( 32.90, 60.44) --
	( 33.32, 60.63) --
	( 33.73, 60.79) --
	( 34.14, 61.02) --
	( 34.56, 61.17) --
	( 34.97, 61.34) --
	( 35.39, 61.51) --
	( 35.80, 61.71) --
	( 36.21, 61.96) --
	( 36.63, 62.21) --
	( 37.04, 62.42) --
	( 37.46, 62.70) --
	( 37.87, 62.93) --
	( 38.28, 63.18) --
	( 38.70, 63.38) --
	( 39.11, 63.58) --
	( 39.53, 63.75) --
	( 39.94, 63.92) --
	( 40.35, 64.07) --
	( 40.77, 64.24) --
	( 41.18, 64.43) --
	( 41.60, 64.55) --
	( 42.01, 64.69) --
	( 42.42, 64.81) --
	( 42.84, 64.92) --
	( 43.25, 65.01) --
	( 43.67, 65.18) --
	( 44.08, 65.31) --
	( 44.49, 65.48) --
	( 44.91, 65.62) --
	( 45.32, 65.68) --
	( 45.74, 65.83) --
	( 46.15, 65.93) --
	( 46.56, 66.04) --
	( 46.98, 66.14) --
	( 47.39, 66.24) --
	( 47.81, 66.35) --
	( 48.22, 66.43) --
	( 48.63, 66.51) --
	( 49.05, 66.55) --
	( 49.46, 66.60) --
	( 49.87, 66.67) --
	( 50.29, 66.74) --
	( 50.70, 66.73) --
	( 51.12, 66.80) --
	( 51.53, 66.87) --
	( 51.94, 66.92) --
	( 52.36, 66.95) --
	( 52.77, 66.98) --
	( 53.19, 67.02) --
	( 53.60, 67.04) --
	( 54.01, 67.13) --
	( 54.43, 67.25) --
	( 54.84, 67.31) --
	( 55.26, 67.43) --
	( 55.67, 67.56) --
	( 56.08, 67.67) --
	( 56.50, 67.75) --
	( 56.91, 67.85) --
	( 57.33, 67.97) --
	( 57.74, 68.09) --
	( 58.15, 68.24) --
	( 58.57, 68.40) --
	( 58.98, 68.54) --
	( 59.40, 68.70) --
	( 59.81, 68.82) --
	( 60.22, 68.91) --
	( 60.64, 69.04) --
	( 61.05, 69.21) --
	( 61.47, 69.33) --
	( 61.88, 69.54) --
	( 62.29, 69.78) --
	( 62.71, 69.98) --
	( 63.12, 70.20) --
	( 63.53, 70.31) --
	( 63.95, 70.46) --
	( 64.36, 70.66) --
	( 64.78, 70.80) --
	( 65.19, 70.97) --
	( 65.60, 71.19) --
	( 66.02, 71.37) --
	( 66.43, 71.59) --
	( 66.85, 71.86) --
	( 67.26, 72.06) --
	( 67.67, 72.26) --
	( 68.09, 72.46) --
	( 68.50, 72.71) --
	( 68.92, 73.00) --
	( 69.33, 73.29) --
	( 69.74, 73.51) --
	( 70.16, 73.76) --
	( 70.57, 74.18) --
	( 70.99, 74.61) --
	( 71.40, 75.04) --
	( 71.81, 75.48) --
	( 72.23, 75.93) --
	( 72.64, 76.39) --
	( 73.06, 76.86) --
	( 73.47, 77.33) --
	( 73.88, 77.66) --
	( 74.30, 77.96) --
	( 74.71, 78.26) --
	( 75.13, 78.56) --
	( 75.54, 78.87) --
	( 75.95, 79.17) --
	( 76.37, 79.48) --
	( 76.78, 79.78) --
	( 77.20, 80.09) --
	( 77.61, 80.40) --
	( 78.02, 80.86) --
	( 78.44, 81.35) --
	( 78.85, 81.85) --
	( 79.26, 82.36) --
	( 79.68, 82.87) --
	( 80.09, 83.38) --
	( 80.51, 83.91) --
	( 80.92, 84.43) --
	( 81.33, 84.96) --
	( 81.75, 85.45) --
	( 82.16, 85.94) --
	( 82.58, 86.44) --
	( 82.99, 86.94) --
	( 83.40, 87.58) --
	( 83.82, 88.24) --
	( 84.23, 88.86) --
	( 84.65, 89.44) --
	( 85.06, 90.02) --
	( 85.47, 90.61) --
	( 85.89, 91.21) --
	( 86.30, 91.80) --
	( 86.72, 92.41) --
	( 87.13, 93.02) --
	( 87.54, 93.63) --
	( 87.96, 94.25) --
	( 88.37, 94.87) --
	( 88.79, 95.50) --
	( 89.20, 96.13) --
	( 89.61, 96.77) --
	( 90.03, 97.42) --
	( 90.44, 98.06) --
	( 90.86, 98.71) --
	( 91.27, 99.37) --
	( 91.68,100.03) --
	( 92.10,100.70) --
	( 92.51,101.37) --
	( 92.93,102.04) --
	( 93.34,102.72) --
	( 93.75,103.40) --
	( 94.17,104.09) --
	( 94.58,104.78) --
	( 94.99,105.47) --
	( 95.41,106.17) --
	( 95.82,106.87) --
	( 96.24,107.68) --
	( 96.65,108.53) --
	( 97.06,109.39) --
	( 97.48,110.26) --
	( 97.89,111.14) --
	( 98.31,112.02) --
	( 98.72,112.91) --
	( 99.13,113.81) --
	( 99.55,114.71) --
	( 99.96,115.62) --
	(100.38,116.54) --
	(100.79,117.46);
\end{scope}
\end{tikzpicture}

%% file: out-pima-plot-x6.tikz
\begin{tikzpicture}[x=1pt,y=1pt]
\definecolor[named]{drawColor}{rgb}{0.00,0.00,0.00}
\definecolor[named]{fillColor}{rgb}{1.00,1.00,1.00}
\fill[color=fillColor,fill opacity=0.00,] (0,0) rectangle (108.41,130.09);
\begin{scope}
\path[clip] (  0.00,  0.00) rectangle (108.41,130.09);
\definecolor[named]{drawColor}{rgb}{0.00,0.00,0.00}

\draw[color=drawColor,line cap=round,line join=round,fill opacity=0.00,] ( 15.42, 33.12) -- (103.61, 33.12);

\draw[color=drawColor,line cap=round,line join=round,fill opacity=0.00,] ( 15.42, 33.12) -- ( 15.42, 29.52);

\draw[color=drawColor,line cap=round,line join=round,fill opacity=0.00,] ( 33.06, 33.12) -- ( 33.06, 29.52);

\draw[color=drawColor,line cap=round,line join=round,fill opacity=0.00,] ( 50.69, 33.12) -- ( 50.69, 29.52);

\draw[color=drawColor,line cap=round,line join=round,fill opacity=0.00,] ( 68.33, 33.12) -- ( 68.33, 29.52);

\draw[color=drawColor,line cap=round,line join=round,fill opacity=0.00,] ( 85.97, 33.12) -- ( 85.97, 29.52);

\draw[color=drawColor,line cap=round,line join=round,fill opacity=0.00,] (103.61, 33.12) -- (103.61, 29.52);

\node[color=drawColor,anchor=base,inner sep=0pt, outer sep=0pt, scale=  0.60] at ( 15.42, 18.72) {0.0};

\node[color=drawColor,anchor=base,inner sep=0pt, outer sep=0pt, scale=  0.60] at ( 33.06, 18.72) {0.5};

\node[color=drawColor,anchor=base,inner sep=0pt, outer sep=0pt, scale=  0.60] at ( 50.69, 18.72) {1.0};

\node[color=drawColor,anchor=base,inner sep=0pt, outer sep=0pt, scale=  0.60] at ( 68.33, 18.72) {1.5};

\node[color=drawColor,anchor=base,inner sep=0pt, outer sep=0pt, scale=  0.60] at ( 85.97, 18.72) {2.0};

\node[color=drawColor,anchor=base,inner sep=0pt, outer sep=0pt, scale=  0.60] at (103.61, 18.72) {2.5};

\draw[color=drawColor,line cap=round,line join=round,fill opacity=0.00,] ( 15.12, 36.15) -- ( 15.12,111.93);

\draw[color=drawColor,line cap=round,line join=round,fill opacity=0.00,] ( 15.12, 36.15) -- ( 11.52, 36.15);

\draw[color=drawColor,line cap=round,line join=round,fill opacity=0.00,] ( 15.12, 48.78) -- ( 11.52, 48.78);

\draw[color=drawColor,line cap=round,line join=round,fill opacity=0.00,] ( 15.12, 61.41) -- ( 11.52, 61.41);

\draw[color=drawColor,line cap=round,line join=round,fill opacity=0.00,] ( 15.12, 74.04) -- ( 11.52, 74.04);

\draw[color=drawColor,line cap=round,line join=round,fill opacity=0.00,] ( 15.12, 86.67) -- ( 11.52, 86.67);

\draw[color=drawColor,line cap=round,line join=round,fill opacity=0.00,] ( 15.12, 99.30) -- ( 11.52, 99.30);

\draw[color=drawColor,line cap=round,line join=round,fill opacity=0.00,] ( 15.12,111.93) -- ( 11.52,111.93);

\node[color=drawColor,anchor=base east,inner sep=0pt, outer sep=0pt, scale=  0.60] at (  7.92, 34.09) {-2};

\node[color=drawColor,anchor=base east,inner sep=0pt, outer sep=0pt, scale=  0.60] at (  7.92, 46.72) {-1};

\node[color=drawColor,anchor=base east,inner sep=0pt, outer sep=0pt, scale=  0.60] at (  7.92, 59.35) {0};

\node[color=drawColor,anchor=base east,inner sep=0pt, outer sep=0pt, scale=  0.60] at (  7.92, 71.98) {1};

\node[color=drawColor,anchor=base east,inner sep=0pt, outer sep=0pt, scale=  0.60] at (  7.92, 84.61) {2};

\node[color=drawColor,anchor=base east,inner sep=0pt, outer sep=0pt, scale=  0.60] at (  7.92, 97.24) {3};

\node[color=drawColor,anchor=base east,inner sep=0pt, outer sep=0pt, scale=  0.60] at (  7.92,109.87) {4};

\draw[color=drawColor,line cap=round,line join=round,fill opacity=0.00,] ( 15.12, 33.12) --
	(104.08, 33.12) --
	(104.08,114.97) --
	( 15.12,114.97) --
	( 15.12, 33.12);
\end{scope}
\begin{scope}
\path[clip] (  0.00,  0.00) rectangle (108.41,130.09);
\definecolor[named]{drawColor}{rgb}{0.00,0.00,0.00}

\node[color=drawColor,anchor=base,inner sep=0pt, outer sep=0pt, scale=  0.72] at ( 59.60,120.04) {\bfseries $d_{6}= 2$};

\node[color=drawColor,anchor=base,inner sep=0pt, outer sep=0pt, scale=  0.72] at ( 59.60,  4.32) {$x_{6}$};
\end{scope}
\begin{scope}
\path[clip] ( 15.12, 33.12) rectangle (104.08,114.97);
\definecolor[named]{drawColor}{rgb}{0.00,0.00,0.00}

\draw[color=drawColor,line width= 0.8pt,line cap=round,line join=round,fill opacity=0.00,] ( 18.42, 53.12) --
	( 18.83, 53.38) --
	( 19.24, 53.65) --
	( 19.66, 53.92) --
	( 20.07, 54.19) --
	( 20.48, 54.45) --
	( 20.90, 54.72) --
	( 21.31, 54.98) --
	( 21.73, 55.25) --
	( 22.14, 55.52) --
	( 22.55, 55.78) --
	( 22.97, 56.05) --
	( 23.38, 56.31) --
	( 23.80, 56.58) --
	( 24.21, 56.85) --
	( 24.62, 57.11) --
	( 25.04, 57.38) --
	( 25.45, 57.64) --
	( 25.87, 57.90) --
	( 26.28, 58.17) --
	( 26.69, 58.43) --
	( 27.11, 58.69) --
	( 27.52, 58.95) --
	( 27.94, 59.21) --
	( 28.35, 59.46) --
	( 28.76, 59.72) --
	( 29.18, 59.97) --
	( 29.59, 60.22) --
	( 30.01, 60.47) --
	( 30.42, 60.72) --
	( 30.83, 60.97) --
	( 31.25, 61.21) --
	( 31.66, 61.45) --
	( 32.08, 61.70) --
	( 32.49, 61.93) --
	( 32.90, 62.17) --
	( 33.32, 62.40) --
	( 33.73, 62.64) --
	( 34.14, 62.87) --
	( 34.56, 63.09) --
	( 34.97, 63.32) --
	( 35.39, 63.54) --
	( 35.80, 63.76) --
	( 36.21, 63.98) --
	( 36.63, 64.20) --
	( 37.04, 64.41) --
	( 37.46, 64.63) --
	( 37.87, 64.84) --
	( 38.28, 65.04) --
	( 38.70, 65.25) --
	( 39.11, 65.45) --
	( 39.53, 65.65) --
	( 39.94, 65.85) --
	( 40.35, 66.04) --
	( 40.77, 66.23) --
	( 41.18, 66.42) --
	( 41.60, 66.61) --
	( 42.01, 66.80) --
	( 42.42, 66.98) --
	( 42.84, 67.16) --
	( 43.25, 67.33) --
	( 43.67, 67.51) --
	( 44.08, 67.68) --
	( 44.49, 67.85) --
	( 44.91, 68.02) --
	( 45.32, 68.18) --
	( 45.74, 68.34) --
	( 46.15, 68.50) --
	( 46.56, 68.66) --
	( 46.98, 68.81) --
	( 47.39, 68.96) --
	( 47.81, 69.11) --
	( 48.22, 69.26) --
	( 48.63, 69.40) --
	( 49.05, 69.55) --
	( 49.46, 69.68) --
	( 49.87, 69.82) --
	( 50.29, 69.96) --
	( 50.70, 70.09) --
	( 51.12, 70.22) --
	( 51.53, 70.35) --
	( 51.94, 70.47) --
	( 52.36, 70.60) --
	( 52.77, 70.72) --
	( 53.19, 70.84) --
	( 53.60, 70.95) --
	( 54.01, 71.07) --
	( 54.43, 71.18) --
	( 54.84, 71.29) --
	( 55.26, 71.39) --
	( 55.67, 71.50) --
	( 56.08, 71.60) --
	( 56.50, 71.70) --
	( 56.91, 71.80) --
	( 57.33, 71.90) --
	( 57.74, 71.99) --
	( 58.15, 72.09) --
	( 58.57, 72.18) --
	( 58.98, 72.27) --
	( 59.40, 72.35) --
	( 59.81, 72.44) --
	( 60.22, 72.52) --
	( 60.64, 72.60) --
	( 61.05, 72.68) --
	( 61.47, 72.75) --
	( 61.88, 72.83) --
	( 62.29, 72.90) --
	( 62.71, 72.97) --
	( 63.12, 73.04) --
	( 63.53, 73.10) --
	( 63.95, 73.17) --
	( 64.36, 73.23) --
	( 64.78, 73.29) --
	( 65.19, 73.35) --
	( 65.60, 73.41) --
	( 66.02, 73.46) --
	( 66.43, 73.51) --
	( 66.85, 73.57) --
	( 67.26, 73.62) --
	( 67.67, 73.66) --
	( 68.09, 73.71) --
	( 68.50, 73.75) --
	( 68.92, 73.80) --
	( 69.33, 73.84) --
	( 69.74, 73.88) --
	( 70.16, 73.91) --
	( 70.57, 73.95) --
	( 70.99, 73.98) --
	( 71.40, 74.01) --
	( 71.81, 74.05) --
	( 72.23, 74.07) --
	( 72.64, 74.10) --
	( 73.06, 74.13) --
	( 73.47, 74.15) --
	( 73.88, 74.17) --
	( 74.30, 74.19) --
	( 74.71, 74.21) --
	( 75.13, 74.23) --
	( 75.54, 74.25) --
	( 75.95, 74.26) --
	( 76.37, 74.28) --
	( 76.78, 74.29) --
	( 77.20, 74.30) --
	( 77.61, 74.31) --
	( 78.02, 74.31) --
	( 78.44, 74.32) --
	( 78.85, 74.33) --
	( 79.26, 74.33) --
	( 79.68, 74.33) --
	( 80.09, 74.33) --
	( 80.51, 74.33) --
	( 80.92, 74.33) --
	( 81.33, 74.32) --
	( 81.75, 74.32) --
	( 82.16, 74.31) --
	( 82.58, 74.30) --
	( 82.99, 74.30) --
	( 83.40, 74.29) --
	( 83.82, 74.27) --
	( 84.23, 74.26) --
	( 84.65, 74.25) --
	( 85.06, 74.23) --
	( 85.47, 74.22) --
	( 85.89, 74.20) --
	( 86.30, 74.18) --
	( 86.72, 74.16) --
	( 87.13, 74.14) --
	( 87.54, 74.12) --
	( 87.96, 74.09) --
	( 88.37, 74.07) --
	( 88.79, 74.04) --
	( 89.20, 74.02) --
	( 89.61, 73.99) --
	( 90.03, 73.96) --
	( 90.44, 73.93) --
	( 90.86, 73.90) --
	( 91.27, 73.87) --
	( 91.68, 73.84) --
	( 92.10, 73.80) --
	( 92.51, 73.77) --
	( 92.93, 73.73) --
	( 93.34, 73.69) --
	( 93.75, 73.66) --
	( 94.17, 73.62) --
	( 94.58, 73.58) --
	( 94.99, 73.54) --
	( 95.41, 73.50) --
	( 95.82, 73.46) --
	( 96.24, 73.41) --
	( 96.65, 73.37) --
	( 97.06, 73.32) --
	( 97.48, 73.28) --
	( 97.89, 73.23) --
	( 98.31, 73.19) --
	( 98.72, 73.14) --
	( 99.13, 73.09) --
	( 99.55, 73.04) --
	( 99.96, 72.99) --
	(100.38, 72.94) --
	(100.79, 72.89);

\draw[color=drawColor,line width= 0.8pt,dash pattern=on 4pt off 4pt ,line cap=round,line join=round,fill opacity=0.00,] ( 18.42, 47.84) --
	( 18.83, 48.33) --
	( 19.24, 48.84) --
	( 19.66, 49.28) --
	( 20.07, 49.67) --
	( 20.48, 50.19) --
	( 20.90, 50.69) --
	( 21.31, 51.13) --
	( 21.73, 51.60) --
	( 22.14, 52.06) --
	( 22.55, 52.52) --
	( 22.97, 52.98) --
	( 23.38, 53.40) --
	( 23.80, 53.83) --
	( 24.21, 54.28) --
	( 24.62, 54.70) --
	( 25.04, 55.11) --
	( 25.45, 55.49) --
	( 25.87, 55.89) --
	( 26.28, 56.22) --
	( 26.69, 56.56) --
	( 27.11, 56.91) --
	( 27.52, 57.29) --
	( 27.94, 57.60) --
	( 28.35, 57.90) --
	( 28.76, 58.18) --
	( 29.18, 58.45) --
	( 29.59, 58.69) --
	( 30.01, 58.94) --
	( 30.42, 59.16) --
	( 30.83, 59.38) --
	( 31.25, 59.63) --
	( 31.66, 59.85) --
	( 32.08, 60.03) --
	( 32.49, 60.21) --
	( 32.90, 60.37) --
	( 33.32, 60.55) --
	( 33.73, 60.71) --
	( 34.14, 60.86) --
	( 34.56, 61.02) --
	( 34.97, 61.18) --
	( 35.39, 61.37) --
	( 35.80, 61.52) --
	( 36.21, 61.68) --
	( 36.63, 61.83) --
	( 37.04, 61.98) --
	( 37.46, 62.14) --
	( 37.87, 62.29) --
	( 38.28, 62.45) --
	( 38.70, 62.61) --
	( 39.11, 62.74) --
	( 39.53, 62.87) --
	( 39.94, 63.00) --
	( 40.35, 63.14) --
	( 40.77, 63.28) --
	( 41.18, 63.38) --
	( 41.60, 63.49) --
	( 42.01, 63.60) --
	( 42.42, 63.71) --
	( 42.84, 63.80) --
	( 43.25, 63.89) --
	( 43.67, 63.97) --
	( 44.08, 64.04) --
	( 44.49, 64.15) --
	( 44.91, 64.22) --
	( 45.32, 64.32) --
	( 45.74, 64.39) --
	( 46.15, 64.47) --
	( 46.56, 64.55) --
	( 46.98, 64.63) --
	( 47.39, 64.69) --
	( 47.81, 64.74) --
	( 48.22, 64.78) --
	( 48.63, 64.84) --
	( 49.05, 64.91) --
	( 49.46, 64.99) --
	( 49.87, 65.05) --
	( 50.29, 65.08) --
	( 50.70, 65.16) --
	( 51.12, 65.22) --
	( 51.53, 65.28) --
	( 51.94, 65.32) --
	( 52.36, 65.37) --
	( 52.77, 65.41) --
	( 53.19, 65.45) --
	( 53.60, 65.47) --
	( 54.01, 65.50) --
	( 54.43, 65.53) --
	( 54.84, 65.56) --
	( 55.26, 65.61) --
	( 55.67, 65.64) --
	( 56.08, 65.63) --
	( 56.50, 65.72) --
	( 56.91, 65.74) --
	( 57.33, 65.72) --
	( 57.74, 65.72) --
	( 58.15, 65.72) --
	( 58.57, 65.72) --
	( 58.98, 65.75) --
	( 59.40, 65.73) --
	( 59.81, 65.69) --
	( 60.22, 65.66) --
	( 60.64, 65.65) --
	( 61.05, 65.64) --
	( 61.47, 65.65) --
	( 61.88, 65.61) --
	( 62.29, 65.59) --
	( 62.71, 65.55) --
	( 63.12, 65.51) --
	( 63.53, 65.46) --
	( 63.95, 65.39) --
	( 64.36, 65.33) --
	( 64.78, 65.25) --
	( 65.19, 65.21) --
	( 65.60, 65.14) --
	( 66.02, 65.07) --
	( 66.43, 65.00) --
	( 66.85, 64.92) --
	( 67.26, 64.85) --
	( 67.67, 64.72) --
	( 68.09, 64.61) --
	( 68.50, 64.52) --
	( 68.92, 64.39) --
	( 69.33, 64.29) --
	( 69.74, 64.15) --
	( 70.16, 64.01) --
	( 70.57, 63.87) --
	( 70.99, 63.71) --
	( 71.40, 63.60) --
	( 71.81, 63.45) --
	( 72.23, 63.33) --
	( 72.64, 63.15) --
	( 73.06, 63.01) --
	( 73.47, 62.87) --
	( 73.88, 62.75) --
	( 74.30, 62.59) --
	( 74.71, 62.41) --
	( 75.13, 62.24) --
	( 75.54, 62.07) --
	( 75.95, 61.95) --
	( 76.37, 61.79) --
	( 76.78, 61.62) --
	( 77.20, 61.40) --
	( 77.61, 61.18) --
	( 78.02, 60.99) --
	( 78.44, 60.78) --
	( 78.85, 60.61) --
	( 79.26, 60.41) --
	( 79.68, 60.16) --
	( 80.09, 59.91) --
	( 80.51, 59.66) --
	( 80.92, 59.42) --
	( 81.33, 59.20) --
	( 81.75, 58.97) --
	( 82.16, 58.67) --
	( 82.58, 58.40) --
	( 82.99, 58.19) --
	( 83.40, 57.95) --
	( 83.82, 57.68) --
	( 84.23, 57.41) --
	( 84.65, 57.10) --
	( 85.06, 56.84) --
	( 85.47, 56.58) --
	( 85.89, 56.32) --
	( 86.30, 56.08) --
	( 86.72, 55.77) --
	( 87.13, 55.53) --
	( 87.54, 55.27) --
	( 87.96, 54.94) --
	( 88.37, 54.68) --
	( 88.79, 54.38) --
	( 89.20, 54.07) --
	( 89.61, 53.79) --
	( 90.03, 53.51) --
	( 90.44, 53.25) --
	( 90.86, 52.84) --
	( 91.27, 52.54) --
	( 91.68, 52.25) --
	( 92.10, 51.92) --
	( 92.51, 51.65) --
	( 92.93, 51.31) --
	( 93.34, 50.99) --
	( 93.75, 50.69) --
	( 94.17, 50.32) --
	( 94.58, 49.96) --
	( 94.99, 49.64) --
	( 95.41, 49.38) --
	( 95.82, 48.98) --
	( 96.24, 48.66) --
	( 96.65, 48.27) --
	( 97.06, 47.86) --
	( 97.48, 47.50) --
	( 97.89, 47.12) --
	( 98.31, 46.65) --
	( 98.72, 46.28) --
	( 99.13, 45.89) --
	( 99.55, 45.54) --
	( 99.96, 45.17) --
	(100.38, 44.81) --
	(100.79, 44.51);

\draw[color=drawColor,line width= 0.8pt,dash pattern=on 4pt off 4pt ,line cap=round,line join=round,fill opacity=0.00,] ( 18.42, 58.48) --
	( 18.83, 58.54) --
	( 19.24, 58.58) --
	( 19.66, 58.62) --
	( 20.07, 58.64) --
	( 20.48, 58.69) --
	( 20.90, 58.76) --
	( 21.31, 58.80) --
	( 21.73, 58.86) --
	( 22.14, 58.94) --
	( 22.55, 59.01) --
	( 22.97, 59.10) --
	( 23.38, 59.18) --
	( 23.80, 59.27) --
	( 24.21, 59.36) --
	( 24.62, 59.48) --
	( 25.04, 59.59) --
	( 25.45, 59.71) --
	( 25.87, 59.85) --
	( 26.28, 59.99) --
	( 26.69, 60.16) --
	( 27.11, 60.35) --
	( 27.52, 60.55) --
	( 27.94, 60.77) --
	( 28.35, 61.00) --
	( 28.76, 61.24) --
	( 29.18, 61.48) --
	( 29.59, 61.73) --
	( 30.01, 61.99) --
	( 30.42, 62.25) --
	( 30.83, 62.54) --
	( 31.25, 62.79) --
	( 31.66, 63.09) --
	( 32.08, 63.38) --
	( 32.49, 63.67) --
	( 32.90, 63.98) --
	( 33.32, 64.29) --
	( 33.73, 64.59) --
	( 34.14, 64.88) --
	( 34.56, 65.16) --
	( 34.97, 65.45) --
	( 35.39, 65.74) --
	( 35.80, 66.00) --
	( 36.21, 66.26) --
	( 36.63, 66.53) --
	( 37.04, 66.81) --
	( 37.46, 67.06) --
	( 37.87, 67.33) --
	( 38.28, 67.59) --
	( 38.70, 67.84) --
	( 39.11, 68.10) --
	( 39.53, 68.37) --
	( 39.94, 68.64) --
	( 40.35, 68.90) --
	( 40.77, 69.18) --
	( 41.18, 69.42) --
	( 41.60, 69.66) --
	( 42.01, 69.92) --
	( 42.42, 70.19) --
	( 42.84, 70.45) --
	( 43.25, 70.69) --
	( 43.67, 70.95) --
	( 44.08, 71.19) --
	( 44.49, 71.47) --
	( 44.91, 71.73) --
	( 45.32, 71.99) --
	( 45.74, 72.24) --
	( 46.15, 72.46) --
	( 46.56, 72.67) --
	( 46.98, 72.96) --
	( 47.39, 73.25) --
	( 47.81, 73.50) --
	( 48.22, 73.72) --
	( 48.63, 73.93) --
	( 49.05, 74.21) --
	( 49.46, 74.42) --
	( 49.87, 74.62) --
	( 50.29, 74.83) --
	( 50.70, 75.06) --
	( 51.12, 75.26) --
	( 51.53, 75.46) --
	( 51.94, 75.65) --
	( 52.36, 75.83) --
	( 52.77, 76.03) --
	( 53.19, 76.28) --
	( 53.60, 76.51) --
	( 54.01, 76.72) --
	( 54.43, 76.88) --
	( 54.84, 77.05) --
	( 55.26, 77.26) --
	( 55.67, 77.50) --
	( 56.08, 77.75) --
	( 56.50, 78.00) --
	( 56.91, 78.24) --
	( 57.33, 78.48) --
	( 57.74, 78.69) --
	( 58.15, 78.91) --
	( 58.57, 79.10) --
	( 58.98, 79.30) --
	( 59.40, 79.51) --
	( 59.81, 79.74) --
	( 60.22, 79.90) --
	( 60.64, 80.04) --
	( 61.05, 80.23) --
	( 61.47, 80.46) --
	( 61.88, 80.65) --
	( 62.29, 80.80) --
	( 62.71, 80.97) --
	( 63.12, 81.10) --
	( 63.53, 81.23) --
	( 63.95, 81.35) --
	( 64.36, 81.50) --
	( 64.78, 81.62) --
	( 65.19, 81.82) --
	( 65.60, 81.98) --
	( 66.02, 82.24) --
	( 66.43, 82.40) --
	( 66.85, 82.64) --
	( 67.26, 82.81) --
	( 67.67, 83.01) --
	( 68.09, 83.22) --
	( 68.50, 83.36) --
	( 68.92, 83.66) --
	( 69.33, 83.86) --
	( 69.74, 84.04) --
	( 70.16, 84.23) --
	( 70.57, 84.45) --
	( 70.99, 84.72) --
	( 71.40, 84.99) --
	( 71.81, 85.30) --
	( 72.23, 85.49) --
	( 72.64, 85.67) --
	( 73.06, 85.92) --
	( 73.47, 86.13) --
	( 73.88, 86.37) --
	( 74.30, 86.54) --
	( 74.71, 86.70) --
	( 75.13, 86.90) --
	( 75.54, 87.08) --
	( 75.95, 87.34) --
	( 76.37, 87.66) --
	( 76.78, 87.84) --
	( 77.20, 88.10) --
	( 77.61, 88.38) --
	( 78.02, 88.66) --
	( 78.44, 88.94) --
	( 78.85, 89.24) --
	( 79.26, 89.50) --
	( 79.68, 89.77) --
	( 80.09, 90.01) --
	( 80.51, 90.22) --
	( 80.92, 90.45) --
	( 81.33, 90.69) --
	( 81.75, 90.90) --
	( 82.16, 91.18) --
	( 82.58, 91.41) --
	( 82.99, 91.68) --
	( 83.40, 92.02) --
	( 83.82, 92.32) --
	( 84.23, 92.51) --
	( 84.65, 92.77) --
	( 85.06, 93.06) --
	( 85.47, 93.37) --
	( 85.89, 93.61) --
	( 86.30, 93.79) --
	( 86.72, 93.96) --
	( 87.13, 94.13) --
	( 87.54, 94.31) --
	( 87.96, 94.61) --
	( 88.37, 94.92) --
	( 88.79, 95.16) --
	( 89.20, 95.42) --
	( 89.61, 95.70) --
	( 90.03, 95.97) --
	( 90.44, 96.18) --
	( 90.86, 96.55) --
	( 91.27, 96.89) --
	( 91.68, 97.16) --
	( 92.10, 97.44) --
	( 92.51, 97.73) --
	( 92.93, 98.03) --
	( 93.34, 98.32) --
	( 93.75, 98.66) --
	( 94.17, 98.99) --
	( 94.58, 99.31) --
	( 94.99, 99.51) --
	( 95.41, 99.74) --
	( 95.82,100.08) --
	( 96.24,100.41) --
	( 96.65,100.71) --
	( 97.06,100.95) --
	( 97.48,101.23) --
	( 97.89,101.53) --
	( 98.31,101.86) --
	( 98.72,102.15) --
	( 99.13,102.49) --
	( 99.55,102.79) --
	( 99.96,103.15) --
	(100.38,103.48) --
	(100.79,103.84);

\draw[color=drawColor,line width= 0.8pt,dash pattern=on 1pt off 3pt ,line cap=round,line join=round,fill opacity=0.00,] ( 18.42, 45.79) --
	( 18.83, 46.34) --
	( 19.24, 46.92) --
	( 19.66, 47.49) --
	( 20.07, 48.06) --
	( 20.48, 48.60) --
	( 20.90, 49.15) --
	( 21.31, 49.72) --
	( 21.73, 50.23) --
	( 22.14, 50.82) --
	( 22.55, 51.38) --
	( 22.97, 51.88) --
	( 23.38, 52.43) --
	( 23.80, 52.96) --
	( 24.21, 53.46) --
	( 24.62, 53.95) --
	( 25.04, 54.41) --
	( 25.45, 54.86) --
	( 25.87, 55.34) --
	( 26.28, 55.71) --
	( 26.69, 56.10) --
	( 27.11, 56.51) --
	( 27.52, 56.81) --
	( 27.94, 57.09) --
	( 28.35, 57.39) --
	( 28.76, 57.68) --
	( 29.18, 57.95) --
	( 29.59, 58.21) --
	( 30.01, 58.49) --
	( 30.42, 58.72) --
	( 30.83, 58.92) --
	( 31.25, 59.07) --
	( 31.66, 59.21) --
	( 32.08, 59.39) --
	( 32.49, 59.57) --
	( 32.90, 59.73) --
	( 33.32, 59.87) --
	( 33.73, 60.04) --
	( 34.14, 60.16) --
	( 34.56, 60.31) --
	( 34.97, 60.40) --
	( 35.39, 60.58) --
	( 35.80, 60.72) --
	( 36.21, 60.85) --
	( 36.63, 61.01) --
	( 37.04, 61.16) --
	( 37.46, 61.32) --
	( 37.87, 61.44) --
	( 38.28, 61.55) --
	( 38.70, 61.56) --
	( 39.11, 61.65) --
	( 39.53, 61.74) --
	( 39.94, 61.89) --
	( 40.35, 61.96) --
	( 40.77, 62.02) --
	( 41.18, 62.10) --
	( 41.60, 62.18) --
	( 42.01, 62.26) --
	( 42.42, 62.34) --
	( 42.84, 62.39) --
	( 43.25, 62.47) --
	( 43.67, 62.57) --
	( 44.08, 62.63) --
	( 44.49, 62.73) --
	( 44.91, 62.76) --
	( 45.32, 62.75) --
	( 45.74, 62.80) --
	( 46.15, 62.80) --
	( 46.56, 62.85) --
	( 46.98, 62.91) --
	( 47.39, 63.00) --
	( 47.81, 63.08) --
	( 48.22, 63.10) --
	( 48.63, 63.13) --
	( 49.05, 63.14) --
	( 49.46, 63.16) --
	( 49.87, 63.17) --
	( 50.29, 63.28) --
	( 50.70, 63.42) --
	( 51.12, 63.52) --
	( 51.53, 63.46) --
	( 51.94, 63.49) --
	( 52.36, 63.57) --
	( 52.77, 63.56) --
	( 53.19, 63.50) --
	( 53.60, 63.43) --
	( 54.01, 63.43) --
	( 54.43, 63.45) --
	( 54.84, 63.48) --
	( 55.26, 63.45) --
	( 55.67, 63.45) --
	( 56.08, 63.50) --
	( 56.50, 63.48) --
	( 56.91, 63.47) --
	( 57.33, 63.43) --
	( 57.74, 63.42) --
	( 58.15, 63.43) --
	( 58.57, 63.40) --
	( 58.98, 63.35) --
	( 59.40, 63.29) --
	( 59.81, 63.24) --
	( 60.22, 63.16) --
	( 60.64, 63.08) --
	( 61.05, 63.00) --
	( 61.47, 62.98) --
	( 61.88, 62.93) --
	( 62.29, 62.87) --
	( 62.71, 62.75) --
	( 63.12, 62.63) --
	( 63.53, 62.60) --
	( 63.95, 62.49) --
	( 64.36, 62.46) --
	( 64.78, 62.32) --
	( 65.19, 62.22) --
	( 65.60, 62.17) --
	( 66.02, 62.08) --
	( 66.43, 62.02) --
	( 66.85, 61.86) --
	( 67.26, 61.71) --
	( 67.67, 61.50) --
	( 68.09, 61.36) --
	( 68.50, 61.18) --
	( 68.92, 61.07) --
	( 69.33, 60.94) --
	( 69.74, 60.68) --
	( 70.16, 60.45) --
	( 70.57, 60.39) --
	( 70.99, 60.29) --
	( 71.40, 60.10) --
	( 71.81, 59.89) --
	( 72.23, 59.79) --
	( 72.64, 59.63) --
	( 73.06, 59.38) --
	( 73.47, 59.13) --
	( 73.88, 58.87) --
	( 74.30, 58.61) --
	( 74.71, 58.35) --
	( 75.13, 58.13) --
	( 75.54, 57.82) --
	( 75.95, 57.55) --
	( 76.37, 57.38) --
	( 76.78, 57.19) --
	( 77.20, 56.91) --
	( 77.61, 56.62) --
	( 78.02, 56.44) --
	( 78.44, 56.26) --
	( 78.85, 56.08) --
	( 79.26, 55.80) --
	( 79.68, 55.43) --
	( 80.09, 55.06) --
	( 80.51, 54.83) --
	( 80.92, 54.49) --
	( 81.33, 54.11) --
	( 81.75, 53.68) --
	( 82.16, 53.37) --
	( 82.58, 52.99) --
	( 82.99, 52.65) --
	( 83.40, 52.30) --
	( 83.82, 52.02) --
	( 84.23, 51.59) --
	( 84.65, 51.23) --
	( 85.06, 50.90) --
	( 85.47, 50.56) --
	( 85.89, 50.23) --
	( 86.30, 49.89) --
	( 86.72, 49.55) --
	( 87.13, 49.20) --
	( 87.54, 48.86) --
	( 87.96, 48.51) --
	( 88.37, 48.16) --
	( 88.79, 47.80) --
	( 89.20, 47.42) --
	( 89.61, 47.05) --
	( 90.03, 46.61) --
	( 90.44, 46.19) --
	( 90.86, 45.73) --
	( 91.27, 45.26) --
	( 91.68, 44.98) --
	( 92.10, 44.60) --
	( 92.51, 44.16) --
	( 92.93, 43.72) --
	( 93.34, 43.25) --
	( 93.75, 42.77) --
	( 94.17, 42.36) --
	( 94.58, 41.90) --
	( 94.99, 41.43) --
	( 95.41, 40.95) --
	( 95.82, 40.47) --
	( 96.24, 39.99) --
	( 96.65, 39.50) --
	( 97.06, 39.11) --
	( 97.48, 38.71) --
	( 97.89, 38.31) --
	( 98.31, 37.92) --
	( 98.72, 37.52) --
	( 99.13, 37.13) --
	( 99.55, 36.70) --
	( 99.96, 36.16) --
	(100.38, 35.67) --
	(100.79, 35.33);

\draw[color=drawColor,line width= 0.8pt,dash pattern=on 1pt off 3pt ,line cap=round,line join=round,fill opacity=0.00,] ( 18.42, 60.16) --
	( 18.83, 60.16) --
	( 19.24, 60.13) --
	( 19.66, 60.14) --
	( 20.07, 60.12) --
	( 20.48, 60.10) --
	( 20.90, 60.10) --
	( 21.31, 60.12) --
	( 21.73, 60.15) --
	( 22.14, 60.13) --
	( 22.55, 60.19) --
	( 22.97, 60.09) --
	( 23.38, 60.09) --
	( 23.80, 60.12) --
	( 24.21, 60.21) --
	( 24.62, 60.26) --
	( 25.04, 60.37) --
	( 25.45, 60.50) --
	( 25.87, 60.58) --
	( 26.28, 60.70) --
	( 26.69, 60.81) --
	( 27.11, 60.98) --
	( 27.52, 61.19) --
	( 27.94, 61.38) --
	( 28.35, 61.56) --
	( 28.76, 61.79) --
	( 29.18, 62.02) --
	( 29.59, 62.29) --
	( 30.01, 62.52) --
	( 30.42, 62.81) --
	( 30.83, 63.07) --
	( 31.25, 63.39) --
	( 31.66, 63.70) --
	( 32.08, 63.97) --
	( 32.49, 64.27) --
	( 32.90, 64.59) --
	( 33.32, 64.86) --
	( 33.73, 65.12) --
	( 34.14, 65.43) --
	( 34.56, 65.77) --
	( 34.97, 66.06) --
	( 35.39, 66.34) --
	( 35.80, 66.64) --
	( 36.21, 66.94) --
	( 36.63, 67.24) --
	( 37.04, 67.53) --
	( 37.46, 67.80) --
	( 37.87, 68.08) --
	( 38.28, 68.42) --
	( 38.70, 68.73) --
	( 39.11, 69.02) --
	( 39.53, 69.31) --
	( 39.94, 69.62) --
	( 40.35, 69.86) --
	( 40.77, 70.19) --
	( 41.18, 70.49) --
	( 41.60, 70.76) --
	( 42.01, 71.04) --
	( 42.42, 71.26) --
	( 42.84, 71.55) --
	( 43.25, 71.92) --
	( 43.67, 72.23) --
	( 44.08, 72.52) --
	( 44.49, 72.79) --
	( 44.91, 73.09) --
	( 45.32, 73.38) --
	( 45.74, 73.66) --
	( 46.15, 73.93) --
	( 46.56, 74.23) --
	( 46.98, 74.49) --
	( 47.39, 74.71) --
	( 47.81, 74.92) --
	( 48.22, 75.16) --
	( 48.63, 75.41) --
	( 49.05, 75.69) --
	( 49.46, 75.97) --
	( 49.87, 76.22) --
	( 50.29, 76.47) --
	( 50.70, 76.69) --
	( 51.12, 76.91) --
	( 51.53, 77.12) --
	( 51.94, 77.33) --
	( 52.36, 77.55) --
	( 52.77, 77.77) --
	( 53.19, 77.95) --
	( 53.60, 78.16) --
	( 54.01, 78.35) --
	( 54.43, 78.52) --
	( 54.84, 78.78) --
	( 55.26, 79.01) --
	( 55.67, 79.22) --
	( 56.08, 79.52) --
	( 56.50, 79.77) --
	( 56.91, 80.01) --
	( 57.33, 80.26) --
	( 57.74, 80.49) --
	( 58.15, 80.73) --
	( 58.57, 80.96) --
	( 58.98, 81.11) --
	( 59.40, 81.28) --
	( 59.81, 81.52) --
	( 60.22, 81.75) --
	( 60.64, 82.00) --
	( 61.05, 82.25) --
	( 61.47, 82.49) --
	( 61.88, 82.64) --
	( 62.29, 82.90) --
	( 62.71, 83.12) --
	( 63.12, 83.37) --
	( 63.53, 83.61) --
	( 63.95, 83.85) --
	( 64.36, 84.10) --
	( 64.78, 84.32) --
	( 65.19, 84.58) --
	( 65.60, 84.90) --
	( 66.02, 85.22) --
	( 66.43, 85.46) --
	( 66.85, 85.65) --
	( 67.26, 85.81) --
	( 67.67, 85.94) --
	( 68.09, 86.16) --
	( 68.50, 86.38) --
	( 68.92, 86.60) --
	( 69.33, 86.82) --
	( 69.74, 87.04) --
	( 70.16, 87.36) --
	( 70.57, 87.67) --
	( 70.99, 88.10) --
	( 71.40, 88.43) --
	( 71.81, 88.78) --
	( 72.23, 89.14) --
	( 72.64, 89.50) --
	( 73.06, 89.86) --
	( 73.47, 90.17) --
	( 73.88, 90.45) --
	( 74.30, 90.72) --
	( 74.71, 90.92) --
	( 75.13, 91.28) --
	( 75.54, 91.58) --
	( 75.95, 91.92) --
	( 76.37, 92.28) --
	( 76.78, 92.64) --
	( 77.20, 93.01) --
	( 77.61, 93.38) --
	( 78.02, 93.75) --
	( 78.44, 94.03) --
	( 78.85, 94.31) --
	( 79.26, 94.59) --
	( 79.68, 94.86) --
	( 80.09, 95.24) --
	( 80.51, 95.62) --
	( 80.92, 96.00) --
	( 81.33, 96.38) --
	( 81.75, 96.67) --
	( 82.16, 96.94) --
	( 82.58, 97.27) --
	( 82.99, 97.67) --
	( 83.40, 98.07) --
	( 83.82, 98.47) --
	( 84.23, 98.91) --
	( 84.65, 99.28) --
	( 85.06, 99.69) --
	( 85.47,100.10) --
	( 85.89,100.37) --
	( 86.30,100.66) --
	( 86.72,100.98) --
	( 87.13,101.31) --
	( 87.54,101.88) --
	( 87.96,102.42) --
	( 88.37,102.80) --
	( 88.79,103.10) --
	( 89.20,103.37) --
	( 89.61,103.70) --
	( 90.03,104.15) --
	( 90.44,104.58) --
	( 90.86,104.89) --
	( 91.27,105.20) --
	( 91.68,105.51) --
	( 92.10,105.82) --
	( 92.51,106.13) --
	( 92.93,106.44) --
	( 93.34,106.81) --
	( 93.75,107.09) --
	( 94.17,107.41) --
	( 94.58,107.83) --
	( 94.99,108.27) --
	( 95.41,108.70) --
	( 95.82,109.14) --
	( 96.24,109.48) --
	( 96.65,109.77) --
	( 97.06,110.07) --
	( 97.48,110.56) --
	( 97.89,111.05) --
	( 98.31,111.50) --
	( 98.72,111.91) --
	( 99.13,112.26) --
	( 99.55,112.61) --
	( 99.96,112.95) --
	(100.38,113.37) --
	(100.79,113.85);
\end{scope}
\end{tikzpicture}

%% file: out-pima-plot-x7.tikz
\begin{tikzpicture}[x=1pt,y=1pt]
\definecolor[named]{drawColor}{rgb}{0.00,0.00,0.00}
\definecolor[named]{fillColor}{rgb}{1.00,1.00,1.00}
\fill[color=fillColor,fill opacity=0.00,] (0,0) rectangle (108.41,130.09);
\begin{scope}
\path[clip] (  0.00,  0.00) rectangle (108.41,130.09);
\definecolor[named]{drawColor}{rgb}{0.47,0.64,0.37}
\definecolor[named]{fillColor}{rgb}{0.56,0.93,0.17}
\definecolor[named]{drawColor}{rgb}{0.00,0.00,0.00}

\draw[color=drawColor,line cap=round,line join=round,fill opacity=0.00,] ( 17.04, 33.12) -- ( 99.42, 33.12);

\draw[color=drawColor,line cap=round,line join=round,fill opacity=0.00,] ( 17.04, 33.12) -- ( 17.04, 29.52);

\draw[color=drawColor,line cap=round,line join=round,fill opacity=0.00,] ( 30.77, 33.12) -- ( 30.77, 29.52);

\draw[color=drawColor,line cap=round,line join=round,fill opacity=0.00,] ( 44.50, 33.12) -- ( 44.50, 29.52);

\draw[color=drawColor,line cap=round,line join=round,fill opacity=0.00,] ( 58.23, 33.12) -- ( 58.23, 29.52);

\draw[color=drawColor,line cap=round,line join=round,fill opacity=0.00,] ( 71.96, 33.12) -- ( 71.96, 29.52);

\draw[color=drawColor,line cap=round,line join=round,fill opacity=0.00,] ( 85.69, 33.12) -- ( 85.69, 29.52);

\draw[color=drawColor,line cap=round,line join=round,fill opacity=0.00,] ( 99.42, 33.12) -- ( 99.42, 29.52);

\node[color=drawColor,anchor=base,inner sep=0pt, outer sep=0pt, scale=  0.60] at ( 17.04, 18.72) {20};

\node[color=drawColor,anchor=base,inner sep=0pt, outer sep=0pt, scale=  0.60] at ( 30.77, 18.72) {30};

\node[color=drawColor,anchor=base,inner sep=0pt, outer sep=0pt, scale=  0.60] at ( 44.50, 18.72) {40};

\node[color=drawColor,anchor=base,inner sep=0pt, outer sep=0pt, scale=  0.60] at ( 58.23, 18.72) {50};

\node[color=drawColor,anchor=base,inner sep=0pt, outer sep=0pt, scale=  0.60] at ( 71.96, 18.72) {60};

\node[color=drawColor,anchor=base,inner sep=0pt, outer sep=0pt, scale=  0.60] at ( 85.69, 18.72) {70};

\node[color=drawColor,anchor=base,inner sep=0pt, outer sep=0pt, scale=  0.60] at ( 99.42, 18.72) {80};

\draw[color=drawColor,line cap=round,line join=round,fill opacity=0.00,] ( 15.12, 36.15) -- ( 15.12,111.93);

\draw[color=drawColor,line cap=round,line join=round,fill opacity=0.00,] ( 15.12, 36.15) -- ( 11.52, 36.15);

\draw[color=drawColor,line cap=round,line join=round,fill opacity=0.00,] ( 15.12, 48.78) -- ( 11.52, 48.78);

\draw[color=drawColor,line cap=round,line join=round,fill opacity=0.00,] ( 15.12, 61.41) -- ( 11.52, 61.41);

\draw[color=drawColor,line cap=round,line join=round,fill opacity=0.00,] ( 15.12, 74.04) -- ( 11.52, 74.04);

\draw[color=drawColor,line cap=round,line join=round,fill opacity=0.00,] ( 15.12, 86.67) -- ( 11.52, 86.67);

\draw[color=drawColor,line cap=round,line join=round,fill opacity=0.00,] ( 15.12, 99.30) -- ( 11.52, 99.30);

\draw[color=drawColor,line cap=round,line join=round,fill opacity=0.00,] ( 15.12,111.93) -- ( 11.52,111.93);

\node[color=drawColor,anchor=base east,inner sep=0pt, outer sep=0pt, scale=  0.60] at (  7.92, 34.09) {-8};

\node[color=drawColor,anchor=base east,inner sep=0pt, outer sep=0pt, scale=  0.60] at (  7.92, 46.72) {-6};

\node[color=drawColor,anchor=base east,inner sep=0pt, outer sep=0pt, scale=  0.60] at (  7.92, 59.35) {-4};

\node[color=drawColor,anchor=base east,inner sep=0pt, outer sep=0pt, scale=  0.60] at (  7.92, 71.98) {-2};

\node[color=drawColor,anchor=base east,inner sep=0pt, outer sep=0pt, scale=  0.60] at (  7.92, 84.61) {0};

\node[color=drawColor,anchor=base east,inner sep=0pt, outer sep=0pt, scale=  0.60] at (  7.92, 97.24) {2};

\node[color=drawColor,anchor=base east,inner sep=0pt, outer sep=0pt, scale=  0.60] at (  7.92,109.87) {4};

\draw[color=drawColor,line cap=round,line join=round,fill opacity=0.00,] ( 15.12, 33.12) --
	(104.08, 33.12) --
	(104.08,114.97) --
	( 15.12,114.97) --
	( 15.12, 33.12);
\end{scope}
\begin{scope}
\path[clip] (  0.00,  0.00) rectangle (108.41,130.09);
\definecolor[named]{drawColor}{rgb}{0.47,0.64,0.37}
\definecolor[named]{fillColor}{rgb}{0.56,0.93,0.17}
\definecolor[named]{drawColor}{rgb}{0.00,0.00,0.00}

\node[color=drawColor,anchor=base,inner sep=0pt, outer sep=0pt, scale=  0.72] at ( 59.60,120.04) {\bfseries $d_{7}= 4$};

\node[color=drawColor,anchor=base,inner sep=0pt, outer sep=0pt, scale=  0.72] at ( 59.60,  4.32) {$x_{7}$};
\end{scope}
\begin{scope}
\path[clip] ( 15.12, 33.12) rectangle (104.08,114.97);
\definecolor[named]{drawColor}{rgb}{0.47,0.64,0.37}
\definecolor[named]{fillColor}{rgb}{0.56,0.93,0.17}
\definecolor[named]{drawColor}{rgb}{0.00,0.00,0.00}

\draw[color=drawColor,line width= 0.8pt,line cap=round,line join=round,fill opacity=0.00,] ( 18.42, 79.52) --
	( 18.83, 79.93) --
	( 19.24, 80.33) --
	( 19.66, 80.73) --
	( 20.07, 81.12) --
	( 20.48, 81.50) --
	( 20.90, 81.88) --
	( 21.31, 82.24) --
	( 21.73, 82.60) --
	( 22.14, 82.96) --
	( 22.55, 83.30) --
	( 22.97, 83.64) --
	( 23.38, 83.97) --
	( 23.80, 84.30) --
	( 24.21, 84.61) --
	( 24.62, 84.92) --
	( 25.04, 85.22) --
	( 25.45, 85.51) --
	( 25.87, 85.80) --
	( 26.28, 86.07) --
	( 26.69, 86.34) --
	( 27.11, 86.60) --
	( 27.52, 86.85) --
	( 27.94, 87.09) --
	( 28.35, 87.32) --
	( 28.76, 87.54) --
	( 29.18, 87.75) --
	( 29.59, 87.96) --
	( 30.01, 88.15) --
	( 30.42, 88.34) --
	( 30.83, 88.52) --
	( 31.25, 88.68) --
	( 31.66, 88.84) --
	( 32.08, 88.99) --
	( 32.49, 89.13) --
	( 32.90, 89.26) --
	( 33.32, 89.39) --
	( 33.73, 89.51) --
	( 34.14, 89.63) --
	( 34.56, 89.74) --
	( 34.97, 89.84) --
	( 35.39, 89.94) --
	( 35.80, 90.04) --
	( 36.21, 90.14) --
	( 36.63, 90.23) --
	( 37.04, 90.32) --
	( 37.46, 90.41) --
	( 37.87, 90.49) --
	( 38.28, 90.58) --
	( 38.70, 90.67) --
	( 39.11, 90.76) --
	( 39.53, 90.85) --
	( 39.94, 90.94) --
	( 40.35, 91.03) --
	( 40.77, 91.13) --
	( 41.18, 91.23) --
	( 41.60, 91.33) --
	( 42.01, 91.44) --
	( 42.42, 91.56) --
	( 42.84, 91.68) --
	( 43.25, 91.80) --
	( 43.67, 91.93) --
	( 44.08, 92.07) --
	( 44.49, 92.21) --
	( 44.91, 92.36) --
	( 45.32, 92.51) --
	( 45.74, 92.66) --
	( 46.15, 92.82) --
	( 46.56, 92.97) --
	( 46.98, 93.13) --
	( 47.39, 93.29) --
	( 47.81, 93.44) --
	( 48.22, 93.59) --
	( 48.63, 93.74) --
	( 49.05, 93.89) --
	( 49.46, 94.04) --
	( 49.87, 94.17) --
	( 50.29, 94.31) --
	( 50.70, 94.43) --
	( 51.12, 94.56) --
	( 51.53, 94.67) --
	( 51.94, 94.77) --
	( 52.36, 94.87) --
	( 52.77, 94.95) --
	( 53.19, 95.03) --
	( 53.60, 95.09) --
	( 54.01, 95.14) --
	( 54.43, 95.18) --
	( 54.84, 95.21) --
	( 55.26, 95.22) --
	( 55.67, 95.22) --
	( 56.08, 95.20) --
	( 56.50, 95.16) --
	( 56.91, 95.12) --
	( 57.33, 95.06) --
	( 57.74, 94.98) --
	( 58.15, 94.89) --
	( 58.57, 94.80) --
	( 58.98, 94.68) --
	( 59.40, 94.56) --
	( 59.81, 94.43) --
	( 60.22, 94.29) --
	( 60.64, 94.14) --
	( 61.05, 93.97) --
	( 61.47, 93.80) --
	( 61.88, 93.62) --
	( 62.29, 93.44) --
	( 62.71, 93.25) --
	( 63.12, 93.05) --
	( 63.53, 92.84) --
	( 63.95, 92.63) --
	( 64.36, 92.41) --
	( 64.78, 92.19) --
	( 65.19, 91.96) --
	( 65.60, 91.73) --
	( 66.02, 91.50) --
	( 66.43, 91.26) --
	( 66.85, 91.03) --
	( 67.26, 90.79) --
	( 67.67, 90.55) --
	( 68.09, 90.30) --
	( 68.50, 90.06) --
	( 68.92, 89.82) --
	( 69.33, 89.58) --
	( 69.74, 89.33) --
	( 70.16, 89.09) --
	( 70.57, 88.84) --
	( 70.99, 88.60) --
	( 71.40, 88.35) --
	( 71.81, 88.11) --
	( 72.23, 87.86) --
	( 72.64, 87.61) --
	( 73.06, 87.36) --
	( 73.47, 87.12) --
	( 73.88, 86.87) --
	( 74.30, 86.62) --
	( 74.71, 86.37) --
	( 75.13, 86.12) --
	( 75.54, 85.87) --
	( 75.95, 85.62) --
	( 76.37, 85.36) --
	( 76.78, 85.11) --
	( 77.20, 84.86) --
	( 77.61, 84.61) --
	( 78.02, 84.35) --
	( 78.44, 84.10) --
	( 78.85, 83.84) --
	( 79.26, 83.59) --
	( 79.68, 83.33) --
	( 80.09, 83.08) --
	( 80.51, 82.82) --
	( 80.92, 82.56) --
	( 81.33, 82.31) --
	( 81.75, 82.05) --
	( 82.16, 81.79) --
	( 82.58, 81.53) --
	( 82.99, 81.27) --
	( 83.40, 81.01) --
	( 83.82, 80.76) --
	( 84.23, 80.50) --
	( 84.65, 80.23) --
	( 85.06, 79.97) --
	( 85.47, 79.71) --
	( 85.89, 79.45) --
	( 86.30, 79.19) --
	( 86.72, 78.93) --
	( 87.13, 78.66) --
	( 87.54, 78.40) --
	( 87.96, 78.14) --
	( 88.37, 77.87) --
	( 88.79, 77.61) --
	( 89.20, 77.34) --
	( 89.61, 77.08) --
	( 90.03, 76.81) --
	( 90.44, 76.55) --
	( 90.86, 76.28) --
	( 91.27, 76.01) --
	( 91.68, 75.75) --
	( 92.10, 75.48) --
	( 92.51, 75.21) --
	( 92.93, 74.94) --
	( 93.34, 74.67) --
	( 93.75, 74.40) --
	( 94.17, 74.14) --
	( 94.58, 73.87) --
	( 94.99, 73.60) --
	( 95.41, 73.33) --
	( 95.82, 73.06) --
	( 96.24, 72.79) --
	( 96.65, 72.51) --
	( 97.06, 72.24) --
	( 97.48, 71.97) --
	( 97.89, 71.70) --
	( 98.31, 71.43) --
	( 98.72, 71.15) --
	( 99.13, 70.88) --
	( 99.55, 70.61) --
	( 99.96, 70.33) --
	(100.38, 70.06) --
	(100.79, 69.79);

\draw[color=drawColor,line width= 0.8pt,dash pattern=on 4pt off 4pt ,line cap=round,line join=round,fill opacity=0.00,] ( 18.42, 76.33) --
	( 18.83, 77.01) --
	( 19.24, 77.67) --
	( 19.66, 78.27) --
	( 20.07, 78.90) --
	( 20.48, 79.48) --
	( 20.90, 80.04) --
	( 21.31, 80.58) --
	( 21.73, 81.06) --
	( 22.14, 81.56) --
	( 22.55, 82.03) --
	( 22.97, 82.45) --
	( 23.38, 82.81) --
	( 23.80, 83.13) --
	( 24.21, 83.44) --
	( 24.62, 83.71) --
	( 25.04, 83.94) --
	( 25.45, 84.16) --
	( 25.87, 84.40) --
	( 26.28, 84.58) --
	( 26.69, 84.77) --
	( 27.11, 84.95) --
	( 27.52, 85.15) --
	( 27.94, 85.32) --
	( 28.35, 85.50) --
	( 28.76, 85.68) --
	( 29.18, 85.85) --
	( 29.59, 86.01) --
	( 30.01, 86.16) --
	( 30.42, 86.31) --
	( 30.83, 86.46) --
	( 31.25, 86.61) --
	( 31.66, 86.75) --
	( 32.08, 86.89) --
	( 32.49, 87.02) --
	( 32.90, 87.15) --
	( 33.32, 87.27) --
	( 33.73, 87.40) --
	( 34.14, 87.51) --
	( 34.56, 87.60) --
	( 34.97, 87.68) --
	( 35.39, 87.76) --
	( 35.80, 87.84) --
	( 36.21, 87.88) --
	( 36.63, 87.96) --
	( 37.04, 88.00) --
	( 37.46, 88.08) --
	( 37.87, 88.13) --
	( 38.28, 88.19) --
	( 38.70, 88.22) --
	( 39.11, 88.25) --
	( 39.53, 88.28) --
	( 39.94, 88.33) --
	( 40.35, 88.39) --
	( 40.77, 88.47) --
	( 41.18, 88.55) --
	( 41.60, 88.63) --
	( 42.01, 88.73) --
	( 42.42, 88.81) --
	( 42.84, 88.94) --
	( 43.25, 89.09) --
	( 43.67, 89.22) --
	( 44.08, 89.36) --
	( 44.49, 89.54) --
	( 44.91, 89.72) --
	( 45.32, 89.90) --
	( 45.74, 90.08) --
	( 46.15, 90.26) --
	( 46.56, 90.41) --
	( 46.98, 90.57) --
	( 47.39, 90.73) --
	( 47.81, 90.87) --
	( 48.22, 91.00) --
	( 48.63, 91.12) --
	( 49.05, 91.24) --
	( 49.46, 91.36) --
	( 49.87, 91.48) --
	( 50.29, 91.54) --
	( 50.70, 91.61) --
	( 51.12, 91.67) --
	( 51.53, 91.72) --
	( 51.94, 91.75) --
	( 52.36, 91.77) --
	( 52.77, 91.79) --
	( 53.19, 91.82) --
	( 53.60, 91.81) --
	( 54.01, 91.80) --
	( 54.43, 91.78) --
	( 54.84, 91.74) --
	( 55.26, 91.71) --
	( 55.67, 91.65) --
	( 56.08, 91.59) --
	( 56.50, 91.53) --
	( 56.91, 91.47) --
	( 57.33, 91.39) --
	( 57.74, 91.29) --
	( 58.15, 91.18) --
	( 58.57, 91.07) --
	( 58.98, 90.94) --
	( 59.40, 90.82) --
	( 59.81, 90.67) --
	( 60.22, 90.48) --
	( 60.64, 90.32) --
	( 61.05, 90.10) --
	( 61.47, 89.92) --
	( 61.88, 89.73) --
	( 62.29, 89.50) --
	( 62.71, 89.26) --
	( 63.12, 89.02) --
	( 63.53, 88.80) --
	( 63.95, 88.52) --
	( 64.36, 88.26) --
	( 64.78, 87.95) --
	( 65.19, 87.65) --
	( 65.60, 87.32) --
	( 66.02, 87.02) --
	( 66.43, 86.68) --
	( 66.85, 86.34) --
	( 67.26, 86.02) --
	( 67.67, 85.67) --
	( 68.09, 85.30) --
	( 68.50, 84.97) --
	( 68.92, 84.58) --
	( 69.33, 84.23) --
	( 69.74, 83.94) --
	( 70.16, 83.58) --
	( 70.57, 83.24) --
	( 70.99, 82.90) --
	( 71.40, 82.57) --
	( 71.81, 82.24) --
	( 72.23, 81.89) --
	( 72.64, 81.47) --
	( 73.06, 81.07) --
	( 73.47, 80.69) --
	( 73.88, 80.34) --
	( 74.30, 79.92) --
	( 74.71, 79.51) --
	( 75.13, 78.93) --
	( 75.54, 78.47) --
	( 75.95, 78.04) --
	( 76.37, 77.61) --
	( 76.78, 77.25) --
	( 77.20, 76.81) --
	( 77.61, 76.32) --
	( 78.02, 75.85) --
	( 78.44, 75.34) --
	( 78.85, 74.82) --
	( 79.26, 74.27) --
	( 79.68, 73.65) --
	( 80.09, 73.05) --
	( 80.51, 72.53) --
	( 80.92, 71.88) --
	( 81.33, 71.17) --
	( 81.75, 70.57) --
	( 82.16, 69.98) --
	( 82.58, 69.30) --
	( 82.99, 68.49) --
	( 83.40, 67.66) --
	( 83.82, 67.09) --
	( 84.23, 66.36) --
	( 84.65, 65.55) --
	( 85.06, 64.93) --
	( 85.47, 64.33) --
	( 85.89, 63.48) --
	( 86.30, 62.85) --
	( 86.72, 62.00) --
	( 87.13, 61.14) --
	( 87.54, 60.41) --
	( 87.96, 59.56) --
	( 88.37, 58.90) --
	( 88.79, 58.19) --
	( 89.20, 57.37) --
	( 89.61, 56.49) --
	( 90.03, 55.71) --
	( 90.44, 54.85) --
	( 90.86, 53.79) --
	( 91.27, 53.00) --
	( 91.68, 51.98) --
	( 92.10, 51.22) --
	( 92.51, 50.29) --
	( 92.93, 49.32) --
	( 93.34, 48.40) --
	( 93.75, 47.45) --
	( 94.17, 46.31) --
	( 94.58, 45.42) --
	( 94.99, 44.52) --
	( 95.41, 43.62) --
	( 95.82, 42.71) --
	( 96.24, 41.79) --
	( 96.65, 40.87) --
	( 97.06, 39.90) --
	( 97.48, 38.72) --
	( 97.89, 37.60) --
	( 98.31, 36.57) --
	( 98.72, 35.64) --
	( 99.13, 34.71) --
	( 99.55, 33.77) --
	( 99.96, 32.83) --
	(100.38, 31.89) --
	(100.79, 30.93);

\draw[color=drawColor,line width= 0.8pt,dash pattern=on 4pt off 4pt ,line cap=round,line join=round,fill opacity=0.00,] ( 18.42, 82.51) --
	( 18.83, 82.68) --
	( 19.24, 82.84) --
	( 19.66, 83.02) --
	( 20.07, 83.23) --
	( 20.48, 83.41) --
	( 20.90, 83.61) --
	( 21.31, 83.82) --
	( 21.73, 84.05) --
	( 22.14, 84.29) --
	( 22.55, 84.54) --
	( 22.97, 84.82) --
	( 23.38, 85.13) --
	( 23.80, 85.47) --
	( 24.21, 85.82) --
	( 24.62, 86.17) --
	( 25.04, 86.51) --
	( 25.45, 86.86) --
	( 25.87, 87.23) --
	( 26.28, 87.56) --
	( 26.69, 87.92) --
	( 27.11, 88.25) --
	( 27.52, 88.58) --
	( 27.94, 88.93) --
	( 28.35, 89.23) --
	( 28.76, 89.51) --
	( 29.18, 89.77) --
	( 29.59, 90.01) --
	( 30.01, 90.23) --
	( 30.42, 90.45) --
	( 30.83, 90.63) --
	( 31.25, 90.80) --
	( 31.66, 90.99) --
	( 32.08, 91.14) --
	( 32.49, 91.30) --
	( 32.90, 91.46) --
	( 33.32, 91.61) --
	( 33.73, 91.71) --
	( 34.14, 91.85) --
	( 34.56, 91.96) --
	( 34.97, 92.09) --
	( 35.39, 92.19) --
	( 35.80, 92.32) --
	( 36.21, 92.41) --
	( 36.63, 92.54) --
	( 37.04, 92.62) --
	( 37.46, 92.74) --
	( 37.87, 92.89) --
	( 38.28, 93.00) --
	( 38.70, 93.14) --
	( 39.11, 93.25) --
	( 39.53, 93.38) --
	( 39.94, 93.54) --
	( 40.35, 93.64) --
	( 40.77, 93.77) --
	( 41.18, 93.91) --
	( 41.60, 94.04) --
	( 42.01, 94.17) --
	( 42.42, 94.26) --
	( 42.84, 94.40) --
	( 43.25, 94.51) --
	( 43.67, 94.62) --
	( 44.08, 94.74) --
	( 44.49, 94.85) --
	( 44.91, 94.99) --
	( 45.32, 95.13) --
	( 45.74, 95.27) --
	( 46.15, 95.41) --
	( 46.56, 95.55) --
	( 46.98, 95.70) --
	( 47.39, 95.85) --
	( 47.81, 96.02) --
	( 48.22, 96.19) --
	( 48.63, 96.37) --
	( 49.05, 96.55) --
	( 49.46, 96.76) --
	( 49.87, 96.94) --
	( 50.29, 97.12) --
	( 50.70, 97.29) --
	( 51.12, 97.45) --
	( 51.53, 97.62) --
	( 51.94, 97.79) --
	( 52.36, 97.98) --
	( 52.77, 98.14) --
	( 53.19, 98.27) --
	( 53.60, 98.40) --
	( 54.01, 98.53) --
	( 54.43, 98.64) --
	( 54.84, 98.73) --
	( 55.26, 98.81) --
	( 55.67, 98.86) --
	( 56.08, 98.86) --
	( 56.50, 98.88) --
	( 56.91, 98.89) --
	( 57.33, 98.86) --
	( 57.74, 98.78) --
	( 58.15, 98.74) --
	( 58.57, 98.64) --
	( 58.98, 98.53) --
	( 59.40, 98.44) --
	( 59.81, 98.28) --
	( 60.22, 98.19) --
	( 60.64, 98.06) --
	( 61.05, 97.88) --
	( 61.47, 97.67) --
	( 61.88, 97.56) --
	( 62.29, 97.37) --
	( 62.71, 97.23) --
	( 63.12, 97.09) --
	( 63.53, 96.89) --
	( 63.95, 96.76) --
	( 64.36, 96.59) --
	( 64.78, 96.43) --
	( 65.19, 96.29) --
	( 65.60, 96.09) --
	( 66.02, 95.95) --
	( 66.43, 95.77) --
	( 66.85, 95.56) --
	( 67.26, 95.39) --
	( 67.67, 95.21) --
	( 68.09, 95.11) --
	( 68.50, 94.96) --
	( 68.92, 94.88) --
	( 69.33, 94.73) --
	( 69.74, 94.55) --
	( 70.16, 94.41) --
	( 70.57, 94.27) --
	( 70.99, 94.11) --
	( 71.40, 94.02) --
	( 71.81, 93.86) --
	( 72.23, 93.72) --
	( 72.64, 93.58) --
	( 73.06, 93.46) --
	( 73.47, 93.34) --
	( 73.88, 93.25) --
	( 74.30, 93.12) --
	( 74.71, 93.06) --
	( 75.13, 92.99) --
	( 75.54, 92.90) --
	( 75.95, 92.77) --
	( 76.37, 92.67) --
	( 76.78, 92.62) --
	( 77.20, 92.53) --
	( 77.61, 92.38) --
	( 78.02, 92.28) --
	( 78.44, 92.22) --
	( 78.85, 92.14) --
	( 79.26, 92.09) --
	( 79.68, 92.05) --
	( 80.09, 92.03) --
	( 80.51, 91.99) --
	( 80.92, 91.95) --
	( 81.33, 91.99) --
	( 81.75, 92.00) --
	( 82.16, 91.95) --
	( 82.58, 91.95) --
	( 82.99, 91.99) --
	( 83.40, 92.02) --
	( 83.82, 92.06) --
	( 84.23, 92.12) --
	( 84.65, 92.16) --
	( 85.06, 92.24) --
	( 85.47, 92.30) --
	( 85.89, 92.37) --
	( 86.30, 92.38) --
	( 86.72, 92.48) --
	( 87.13, 92.57) --
	( 87.54, 92.70) --
	( 87.96, 92.87) --
	( 88.37, 92.97) --
	( 88.79, 93.08) --
	( 89.20, 93.14) --
	( 89.61, 93.27) --
	( 90.03, 93.48) --
	( 90.44, 93.70) --
	( 90.86, 93.90) --
	( 91.27, 94.04) --
	( 91.68, 94.20) --
	( 92.10, 94.38) --
	( 92.51, 94.57) --
	( 92.93, 94.79) --
	( 93.34, 94.98) --
	( 93.75, 95.24) --
	( 94.17, 95.47) --
	( 94.58, 95.76) --
	( 94.99, 96.03) --
	( 95.41, 96.27) --
	( 95.82, 96.52) --
	( 96.24, 96.78) --
	( 96.65, 97.03) --
	( 97.06, 97.29) --
	( 97.48, 97.55) --
	( 97.89, 97.83) --
	( 98.31, 98.08) --
	( 98.72, 98.35) --
	( 99.13, 98.64) --
	( 99.55, 98.93) --
	( 99.96, 99.22) --
	(100.38, 99.52) --
	(100.79, 99.86);

\draw[color=drawColor,line width= 0.8pt,dash pattern=on 1pt off 3pt ,line cap=round,line join=round,fill opacity=0.00,] ( 18.42, 74.90) --
	( 18.83, 75.70) --
	( 19.24, 76.46) --
	( 19.66, 77.18) --
	( 20.07, 77.89) --
	( 20.48, 78.58) --
	( 20.90, 79.22) --
	( 21.31, 79.85) --
	( 21.73, 80.41) --
	( 22.14, 81.04) --
	( 22.55, 81.47) --
	( 22.97, 81.86) --
	( 23.38, 82.20) --
	( 23.80, 82.54) --
	( 24.21, 82.86) --
	( 24.62, 83.14) --
	( 25.04, 83.35) --
	( 25.45, 83.57) --
	( 25.87, 83.77) --
	( 26.28, 83.92) --
	( 26.69, 84.05) --
	( 27.11, 84.26) --
	( 27.52, 84.39) --
	( 27.94, 84.52) --
	( 28.35, 84.67) --
	( 28.76, 84.81) --
	( 29.18, 84.94) --
	( 29.59, 85.07) --
	( 30.01, 85.22) --
	( 30.42, 85.37) --
	( 30.83, 85.48) --
	( 31.25, 85.63) --
	( 31.66, 85.77) --
	( 32.08, 85.91) --
	( 32.49, 86.05) --
	( 32.90, 86.19) --
	( 33.32, 86.31) --
	( 33.73, 86.42) --
	( 34.14, 86.48) --
	( 34.56, 86.59) --
	( 34.97, 86.64) --
	( 35.39, 86.75) --
	( 35.80, 86.80) --
	( 36.21, 86.83) --
	( 36.63, 86.83) --
	( 37.04, 86.91) --
	( 37.46, 86.99) --
	( 37.87, 87.04) --
	( 38.28, 87.04) --
	( 38.70, 87.05) --
	( 39.11, 87.09) --
	( 39.53, 87.17) --
	( 39.94, 87.20) --
	( 40.35, 87.25) --
	( 40.77, 87.31) --
	( 41.18, 87.39) --
	( 41.60, 87.49) --
	( 42.01, 87.60) --
	( 42.42, 87.72) --
	( 42.84, 87.81) --
	( 43.25, 87.90) --
	( 43.67, 88.02) --
	( 44.08, 88.23) --
	( 44.49, 88.41) --
	( 44.91, 88.59) --
	( 45.32, 88.78) --
	( 45.74, 88.95) --
	( 46.15, 89.15) --
	( 46.56, 89.36) --
	( 46.98, 89.47) --
	( 47.39, 89.64) --
	( 47.81, 89.71) --
	( 48.22, 89.82) --
	( 48.63, 89.99) --
	( 49.05, 90.04) --
	( 49.46, 90.07) --
	( 49.87, 90.10) --
	( 50.29, 90.17) --
	( 50.70, 90.19) --
	( 51.12, 90.20) --
	( 51.53, 90.22) --
	( 51.94, 90.23) --
	( 52.36, 90.20) --
	( 52.77, 90.23) --
	( 53.19, 90.22) --
	( 53.60, 90.11) --
	( 54.01, 90.09) --
	( 54.43, 90.09) --
	( 54.84, 90.10) --
	( 55.26, 90.08) --
	( 55.67, 89.99) --
	( 56.08, 89.90) --
	( 56.50, 89.83) --
	( 56.91, 89.76) --
	( 57.33, 89.63) --
	( 57.74, 89.54) --
	( 58.15, 89.48) --
	( 58.57, 89.36) --
	( 58.98, 89.17) --
	( 59.40, 89.01) --
	( 59.81, 88.79) --
	( 60.22, 88.56) --
	( 60.64, 88.43) --
	( 61.05, 88.36) --
	( 61.47, 88.23) --
	( 61.88, 88.00) --
	( 62.29, 87.78) --
	( 62.71, 87.50) --
	( 63.12, 87.30) --
	( 63.53, 86.99) --
	( 63.95, 86.77) --
	( 64.36, 86.43) --
	( 64.78, 86.12) --
	( 65.19, 85.78) --
	( 65.60, 85.50) --
	( 66.02, 85.22) --
	( 66.43, 84.87) --
	( 66.85, 84.48) --
	( 67.26, 84.19) --
	( 67.67, 83.76) --
	( 68.09, 83.34) --
	( 68.50, 82.98) --
	( 68.92, 82.67) --
	( 69.33, 82.24) --
	( 69.74, 81.80) --
	( 70.16, 81.41) --
	( 70.57, 81.12) --
	( 70.99, 80.79) --
	( 71.40, 80.32) --
	( 71.81, 79.80) --
	( 72.23, 79.27) --
	( 72.64, 78.74) --
	( 73.06, 78.19) --
	( 73.47, 77.64) --
	( 73.88, 77.20) --
	( 74.30, 76.83) --
	( 74.71, 76.33) --
	( 75.13, 75.68) --
	( 75.54, 75.02) --
	( 75.95, 74.35) --
	( 76.37, 73.67) --
	( 76.78, 72.99) --
	( 77.20, 72.30) --
	( 77.61, 71.63) --
	( 78.02, 70.99) --
	( 78.44, 70.34) --
	( 78.85, 69.68) --
	( 79.26, 69.01) --
	( 79.68, 68.34) --
	( 80.09, 67.66) --
	( 80.51, 66.98) --
	( 80.92, 66.29) --
	( 81.33, 65.59) --
	( 81.75, 64.89) --
	( 82.16, 64.18) --
	( 82.58, 63.42) --
	( 82.99, 62.44) --
	( 83.40, 61.45) --
	( 83.82, 60.45) --
	( 84.23, 59.44) --
	( 84.65, 58.41) --
	( 85.06, 57.38) --
	( 85.47, 56.33) --
	( 85.89, 55.28) --
	( 86.30, 54.21) --
	( 86.72, 53.13) --
	( 87.13, 52.04) --
	( 87.54, 50.95) --
	( 87.96, 49.84) --
	( 88.37, 48.72) --
	( 88.79, 47.59) --
	( 89.20, 46.45) --
	( 89.61, 45.30) --
	( 90.03, 44.14) --
	( 90.44, 42.98) --
	( 90.86, 41.80) --
	( 91.27, 40.61) --
	( 91.68, 39.42) --
	( 92.10, 38.21) --
	( 92.51, 37.00) --
	( 92.93, 35.77) --
	( 93.34, 34.54) --
	( 93.75, 33.30) --
	( 94.17, 32.06) --
	( 94.58, 30.80) --
	( 94.99, 29.53) --
	( 95.41, 28.26) --
	( 95.82, 26.98) --
	( 96.24, 25.69) --
	( 96.65, 24.39) --
	( 97.06, 23.09) --
	( 97.48, 21.78) --
	( 97.89, 20.46) --
	( 98.31, 19.13) --
	( 98.72, 17.80) --
	( 99.13, 16.45) --
	( 99.55, 15.11) --
	( 99.96, 13.75) --
	(100.38, 12.39) --
	(100.79, 11.02);

\draw[color=drawColor,line width= 0.8pt,dash pattern=on 1pt off 3pt ,line cap=round,line join=round,fill opacity=0.00,] ( 18.42, 83.95) --
	( 18.83, 84.02) --
	( 19.24, 84.00) --
	( 19.66, 84.00) --
	( 20.07, 84.03) --
	( 20.48, 84.15) --
	( 20.90, 84.34) --
	( 21.31, 84.44) --
	( 21.73, 84.62) --
	( 22.14, 84.87) --
	( 22.55, 85.12) --
	( 22.97, 85.36) --
	( 23.38, 85.68) --
	( 23.80, 86.03) --
	( 24.21, 86.34) --
	( 24.62, 86.69) --
	( 25.04, 87.09) --
	( 25.45, 87.52) --
	( 25.87, 87.91) --
	( 26.28, 88.28) --
	( 26.69, 88.65) --
	( 27.11, 88.98) --
	( 27.52, 89.32) --
	( 27.94, 89.64) --
	( 28.35, 89.97) --
	( 28.76, 90.31) --
	( 29.18, 90.64) --
	( 29.59, 90.89) --
	( 30.01, 91.10) --
	( 30.42, 91.41) --
	( 30.83, 91.70) --
	( 31.25, 91.92) --
	( 31.66, 92.07) --
	( 32.08, 92.20) --
	( 32.49, 92.35) --
	( 32.90, 92.53) --
	( 33.32, 92.71) --
	( 33.73, 92.84) --
	( 34.14, 92.96) --
	( 34.56, 93.05) --
	( 34.97, 93.17) --
	( 35.39, 93.33) --
	( 35.80, 93.46) --
	( 36.21, 93.53) --
	( 36.63, 93.65) --
	( 37.04, 93.76) --
	( 37.46, 93.89) --
	( 37.87, 94.07) --
	( 38.28, 94.25) --
	( 38.70, 94.37) --
	( 39.11, 94.44) --
	( 39.53, 94.47) --
	( 39.94, 94.56) --
	( 40.35, 94.68) --
	( 40.77, 94.80) --
	( 41.18, 94.94) --
	( 41.60, 95.13) --
	( 42.01, 95.34) --
	( 42.42, 95.52) --
	( 42.84, 95.66) --
	( 43.25, 95.78) --
	( 43.67, 95.90) --
	( 44.08, 96.00) --
	( 44.49, 96.12) --
	( 44.91, 96.24) --
	( 45.32, 96.36) --
	( 45.74, 96.46) --
	( 46.15, 96.58) --
	( 46.56, 96.78) --
	( 46.98, 96.96) --
	( 47.39, 97.10) --
	( 47.81, 97.18) --
	( 48.22, 97.30) --
	( 48.63, 97.52) --
	( 49.05, 97.76) --
	( 49.46, 98.01) --
	( 49.87, 98.25) --
	( 50.29, 98.47) --
	( 50.70, 98.68) --
	( 51.12, 98.88) --
	( 51.53, 99.11) --
	( 51.94, 99.44) --
	( 52.36, 99.70) --
	( 52.77, 99.89) --
	( 53.19, 99.95) --
	( 53.60,100.09) --
	( 54.01,100.11) --
	( 54.43,100.14) --
	( 54.84,100.30) --
	( 55.26,100.43) --
	( 55.67,100.53) --
	( 56.08,100.59) --
	( 56.50,100.60) --
	( 56.91,100.62) --
	( 57.33,100.60) --
	( 57.74,100.53) --
	( 58.15,100.47) --
	( 58.57,100.45) --
	( 58.98,100.41) --
	( 59.40,100.34) --
	( 59.81,100.25) --
	( 60.22,100.14) --
	( 60.64,100.01) --
	( 61.05, 99.86) --
	( 61.47, 99.68) --
	( 61.88, 99.50) --
	( 62.29, 99.35) --
	( 62.71, 99.23) --
	( 63.12, 99.06) --
	( 63.53, 98.88) --
	( 63.95, 98.70) --
	( 64.36, 98.63) --
	( 64.78, 98.52) --
	( 65.19, 98.38) --
	( 65.60, 98.18) --
	( 66.02, 98.12) --
	( 66.43, 97.96) --
	( 66.85, 97.88) --
	( 67.26, 97.72) --
	( 67.67, 97.61) --
	( 68.09, 97.48) --
	( 68.50, 97.34) --
	( 68.92, 97.25) --
	( 69.33, 97.11) --
	( 69.74, 96.94) --
	( 70.16, 96.76) --
	( 70.57, 96.64) --
	( 70.99, 96.52) --
	( 71.40, 96.40) --
	( 71.81, 96.22) --
	( 72.23, 96.15) --
	( 72.64, 96.04) --
	( 73.06, 96.01) --
	( 73.47, 95.89) --
	( 73.88, 95.91) --
	( 74.30, 95.93) --
	( 74.71, 95.87) --
	( 75.13, 95.87) --
	( 75.54, 95.75) --
	( 75.95, 95.70) --
	( 76.37, 95.74) --
	( 76.78, 95.67) --
	( 77.20, 95.62) --
	( 77.61, 95.65) --
	( 78.02, 95.68) --
	( 78.44, 95.58) --
	( 78.85, 95.65) --
	( 79.26, 95.72) --
	( 79.68, 95.73) --
	( 80.09, 95.75) --
	( 80.51, 95.64) --
	( 80.92, 95.78) --
	( 81.33, 95.92) --
	( 81.75, 95.97) --
	( 82.16, 96.10) --
	( 82.58, 96.29) --
	( 82.99, 96.31) --
	( 83.40, 96.42) --
	( 83.82, 96.53) --
	( 84.23, 96.73) --
	( 84.65, 96.87) --
	( 85.06, 97.04) --
	( 85.47, 97.22) --
	( 85.89, 97.40) --
	( 86.30, 97.58) --
	( 86.72, 97.77) --
	( 87.13, 97.95) --
	( 87.54, 98.20) --
	( 87.96, 98.41) --
	( 88.37, 98.58) --
	( 88.79, 98.82) --
	( 89.20, 99.04) --
	( 89.61, 99.20) --
	( 90.03, 99.35) --
	( 90.44, 99.66) --
	( 90.86,100.05) --
	( 91.27,100.30) --
	( 91.68,100.56) --
	( 92.10,100.86) --
	( 92.51,101.30) --
	( 92.93,101.75) --
	( 93.34,102.20) --
	( 93.75,102.76) --
	( 94.17,103.17) --
	( 94.58,103.61) --
	( 94.99,104.06) --
	( 95.41,104.54) --
	( 95.82,105.03) --
	( 96.24,105.61) --
	( 96.65,106.20) --
	( 97.06,106.73) --
	( 97.48,107.03) --
	( 97.89,107.51) --
	( 98.31,107.87) --
	( 98.72,108.43) --
	( 99.13,108.73) --
	( 99.55,109.24) --
	( 99.96,109.89) --
	(100.38,110.54) --
	(100.79,111.20);
\end{scope}
\end{tikzpicture}

%% file: out-pima-topInc-plot-x2.tikz
\begin{tikzpicture}[x=1pt,y=1pt]
\definecolor[named]{drawColor}{rgb}{0.00,0.00,0.00}
\definecolor[named]{fillColor}{rgb}{1.00,1.00,1.00}
\fill[color=fillColor,fill opacity=0.00,] (0,0) rectangle (108.41,130.09);
\begin{scope}
\path[clip] (  0.00,  0.00) rectangle (108.41,130.09);
\definecolor[named]{drawColor}{rgb}{0.00,0.00,0.00}

\draw[color=drawColor,line cap=round,line join=round,fill opacity=0.00,] ( 20.72, 33.12) -- (101.37, 33.12);

\draw[color=drawColor,line cap=round,line join=round,fill opacity=0.00,] ( 20.72, 33.12) -- ( 20.72, 29.52);

\draw[color=drawColor,line cap=round,line join=round,fill opacity=0.00,] ( 32.24, 33.12) -- ( 32.24, 29.52);

\draw[color=drawColor,line cap=round,line join=round,fill opacity=0.00,] ( 43.76, 33.12) -- ( 43.76, 29.52);

\draw[color=drawColor,line cap=round,line join=round,fill opacity=0.00,] ( 55.28, 33.12) -- ( 55.28, 29.52);

\draw[color=drawColor,line cap=round,line join=round,fill opacity=0.00,] ( 66.80, 33.12) -- ( 66.80, 29.52);

\draw[color=drawColor,line cap=round,line join=round,fill opacity=0.00,] ( 78.32, 33.12) -- ( 78.32, 29.52);

\draw[color=drawColor,line cap=round,line join=round,fill opacity=0.00,] ( 89.85, 33.12) -- ( 89.85, 29.52);

\draw[color=drawColor,line cap=round,line join=round,fill opacity=0.00,] (101.37, 33.12) -- (101.37, 29.52);

\node[color=drawColor,anchor=base,inner sep=0pt, outer sep=0pt, scale=  0.60] at ( 20.72, 18.72) {60};

\node[color=drawColor,anchor=base,inner sep=0pt, outer sep=0pt, scale=  0.60] at ( 32.24, 18.72) {80};

\node[color=drawColor,anchor=base,inner sep=0pt, outer sep=0pt, scale=  0.60] at ( 55.28, 18.72) {120};

\node[color=drawColor,anchor=base,inner sep=0pt, outer sep=0pt, scale=  0.60] at ( 78.32, 18.72) {160};

\node[color=drawColor,anchor=base,inner sep=0pt, outer sep=0pt, scale=  0.60] at (101.37, 18.72) {200};

\draw[color=drawColor,line cap=round,line join=round,fill opacity=0.00,] ( 15.12, 36.15) -- ( 15.12,111.93);

\draw[color=drawColor,line cap=round,line join=round,fill opacity=0.00,] ( 15.12, 36.15) -- ( 11.52, 36.15);

\draw[color=drawColor,line cap=round,line join=round,fill opacity=0.00,] ( 15.12, 55.10) -- ( 11.52, 55.10);

\draw[color=drawColor,line cap=round,line join=round,fill opacity=0.00,] ( 15.12, 74.04) -- ( 11.52, 74.04);

\draw[color=drawColor,line cap=round,line join=round,fill opacity=0.00,] ( 15.12, 92.99) -- ( 11.52, 92.99);

\draw[color=drawColor,line cap=round,line join=round,fill opacity=0.00,] ( 15.12,111.93) -- ( 11.52,111.93);

\node[color=drawColor,anchor=base east,inner sep=0pt, outer sep=0pt, scale=  0.60] at (  7.92, 34.09) {-4};

\node[color=drawColor,anchor=base east,inner sep=0pt, outer sep=0pt, scale=  0.60] at (  7.92, 53.03) {-2};

\node[color=drawColor,anchor=base east,inner sep=0pt, outer sep=0pt, scale=  0.60] at (  7.92, 71.98) {0};

\node[color=drawColor,anchor=base east,inner sep=0pt, outer sep=0pt, scale=  0.60] at (  7.92, 90.92) {2};

\node[color=drawColor,anchor=base east,inner sep=0pt, outer sep=0pt, scale=  0.60] at (  7.92,109.87) {4};

\draw[color=drawColor,line cap=round,line join=round,fill opacity=0.00,] ( 15.12, 33.12) --
	(104.08, 33.12) --
	(104.08,114.97) --
	( 15.12,114.97) --
	( 15.12, 33.12);
\end{scope}
\begin{scope}
\path[clip] (  0.00,  0.00) rectangle (108.41,130.09);
\definecolor[named]{drawColor}{rgb}{0.00,0.00,0.00}

\node[color=drawColor,anchor=base,inner sep=0pt, outer sep=0pt, scale=  0.72] at ( 59.60,  4.32) {$x_{2}$};
\end{scope}
\begin{scope}
\path[clip] ( 15.12, 33.12) rectangle (104.08,114.97);
\definecolor[named]{drawColor}{rgb}{0.00,0.00,0.00}

\draw[color=drawColor,line width= 0.8pt,line cap=round,line join=round,fill opacity=0.00,] ( 18.42, 51.14) --
	( 18.83, 51.41) --
	( 19.24, 51.68) --
	( 19.66, 51.96) --
	( 20.07, 52.23) --
	( 20.48, 52.50) --
	( 20.90, 52.78) --
	( 21.31, 53.05) --
	( 21.73, 53.32) --
	( 22.14, 53.59) --
	( 22.55, 53.86) --
	( 22.97, 54.13) --
	( 23.38, 54.41) --
	( 23.80, 54.68) --
	( 24.21, 54.95) --
	( 24.62, 55.21) --
	( 25.04, 55.48) --
	( 25.45, 55.75) --
	( 25.87, 56.02) --
	( 26.28, 56.29) --
	( 26.69, 56.56) --
	( 27.11, 56.82) --
	( 27.52, 57.09) --
	( 27.94, 57.36) --
	( 28.35, 57.62) --
	( 28.76, 57.89) --
	( 29.18, 58.16) --
	( 29.59, 58.42) --
	( 30.01, 58.69) --
	( 30.42, 58.95) --
	( 30.83, 59.21) --
	( 31.25, 59.48) --
	( 31.66, 59.74) --
	( 32.08, 60.00) --
	( 32.49, 60.26) --
	( 32.90, 60.53) --
	( 33.32, 60.79) --
	( 33.73, 61.05) --
	( 34.14, 61.31) --
	( 34.56, 61.57) --
	( 34.97, 61.83) --
	( 35.39, 62.09) --
	( 35.80, 62.34) --
	( 36.21, 62.60) --
	( 36.63, 62.86) --
	( 37.04, 63.12) --
	( 37.46, 63.37) --
	( 37.87, 63.63) --
	( 38.28, 63.88) --
	( 38.70, 64.14) --
	( 39.11, 64.39) --
	( 39.53, 64.65) --
	( 39.94, 64.90) --
	( 40.35, 65.15) --
	( 40.77, 65.41) --
	( 41.18, 65.66) --
	( 41.60, 65.91) --
	( 42.01, 66.16) --
	( 42.42, 66.41) --
	( 42.84, 66.66) --
	( 43.25, 66.91) --
	( 43.67, 67.16) --
	( 44.08, 67.40) --
	( 44.49, 67.65) --
	( 44.91, 67.90) --
	( 45.32, 68.14) --
	( 45.74, 68.39) --
	( 46.15, 68.63) --
	( 46.56, 68.88) --
	( 46.98, 69.12) --
	( 47.39, 69.36) --
	( 47.81, 69.60) --
	( 48.22, 69.84) --
	( 48.63, 70.08) --
	( 49.05, 70.32) --
	( 49.46, 70.56) --
	( 49.87, 70.80) --
	( 50.29, 71.04) --
	( 50.70, 71.27) --
	( 51.12, 71.51) --
	( 51.53, 71.74) --
	( 51.94, 71.98) --
	( 52.36, 72.21) --
	( 52.77, 72.44) --
	( 53.19, 72.67) --
	( 53.60, 72.91) --
	( 54.01, 73.14) --
	( 54.43, 73.36) --
	( 54.84, 73.59) --
	( 55.26, 73.82) --
	( 55.67, 74.05) --
	( 56.08, 74.27) --
	( 56.50, 74.50) --
	( 56.91, 74.73) --
	( 57.33, 74.95) --
	( 57.74, 75.18) --
	( 58.15, 75.40) --
	( 58.57, 75.62) --
	( 58.98, 75.85) --
	( 59.40, 76.07) --
	( 59.81, 76.30) --
	( 60.22, 76.52) --
	( 60.64, 76.74) --
	( 61.05, 76.97) --
	( 61.47, 77.19) --
	( 61.88, 77.42) --
	( 62.29, 77.64) --
	( 62.71, 77.87) --
	( 63.12, 78.09) --
	( 63.53, 78.32) --
	( 63.95, 78.54) --
	( 64.36, 78.77) --
	( 64.78, 79.00) --
	( 65.19, 79.23) --
	( 65.60, 79.46) --
	( 66.02, 79.69) --
	( 66.43, 79.92) --
	( 66.85, 80.15) --
	( 67.26, 80.38) --
	( 67.67, 80.61) --
	( 68.09, 80.85) --
	( 68.50, 81.08) --
	( 68.92, 81.32) --
	( 69.33, 81.56) --
	( 69.74, 81.80) --
	( 70.16, 82.03) --
	( 70.57, 82.27) --
	( 70.99, 82.52) --
	( 71.40, 82.76) --
	( 71.81, 83.00) --
	( 72.23, 83.24) --
	( 72.64, 83.49) --
	( 73.06, 83.73) --
	( 73.47, 83.98) --
	( 73.88, 84.22) --
	( 74.30, 84.47) --
	( 74.71, 84.72) --
	( 75.13, 84.96) --
	( 75.54, 85.21) --
	( 75.95, 85.46) --
	( 76.37, 85.71) --
	( 76.78, 85.96) --
	( 77.20, 86.21) --
	( 77.61, 86.45) --
	( 78.02, 86.70) --
	( 78.44, 86.95) --
	( 78.85, 87.20) --
	( 79.26, 87.45) --
	( 79.68, 87.70) --
	( 80.09, 87.95) --
	( 80.51, 88.20) --
	( 80.92, 88.45) --
	( 81.33, 88.70) --
	( 81.75, 88.95) --
	( 82.16, 89.20) --
	( 82.58, 89.44) --
	( 82.99, 89.69) --
	( 83.40, 89.94) --
	( 83.82, 90.19) --
	( 84.23, 90.43) --
	( 84.65, 90.68) --
	( 85.06, 90.93) --
	( 85.47, 91.17) --
	( 85.89, 91.42) --
	( 86.30, 91.67) --
	( 86.72, 91.91) --
	( 87.13, 92.16) --
	( 87.54, 92.40) --
	( 87.96, 92.64) --
	( 88.37, 92.89) --
	( 88.79, 93.13) --
	( 89.20, 93.38) --
	( 89.61, 93.62) --
	( 90.03, 93.86) --
	( 90.44, 94.10) --
	( 90.86, 94.35) --
	( 91.27, 94.59) --
	( 91.68, 94.83) --
	( 92.10, 95.07) --
	( 92.51, 95.31) --
	( 92.93, 95.55) --
	( 93.34, 95.79) --
	( 93.75, 96.03) --
	( 94.17, 96.27) --
	( 94.58, 96.51) --
	( 94.99, 96.75) --
	( 95.41, 96.99) --
	( 95.82, 97.23) --
	( 96.24, 97.47) --
	( 96.65, 97.71) --
	( 97.06, 97.95) --
	( 97.48, 98.19) --
	( 97.89, 98.43) --
	( 98.31, 98.66) --
	( 98.72, 98.90) --
	( 99.13, 99.14) --
	( 99.55, 99.38) --
	( 99.96, 99.61) --
	(100.38, 99.85) --
	(100.79,100.09);

\draw[color=drawColor,line width= 0.8pt,dash pattern=on 4pt off 4pt ,line cap=round,line join=round,fill opacity=0.00,] ( 18.42, 36.35) --
	( 18.83, 37.00) --
	( 19.24, 37.73) --
	( 19.66, 38.39) --
	( 20.07, 39.00) --
	( 20.48, 39.68) --
	( 20.90, 40.37) --
	( 21.31, 41.02) --
	( 21.73, 41.65) --
	( 22.14, 42.26) --
	( 22.55, 42.98) --
	( 22.97, 43.53) --
	( 23.38, 44.00) --
	( 23.80, 44.55) --
	( 24.21, 45.10) --
	( 24.62, 45.68) --
	( 25.04, 46.21) --
	( 25.45, 46.74) --
	( 25.87, 47.28) --
	( 26.28, 47.80) --
	( 26.69, 48.40) --
	( 27.11, 48.98) --
	( 27.52, 49.50) --
	( 27.94, 50.03) --
	( 28.35, 50.60) --
	( 28.76, 51.12) --
	( 29.18, 51.56) --
	( 29.59, 51.99) --
	( 30.01, 52.57) --
	( 30.42, 53.05) --
	( 30.83, 53.56) --
	( 31.25, 54.09) --
	( 31.66, 54.57) --
	( 32.08, 55.04) --
	( 32.49, 55.50) --
	( 32.90, 55.93) --
	( 33.32, 56.39) --
	( 33.73, 56.83) --
	( 34.14, 57.27) --
	( 34.56, 57.71) --
	( 34.97, 58.16) --
	( 35.39, 58.60) --
	( 35.80, 58.97) --
	( 36.21, 59.35) --
	( 36.63, 59.72) --
	( 37.04, 60.08) --
	( 37.46, 60.44) --
	( 37.87, 60.79) --
	( 38.28, 61.15) --
	( 38.70, 61.50) --
	( 39.11, 61.84) --
	( 39.53, 62.18) --
	( 39.94, 62.50) --
	( 40.35, 62.81) --
	( 40.77, 63.12) --
	( 41.18, 63.45) --
	( 41.60, 63.80) --
	( 42.01, 64.12) --
	( 42.42, 64.42) --
	( 42.84, 64.72) --
	( 43.25, 65.05) --
	( 43.67, 65.35) --
	( 44.08, 65.62) --
	( 44.49, 65.91) --
	( 44.91, 66.20) --
	( 45.32, 66.48) --
	( 45.74, 66.77) --
	( 46.15, 67.04) --
	( 46.56, 67.31) --
	( 46.98, 67.59) --
	( 47.39, 67.87) --
	( 47.81, 68.13) --
	( 48.22, 68.38) --
	( 48.63, 68.65) --
	( 49.05, 68.91) --
	( 49.46, 69.14) --
	( 49.87, 69.35) --
	( 50.29, 69.56) --
	( 50.70, 69.77) --
	( 51.12, 69.99) --
	( 51.53, 70.20) --
	( 51.94, 70.41) --
	( 52.36, 70.62) --
	( 52.77, 70.84) --
	( 53.19, 71.04) --
	( 53.60, 71.25) --
	( 54.01, 71.46) --
	( 54.43, 71.67) --
	( 54.84, 71.88) --
	( 55.26, 72.09) --
	( 55.67, 72.29) --
	( 56.08, 72.50) --
	( 56.50, 72.72) --
	( 56.91, 72.91) --
	( 57.33, 73.11) --
	( 57.74, 73.31) --
	( 58.15, 73.51) --
	( 58.57, 73.70) --
	( 58.98, 73.91) --
	( 59.40, 74.11) --
	( 59.81, 74.30) --
	( 60.22, 74.47) --
	( 60.64, 74.66) --
	( 61.05, 74.83) --
	( 61.47, 75.00) --
	( 61.88, 75.17) --
	( 62.29, 75.34) --
	( 62.71, 75.52) --
	( 63.12, 75.68) --
	( 63.53, 75.85) --
	( 63.95, 76.00) --
	( 64.36, 76.16) --
	( 64.78, 76.34) --
	( 65.19, 76.50) --
	( 65.60, 76.69) --
	( 66.02, 76.91) --
	( 66.43, 77.11) --
	( 66.85, 77.31) --
	( 67.26, 77.54) --
	( 67.67, 77.75) --
	( 68.09, 77.96) --
	( 68.50, 78.17) --
	( 68.92, 78.41) --
	( 69.33, 78.63) --
	( 69.74, 78.85) --
	( 70.16, 79.09) --
	( 70.57, 79.34) --
	( 70.99, 79.58) --
	( 71.40, 79.82) --
	( 71.81, 80.08) --
	( 72.23, 80.33) --
	( 72.64, 80.52) --
	( 73.06, 80.76) --
	( 73.47, 80.97) --
	( 73.88, 81.17) --
	( 74.30, 81.40) --
	( 74.71, 81.63) --
	( 75.13, 81.89) --
	( 75.54, 82.13) --
	( 75.95, 82.34) --
	( 76.37, 82.56) --
	( 76.78, 82.78) --
	( 77.20, 82.99) --
	( 77.61, 83.19) --
	( 78.02, 83.40) --
	( 78.44, 83.62) --
	( 78.85, 83.82) --
	( 79.26, 84.03) --
	( 79.68, 84.24) --
	( 80.09, 84.43) --
	( 80.51, 84.62) --
	( 80.92, 84.82) --
	( 81.33, 85.00) --
	( 81.75, 85.20) --
	( 82.16, 85.39) --
	( 82.58, 85.59) --
	( 82.99, 85.79) --
	( 83.40, 85.99) --
	( 83.82, 86.17) --
	( 84.23, 86.36) --
	( 84.65, 86.53) --
	( 85.06, 86.72) --
	( 85.47, 86.90) --
	( 85.89, 87.10) --
	( 86.30, 87.28) --
	( 86.72, 87.46) --
	( 87.13, 87.65) --
	( 87.54, 87.82) --
	( 87.96, 88.01) --
	( 88.37, 88.19) --
	( 88.79, 88.37) --
	( 89.20, 88.54) --
	( 89.61, 88.70) --
	( 90.03, 88.85) --
	( 90.44, 88.99) --
	( 90.86, 89.16) --
	( 91.27, 89.32) --
	( 91.68, 89.47) --
	( 92.10, 89.62) --
	( 92.51, 89.78) --
	( 92.93, 89.93) --
	( 93.34, 90.05) --
	( 93.75, 90.18) --
	( 94.17, 90.34) --
	( 94.58, 90.47) --
	( 94.99, 90.60) --
	( 95.41, 90.73) --
	( 95.82, 90.84) --
	( 96.24, 90.95) --
	( 96.65, 91.09) --
	( 97.06, 91.22) --
	( 97.48, 91.34) --
	( 97.89, 91.42) --
	( 98.31, 91.52) --
	( 98.72, 91.61) --
	( 99.13, 91.70) --
	( 99.55, 91.78) --
	( 99.96, 91.88) --
	(100.38, 91.92) --
	(100.79, 92.05);

\draw[color=drawColor,line width= 0.8pt,dash pattern=on 4pt off 4pt ,line cap=round,line join=round,fill opacity=0.00,] ( 18.42, 60.52) --
	( 18.83, 60.60) --
	( 19.24, 60.64) --
	( 19.66, 60.73) --
	( 20.07, 60.81) --
	( 20.48, 60.89) --
	( 20.90, 60.97) --
	( 21.31, 61.05) --
	( 21.73, 61.16) --
	( 22.14, 61.24) --
	( 22.55, 61.34) --
	( 22.97, 61.39) --
	( 23.38, 61.50) --
	( 23.80, 61.63) --
	( 24.21, 61.72) --
	( 24.62, 61.80) --
	( 25.04, 61.91) --
	( 25.45, 62.03) --
	( 25.87, 62.13) --
	( 26.28, 62.22) --
	( 26.69, 62.33) --
	( 27.11, 62.47) --
	( 27.52, 62.58) --
	( 27.94, 62.70) --
	( 28.35, 62.84) --
	( 28.76, 62.94) --
	( 29.18, 63.06) --
	( 29.59, 63.23) --
	( 30.01, 63.36) --
	( 30.42, 63.52) --
	( 30.83, 63.67) --
	( 31.25, 63.82) --
	( 31.66, 63.94) --
	( 32.08, 64.09) --
	( 32.49, 64.23) --
	( 32.90, 64.41) --
	( 33.32, 64.57) --
	( 33.73, 64.73) --
	( 34.14, 64.90) --
	( 34.56, 65.07) --
	( 34.97, 65.22) --
	( 35.39, 65.38) --
	( 35.80, 65.55) --
	( 36.21, 65.73) --
	( 36.63, 65.91) --
	( 37.04, 66.10) --
	( 37.46, 66.27) --
	( 37.87, 66.44) --
	( 38.28, 66.61) --
	( 38.70, 66.78) --
	( 39.11, 66.95) --
	( 39.53, 67.11) --
	( 39.94, 67.31) --
	( 40.35, 67.49) --
	( 40.77, 67.67) --
	( 41.18, 67.87) --
	( 41.60, 68.06) --
	( 42.01, 68.28) --
	( 42.42, 68.48) --
	( 42.84, 68.66) --
	( 43.25, 68.88) --
	( 43.67, 69.10) --
	( 44.08, 69.31) --
	( 44.49, 69.53) --
	( 44.91, 69.75) --
	( 45.32, 69.97) --
	( 45.74, 70.21) --
	( 46.15, 70.47) --
	( 46.56, 70.71) --
	( 46.98, 70.97) --
	( 47.39, 71.25) --
	( 47.81, 71.54) --
	( 48.22, 71.80) --
	( 48.63, 72.06) --
	( 49.05, 72.32) --
	( 49.46, 72.60) --
	( 49.87, 72.88) --
	( 50.29, 73.17) --
	( 50.70, 73.45) --
	( 51.12, 73.71) --
	( 51.53, 73.96) --
	( 51.94, 74.19) --
	( 52.36, 74.43) --
	( 52.77, 74.66) --
	( 53.19, 74.92) --
	( 53.60, 75.13) --
	( 54.01, 75.36) --
	( 54.43, 75.56) --
	( 54.84, 75.74) --
	( 55.26, 75.97) --
	( 55.67, 76.18) --
	( 56.08, 76.40) --
	( 56.50, 76.61) --
	( 56.91, 76.83) --
	( 57.33, 77.03) --
	( 57.74, 77.25) --
	( 58.15, 77.46) --
	( 58.57, 77.68) --
	( 58.98, 77.86) --
	( 59.40, 78.08) --
	( 59.81, 78.30) --
	( 60.22, 78.55) --
	( 60.64, 78.78) --
	( 61.05, 79.00) --
	( 61.47, 79.23) --
	( 61.88, 79.48) --
	( 62.29, 79.68) --
	( 62.71, 79.92) --
	( 63.12, 80.15) --
	( 63.53, 80.37) --
	( 63.95, 80.63) --
	( 64.36, 80.86) --
	( 64.78, 81.11) --
	( 65.19, 81.36) --
	( 65.60, 81.60) --
	( 66.02, 81.86) --
	( 66.43, 82.12) --
	( 66.85, 82.36) --
	( 67.26, 82.62) --
	( 67.67, 82.89) --
	( 68.09, 83.16) --
	( 68.50, 83.42) --
	( 68.92, 83.70) --
	( 69.33, 83.95) --
	( 69.74, 84.23) --
	( 70.16, 84.52) --
	( 70.57, 84.82) --
	( 70.99, 85.09) --
	( 71.40, 85.37) --
	( 71.81, 85.67) --
	( 72.23, 85.95) --
	( 72.64, 86.22) --
	( 73.06, 86.50) --
	( 73.47, 86.79) --
	( 73.88, 87.07) --
	( 74.30, 87.36) --
	( 74.71, 87.66) --
	( 75.13, 87.92) --
	( 75.54, 88.21) --
	( 75.95, 88.51) --
	( 76.37, 88.82) --
	( 76.78, 89.11) --
	( 77.20, 89.40) --
	( 77.61, 89.70) --
	( 78.02, 90.00) --
	( 78.44, 90.29) --
	( 78.85, 90.57) --
	( 79.26, 90.88) --
	( 79.68, 91.19) --
	( 80.09, 91.51) --
	( 80.51, 91.81) --
	( 80.92, 92.14) --
	( 81.33, 92.45) --
	( 81.75, 92.77) --
	( 82.16, 93.06) --
	( 82.58, 93.38) --
	( 82.99, 93.70) --
	( 83.40, 93.99) --
	( 83.82, 94.29) --
	( 84.23, 94.60) --
	( 84.65, 94.91) --
	( 85.06, 95.26) --
	( 85.47, 95.59) --
	( 85.89, 95.91) --
	( 86.30, 96.24) --
	( 86.72, 96.54) --
	( 87.13, 96.83) --
	( 87.54, 97.11) --
	( 87.96, 97.41) --
	( 88.37, 97.75) --
	( 88.79, 98.05) --
	( 89.20, 98.36) --
	( 89.61, 98.67) --
	( 90.03, 98.98) --
	( 90.44, 99.32) --
	( 90.86, 99.67) --
	( 91.27,100.01) --
	( 91.68,100.36) --
	( 92.10,100.72) --
	( 92.51,101.08) --
	( 92.93,101.43) --
	( 93.34,101.81) --
	( 93.75,102.18) --
	( 94.17,102.51) --
	( 94.58,102.91) --
	( 94.99,103.27) --
	( 95.41,103.62) --
	( 95.82,104.03) --
	( 96.24,104.42) --
	( 96.65,104.80) --
	( 97.06,105.19) --
	( 97.48,105.64) --
	( 97.89,106.07) --
	( 98.31,106.53) --
	( 98.72,107.03) --
	( 99.13,107.38) --
	( 99.55,107.85) --
	( 99.96,108.30) --
	(100.38,108.74) --
	(100.79,109.11);

\draw[color=drawColor,line width= 0.8pt,dash pattern=on 1pt off 3pt ,line cap=round,line join=round,fill opacity=0.00,] ( 18.42, 22.62) --
	( 18.83, 23.71) --
	( 19.24, 24.80) --
	( 19.66, 25.87) --
	( 20.07, 26.94) --
	( 20.48, 27.99) --
	( 20.90, 29.04) --
	( 21.31, 30.07) --
	( 21.73, 31.10) --
	( 22.14, 32.11) --
	( 22.55, 33.11) --
	( 22.97, 34.11) --
	( 23.38, 35.09) --
	( 23.80, 36.06) --
	( 24.21, 37.02) --
	( 24.62, 37.96) --
	( 25.04, 38.90) --
	( 25.45, 39.80) --
	( 25.87, 40.57) --
	( 26.28, 41.45) --
	( 26.69, 42.37) --
	( 27.11, 43.07) --
	( 27.52, 43.88) --
	( 27.94, 44.61) --
	( 28.35, 45.22) --
	( 28.76, 45.84) --
	( 29.18, 46.50) --
	( 29.59, 47.31) --
	( 30.01, 48.05) --
	( 30.42, 48.86) --
	( 30.83, 49.63) --
	( 31.25, 50.27) --
	( 31.66, 50.85) --
	( 32.08, 51.46) --
	( 32.49, 52.23) --
	( 32.90, 52.76) --
	( 33.32, 53.46) --
	( 33.73, 54.07) --
	( 34.14, 54.62) --
	( 34.56, 55.14) --
	( 34.97, 55.87) --
	( 35.39, 56.35) --
	( 35.80, 56.75) --
	( 36.21, 57.28) --
	( 36.63, 57.75) --
	( 37.04, 58.23) --
	( 37.46, 58.72) --
	( 37.87, 59.18) --
	( 38.28, 59.62) --
	( 38.70, 60.09) --
	( 39.11, 60.63) --
	( 39.53, 61.07) --
	( 39.94, 61.47) --
	( 40.35, 61.83) --
	( 40.77, 62.20) --
	( 41.18, 62.63) --
	( 41.60, 63.03) --
	( 42.01, 63.43) --
	( 42.42, 63.75) --
	( 42.84, 64.04) --
	( 43.25, 64.33) --
	( 43.67, 64.62) --
	( 44.08, 64.91) --
	( 44.49, 65.25) --
	( 44.91, 65.60) --
	( 45.32, 65.92) --
	( 45.74, 66.16) --
	( 46.15, 66.49) --
	( 46.56, 66.71) --
	( 46.98, 66.92) --
	( 47.39, 67.18) --
	( 47.81, 67.41) --
	( 48.22, 67.64) --
	( 48.63, 67.87) --
	( 49.05, 68.03) --
	( 49.46, 68.21) --
	( 49.87, 68.51) --
	( 50.29, 68.69) --
	( 50.70, 68.90) --
	( 51.12, 69.11) --
	( 51.53, 69.32) --
	( 51.94, 69.56) --
	( 52.36, 69.75) --
	( 52.77, 69.91) --
	( 53.19, 70.13) --
	( 53.60, 70.30) --
	( 54.01, 70.45) --
	( 54.43, 70.64) --
	( 54.84, 70.85) --
	( 55.26, 71.04) --
	( 55.67, 71.26) --
	( 56.08, 71.47) --
	( 56.50, 71.66) --
	( 56.91, 71.86) --
	( 57.33, 72.11) --
	( 57.74, 72.31) --
	( 58.15, 72.49) --
	( 58.57, 72.62) --
	( 58.98, 72.76) --
	( 59.40, 72.88) --
	( 59.81, 73.05) --
	( 60.22, 73.21) --
	( 60.64, 73.31) --
	( 61.05, 73.52) --
	( 61.47, 73.65) --
	( 61.88, 73.74) --
	( 62.29, 73.90) --
	( 62.71, 74.04) --
	( 63.12, 74.24) --
	( 63.53, 74.40) --
	( 63.95, 74.53) --
	( 64.36, 74.68) --
	( 64.78, 74.82) --
	( 65.19, 74.95) --
	( 65.60, 75.10) --
	( 66.02, 75.27) --
	( 66.43, 75.46) --
	( 66.85, 75.68) --
	( 67.26, 75.86) --
	( 67.67, 76.04) --
	( 68.09, 76.24) --
	( 68.50, 76.44) --
	( 68.92, 76.63) --
	( 69.33, 76.85) --
	( 69.74, 77.09) --
	( 70.16, 77.32) --
	( 70.57, 77.56) --
	( 70.99, 77.84) --
	( 71.40, 78.11) --
	( 71.81, 78.42) --
	( 72.23, 78.67) --
	( 72.64, 78.96) --
	( 73.06, 79.17) --
	( 73.47, 79.46) --
	( 73.88, 79.78) --
	( 74.30, 80.02) --
	( 74.71, 80.24) --
	( 75.13, 80.48) --
	( 75.54, 80.69) --
	( 75.95, 80.99) --
	( 76.37, 81.19) --
	( 76.78, 81.36) --
	( 77.20, 81.62) --
	( 77.61, 81.82) --
	( 78.02, 82.00) --
	( 78.44, 82.21) --
	( 78.85, 82.43) --
	( 79.26, 82.58) --
	( 79.68, 82.77) --
	( 80.09, 82.97) --
	( 80.51, 83.17) --
	( 80.92, 83.36) --
	( 81.33, 83.59) --
	( 81.75, 83.80) --
	( 82.16, 83.99) --
	( 82.58, 84.15) --
	( 82.99, 84.32) --
	( 83.40, 84.48) --
	( 83.82, 84.64) --
	( 84.23, 84.81) --
	( 84.65, 85.00) --
	( 85.06, 85.15) --
	( 85.47, 85.31) --
	( 85.89, 85.48) --
	( 86.30, 85.63) --
	( 86.72, 85.79) --
	( 87.13, 85.94) --
	( 87.54, 86.05) --
	( 87.96, 86.18) --
	( 88.37, 86.33) --
	( 88.79, 86.50) --
	( 89.20, 86.64) --
	( 89.61, 86.79) --
	( 90.03, 86.91) --
	( 90.44, 87.07) --
	( 90.86, 87.10) --
	( 91.27, 87.19) --
	( 91.68, 87.29) --
	( 92.10, 87.39) --
	( 92.51, 87.52) --
	( 92.93, 87.66) --
	( 93.34, 87.80) --
	( 93.75, 87.87) --
	( 94.17, 87.92) --
	( 94.58, 88.01) --
	( 94.99, 88.08) --
	( 95.41, 88.17) --
	( 95.82, 88.22) --
	( 96.24, 88.31) --
	( 96.65, 88.31) --
	( 97.06, 88.43) --
	( 97.48, 88.43) --
	( 97.89, 88.42) --
	( 98.31, 88.40) --
	( 98.72, 88.27) --
	( 99.13, 88.22) --
	( 99.55, 88.21) --
	( 99.96, 88.24) --
	(100.38, 88.07) --
	(100.79, 87.87);

\draw[color=drawColor,line width= 0.8pt,dash pattern=on 1pt off 3pt ,line cap=round,line join=round,fill opacity=0.00,] ( 18.42, 66.35) --
	( 18.83, 66.26) --
	( 19.24, 66.11) --
	( 19.66, 65.96) --
	( 20.07, 65.80) --
	( 20.48, 65.68) --
	( 20.90, 65.58) --
	( 21.31, 65.45) --
	( 21.73, 65.41) --
	( 22.14, 65.35) --
	( 22.55, 65.19) --
	( 22.97, 65.15) --
	( 23.38, 65.20) --
	( 23.80, 65.21) --
	( 24.21, 65.16) --
	( 24.62, 65.07) --
	( 25.04, 65.03) --
	( 25.45, 64.90) --
	( 25.87, 64.94) --
	( 26.28, 65.01) --
	( 26.69, 65.01) --
	( 27.11, 65.00) --
	( 27.52, 65.03) --
	( 27.94, 65.11) --
	( 28.35, 65.18) --
	( 28.76, 65.23) --
	( 29.18, 65.29) --
	( 29.59, 65.27) --
	( 30.01, 65.37) --
	( 30.42, 65.45) --
	( 30.83, 65.54) --
	( 31.25, 65.62) --
	( 31.66, 65.73) --
	( 32.08, 65.82) --
	( 32.49, 65.92) --
	( 32.90, 66.01) --
	( 33.32, 66.09) --
	( 33.73, 66.20) --
	( 34.14, 66.34) --
	( 34.56, 66.46) --
	( 34.97, 66.59) --
	( 35.39, 66.73) --
	( 35.80, 66.90) --
	( 36.21, 67.04) --
	( 36.63, 67.13) --
	( 37.04, 67.26) --
	( 37.46, 67.42) --
	( 37.87, 67.55) --
	( 38.28, 67.73) --
	( 38.70, 67.91) --
	( 39.11, 68.05) --
	( 39.53, 68.21) --
	( 39.94, 68.43) --
	( 40.35, 68.60) --
	( 40.77, 68.76) --
	( 41.18, 68.89) --
	( 41.60, 69.02) --
	( 42.01, 69.27) --
	( 42.42, 69.40) --
	( 42.84, 69.62) --
	( 43.25, 69.83) --
	( 43.67, 70.07) --
	( 44.08, 70.32) --
	( 44.49, 70.56) --
	( 44.91, 70.82) --
	( 45.32, 71.14) --
	( 45.74, 71.45) --
	( 46.15, 71.76) --
	( 46.56, 72.05) --
	( 46.98, 72.34) --
	( 47.39, 72.61) --
	( 47.81, 72.87) --
	( 48.22, 73.16) --
	( 48.63, 73.38) --
	( 49.05, 73.66) --
	( 49.46, 73.95) --
	( 49.87, 74.22) --
	( 50.29, 74.45) --
	( 50.70, 74.68) --
	( 51.12, 74.94) --
	( 51.53, 75.18) --
	( 51.94, 75.40) --
	( 52.36, 75.64) --
	( 52.77, 75.94) --
	( 53.19, 76.23) --
	( 53.60, 76.42) --
	( 54.01, 76.69) --
	( 54.43, 76.89) --
	( 54.84, 77.13) --
	( 55.26, 77.35) --
	( 55.67, 77.58) --
	( 56.08, 77.76) --
	( 56.50, 77.92) --
	( 56.91, 78.09) --
	( 57.33, 78.34) --
	( 57.74, 78.46) --
	( 58.15, 78.64) --
	( 58.57, 78.90) --
	( 58.98, 79.07) --
	( 59.40, 79.25) --
	( 59.81, 79.43) --
	( 60.22, 79.63) --
	( 60.64, 79.81) --
	( 61.05, 80.05) --
	( 61.47, 80.30) --
	( 61.88, 80.49) --
	( 62.29, 80.71) --
	( 62.71, 80.93) --
	( 63.12, 81.23) --
	( 63.53, 81.52) --
	( 63.95, 81.80) --
	( 64.36, 82.09) --
	( 64.78, 82.36) --
	( 65.19, 82.56) --
	( 65.60, 82.85) --
	( 66.02, 83.08) --
	( 66.43, 83.31) --
	( 66.85, 83.55) --
	( 67.26, 83.80) --
	( 67.67, 84.06) --
	( 68.09, 84.29) --
	( 68.50, 84.54) --
	( 68.92, 84.80) --
	( 69.33, 85.06) --
	( 69.74, 85.32) --
	( 70.16, 85.58) --
	( 70.57, 85.85) --
	( 70.99, 86.10) --
	( 71.40, 86.37) --
	( 71.81, 86.61) --
	( 72.23, 86.84) --
	( 72.64, 87.12) --
	( 73.06, 87.37) --
	( 73.47, 87.77) --
	( 73.88, 88.14) --
	( 74.30, 88.39) --
	( 74.71, 88.64) --
	( 75.13, 88.97) --
	( 75.54, 89.37) --
	( 75.95, 89.78) --
	( 76.37, 90.18) --
	( 76.78, 90.57) --
	( 77.20, 90.96) --
	( 77.61, 91.35) --
	( 78.02, 91.73) --
	( 78.44, 92.02) --
	( 78.85, 92.28) --
	( 79.26, 92.62) --
	( 79.68, 92.91) --
	( 80.09, 93.25) --
	( 80.51, 93.58) --
	( 80.92, 93.90) --
	( 81.33, 94.39) --
	( 81.75, 94.74) --
	( 82.16, 95.06) --
	( 82.58, 95.47) --
	( 82.99, 95.82) --
	( 83.40, 96.07) --
	( 83.82, 96.48) --
	( 84.23, 96.89) --
	( 84.65, 97.18) --
	( 85.06, 97.41) --
	( 85.47, 97.62) --
	( 85.89, 97.85) --
	( 86.30, 98.25) --
	( 86.72, 98.47) --
	( 87.13, 98.76) --
	( 87.54, 98.98) --
	( 87.96, 99.27) --
	( 88.37, 99.64) --
	( 88.79, 99.98) --
	( 89.20,100.41) --
	( 89.61,100.79) --
	( 90.03,101.13) --
	( 90.44,101.47) --
	( 90.86,101.70) --
	( 91.27,102.06) --
	( 91.68,102.51) --
	( 92.10,102.96) --
	( 92.51,103.41) --
	( 92.93,103.84) --
	( 93.34,104.24) --
	( 93.75,104.77) --
	( 94.17,105.18) --
	( 94.58,105.78) --
	( 94.99,106.44) --
	( 95.41,107.04) --
	( 95.82,107.39) --
	( 96.24,107.71) --
	( 96.65,108.19) --
	( 97.06,108.66) --
	( 97.48,109.06) --
	( 97.89,109.49) --
	( 98.31,109.90) --
	( 98.72,110.32) --
	( 99.13,110.88) --
	( 99.55,111.25) --
	( 99.96,111.67) --
	(100.38,112.14) --
	(100.79,112.75);
\end{scope}
\end{tikzpicture}

%% file: out-pima-topInc-plot-x5.tikz
\begin{tikzpicture}[x=1pt,y=1pt]
\definecolor[named]{drawColor}{rgb}{0.00,0.00,0.00}
\definecolor[named]{fillColor}{rgb}{1.00,1.00,1.00}
\fill[color=fillColor,fill opacity=0.00,] (0,0) rectangle (108.41,130.09);
\begin{scope}
\path[clip] (  0.00,  0.00) rectangle (108.41,130.09);
\definecolor[named]{drawColor}{rgb}{0.00,0.00,0.00}

\draw[color=drawColor,line cap=round,line join=round,fill opacity=0.00,] ( 21.45, 33.12) -- ( 88.83, 33.12);

\draw[color=drawColor,line cap=round,line join=round,fill opacity=0.00,] ( 21.45, 33.12) -- ( 21.45, 29.52);

\draw[color=drawColor,line cap=round,line join=round,fill opacity=0.00,] ( 38.29, 33.12) -- ( 38.29, 29.52);

\draw[color=drawColor,line cap=round,line join=round,fill opacity=0.00,] ( 55.14, 33.12) -- ( 55.14, 29.52);

\draw[color=drawColor,line cap=round,line join=round,fill opacity=0.00,] ( 71.98, 33.12) -- ( 71.98, 29.52);

\draw[color=drawColor,line cap=round,line join=round,fill opacity=0.00,] ( 88.83, 33.12) -- ( 88.83, 29.52);

\node[color=drawColor,anchor=base,inner sep=0pt, outer sep=0pt, scale=  0.60] at ( 21.45, 18.72) {20};

\node[color=drawColor,anchor=base,inner sep=0pt, outer sep=0pt, scale=  0.60] at ( 38.29, 18.72) {30};

\node[color=drawColor,anchor=base,inner sep=0pt, outer sep=0pt, scale=  0.60] at ( 55.14, 18.72) {40};

\node[color=drawColor,anchor=base,inner sep=0pt, outer sep=0pt, scale=  0.60] at ( 71.98, 18.72) {50};

\node[color=drawColor,anchor=base,inner sep=0pt, outer sep=0pt, scale=  0.60] at ( 88.83, 18.72) {60};

\draw[color=drawColor,line cap=round,line join=round,fill opacity=0.00,] ( 15.12, 36.15) -- ( 15.12,111.93);

\draw[color=drawColor,line cap=round,line join=round,fill opacity=0.00,] ( 15.12, 36.15) -- ( 11.52, 36.15);

\draw[color=drawColor,line cap=round,line join=round,fill opacity=0.00,] ( 15.12, 48.78) -- ( 11.52, 48.78);

\draw[color=drawColor,line cap=round,line join=round,fill opacity=0.00,] ( 15.12, 61.41) -- ( 11.52, 61.41);

\draw[color=drawColor,line cap=round,line join=round,fill opacity=0.00,] ( 15.12, 74.04) -- ( 11.52, 74.04);

\draw[color=drawColor,line cap=round,line join=round,fill opacity=0.00,] ( 15.12, 86.67) -- ( 11.52, 86.67);

\draw[color=drawColor,line cap=round,line join=round,fill opacity=0.00,] ( 15.12, 99.30) -- ( 11.52, 99.30);

\draw[color=drawColor,line cap=round,line join=round,fill opacity=0.00,] ( 15.12,111.93) -- ( 11.52,111.93);

\node[color=drawColor,anchor=base east,inner sep=0pt, outer sep=0pt, scale=  0.60] at (  7.92, 34.09) {-4};

\node[color=drawColor,anchor=base east,inner sep=0pt, outer sep=0pt, scale=  0.60] at (  7.92, 46.72) {-2};

\node[color=drawColor,anchor=base east,inner sep=0pt, outer sep=0pt, scale=  0.60] at (  7.92, 59.35) {0};

\node[color=drawColor,anchor=base east,inner sep=0pt, outer sep=0pt, scale=  0.60] at (  7.92, 71.98) {2};

\node[color=drawColor,anchor=base east,inner sep=0pt, outer sep=0pt, scale=  0.60] at (  7.92, 84.61) {4};

\node[color=drawColor,anchor=base east,inner sep=0pt, outer sep=0pt, scale=  0.60] at (  7.92, 97.24) {6};

\node[color=drawColor,anchor=base east,inner sep=0pt, outer sep=0pt, scale=  0.60] at (  7.92,109.87) {8};

\draw[color=drawColor,line cap=round,line join=round,fill opacity=0.00,] ( 15.12, 33.12) --
	(104.08, 33.12) --
	(104.08,114.97) --
	( 15.12,114.97) --
	( 15.12, 33.12);
\end{scope}
\begin{scope}
\path[clip] (  0.00,  0.00) rectangle (108.41,130.09);
\definecolor[named]{drawColor}{rgb}{0.00,0.00,0.00}

\node[color=drawColor,anchor=base,inner sep=0pt, outer sep=0pt, scale=  0.72] at ( 59.60,  4.32) {$x_{5}$};
\end{scope}
\begin{scope}
\path[clip] ( 15.12, 33.12) rectangle (104.08,114.97);
\definecolor[named]{drawColor}{rgb}{0.00,0.00,0.00}

\draw[color=drawColor,line width= 0.8pt,line cap=round,line join=round,fill opacity=0.00,] ( 18.42, 47.37) --
	( 18.83, 47.70) --
	( 19.24, 48.03) --
	( 19.66, 48.36) --
	( 20.07, 48.68) --
	( 20.48, 49.01) --
	( 20.90, 49.33) --
	( 21.31, 49.65) --
	( 21.73, 49.97) --
	( 22.14, 50.29) --
	( 22.55, 50.61) --
	( 22.97, 50.93) --
	( 23.38, 51.24) --
	( 23.80, 51.55) --
	( 24.21, 51.87) --
	( 24.62, 52.18) --
	( 25.04, 52.49) --
	( 25.45, 52.79) --
	( 25.87, 53.10) --
	( 26.28, 53.40) --
	( 26.69, 53.71) --
	( 27.11, 54.01) --
	( 27.52, 54.31) --
	( 27.94, 54.61) --
	( 28.35, 54.90) --
	( 28.76, 55.20) --
	( 29.18, 55.49) --
	( 29.59, 55.78) --
	( 30.01, 56.07) --
	( 30.42, 56.36) --
	( 30.83, 56.65) --
	( 31.25, 56.93) --
	( 31.66, 57.21) --
	( 32.08, 57.49) --
	( 32.49, 57.77) --
	( 32.90, 58.04) --
	( 33.32, 58.32) --
	( 33.73, 58.58) --
	( 34.14, 58.85) --
	( 34.56, 59.11) --
	( 34.97, 59.36) --
	( 35.39, 59.61) --
	( 35.80, 59.85) --
	( 36.21, 60.09) --
	( 36.63, 60.32) --
	( 37.04, 60.55) --
	( 37.46, 60.77) --
	( 37.87, 60.98) --
	( 38.28, 61.18) --
	( 38.70, 61.37) --
	( 39.11, 61.56) --
	( 39.53, 61.73) --
	( 39.94, 61.90) --
	( 40.35, 62.06) --
	( 40.77, 62.21) --
	( 41.18, 62.35) --
	( 41.60, 62.48) --
	( 42.01, 62.61) --
	( 42.42, 62.73) --
	( 42.84, 62.84) --
	( 43.25, 62.94) --
	( 43.67, 63.04) --
	( 44.08, 63.14) --
	( 44.49, 63.22) --
	( 44.91, 63.30) --
	( 45.32, 63.38) --
	( 45.74, 63.45) --
	( 46.15, 63.52) --
	( 46.56, 63.58) --
	( 46.98, 63.64) --
	( 47.39, 63.70) --
	( 47.81, 63.75) --
	( 48.22, 63.81) --
	( 48.63, 63.85) --
	( 49.05, 63.90) --
	( 49.46, 63.95) --
	( 49.87, 63.99) --
	( 50.29, 64.03) --
	( 50.70, 64.07) --
	( 51.12, 64.11) --
	( 51.53, 64.15) --
	( 51.94, 64.19) --
	( 52.36, 64.23) --
	( 52.77, 64.27) --
	( 53.19, 64.31) --
	( 53.60, 64.35) --
	( 54.01, 64.39) --
	( 54.43, 64.44) --
	( 54.84, 64.48) --
	( 55.26, 64.52) --
	( 55.67, 64.57) --
	( 56.08, 64.62) --
	( 56.50, 64.67) --
	( 56.91, 64.73) --
	( 57.33, 64.79) --
	( 57.74, 64.84) --
	( 58.15, 64.91) --
	( 58.57, 64.97) --
	( 58.98, 65.04) --
	( 59.40, 65.11) --
	( 59.81, 65.18) --
	( 60.22, 65.26) --
	( 60.64, 65.34) --
	( 61.05, 65.42) --
	( 61.47, 65.50) --
	( 61.88, 65.58) --
	( 62.29, 65.67) --
	( 62.71, 65.76) --
	( 63.12, 65.86) --
	( 63.53, 65.95) --
	( 63.95, 66.05) --
	( 64.36, 66.15) --
	( 64.78, 66.25) --
	( 65.19, 66.35) --
	( 65.60, 66.46) --
	( 66.02, 66.57) --
	( 66.43, 66.68) --
	( 66.85, 66.80) --
	( 67.26, 66.91) --
	( 67.67, 67.03) --
	( 68.09, 67.15) --
	( 68.50, 67.27) --
	( 68.92, 67.40) --
	( 69.33, 67.52) --
	( 69.74, 67.65) --
	( 70.16, 67.79) --
	( 70.57, 67.92) --
	( 70.99, 68.05) --
	( 71.40, 68.19) --
	( 71.81, 68.33) --
	( 72.23, 68.47) --
	( 72.64, 68.62) --
	( 73.06, 68.76) --
	( 73.47, 68.91) --
	( 73.88, 69.06) --
	( 74.30, 69.21) --
	( 74.71, 69.37) --
	( 75.13, 69.52) --
	( 75.54, 69.68) --
	( 75.95, 69.84) --
	( 76.37, 70.00) --
	( 76.78, 70.16) --
	( 77.20, 70.33) --
	( 77.61, 70.49) --
	( 78.02, 70.66) --
	( 78.44, 70.83) --
	( 78.85, 71.01) --
	( 79.26, 71.18) --
	( 79.68, 71.36) --
	( 80.09, 71.53) --
	( 80.51, 71.71) --
	( 80.92, 71.89) --
	( 81.33, 72.08) --
	( 81.75, 72.26) --
	( 82.16, 72.45) --
	( 82.58, 72.63) --
	( 82.99, 72.82) --
	( 83.40, 73.01) --
	( 83.82, 73.21) --
	( 84.23, 73.40) --
	( 84.65, 73.59) --
	( 85.06, 73.79) --
	( 85.47, 73.99) --
	( 85.89, 74.19) --
	( 86.30, 74.39) --
	( 86.72, 74.59) --
	( 87.13, 74.80) --
	( 87.54, 75.00) --
	( 87.96, 75.21) --
	( 88.37, 75.42) --
	( 88.79, 75.63) --
	( 89.20, 75.84) --
	( 89.61, 76.05) --
	( 90.03, 76.27) --
	( 90.44, 76.48) --
	( 90.86, 76.70) --
	( 91.27, 76.92) --
	( 91.68, 77.14) --
	( 92.10, 77.36) --
	( 92.51, 77.58) --
	( 92.93, 77.80) --
	( 93.34, 78.03) --
	( 93.75, 78.25) --
	( 94.17, 78.48) --
	( 94.58, 78.70) --
	( 94.99, 78.93) --
	( 95.41, 79.16) --
	( 95.82, 79.39) --
	( 96.24, 79.62) --
	( 96.65, 79.86) --
	( 97.06, 80.09) --
	( 97.48, 80.33) --
	( 97.89, 80.56) --
	( 98.31, 80.80) --
	( 98.72, 81.04) --
	( 99.13, 81.28) --
	( 99.55, 81.52) --
	( 99.96, 81.76) --
	(100.38, 82.00) --
	(100.79, 82.24);

\draw[color=drawColor,line width= 0.8pt,dash pattern=on 4pt off 4pt ,line cap=round,line join=round,fill opacity=0.00,] ( 18.42, 31.52) --
	( 18.83, 32.53) --
	( 19.24, 33.58) --
	( 19.66, 34.50) --
	( 20.07, 35.45) --
	( 20.48, 36.34) --
	( 20.90, 37.17) --
	( 21.31, 38.00) --
	( 21.73, 38.91) --
	( 22.14, 39.74) --
	( 22.55, 40.64) --
	( 22.97, 41.40) --
	( 23.38, 42.12) --
	( 23.80, 42.92) --
	( 24.21, 43.73) --
	( 24.62, 44.57) --
	( 25.04, 45.32) --
	( 25.45, 46.03) --
	( 25.87, 46.69) --
	( 26.28, 47.38) --
	( 26.69, 48.06) --
	( 27.11, 48.67) --
	( 27.52, 49.26) --
	( 27.94, 49.84) --
	( 28.35, 50.42) --
	( 28.76, 50.99) --
	( 29.18, 51.55) --
	( 29.59, 52.09) --
	( 30.01, 52.67) --
	( 30.42, 53.23) --
	( 30.83, 53.72) --
	( 31.25, 54.23) --
	( 31.66, 54.67) --
	( 32.08, 55.14) --
	( 32.49, 55.59) --
	( 32.90, 56.05) --
	( 33.32, 56.49) --
	( 33.73, 56.86) --
	( 34.14, 57.18) --
	( 34.56, 57.51) --
	( 34.97, 57.78) --
	( 35.39, 58.07) --
	( 35.80, 58.32) --
	( 36.21, 58.55) --
	( 36.63, 58.78) --
	( 37.04, 58.99) --
	( 37.46, 59.18) --
	( 37.87, 59.37) --
	( 38.28, 59.55) --
	( 38.70, 59.73) --
	( 39.11, 59.90) --
	( 39.53, 60.07) --
	( 39.94, 60.23) --
	( 40.35, 60.38) --
	( 40.77, 60.53) --
	( 41.18, 60.68) --
	( 41.60, 60.83) --
	( 42.01, 60.98) --
	( 42.42, 61.13) --
	( 42.84, 61.28) --
	( 43.25, 61.43) --
	( 43.67, 61.50) --
	( 44.08, 61.58) --
	( 44.49, 61.65) --
	( 44.91, 61.73) --
	( 45.32, 61.80) --
	( 45.74, 61.87) --
	( 46.15, 61.93) --
	( 46.56, 62.00) --
	( 46.98, 62.06) --
	( 47.39, 62.11) --
	( 47.81, 62.15) --
	( 48.22, 62.20) --
	( 48.63, 62.22) --
	( 49.05, 62.25) --
	( 49.46, 62.28) --
	( 49.87, 62.31) --
	( 50.29, 62.33) --
	( 50.70, 62.34) --
	( 51.12, 62.35) --
	( 51.53, 62.36) --
	( 51.94, 62.37) --
	( 52.36, 62.37) --
	( 52.77, 62.39) --
	( 53.19, 62.40) --
	( 53.60, 62.39) --
	( 54.01, 62.39) --
	( 54.43, 62.37) --
	( 54.84, 62.33) --
	( 55.26, 62.29) --
	( 55.67, 62.27) --
	( 56.08, 62.24) --
	( 56.50, 62.20) --
	( 56.91, 62.18) --
	( 57.33, 62.17) --
	( 57.74, 62.16) --
	( 58.15, 62.14) --
	( 58.57, 62.11) --
	( 58.98, 62.10) --
	( 59.40, 62.10) --
	( 59.81, 62.12) --
	( 60.22, 62.12) --
	( 60.64, 62.12) --
	( 61.05, 62.15) --
	( 61.47, 62.20) --
	( 61.88, 62.21) --
	( 62.29, 62.26) --
	( 62.71, 62.29) --
	( 63.12, 62.29) --
	( 63.53, 62.35) --
	( 63.95, 62.40) --
	( 64.36, 62.44) --
	( 64.78, 62.51) --
	( 65.19, 62.57) --
	( 65.60, 62.65) --
	( 66.02, 62.73) --
	( 66.43, 62.78) --
	( 66.85, 62.83) --
	( 67.26, 62.89) --
	( 67.67, 62.93) --
	( 68.09, 63.02) --
	( 68.50, 63.08) --
	( 68.92, 63.13) --
	( 69.33, 63.18) --
	( 69.74, 63.23) --
	( 70.16, 63.28) --
	( 70.57, 63.32) --
	( 70.99, 63.38) --
	( 71.40, 63.43) --
	( 71.81, 63.49) --
	( 72.23, 63.58) --
	( 72.64, 63.61) --
	( 73.06, 63.66) --
	( 73.47, 63.70) --
	( 73.88, 63.75) --
	( 74.30, 63.80) --
	( 74.71, 63.82) --
	( 75.13, 63.81) --
	( 75.54, 63.85) --
	( 75.95, 63.91) --
	( 76.37, 63.99) --
	( 76.78, 64.03) --
	( 77.20, 64.07) --
	( 77.61, 64.07) --
	( 78.02, 64.05) --
	( 78.44, 64.11) --
	( 78.85, 64.13) --
	( 79.26, 64.17) --
	( 79.68, 64.15) --
	( 80.09, 64.14) --
	( 80.51, 64.15) --
	( 80.92, 64.15) --
	( 81.33, 64.19) --
	( 81.75, 64.17) --
	( 82.16, 64.16) --
	( 82.58, 64.18) --
	( 82.99, 64.22) --
	( 83.40, 64.23) --
	( 83.82, 64.21) --
	( 84.23, 64.23) --
	( 84.65, 64.22) --
	( 85.06, 64.19) --
	( 85.47, 64.15) --
	( 85.89, 64.17) --
	( 86.30, 64.20) --
	( 86.72, 64.20) --
	( 87.13, 64.21) --
	( 87.54, 64.15) --
	( 87.96, 64.13) --
	( 88.37, 64.08) --
	( 88.79, 64.02) --
	( 89.20, 63.95) --
	( 89.61, 63.93) --
	( 90.03, 63.90) --
	( 90.44, 63.87) --
	( 90.86, 63.80) --
	( 91.27, 63.77) --
	( 91.68, 63.74) --
	( 92.10, 63.68) --
	( 92.51, 63.64) --
	( 92.93, 63.62) --
	( 93.34, 63.47) --
	( 93.75, 63.40) --
	( 94.17, 63.32) --
	( 94.58, 63.29) --
	( 94.99, 63.23) --
	( 95.41, 63.18) --
	( 95.82, 63.11) --
	( 96.24, 63.04) --
	( 96.65, 62.96) --
	( 97.06, 62.91) --
	( 97.48, 62.81) --
	( 97.89, 62.76) --
	( 98.31, 62.70) --
	( 98.72, 62.61) --
	( 99.13, 62.54) --
	( 99.55, 62.47) --
	( 99.96, 62.42) --
	(100.38, 62.37) --
	(100.79, 62.32);

\draw[color=drawColor,line width= 0.8pt,dash pattern=on 4pt off 4pt ,line cap=round,line join=round,fill opacity=0.00,] ( 18.42, 56.84) --
	( 18.83, 56.89) --
	( 19.24, 56.95) --
	( 19.66, 56.98) --
	( 20.07, 57.03) --
	( 20.48, 57.08) --
	( 20.90, 57.15) --
	( 21.31, 57.20) --
	( 21.73, 57.23) --
	( 22.14, 57.30) --
	( 22.55, 57.37) --
	( 22.97, 57.44) --
	( 23.38, 57.53) --
	( 23.80, 57.61) --
	( 24.21, 57.68) --
	( 24.62, 57.75) --
	( 25.04, 57.83) --
	( 25.45, 57.89) --
	( 25.87, 57.97) --
	( 26.28, 58.06) --
	( 26.69, 58.15) --
	( 27.11, 58.23) --
	( 27.52, 58.32) --
	( 27.94, 58.42) --
	( 28.35, 58.50) --
	( 28.76, 58.60) --
	( 29.18, 58.68) --
	( 29.59, 58.77) --
	( 30.01, 58.89) --
	( 30.42, 59.00) --
	( 30.83, 59.11) --
	( 31.25, 59.22) --
	( 31.66, 59.36) --
	( 32.08, 59.51) --
	( 32.49, 59.67) --
	( 32.90, 59.84) --
	( 33.32, 60.04) --
	( 33.73, 60.29) --
	( 34.14, 60.53) --
	( 34.56, 60.81) --
	( 34.97, 61.10) --
	( 35.39, 61.43) --
	( 35.80, 61.78) --
	( 36.21, 62.10) --
	( 36.63, 62.40) --
	( 37.04, 62.73) --
	( 37.46, 63.03) --
	( 37.87, 63.30) --
	( 38.28, 63.60) --
	( 38.70, 63.85) --
	( 39.11, 64.07) --
	( 39.53, 64.29) --
	( 39.94, 64.47) --
	( 40.35, 64.64) --
	( 40.77, 64.79) --
	( 41.18, 64.92) --
	( 41.60, 65.02) --
	( 42.01, 65.08) --
	( 42.42, 65.14) --
	( 42.84, 65.19) --
	( 43.25, 65.24) --
	( 43.67, 65.30) --
	( 44.08, 65.35) --
	( 44.49, 65.43) --
	( 44.91, 65.49) --
	( 45.32, 65.54) --
	( 45.74, 65.59) --
	( 46.15, 65.65) --
	( 46.56, 65.69) --
	( 46.98, 65.73) --
	( 47.39, 65.78) --
	( 47.81, 65.83) --
	( 48.22, 65.87) --
	( 48.63, 65.90) --
	( 49.05, 65.94) --
	( 49.46, 65.96) --
	( 49.87, 65.97) --
	( 50.29, 65.98) --
	( 50.70, 66.03) --
	( 51.12, 66.05) --
	( 51.53, 66.07) --
	( 51.94, 66.10) --
	( 52.36, 66.15) --
	( 52.77, 66.19) --
	( 53.19, 66.24) --
	( 53.60, 66.30) --
	( 54.01, 66.36) --
	( 54.43, 66.46) --
	( 54.84, 66.53) --
	( 55.26, 66.65) --
	( 55.67, 66.74) --
	( 56.08, 66.87) --
	( 56.50, 66.97) --
	( 56.91, 67.09) --
	( 57.33, 67.21) --
	( 57.74, 67.35) --
	( 58.15, 67.46) --
	( 58.57, 67.56) --
	( 58.98, 67.68) --
	( 59.40, 67.83) --
	( 59.81, 67.98) --
	( 60.22, 68.11) --
	( 60.64, 68.25) --
	( 61.05, 68.39) --
	( 61.47, 68.54) --
	( 61.88, 68.70) --
	( 62.29, 68.85) --
	( 62.71, 69.01) --
	( 63.12, 69.16) --
	( 63.53, 69.33) --
	( 63.95, 69.48) --
	( 64.36, 69.61) --
	( 64.78, 69.79) --
	( 65.19, 69.97) --
	( 65.60, 70.15) --
	( 66.02, 70.32) --
	( 66.43, 70.50) --
	( 66.85, 70.67) --
	( 67.26, 70.83) --
	( 67.67, 71.01) --
	( 68.09, 71.20) --
	( 68.50, 71.40) --
	( 68.92, 71.59) --
	( 69.33, 71.78) --
	( 69.74, 71.99) --
	( 70.16, 72.23) --
	( 70.57, 72.51) --
	( 70.99, 72.80) --
	( 71.40, 73.00) --
	( 71.81, 73.29) --
	( 72.23, 73.55) --
	( 72.64, 73.81) --
	( 73.06, 74.11) --
	( 73.47, 74.36) --
	( 73.88, 74.66) --
	( 74.30, 75.00) --
	( 74.71, 75.35) --
	( 75.13, 75.68) --
	( 75.54, 76.03) --
	( 75.95, 76.35) --
	( 76.37, 76.69) --
	( 76.78, 77.01) --
	( 77.20, 77.36) --
	( 77.61, 77.69) --
	( 78.02, 78.06) --
	( 78.44, 78.50) --
	( 78.85, 78.91) --
	( 79.26, 79.32) --
	( 79.68, 79.73) --
	( 80.09, 80.19) --
	( 80.51, 80.75) --
	( 80.92, 81.18) --
	( 81.33, 81.63) --
	( 81.75, 82.07) --
	( 82.16, 82.52) --
	( 82.58, 83.07) --
	( 82.99, 83.55) --
	( 83.40, 84.08) --
	( 83.82, 84.54) --
	( 84.23, 85.06) --
	( 84.65, 85.59) --
	( 85.06, 86.17) --
	( 85.47, 86.75) --
	( 85.89, 87.29) --
	( 86.30, 87.76) --
	( 86.72, 88.35) --
	( 87.13, 88.93) --
	( 87.54, 89.54) --
	( 87.96, 90.14) --
	( 88.37, 90.75) --
	( 88.79, 91.42) --
	( 89.20, 92.16) --
	( 89.61, 92.79) --
	( 90.03, 93.38) --
	( 90.44, 94.04) --
	( 90.86, 94.75) --
	( 91.27, 95.45) --
	( 91.68, 96.15) --
	( 92.10, 96.83) --
	( 92.51, 97.43) --
	( 92.93, 98.06) --
	( 93.34, 98.86) --
	( 93.75, 99.60) --
	( 94.17,100.39) --
	( 94.58,101.14) --
	( 94.99,101.94) --
	( 95.41,102.58) --
	( 95.82,103.40) --
	( 96.24,104.09) --
	( 96.65,104.80) --
	( 97.06,105.56) --
	( 97.48,106.33) --
	( 97.89,107.21) --
	( 98.31,107.98) --
	( 98.72,108.82) --
	( 99.13,109.57) --
	( 99.55,110.39) --
	( 99.96,111.18) --
	(100.38,111.98) --
	(100.79,112.78);

\draw[color=drawColor,line width= 0.8pt,dash pattern=on 1pt off 3pt ,line cap=round,line join=round,fill opacity=0.00,] ( 18.42, 20.44) --
	( 18.83, 22.04) --
	( 19.24, 23.62) --
	( 19.66, 25.14) --
	( 20.07, 26.39) --
	( 20.48, 27.63) --
	( 20.90, 28.86) --
	( 21.31, 30.07) --
	( 21.73, 31.34) --
	( 22.14, 33.11) --
	( 22.55, 34.43) --
	( 22.97, 35.77) --
	( 23.38, 36.94) --
	( 23.80, 38.10) --
	( 24.21, 39.22) --
	( 24.62, 40.34) --
	( 25.04, 41.35) --
	( 25.45, 42.19) --
	( 25.87, 43.03) --
	( 26.28, 43.86) --
	( 26.69, 44.76) --
	( 27.11, 45.60) --
	( 27.52, 46.55) --
	( 27.94, 47.56) --
	( 28.35, 48.32) --
	( 28.76, 49.07) --
	( 29.18, 49.81) --
	( 29.59, 50.50) --
	( 30.01, 51.11) --
	( 30.42, 51.79) --
	( 30.83, 52.38) --
	( 31.25, 52.92) --
	( 31.66, 53.51) --
	( 32.08, 54.03) --
	( 32.49, 54.49) --
	( 32.90, 54.98) --
	( 33.32, 55.48) --
	( 33.73, 55.93) --
	( 34.14, 56.28) --
	( 34.56, 56.71) --
	( 34.97, 57.05) --
	( 35.39, 57.35) --
	( 35.80, 57.66) --
	( 36.21, 57.93) --
	( 36.63, 58.19) --
	( 37.04, 58.45) --
	( 37.46, 58.65) --
	( 37.87, 58.85) --
	( 38.28, 59.07) --
	( 38.70, 59.22) --
	( 39.11, 59.37) --
	( 39.53, 59.54) --
	( 39.94, 59.76) --
	( 40.35, 59.96) --
	( 40.77, 60.12) --
	( 41.18, 60.25) --
	( 41.60, 60.41) --
	( 42.01, 60.58) --
	( 42.42, 60.76) --
	( 42.84, 60.88) --
	( 43.25, 60.99) --
	( 43.67, 61.07) --
	( 44.08, 61.17) --
	( 44.49, 61.25) --
	( 44.91, 61.32) --
	( 45.32, 61.39) --
	( 45.74, 61.40) --
	( 46.15, 61.45) --
	( 46.56, 61.47) --
	( 46.98, 61.48) --
	( 47.39, 61.47) --
	( 47.81, 61.49) --
	( 48.22, 61.46) --
	( 48.63, 61.47) --
	( 49.05, 61.50) --
	( 49.46, 61.54) --
	( 49.87, 61.50) --
	( 50.29, 61.50) --
	( 50.70, 61.51) --
	( 51.12, 61.57) --
	( 51.53, 61.59) --
	( 51.94, 61.59) --
	( 52.36, 61.55) --
	( 52.77, 61.54) --
	( 53.19, 61.53) --
	( 53.60, 61.49) --
	( 54.01, 61.46) --
	( 54.43, 61.39) --
	( 54.84, 61.35) --
	( 55.26, 61.27) --
	( 55.67, 61.19) --
	( 56.08, 61.09) --
	( 56.50, 61.06) --
	( 56.91, 60.98) --
	( 57.33, 60.97) --
	( 57.74, 60.91) --
	( 58.15, 60.87) --
	( 58.57, 60.81) --
	( 58.98, 60.75) --
	( 59.40, 60.72) --
	( 59.81, 60.70) --
	( 60.22, 60.72) --
	( 60.64, 60.71) --
	( 61.05, 60.68) --
	( 61.47, 60.65) --
	( 61.88, 60.67) --
	( 62.29, 60.67) --
	( 62.71, 60.68) --
	( 63.12, 60.69) --
	( 63.53, 60.76) --
	( 63.95, 60.78) --
	( 64.36, 60.77) --
	( 64.78, 60.76) --
	( 65.19, 60.77) --
	( 65.60, 60.78) --
	( 66.02, 60.80) --
	( 66.43, 60.88) --
	( 66.85, 60.93) --
	( 67.26, 60.96) --
	( 67.67, 61.00) --
	( 68.09, 61.08) --
	( 68.50, 61.12) --
	( 68.92, 61.13) --
	( 69.33, 61.13) --
	( 69.74, 61.15) --
	( 70.16, 61.16) --
	( 70.57, 61.22) --
	( 70.99, 61.24) --
	( 71.40, 61.30) --
	( 71.81, 61.33) --
	( 72.23, 61.33) --
	( 72.64, 61.32) --
	( 73.06, 61.27) --
	( 73.47, 61.30) --
	( 73.88, 61.28) --
	( 74.30, 61.27) --
	( 74.71, 61.29) --
	( 75.13, 61.19) --
	( 75.54, 61.14) --
	( 75.95, 61.12) --
	( 76.37, 61.10) --
	( 76.78, 61.03) --
	( 77.20, 61.03) --
	( 77.61, 60.98) --
	( 78.02, 61.01) --
	( 78.44, 60.96) --
	( 78.85, 60.87) --
	( 79.26, 60.84) --
	( 79.68, 60.83) --
	( 80.09, 60.76) --
	( 80.51, 60.73) --
	( 80.92, 60.76) --
	( 81.33, 60.61) --
	( 81.75, 60.45) --
	( 82.16, 60.34) --
	( 82.58, 60.20) --
	( 82.99, 60.22) --
	( 83.40, 60.10) --
	( 83.82, 60.08) --
	( 84.23, 59.88) --
	( 84.65, 59.75) --
	( 85.06, 59.63) --
	( 85.47, 59.47) --
	( 85.89, 59.28) --
	( 86.30, 59.10) --
	( 86.72, 59.02) --
	( 87.13, 58.98) --
	( 87.54, 58.85) --
	( 87.96, 58.71) --
	( 88.37, 58.67) --
	( 88.79, 58.42) --
	( 89.20, 58.26) --
	( 89.61, 58.21) --
	( 90.03, 58.12) --
	( 90.44, 57.98) --
	( 90.86, 57.87) --
	( 91.27, 57.72) --
	( 91.68, 57.56) --
	( 92.10, 57.42) --
	( 92.51, 57.28) --
	( 92.93, 57.15) --
	( 93.34, 57.01) --
	( 93.75, 56.80) --
	( 94.17, 56.56) --
	( 94.58, 56.31) --
	( 94.99, 56.08) --
	( 95.41, 55.89) --
	( 95.82, 55.65) --
	( 96.24, 55.51) --
	( 96.65, 55.32) --
	( 97.06, 55.12) --
	( 97.48, 54.87) --
	( 97.89, 54.61) --
	( 98.31, 54.34) --
	( 98.72, 54.07) --
	( 99.13, 53.80) --
	( 99.55, 53.53) --
	( 99.96, 53.26) --
	(100.38, 52.98) --
	(100.79, 52.71);

\draw[color=drawColor,line width= 0.8pt,dash pattern=on 1pt off 3pt ,line cap=round,line join=round,fill opacity=0.00,] ( 18.42, 59.66) --
	( 18.83, 59.62) --
	( 19.24, 59.39) --
	( 19.66, 59.35) --
	( 20.07, 59.25) --
	( 20.48, 59.24) --
	( 20.90, 59.22) --
	( 21.31, 59.18) --
	( 21.73, 59.17) --
	( 22.14, 59.15) --
	( 22.55, 59.12) --
	( 22.97, 59.08) --
	( 23.38, 59.04) --
	( 23.80, 59.03) --
	( 24.21, 59.07) --
	( 24.62, 59.12) --
	( 25.04, 59.14) --
	( 25.45, 59.18) --
	( 25.87, 59.23) --
	( 26.28, 59.28) --
	( 26.69, 59.34) --
	( 27.11, 59.39) --
	( 27.52, 59.44) --
	( 27.94, 59.49) --
	( 28.35, 59.55) --
	( 28.76, 59.60) --
	( 29.18, 59.66) --
	( 29.59, 59.74) --
	( 30.01, 59.80) --
	( 30.42, 59.88) --
	( 30.83, 60.01) --
	( 31.25, 60.09) --
	( 31.66, 60.21) --
	( 32.08, 60.40) --
	( 32.49, 60.49) --
	( 32.90, 60.71) --
	( 33.32, 60.89) --
	( 33.73, 61.21) --
	( 34.14, 61.42) --
	( 34.56, 61.76) --
	( 34.97, 62.09) --
	( 35.39, 62.41) --
	( 35.80, 62.73) --
	( 36.21, 63.03) --
	( 36.63, 63.40) --
	( 37.04, 63.75) --
	( 37.46, 64.13) --
	( 37.87, 64.41) --
	( 38.28, 64.73) --
	( 38.70, 65.05) --
	( 39.11, 65.35) --
	( 39.53, 65.54) --
	( 39.94, 65.82) --
	( 40.35, 65.99) --
	( 40.77, 66.12) --
	( 41.18, 66.24) --
	( 41.60, 66.32) --
	( 42.01, 66.39) --
	( 42.42, 66.37) --
	( 42.84, 66.35) --
	( 43.25, 66.41) --
	( 43.67, 66.40) --
	( 44.08, 66.46) --
	( 44.49, 66.49) --
	( 44.91, 66.52) --
	( 45.32, 66.53) --
	( 45.74, 66.58) --
	( 46.15, 66.63) --
	( 46.56, 66.71) --
	( 46.98, 66.71) --
	( 47.39, 66.74) --
	( 47.81, 66.77) --
	( 48.22, 66.84) --
	( 48.63, 66.85) --
	( 49.05, 66.88) --
	( 49.46, 66.91) --
	( 49.87, 66.93) --
	( 50.29, 66.93) --
	( 50.70, 66.95) --
	( 51.12, 66.92) --
	( 51.53, 66.91) --
	( 51.94, 66.94) --
	( 52.36, 66.95) --
	( 52.77, 66.97) --
	( 53.19, 67.02) --
	( 53.60, 67.10) --
	( 54.01, 67.20) --
	( 54.43, 67.39) --
	( 54.84, 67.62) --
	( 55.26, 67.73) --
	( 55.67, 67.77) --
	( 56.08, 67.95) --
	( 56.50, 68.08) --
	( 56.91, 68.16) --
	( 57.33, 68.30) --
	( 57.74, 68.43) --
	( 58.15, 68.57) --
	( 58.57, 68.74) --
	( 58.98, 68.94) --
	( 59.40, 69.11) --
	( 59.81, 69.29) --
	( 60.22, 69.50) --
	( 60.64, 69.70) --
	( 61.05, 69.88) --
	( 61.47, 70.08) --
	( 61.88, 70.27) --
	( 62.29, 70.51) --
	( 62.71, 70.71) --
	( 63.12, 70.87) --
	( 63.53, 71.06) --
	( 63.95, 71.25) --
	( 64.36, 71.45) --
	( 64.78, 71.63) --
	( 65.19, 71.82) --
	( 65.60, 72.04) --
	( 66.02, 72.23) --
	( 66.43, 72.42) --
	( 66.85, 72.60) --
	( 67.26, 72.66) --
	( 67.67, 72.72) --
	( 68.09, 72.94) --
	( 68.50, 73.06) --
	( 68.92, 73.18) --
	( 69.33, 73.42) --
	( 69.74, 73.64) --
	( 70.16, 73.98) --
	( 70.57, 74.27) --
	( 70.99, 74.55) --
	( 71.40, 74.75) --
	( 71.81, 74.98) --
	( 72.23, 75.30) --
	( 72.64, 75.62) --
	( 73.06, 76.05) --
	( 73.47, 76.50) --
	( 73.88, 77.06) --
	( 74.30, 77.75) --
	( 74.71, 78.26) --
	( 75.13, 78.71) --
	( 75.54, 79.33) --
	( 75.95, 79.98) --
	( 76.37, 80.64) --
	( 76.78, 81.31) --
	( 77.20, 81.99) --
	( 77.61, 82.69) --
	( 78.02, 83.39) --
	( 78.44, 84.11) --
	( 78.85, 84.83) --
	( 79.26, 85.56) --
	( 79.68, 86.31) --
	( 80.09, 87.06) --
	( 80.51, 87.83) --
	( 80.92, 88.60) --
	( 81.33, 89.39) --
	( 81.75, 90.18) --
	( 82.16, 90.98) --
	( 82.58, 91.69) --
	( 82.99, 92.40) --
	( 83.40, 93.12) --
	( 83.82, 93.84) --
	( 84.23, 94.57) --
	( 84.65, 95.31) --
	( 85.06, 96.05) --
	( 85.47, 96.80) --
	( 85.89, 97.56) --
	( 86.30, 98.32) --
	( 86.72, 99.09) --
	( 87.13, 99.97) --
	( 87.54,100.95) --
	( 87.96,101.93) --
	( 88.37,102.93) --
	( 88.79,103.94) --
	( 89.20,104.96) --
	( 89.61,105.98) --
	( 90.03,107.02) --
	( 90.44,108.07) --
	( 90.86,109.12) --
	( 91.27,110.19) --
	( 91.68,111.26) --
	( 92.10,112.35) --
	( 92.51,113.44) --
	( 92.93,114.55) --
	( 93.34,115.66) --
	( 93.75,116.78) --
	( 94.17,117.82) --
	( 94.58,118.85) --
	( 94.99,119.89) --
	( 95.41,120.93) --
	( 95.82,121.98) --
	( 96.24,123.03) --
	( 96.65,124.09) --
	( 97.06,125.16) --
	( 97.48,126.22) --
	( 97.89,127.30) --
	( 98.31,128.38) --
	( 98.72,129.46) --
	( 98.96,130.09);
\end{scope}
\end{tikzpicture}

%% file: out-pima-topInc-plot-x6.tikz
\begin{tikzpicture}[x=1pt,y=1pt]
\definecolor[named]{drawColor}{rgb}{0.00,0.00,0.00}
\definecolor[named]{fillColor}{rgb}{1.00,1.00,1.00}
\fill[color=fillColor,fill opacity=0.00,] (0,0) rectangle (108.41,130.09);
\begin{scope}
\path[clip] (  0.00,  0.00) rectangle (108.41,130.09);
\definecolor[named]{drawColor}{rgb}{0.00,0.00,0.00}

\draw[color=drawColor,line cap=round,line join=round,fill opacity=0.00,] ( 15.42, 33.12) -- (103.61, 33.12);

\draw[color=drawColor,line cap=round,line join=round,fill opacity=0.00,] ( 15.42, 33.12) -- ( 15.42, 29.52);

\draw[color=drawColor,line cap=round,line join=round,fill opacity=0.00,] ( 33.06, 33.12) -- ( 33.06, 29.52);

\draw[color=drawColor,line cap=round,line join=round,fill opacity=0.00,] ( 50.69, 33.12) -- ( 50.69, 29.52);

\draw[color=drawColor,line cap=round,line join=round,fill opacity=0.00,] ( 68.33, 33.12) -- ( 68.33, 29.52);

\draw[color=drawColor,line cap=round,line join=round,fill opacity=0.00,] ( 85.97, 33.12) -- ( 85.97, 29.52);

\draw[color=drawColor,line cap=round,line join=round,fill opacity=0.00,] (103.61, 33.12) -- (103.61, 29.52);

\node[color=drawColor,anchor=base,inner sep=0pt, outer sep=0pt, scale=  0.60] at ( 15.42, 18.72) {0.0};

\node[color=drawColor,anchor=base,inner sep=0pt, outer sep=0pt, scale=  0.60] at ( 33.06, 18.72) {0.5};

\node[color=drawColor,anchor=base,inner sep=0pt, outer sep=0pt, scale=  0.60] at ( 50.69, 18.72) {1.0};

\node[color=drawColor,anchor=base,inner sep=0pt, outer sep=0pt, scale=  0.60] at ( 68.33, 18.72) {1.5};

\node[color=drawColor,anchor=base,inner sep=0pt, outer sep=0pt, scale=  0.60] at ( 85.97, 18.72) {2.0};

\node[color=drawColor,anchor=base,inner sep=0pt, outer sep=0pt, scale=  0.60] at (103.61, 18.72) {2.5};

\draw[color=drawColor,line cap=round,line join=round,fill opacity=0.00,] ( 15.12, 36.15) -- ( 15.12,111.93);

\draw[color=drawColor,line cap=round,line join=round,fill opacity=0.00,] ( 15.12, 36.15) -- ( 11.52, 36.15);

\draw[color=drawColor,line cap=round,line join=round,fill opacity=0.00,] ( 15.12, 48.78) -- ( 11.52, 48.78);

\draw[color=drawColor,line cap=round,line join=round,fill opacity=0.00,] ( 15.12, 61.41) -- ( 11.52, 61.41);

\draw[color=drawColor,line cap=round,line join=round,fill opacity=0.00,] ( 15.12, 74.04) -- ( 11.52, 74.04);

\draw[color=drawColor,line cap=round,line join=round,fill opacity=0.00,] ( 15.12, 86.67) -- ( 11.52, 86.67);

\draw[color=drawColor,line cap=round,line join=round,fill opacity=0.00,] ( 15.12, 99.30) -- ( 11.52, 99.30);

\draw[color=drawColor,line cap=round,line join=round,fill opacity=0.00,] ( 15.12,111.93) -- ( 11.52,111.93);

\node[color=drawColor,anchor=base east,inner sep=0pt, outer sep=0pt, scale=  0.60] at (  7.92, 34.09) {-2};

\node[color=drawColor,anchor=base east,inner sep=0pt, outer sep=0pt, scale=  0.60] at (  7.92, 46.72) {-1};

\node[color=drawColor,anchor=base east,inner sep=0pt, outer sep=0pt, scale=  0.60] at (  7.92, 59.35) {0};

\node[color=drawColor,anchor=base east,inner sep=0pt, outer sep=0pt, scale=  0.60] at (  7.92, 71.98) {1};

\node[color=drawColor,anchor=base east,inner sep=0pt, outer sep=0pt, scale=  0.60] at (  7.92, 84.61) {2};

\node[color=drawColor,anchor=base east,inner sep=0pt, outer sep=0pt, scale=  0.60] at (  7.92, 97.24) {3};

\node[color=drawColor,anchor=base east,inner sep=0pt, outer sep=0pt, scale=  0.60] at (  7.92,109.87) {4};

\draw[color=drawColor,line cap=round,line join=round,fill opacity=0.00,] ( 15.12, 33.12) --
	(104.08, 33.12) --
	(104.08,114.97) --
	( 15.12,114.97) --
	( 15.12, 33.12);
\end{scope}
\begin{scope}
\path[clip] (  0.00,  0.00) rectangle (108.41,130.09);
\definecolor[named]{drawColor}{rgb}{0.00,0.00,0.00}

\node[color=drawColor,anchor=base,inner sep=0pt, outer sep=0pt, scale=  0.72] at ( 59.60,  4.32) {$x_{6}$};
\end{scope}
\begin{scope}
\path[clip] ( 15.12, 33.12) rectangle (104.08,114.97);
\definecolor[named]{drawColor}{rgb}{0.00,0.00,0.00}

\draw[color=drawColor,line width= 0.8pt,line cap=round,line join=round,fill opacity=0.00,] ( 18.42, 53.11) --
	( 18.83, 53.39) --
	( 19.24, 53.67) --
	( 19.66, 53.94) --
	( 20.07, 54.22) --
	( 20.48, 54.49) --
	( 20.90, 54.77) --
	( 21.31, 55.04) --
	( 21.73, 55.32) --
	( 22.14, 55.59) --
	( 22.55, 55.86) --
	( 22.97, 56.13) --
	( 23.38, 56.40) --
	( 23.80, 56.67) --
	( 24.21, 56.94) --
	( 24.62, 57.20) --
	( 25.04, 57.47) --
	( 25.45, 57.73) --
	( 25.87, 57.99) --
	( 26.28, 58.25) --
	( 26.69, 58.51) --
	( 27.11, 58.77) --
	( 27.52, 59.02) --
	( 27.94, 59.27) --
	( 28.35, 59.52) --
	( 28.76, 59.76) --
	( 29.18, 60.00) --
	( 29.59, 60.24) --
	( 30.01, 60.47) --
	( 30.42, 60.71) --
	( 30.83, 60.94) --
	( 31.25, 61.16) --
	( 31.66, 61.39) --
	( 32.08, 61.61) --
	( 32.49, 61.84) --
	( 32.90, 62.06) --
	( 33.32, 62.27) --
	( 33.73, 62.49) --
	( 34.14, 62.71) --
	( 34.56, 62.92) --
	( 34.97, 63.14) --
	( 35.39, 63.35) --
	( 35.80, 63.57) --
	( 36.21, 63.78) --
	( 36.63, 63.99) --
	( 37.04, 64.20) --
	( 37.46, 64.40) --
	( 37.87, 64.61) --
	( 38.28, 64.82) --
	( 38.70, 65.02) --
	( 39.11, 65.22) --
	( 39.53, 65.42) --
	( 39.94, 65.62) --
	( 40.35, 65.82) --
	( 40.77, 66.01) --
	( 41.18, 66.20) --
	( 41.60, 66.39) --
	( 42.01, 66.58) --
	( 42.42, 66.77) --
	( 42.84, 66.95) --
	( 43.25, 67.14) --
	( 43.67, 67.32) --
	( 44.08, 67.49) --
	( 44.49, 67.67) --
	( 44.91, 67.84) --
	( 45.32, 68.01) --
	( 45.74, 68.18) --
	( 46.15, 68.35) --
	( 46.56, 68.51) --
	( 46.98, 68.67) --
	( 47.39, 68.83) --
	( 47.81, 68.99) --
	( 48.22, 69.15) --
	( 48.63, 69.30) --
	( 49.05, 69.45) --
	( 49.46, 69.60) --
	( 49.87, 69.75) --
	( 50.29, 69.89) --
	( 50.70, 70.03) --
	( 51.12, 70.18) --
	( 51.53, 70.31) --
	( 51.94, 70.45) --
	( 52.36, 70.59) --
	( 52.77, 70.72) --
	( 53.19, 70.85) --
	( 53.60, 70.98) --
	( 54.01, 71.10) --
	( 54.43, 71.23) --
	( 54.84, 71.35) --
	( 55.26, 71.47) --
	( 55.67, 71.59) --
	( 56.08, 71.71) --
	( 56.50, 71.82) --
	( 56.91, 71.94) --
	( 57.33, 72.05) --
	( 57.74, 72.15) --
	( 58.15, 72.26) --
	( 58.57, 72.37) --
	( 58.98, 72.47) --
	( 59.40, 72.57) --
	( 59.81, 72.67) --
	( 60.22, 72.77) --
	( 60.64, 72.86) --
	( 61.05, 72.96) --
	( 61.47, 73.05) --
	( 61.88, 73.14) --
	( 62.29, 73.23) --
	( 62.71, 73.32) --
	( 63.12, 73.40) --
	( 63.53, 73.48) --
	( 63.95, 73.57) --
	( 64.36, 73.64) --
	( 64.78, 73.72) --
	( 65.19, 73.80) --
	( 65.60, 73.87) --
	( 66.02, 73.95) --
	( 66.43, 74.02) --
	( 66.85, 74.09) --
	( 67.26, 74.15) --
	( 67.67, 74.22) --
	( 68.09, 74.28) --
	( 68.50, 74.34) --
	( 68.92, 74.41) --
	( 69.33, 74.46) --
	( 69.74, 74.52) --
	( 70.16, 74.58) --
	( 70.57, 74.63) --
	( 70.99, 74.68) --
	( 71.40, 74.74) --
	( 71.81, 74.78) --
	( 72.23, 74.83) --
	( 72.64, 74.88) --
	( 73.06, 74.92) --
	( 73.47, 74.97) --
	( 73.88, 75.01) --
	( 74.30, 75.05) --
	( 74.71, 75.09) --
	( 75.13, 75.12) --
	( 75.54, 75.16) --
	( 75.95, 75.19) --
	( 76.37, 75.22) --
	( 76.78, 75.25) --
	( 77.20, 75.28) --
	( 77.61, 75.31) --
	( 78.02, 75.34) --
	( 78.44, 75.36) --
	( 78.85, 75.39) --
	( 79.26, 75.41) --
	( 79.68, 75.43) --
	( 80.09, 75.45) --
	( 80.51, 75.47) --
	( 80.92, 75.48) --
	( 81.33, 75.50) --
	( 81.75, 75.51) --
	( 82.16, 75.52) --
	( 82.58, 75.54) --
	( 82.99, 75.54) --
	( 83.40, 75.55) --
	( 83.82, 75.56) --
	( 84.23, 75.57) --
	( 84.65, 75.57) --
	( 85.06, 75.57) --
	( 85.47, 75.57) --
	( 85.89, 75.58) --
	( 86.30, 75.57) --
	( 86.72, 75.57) --
	( 87.13, 75.57) --
	( 87.54, 75.56) --
	( 87.96, 75.56) --
	( 88.37, 75.55) --
	( 88.79, 75.54) --
	( 89.20, 75.53) --
	( 89.61, 75.52) --
	( 90.03, 75.51) --
	( 90.44, 75.50) --
	( 90.86, 75.49) --
	( 91.27, 75.47) --
	( 91.68, 75.45) --
	( 92.10, 75.44) --
	( 92.51, 75.42) --
	( 92.93, 75.40) --
	( 93.34, 75.38) --
	( 93.75, 75.35) --
	( 94.17, 75.33) --
	( 94.58, 75.31) --
	( 94.99, 75.28) --
	( 95.41, 75.26) --
	( 95.82, 75.23) --
	( 96.24, 75.20) --
	( 96.65, 75.17) --
	( 97.06, 75.14) --
	( 97.48, 75.11) --
	( 97.89, 75.08) --
	( 98.31, 75.04) --
	( 98.72, 75.01) --
	( 99.13, 74.97) --
	( 99.55, 74.94) --
	( 99.96, 74.90) --
	(100.38, 74.86) --
	(100.79, 74.82);

\draw[color=drawColor,line width= 0.8pt,dash pattern=on 4pt off 4pt ,line cap=round,line join=round,fill opacity=0.00,] ( 18.42, 45.34) --
	( 18.83, 46.09) --
	( 19.24, 46.81) --
	( 19.66, 47.43) --
	( 20.07, 48.11) --
	( 20.48, 48.84) --
	( 20.90, 49.45) --
	( 21.31, 50.02) --
	( 21.73, 50.69) --
	( 22.14, 51.29) --
	( 22.55, 51.91) --
	( 22.97, 52.47) --
	( 23.38, 53.01) --
	( 23.80, 53.58) --
	( 24.21, 54.10) --
	( 24.62, 54.58) --
	( 25.04, 55.00) --
	( 25.45, 55.42) --
	( 25.87, 55.81) --
	( 26.28, 56.18) --
	( 26.69, 56.51) --
	( 27.11, 56.83) --
	( 27.52, 57.11) --
	( 27.94, 57.38) --
	( 28.35, 57.67) --
	( 28.76, 57.94) --
	( 29.18, 58.24) --
	( 29.59, 58.49) --
	( 30.01, 58.75) --
	( 30.42, 58.98) --
	( 30.83, 59.21) --
	( 31.25, 59.41) --
	( 31.66, 59.61) --
	( 32.08, 59.80) --
	( 32.49, 60.01) --
	( 32.90, 60.15) --
	( 33.32, 60.34) --
	( 33.73, 60.49) --
	( 34.14, 60.63) --
	( 34.56, 60.81) --
	( 34.97, 60.99) --
	( 35.39, 61.19) --
	( 35.80, 61.34) --
	( 36.21, 61.51) --
	( 36.63, 61.69) --
	( 37.04, 61.86) --
	( 37.46, 62.00) --
	( 37.87, 62.16) --
	( 38.28, 62.30) --
	( 38.70, 62.45) --
	( 39.11, 62.58) --
	( 39.53, 62.69) --
	( 39.94, 62.82) --
	( 40.35, 62.92) --
	( 40.77, 63.03) --
	( 41.18, 63.15) --
	( 41.60, 63.26) --
	( 42.01, 63.35) --
	( 42.42, 63.44) --
	( 42.84, 63.54) --
	( 43.25, 63.59) --
	( 43.67, 63.67) --
	( 44.08, 63.77) --
	( 44.49, 63.85) --
	( 44.91, 63.93) --
	( 45.32, 64.01) --
	( 45.74, 64.11) --
	( 46.15, 64.18) --
	( 46.56, 64.30) --
	( 46.98, 64.38) --
	( 47.39, 64.46) --
	( 47.81, 64.52) --
	( 48.22, 64.58) --
	( 48.63, 64.66) --
	( 49.05, 64.73) --
	( 49.46, 64.82) --
	( 49.87, 64.89) --
	( 50.29, 64.94) --
	( 50.70, 65.00) --
	( 51.12, 65.06) --
	( 51.53, 65.11) --
	( 51.94, 65.17) --
	( 52.36, 65.24) --
	( 52.77, 65.29) --
	( 53.19, 65.35) --
	( 53.60, 65.41) --
	( 54.01, 65.49) --
	( 54.43, 65.51) --
	( 54.84, 65.55) --
	( 55.26, 65.57) --
	( 55.67, 65.60) --
	( 56.08, 65.62) --
	( 56.50, 65.64) --
	( 56.91, 65.67) --
	( 57.33, 65.68) --
	( 57.74, 65.72) --
	( 58.15, 65.72) --
	( 58.57, 65.73) --
	( 58.98, 65.73) --
	( 59.40, 65.74) --
	( 59.81, 65.75) --
	( 60.22, 65.75) --
	( 60.64, 65.79) --
	( 61.05, 65.76) --
	( 61.47, 65.79) --
	( 61.88, 65.76) --
	( 62.29, 65.73) --
	( 62.71, 65.70) --
	( 63.12, 65.67) --
	( 63.53, 65.64) --
	( 63.95, 65.62) --
	( 64.36, 65.60) --
	( 64.78, 65.52) --
	( 65.19, 65.45) --
	( 65.60, 65.37) --
	( 66.02, 65.27) --
	( 66.43, 65.18) --
	( 66.85, 65.06) --
	( 67.26, 64.98) --
	( 67.67, 64.87) --
	( 68.09, 64.76) --
	( 68.50, 64.65) --
	( 68.92, 64.51) --
	( 69.33, 64.35) --
	( 69.74, 64.24) --
	( 70.16, 64.07) --
	( 70.57, 63.91) --
	( 70.99, 63.77) --
	( 71.40, 63.56) --
	( 71.81, 63.35) --
	( 72.23, 63.17) --
	( 72.64, 63.04) --
	( 73.06, 62.87) --
	( 73.47, 62.69) --
	( 73.88, 62.49) --
	( 74.30, 62.28) --
	( 74.71, 62.07) --
	( 75.13, 61.84) --
	( 75.54, 61.66) --
	( 75.95, 61.48) --
	( 76.37, 61.29) --
	( 76.78, 61.09) --
	( 77.20, 60.84) --
	( 77.61, 60.60) --
	( 78.02, 60.38) --
	( 78.44, 60.12) --
	( 78.85, 59.89) --
	( 79.26, 59.55) --
	( 79.68, 59.27) --
	( 80.09, 59.02) --
	( 80.51, 58.73) --
	( 80.92, 58.42) --
	( 81.33, 58.12) --
	( 81.75, 57.79) --
	( 82.16, 57.49) --
	( 82.58, 57.15) --
	( 82.99, 56.86) --
	( 83.40, 56.54) --
	( 83.82, 56.21) --
	( 84.23, 55.90) --
	( 84.65, 55.49) --
	( 85.06, 55.14) --
	( 85.47, 54.86) --
	( 85.89, 54.52) --
	( 86.30, 54.11) --
	( 86.72, 53.73) --
	( 87.13, 53.36) --
	( 87.54, 53.05) --
	( 87.96, 52.72) --
	( 88.37, 52.28) --
	( 88.79, 51.88) --
	( 89.20, 51.54) --
	( 89.61, 51.13) --
	( 90.03, 50.68) --
	( 90.44, 50.17) --
	( 90.86, 49.79) --
	( 91.27, 49.40) --
	( 91.68, 48.90) --
	( 92.10, 48.40) --
	( 92.51, 47.99) --
	( 92.93, 47.45) --
	( 93.34, 47.01) --
	( 93.75, 46.58) --
	( 94.17, 46.15) --
	( 94.58, 45.57) --
	( 94.99, 45.07) --
	( 95.41, 44.56) --
	( 95.82, 44.13) --
	( 96.24, 43.68) --
	( 96.65, 43.14) --
	( 97.06, 42.55) --
	( 97.48, 42.07) --
	( 97.89, 41.51) --
	( 98.31, 41.04) --
	( 98.72, 40.29) --
	( 99.13, 39.67) --
	( 99.55, 39.06) --
	( 99.96, 38.40) --
	(100.38, 37.93) --
	(100.79, 37.20);

\draw[color=drawColor,line width= 0.8pt,dash pattern=on 4pt off 4pt ,line cap=round,line join=round,fill opacity=0.00,] ( 18.42, 58.77) --
	( 18.83, 58.76) --
	( 19.24, 58.79) --
	( 19.66, 58.80) --
	( 20.07, 58.85) --
	( 20.48, 58.93) --
	( 20.90, 58.99) --
	( 21.31, 59.02) --
	( 21.73, 59.05) --
	( 22.14, 59.10) --
	( 22.55, 59.16) --
	( 22.97, 59.23) --
	( 23.38, 59.30) --
	( 23.80, 59.40) --
	( 24.21, 59.51) --
	( 24.62, 59.62) --
	( 25.04, 59.75) --
	( 25.45, 59.89) --
	( 25.87, 60.07) --
	( 26.28, 60.29) --
	( 26.69, 60.54) --
	( 27.11, 60.81) --
	( 27.52, 61.10) --
	( 27.94, 61.41) --
	( 28.35, 61.75) --
	( 28.76, 62.04) --
	( 29.18, 62.33) --
	( 29.59, 62.60) --
	( 30.01, 62.89) --
	( 30.42, 63.16) --
	( 30.83, 63.41) --
	( 31.25, 63.72) --
	( 31.66, 64.00) --
	( 32.08, 64.28) --
	( 32.49, 64.60) --
	( 32.90, 64.85) --
	( 33.32, 65.14) --
	( 33.73, 65.42) --
	( 34.14, 65.68) --
	( 34.56, 65.93) --
	( 34.97, 66.19) --
	( 35.39, 66.52) --
	( 35.80, 66.78) --
	( 36.21, 67.08) --
	( 36.63, 67.35) --
	( 37.04, 67.60) --
	( 37.46, 67.82) --
	( 37.87, 68.05) --
	( 38.28, 68.35) --
	( 38.70, 68.62) --
	( 39.11, 68.86) --
	( 39.53, 69.14) --
	( 39.94, 69.38) --
	( 40.35, 69.63) --
	( 40.77, 69.86) --
	( 41.18, 70.07) --
	( 41.60, 70.32) --
	( 42.01, 70.60) --
	( 42.42, 70.83) --
	( 42.84, 71.11) --
	( 43.25, 71.37) --
	( 43.67, 71.64) --
	( 44.08, 71.87) --
	( 44.49, 72.11) --
	( 44.91, 72.38) --
	( 45.32, 72.66) --
	( 45.74, 72.90) --
	( 46.15, 73.13) --
	( 46.56, 73.41) --
	( 46.98, 73.66) --
	( 47.39, 73.91) --
	( 47.81, 74.17) --
	( 48.22, 74.42) --
	( 48.63, 74.64) --
	( 49.05, 74.87) --
	( 49.46, 75.11) --
	( 49.87, 75.32) --
	( 50.29, 75.53) --
	( 50.70, 75.76) --
	( 51.12, 75.92) --
	( 51.53, 76.09) --
	( 51.94, 76.28) --
	( 52.36, 76.51) --
	( 52.77, 76.70) --
	( 53.19, 76.91) --
	( 53.60, 77.10) --
	( 54.01, 77.27) --
	( 54.43, 77.45) --
	( 54.84, 77.64) --
	( 55.26, 77.80) --
	( 55.67, 77.99) --
	( 56.08, 78.19) --
	( 56.50, 78.40) --
	( 56.91, 78.60) --
	( 57.33, 78.77) --
	( 57.74, 78.94) --
	( 58.15, 79.17) --
	( 58.57, 79.38) --
	( 58.98, 79.56) --
	( 59.40, 79.74) --
	( 59.81, 79.98) --
	( 60.22, 80.18) --
	( 60.64, 80.35) --
	( 61.05, 80.53) --
	( 61.47, 80.72) --
	( 61.88, 80.92) --
	( 62.29, 81.09) --
	( 62.71, 81.30) --
	( 63.12, 81.46) --
	( 63.53, 81.68) --
	( 63.95, 81.88) --
	( 64.36, 82.07) --
	( 64.78, 82.29) --
	( 65.19, 82.53) --
	( 65.60, 82.77) --
	( 66.02, 82.99) --
	( 66.43, 83.18) --
	( 66.85, 83.39) --
	( 67.26, 83.57) --
	( 67.67, 83.78) --
	( 68.09, 84.03) --
	( 68.50, 84.27) --
	( 68.92, 84.50) --
	( 69.33, 84.75) --
	( 69.74, 85.00) --
	( 70.16, 85.21) --
	( 70.57, 85.48) --
	( 70.99, 85.74) --
	( 71.40, 86.00) --
	( 71.81, 86.29) --
	( 72.23, 86.53) --
	( 72.64, 86.76) --
	( 73.06, 86.93) --
	( 73.47, 87.17) --
	( 73.88, 87.43) --
	( 74.30, 87.71) --
	( 74.71, 87.95) --
	( 75.13, 88.18) --
	( 75.54, 88.45) --
	( 75.95, 88.71) --
	( 76.37, 88.95) --
	( 76.78, 89.21) --
	( 77.20, 89.47) --
	( 77.61, 89.75) --
	( 78.02, 90.00) --
	( 78.44, 90.26) --
	( 78.85, 90.52) --
	( 79.26, 90.78) --
	( 79.68, 91.03) --
	( 80.09, 91.29) --
	( 80.51, 91.56) --
	( 80.92, 91.84) --
	( 81.33, 92.11) --
	( 81.75, 92.38) --
	( 82.16, 92.65) --
	( 82.58, 92.93) --
	( 82.99, 93.16) --
	( 83.40, 93.41) --
	( 83.82, 93.62) --
	( 84.23, 93.88) --
	( 84.65, 94.15) --
	( 85.06, 94.46) --
	( 85.47, 94.74) --
	( 85.89, 95.01) --
	( 86.30, 95.28) --
	( 86.72, 95.53) --
	( 87.13, 95.81) --
	( 87.54, 96.11) --
	( 87.96, 96.43) --
	( 88.37, 96.71) --
	( 88.79, 96.96) --
	( 89.20, 97.25) --
	( 89.61, 97.51) --
	( 90.03, 97.78) --
	( 90.44, 98.04) --
	( 90.86, 98.30) --
	( 91.27, 98.60) --
	( 91.68, 98.88) --
	( 92.10, 99.21) --
	( 92.51, 99.53) --
	( 92.93, 99.92) --
	( 93.34,100.21) --
	( 93.75,100.51) --
	( 94.17,100.77) --
	( 94.58,101.05) --
	( 94.99,101.31) --
	( 95.41,101.63) --
	( 95.82,101.88) --
	( 96.24,102.14) --
	( 96.65,102.40) --
	( 97.06,102.66) --
	( 97.48,102.92) --
	( 97.89,103.21) --
	( 98.31,103.49) --
	( 98.72,103.80) --
	( 99.13,104.13) --
	( 99.55,104.41) --
	( 99.96,104.76) --
	(100.38,105.11) --
	(100.79,105.46);

\draw[color=drawColor,line width= 0.8pt,dash pattern=on 1pt off 3pt ,line cap=round,line join=round,fill opacity=0.00,] ( 18.42, 39.94) --
	( 18.83, 41.16) --
	( 19.24, 42.36) --
	( 19.66, 43.52) --
	( 20.07, 44.64) --
	( 20.48, 45.77) --
	( 20.90, 46.83) --
	( 21.31, 47.74) --
	( 21.73, 48.50) --
	( 22.14, 49.42) --
	( 22.55, 50.28) --
	( 22.97, 50.97) --
	( 23.38, 51.76) --
	( 23.80, 52.35) --
	( 24.21, 52.95) --
	( 24.62, 53.46) --
	( 25.04, 53.99) --
	( 25.45, 54.40) --
	( 25.87, 54.79) --
	( 26.28, 55.16) --
	( 26.69, 55.47) --
	( 27.11, 55.74) --
	( 27.52, 55.97) --
	( 27.94, 56.20) --
	( 28.35, 56.34) --
	( 28.76, 56.62) --
	( 29.18, 56.83) --
	( 29.59, 57.08) --
	( 30.01, 57.37) --
	( 30.42, 57.69) --
	( 30.83, 57.92) --
	( 31.25, 58.21) --
	( 31.66, 58.42) --
	( 32.08, 58.63) --
	( 32.49, 58.83) --
	( 32.90, 58.93) --
	( 33.32, 58.99) --
	( 33.73, 59.07) --
	( 34.14, 59.10) --
	( 34.56, 59.34) --
	( 34.97, 59.53) --
	( 35.39, 59.71) --
	( 35.80, 59.91) --
	( 36.21, 60.13) --
	( 36.63, 60.36) --
	( 37.04, 60.59) --
	( 37.46, 60.82) --
	( 37.87, 60.94) --
	( 38.28, 61.07) --
	( 38.70, 61.22) --
	( 39.11, 61.39) --
	( 39.53, 61.59) --
	( 39.94, 61.70) --
	( 40.35, 61.81) --
	( 40.77, 61.97) --
	( 41.18, 62.13) --
	( 41.60, 62.26) --
	( 42.01, 62.30) --
	( 42.42, 62.35) --
	( 42.84, 62.39) --
	( 43.25, 62.43) --
	( 43.67, 62.49) --
	( 44.08, 62.56) --
	( 44.49, 62.61) --
	( 44.91, 62.72) --
	( 45.32, 62.82) --
	( 45.74, 62.87) --
	( 46.15, 62.93) --
	( 46.56, 62.95) --
	( 46.98, 62.96) --
	( 47.39, 63.03) --
	( 47.81, 63.06) --
	( 48.22, 63.05) --
	( 48.63, 63.05) --
	( 49.05, 63.10) --
	( 49.46, 63.13) --
	( 49.87, 63.16) --
	( 50.29, 63.21) --
	( 50.70, 63.24) --
	( 51.12, 63.23) --
	( 51.53, 63.26) --
	( 51.94, 63.26) --
	( 52.36, 63.21) --
	( 52.77, 63.19) --
	( 53.19, 63.18) --
	( 53.60, 63.12) --
	( 54.01, 63.18) --
	( 54.43, 63.28) --
	( 54.84, 63.27) --
	( 55.26, 63.25) --
	( 55.67, 63.22) --
	( 56.08, 63.21) --
	( 56.50, 63.17) --
	( 56.91, 63.17) --
	( 57.33, 63.13) --
	( 57.74, 63.10) --
	( 58.15, 63.04) --
	( 58.57, 63.01) --
	( 58.98, 62.99) --
	( 59.40, 62.96) --
	( 59.81, 62.91) --
	( 60.22, 62.88) --
	( 60.64, 62.81) --
	( 61.05, 62.64) --
	( 61.47, 62.53) --
	( 61.88, 62.50) --
	( 62.29, 62.47) --
	( 62.71, 62.41) --
	( 63.12, 62.34) --
	( 63.53, 62.30) --
	( 63.95, 62.24) --
	( 64.36, 62.10) --
	( 64.78, 61.99) --
	( 65.19, 61.85) --
	( 65.60, 61.74) --
	( 66.02, 61.63) --
	( 66.43, 61.50) --
	( 66.85, 61.30) --
	( 67.26, 61.12) --
	( 67.67, 60.85) --
	( 68.09, 60.78) --
	( 68.50, 60.59) --
	( 68.92, 60.43) --
	( 69.33, 60.27) --
	( 69.74, 60.09) --
	( 70.16, 59.93) --
	( 70.57, 59.73) --
	( 70.99, 59.52) --
	( 71.40, 59.29) --
	( 71.81, 59.13) --
	( 72.23, 58.94) --
	( 72.64, 58.74) --
	( 73.06, 58.58) --
	( 73.47, 58.28) --
	( 73.88, 57.96) --
	( 74.30, 57.65) --
	( 74.71, 57.29) --
	( 75.13, 57.01) --
	( 75.54, 56.77) --
	( 75.95, 56.52) --
	( 76.37, 56.24) --
	( 76.78, 55.83) --
	( 77.20, 55.62) --
	( 77.61, 55.33) --
	( 78.02, 54.96) --
	( 78.44, 54.50) --
	( 78.85, 54.04) --
	( 79.26, 53.58) --
	( 79.68, 53.13) --
	( 80.09, 52.67) --
	( 80.51, 52.23) --
	( 80.92, 51.78) --
	( 81.33, 51.35) --
	( 81.75, 50.93) --
	( 82.16, 50.64) --
	( 82.58, 50.46) --
	( 82.99, 50.18) --
	( 83.40, 49.49) --
	( 83.82, 49.09) --
	( 84.23, 48.77) --
	( 84.65, 48.36) --
	( 85.06, 47.89) --
	( 85.47, 47.44) --
	( 85.89, 47.15) --
	( 86.30, 46.61) --
	( 86.72, 46.21) --
	( 87.13, 45.74) --
	( 87.54, 45.28) --
	( 87.96, 44.72) --
	( 88.37, 44.07) --
	( 88.79, 43.49) --
	( 89.20, 42.91) --
	( 89.61, 42.40) --
	( 90.03, 41.78) --
	( 90.44, 41.24) --
	( 90.86, 40.65) --
	( 91.27, 40.06) --
	( 91.68, 39.50) --
	( 92.10, 38.94) --
	( 92.51, 38.26) --
	( 92.93, 37.58) --
	( 93.34, 37.03) --
	( 93.75, 36.38) --
	( 94.17, 35.85) --
	( 94.58, 35.36) --
	( 94.99, 34.87) --
	( 95.41, 34.19) --
	( 95.82, 33.58) --
	( 96.24, 32.82) --
	( 96.65, 32.17) --
	( 97.06, 31.32) --
	( 97.48, 30.70) --
	( 97.89, 29.66) --
	( 98.31, 29.13) --
	( 98.72, 28.28) --
	( 99.13, 27.52) --
	( 99.55, 26.51) --
	( 99.96, 25.49) --
	(100.38, 24.58) --
	(100.79, 23.68);

\draw[color=drawColor,line width= 0.8pt,dash pattern=on 1pt off 3pt ,line cap=round,line join=round,fill opacity=0.00,] ( 18.42, 61.47) --
	( 18.83, 61.22) --
	( 19.24, 60.97) --
	( 19.66, 60.80) --
	( 20.07, 60.49) --
	( 20.48, 60.33) --
	( 20.90, 60.31) --
	( 21.31, 60.26) --
	( 21.73, 60.21) --
	( 22.14, 60.15) --
	( 22.55, 60.17) --
	( 22.97, 60.19) --
	( 23.38, 60.28) --
	( 23.80, 60.34) --
	( 24.21, 60.41) --
	( 24.62, 60.51) --
	( 25.04, 60.54) --
	( 25.45, 60.76) --
	( 25.87, 61.05) --
	( 26.28, 61.37) --
	( 26.69, 61.79) --
	( 27.11, 62.28) --
	( 27.52, 62.72) --
	( 27.94, 63.21) --
	( 28.35, 63.72) --
	( 28.76, 64.00) --
	( 29.18, 64.37) --
	( 29.59, 64.62) --
	( 30.01, 64.85) --
	( 30.42, 65.02) --
	( 30.83, 65.22) --
	( 31.25, 65.38) --
	( 31.66, 65.64) --
	( 32.08, 65.91) --
	( 32.49, 66.15) --
	( 32.90, 66.52) --
	( 33.32, 66.79) --
	( 33.73, 67.03) --
	( 34.14, 67.47) --
	( 34.56, 67.83) --
	( 34.97, 68.11) --
	( 35.39, 68.38) --
	( 35.80, 68.68) --
	( 36.21, 68.90) --
	( 36.63, 69.11) --
	( 37.04, 69.30) --
	( 37.46, 69.53) --
	( 37.87, 69.78) --
	( 38.28, 70.07) --
	( 38.70, 70.36) --
	( 39.11, 70.61) --
	( 39.53, 70.85) --
	( 39.94, 71.14) --
	( 40.35, 71.30) --
	( 40.77, 71.61) --
	( 41.18, 71.89) --
	( 41.60, 72.26) --
	( 42.01, 72.53) --
	( 42.42, 72.88) --
	( 42.84, 73.19) --
	( 43.25, 73.48) --
	( 43.67, 73.76) --
	( 44.08, 74.04) --
	( 44.49, 74.26) --
	( 44.91, 74.55) --
	( 45.32, 74.80) --
	( 45.74, 75.04) --
	( 46.15, 75.33) --
	( 46.56, 75.63) --
	( 46.98, 75.88) --
	( 47.39, 76.15) --
	( 47.81, 76.39) --
	( 48.22, 76.70) --
	( 48.63, 76.91) --
	( 49.05, 77.09) --
	( 49.46, 77.27) --
	( 49.87, 77.44) --
	( 50.29, 77.61) --
	( 50.70, 77.82) --
	( 51.12, 78.01) --
	( 51.53, 78.24) --
	( 51.94, 78.46) --
	( 52.36, 78.70) --
	( 52.77, 78.91) --
	( 53.19, 79.08) --
	( 53.60, 79.32) --
	( 54.01, 79.58) --
	( 54.43, 79.69) --
	( 54.84, 79.87) --
	( 55.26, 80.05) --
	( 55.67, 80.32) --
	( 56.08, 80.59) --
	( 56.50, 80.80) --
	( 56.91, 81.05) --
	( 57.33, 81.22) --
	( 57.74, 81.49) --
	( 58.15, 81.67) --
	( 58.57, 81.87) --
	( 58.98, 82.07) --
	( 59.40, 82.32) --
	( 59.81, 82.60) --
	( 60.22, 82.70) --
	( 60.64, 82.91) --
	( 61.05, 83.18) --
	( 61.47, 83.41) --
	( 61.88, 83.62) --
	( 62.29, 83.83) --
	( 62.71, 84.08) --
	( 63.12, 84.30) --
	( 63.53, 84.54) --
	( 63.95, 84.87) --
	( 64.36, 85.21) --
	( 64.78, 85.53) --
	( 65.19, 85.75) --
	( 65.60, 85.97) --
	( 66.02, 86.17) --
	( 66.43, 86.45) --
	( 66.85, 86.84) --
	( 67.26, 87.22) --
	( 67.67, 87.47) --
	( 68.09, 87.67) --
	( 68.50, 87.88) --
	( 68.92, 88.10) --
	( 69.33, 88.40) --
	( 69.74, 88.71) --
	( 70.16, 89.02) --
	( 70.57, 89.33) --
	( 70.99, 89.64) --
	( 71.40, 89.95) --
	( 71.81, 90.29) --
	( 72.23, 90.60) --
	( 72.64, 90.91) --
	( 73.06, 91.22) --
	( 73.47, 91.53) --
	( 73.88, 91.83) --
	( 74.30, 92.14) --
	( 74.71, 92.45) --
	( 75.13, 92.76) --
	( 75.54, 93.07) --
	( 75.95, 93.38) --
	( 76.37, 93.75) --
	( 76.78, 94.15) --
	( 77.20, 94.46) --
	( 77.61, 94.77) --
	( 78.02, 95.09) --
	( 78.44, 95.36) --
	( 78.85, 95.56) --
	( 79.26, 95.84) --
	( 79.68, 96.02) --
	( 80.09, 96.18) --
	( 80.51, 96.37) --
	( 80.92, 96.81) --
	( 81.33, 97.28) --
	( 81.75, 97.76) --
	( 82.16, 98.17) --
	( 82.58, 98.48) --
	( 82.99, 98.81) --
	( 83.40, 99.12) --
	( 83.82, 99.43) --
	( 84.23, 99.87) --
	( 84.65,100.34) --
	( 85.06,100.82) --
	( 85.47,101.29) --
	( 85.89,101.65) --
	( 86.30,102.02) --
	( 86.72,102.36) --
	( 87.13,102.68) --
	( 87.54,103.00) --
	( 87.96,103.50) --
	( 88.37,103.82) --
	( 88.79,104.15) --
	( 89.20,104.53) --
	( 89.61,104.85) --
	( 90.03,105.17) --
	( 90.44,105.50) --
	( 90.86,105.99) --
	( 91.27,106.48) --
	( 91.68,106.85) --
	( 92.10,107.19) --
	( 92.51,107.51) --
	( 92.93,107.83) --
	( 93.34,108.15) --
	( 93.75,108.47) --
	( 94.17,108.80) --
	( 94.58,109.12) --
	( 94.99,109.44) --
	( 95.41,109.76) --
	( 95.82,110.10) --
	( 96.24,110.48) --
	( 96.65,110.84) --
	( 97.06,111.24) --
	( 97.48,111.72) --
	( 97.89,112.18) --
	( 98.31,112.65) --
	( 98.72,113.40) --
	( 99.13,113.86) --
	( 99.55,114.31) --
	( 99.96,114.76) --
	(100.38,115.20) --
	(100.79,115.54);
\end{scope}
\end{tikzpicture}

%% file: out-pima-topInc-plot-x7.tikz
\begin{tikzpicture}[x=1pt,y=1pt]
\definecolor[named]{drawColor}{rgb}{0.00,0.00,0.00}
\definecolor[named]{fillColor}{rgb}{1.00,1.00,1.00}
\fill[color=fillColor,fill opacity=0.00,] (0,0) rectangle (108.41,130.09);
\begin{scope}
\path[clip] (  0.00,  0.00) rectangle (108.41,130.09);
\definecolor[named]{drawColor}{rgb}{0.00,0.00,0.00}

\draw[color=drawColor,line cap=round,line join=round,fill opacity=0.00,] ( 17.04, 33.12) -- ( 99.42, 33.12);

\draw[color=drawColor,line cap=round,line join=round,fill opacity=0.00,] ( 17.04, 33.12) -- ( 17.04, 29.52);

\draw[color=drawColor,line cap=round,line join=round,fill opacity=0.00,] ( 30.77, 33.12) -- ( 30.77, 29.52);

\draw[color=drawColor,line cap=round,line join=round,fill opacity=0.00,] ( 44.50, 33.12) -- ( 44.50, 29.52);

\draw[color=drawColor,line cap=round,line join=round,fill opacity=0.00,] ( 58.23, 33.12) -- ( 58.23, 29.52);

\draw[color=drawColor,line cap=round,line join=round,fill opacity=0.00,] ( 71.96, 33.12) -- ( 71.96, 29.52);

\draw[color=drawColor,line cap=round,line join=round,fill opacity=0.00,] ( 85.69, 33.12) -- ( 85.69, 29.52);

\draw[color=drawColor,line cap=round,line join=round,fill opacity=0.00,] ( 99.42, 33.12) -- ( 99.42, 29.52);

\node[color=drawColor,anchor=base,inner sep=0pt, outer sep=0pt, scale=  0.60] at ( 17.04, 18.72) {20};

\node[color=drawColor,anchor=base,inner sep=0pt, outer sep=0pt, scale=  0.60] at ( 30.77, 18.72) {30};

\node[color=drawColor,anchor=base,inner sep=0pt, outer sep=0pt, scale=  0.60] at ( 44.50, 18.72) {40};

\node[color=drawColor,anchor=base,inner sep=0pt, outer sep=0pt, scale=  0.60] at ( 58.23, 18.72) {50};

\node[color=drawColor,anchor=base,inner sep=0pt, outer sep=0pt, scale=  0.60] at ( 71.96, 18.72) {60};

\node[color=drawColor,anchor=base,inner sep=0pt, outer sep=0pt, scale=  0.60] at ( 85.69, 18.72) {70};

\node[color=drawColor,anchor=base,inner sep=0pt, outer sep=0pt, scale=  0.60] at ( 99.42, 18.72) {80};

\draw[color=drawColor,line cap=round,line join=round,fill opacity=0.00,] ( 15.12, 36.15) -- ( 15.12,111.93);

\draw[color=drawColor,line cap=round,line join=round,fill opacity=0.00,] ( 15.12, 36.15) -- ( 11.52, 36.15);

\draw[color=drawColor,line cap=round,line join=round,fill opacity=0.00,] ( 15.12, 48.78) -- ( 11.52, 48.78);

\draw[color=drawColor,line cap=round,line join=round,fill opacity=0.00,] ( 15.12, 61.41) -- ( 11.52, 61.41);

\draw[color=drawColor,line cap=round,line join=round,fill opacity=0.00,] ( 15.12, 74.04) -- ( 11.52, 74.04);

\draw[color=drawColor,line cap=round,line join=round,fill opacity=0.00,] ( 15.12, 86.67) -- ( 11.52, 86.67);

\draw[color=drawColor,line cap=round,line join=round,fill opacity=0.00,] ( 15.12, 99.30) -- ( 11.52, 99.30);

\draw[color=drawColor,line cap=round,line join=round,fill opacity=0.00,] ( 15.12,111.93) -- ( 11.52,111.93);

\node[color=drawColor,anchor=base east,inner sep=0pt, outer sep=0pt, scale=  0.60] at (  7.92, 34.09) {-8};

\node[color=drawColor,anchor=base east,inner sep=0pt, outer sep=0pt, scale=  0.60] at (  7.92, 46.72) {-6};

\node[color=drawColor,anchor=base east,inner sep=0pt, outer sep=0pt, scale=  0.60] at (  7.92, 59.35) {-4};

\node[color=drawColor,anchor=base east,inner sep=0pt, outer sep=0pt, scale=  0.60] at (  7.92, 71.98) {-2};

\node[color=drawColor,anchor=base east,inner sep=0pt, outer sep=0pt, scale=  0.60] at (  7.92, 84.61) {0};

\node[color=drawColor,anchor=base east,inner sep=0pt, outer sep=0pt, scale=  0.60] at (  7.92, 97.24) {2};

\node[color=drawColor,anchor=base east,inner sep=0pt, outer sep=0pt, scale=  0.60] at (  7.92,109.87) {4};

\draw[color=drawColor,line cap=round,line join=round,fill opacity=0.00,] ( 15.12, 33.12) --
	(104.08, 33.12) --
	(104.08,114.97) --
	( 15.12,114.97) --
	( 15.12, 33.12);
\end{scope}
\begin{scope}
\path[clip] (  0.00,  0.00) rectangle (108.41,130.09);
\definecolor[named]{drawColor}{rgb}{0.00,0.00,0.00}

\node[color=drawColor,anchor=base,inner sep=0pt, outer sep=0pt, scale=  0.72] at ( 59.60,  4.32) {$x_{7}$};
\end{scope}
\begin{scope}
\path[clip] ( 15.12, 33.12) rectangle (104.08,114.97);
\definecolor[named]{drawColor}{rgb}{0.00,0.00,0.00}

\draw[color=drawColor,line width= 0.8pt,line cap=round,line join=round,fill opacity=0.00,] ( 18.42, 80.00) --
	( 18.83, 80.35) --
	( 19.24, 80.70) --
	( 19.66, 81.04) --
	( 20.07, 81.38) --
	( 20.48, 81.72) --
	( 20.90, 82.05) --
	( 21.31, 82.37) --
	( 21.73, 82.69) --
	( 22.14, 83.00) --
	( 22.55, 83.31) --
	( 22.97, 83.61) --
	( 23.38, 83.91) --
	( 23.80, 84.20) --
	( 24.21, 84.48) --
	( 24.62, 84.76) --
	( 25.04, 85.03) --
	( 25.45, 85.30) --
	( 25.87, 85.56) --
	( 26.28, 85.82) --
	( 26.69, 86.06) --
	( 27.11, 86.31) --
	( 27.52, 86.54) --
	( 27.94, 86.77) --
	( 28.35, 86.99) --
	( 28.76, 87.21) --
	( 29.18, 87.42) --
	( 29.59, 87.62) --
	( 30.01, 87.82) --
	( 30.42, 88.00) --
	( 30.83, 88.18) --
	( 31.25, 88.36) --
	( 31.66, 88.53) --
	( 32.08, 88.69) --
	( 32.49, 88.84) --
	( 32.90, 88.99) --
	( 33.32, 89.14) --
	( 33.73, 89.27) --
	( 34.14, 89.41) --
	( 34.56, 89.54) --
	( 34.97, 89.66) --
	( 35.39, 89.79) --
	( 35.80, 89.91) --
	( 36.21, 90.02) --
	( 36.63, 90.14) --
	( 37.04, 90.25) --
	( 37.46, 90.36) --
	( 37.87, 90.46) --
	( 38.28, 90.57) --
	( 38.70, 90.68) --
	( 39.11, 90.78) --
	( 39.53, 90.89) --
	( 39.94, 90.99) --
	( 40.35, 91.10) --
	( 40.77, 91.21) --
	( 41.18, 91.32) --
	( 41.60, 91.43) --
	( 42.01, 91.54) --
	( 42.42, 91.65) --
	( 42.84, 91.77) --
	( 43.25, 91.89) --
	( 43.67, 92.01) --
	( 44.08, 92.14) --
	( 44.49, 92.27) --
	( 44.91, 92.40) --
	( 45.32, 92.53) --
	( 45.74, 92.66) --
	( 46.15, 92.79) --
	( 46.56, 92.92) --
	( 46.98, 93.05) --
	( 47.39, 93.18) --
	( 47.81, 93.30) --
	( 48.22, 93.43) --
	( 48.63, 93.55) --
	( 49.05, 93.67) --
	( 49.46, 93.78) --
	( 49.87, 93.89) --
	( 50.29, 93.99) --
	( 50.70, 94.09) --
	( 51.12, 94.18) --
	( 51.53, 94.27) --
	( 51.94, 94.35) --
	( 52.36, 94.42) --
	( 52.77, 94.48) --
	( 53.19, 94.53) --
	( 53.60, 94.58) --
	( 54.01, 94.61) --
	( 54.43, 94.64) --
	( 54.84, 94.65) --
	( 55.26, 94.66) --
	( 55.67, 94.65) --
	( 56.08, 94.63) --
	( 56.50, 94.60) --
	( 56.91, 94.55) --
	( 57.33, 94.50) --
	( 57.74, 94.44) --
	( 58.15, 94.36) --
	( 58.57, 94.28) --
	( 58.98, 94.19) --
	( 59.40, 94.09) --
	( 59.81, 93.98) --
	( 60.22, 93.86) --
	( 60.64, 93.74) --
	( 61.05, 93.61) --
	( 61.47, 93.47) --
	( 61.88, 93.32) --
	( 62.29, 93.17) --
	( 62.71, 93.01) --
	( 63.12, 92.85) --
	( 63.53, 92.69) --
	( 63.95, 92.51) --
	( 64.36, 92.34) --
	( 64.78, 92.16) --
	( 65.19, 91.98) --
	( 65.60, 91.79) --
	( 66.02, 91.60) --
	( 66.43, 91.41) --
	( 66.85, 91.22) --
	( 67.26, 91.03) --
	( 67.67, 90.83) --
	( 68.09, 90.64) --
	( 68.50, 90.44) --
	( 68.92, 90.25) --
	( 69.33, 90.05) --
	( 69.74, 89.85) --
	( 70.16, 89.65) --
	( 70.57, 89.46) --
	( 70.99, 89.26) --
	( 71.40, 89.06) --
	( 71.81, 88.86) --
	( 72.23, 88.66) --
	( 72.64, 88.46) --
	( 73.06, 88.26) --
	( 73.47, 88.05) --
	( 73.88, 87.85) --
	( 74.30, 87.65) --
	( 74.71, 87.45) --
	( 75.13, 87.24) --
	( 75.54, 87.04) --
	( 75.95, 86.83) --
	( 76.37, 86.63) --
	( 76.78, 86.42) --
	( 77.20, 86.22) --
	( 77.61, 86.01) --
	( 78.02, 85.81) --
	( 78.44, 85.60) --
	( 78.85, 85.39) --
	( 79.26, 85.18) --
	( 79.68, 84.98) --
	( 80.09, 84.77) --
	( 80.51, 84.56) --
	( 80.92, 84.35) --
	( 81.33, 84.14) --
	( 81.75, 83.93) --
	( 82.16, 83.72) --
	( 82.58, 83.51) --
	( 82.99, 83.30) --
	( 83.40, 83.08) --
	( 83.82, 82.87) --
	( 84.23, 82.66) --
	( 84.65, 82.45) --
	( 85.06, 82.23) --
	( 85.47, 82.02) --
	( 85.89, 81.81) --
	( 86.30, 81.59) --
	( 86.72, 81.38) --
	( 87.13, 81.16) --
	( 87.54, 80.95) --
	( 87.96, 80.73) --
	( 88.37, 80.52) --
	( 88.79, 80.30) --
	( 89.20, 80.08) --
	( 89.61, 79.87) --
	( 90.03, 79.65) --
	( 90.44, 79.43) --
	( 90.86, 79.21) --
	( 91.27, 79.00) --
	( 91.68, 78.78) --
	( 92.10, 78.56) --
	( 92.51, 78.34) --
	( 92.93, 78.12) --
	( 93.34, 77.90) --
	( 93.75, 77.68) --
	( 94.17, 77.46) --
	( 94.58, 77.24) --
	( 94.99, 77.02) --
	( 95.41, 76.80) --
	( 95.82, 76.58) --
	( 96.24, 76.36) --
	( 96.65, 76.13) --
	( 97.06, 75.91) --
	( 97.48, 75.69) --
	( 97.89, 75.47) --
	( 98.31, 75.25) --
	( 98.72, 75.02) --
	( 99.13, 74.80) --
	( 99.55, 74.58) --
	( 99.96, 74.35) --
	(100.38, 74.13) --
	(100.79, 73.91);

\draw[color=drawColor,line width= 0.8pt,dash pattern=on 4pt off 4pt ,line cap=round,line join=round,fill opacity=0.00,] ( 18.42, 76.65) --
	( 18.83, 77.28) --
	( 19.24, 77.91) --
	( 19.66, 78.52) --
	( 20.07, 79.10) --
	( 20.48, 79.70) --
	( 20.90, 80.21) --
	( 21.31, 80.73) --
	( 21.73, 81.21) --
	( 22.14, 81.64) --
	( 22.55, 82.06) --
	( 22.97, 82.42) --
	( 23.38, 82.74) --
	( 23.80, 83.04) --
	( 24.21, 83.33) --
	( 24.62, 83.59) --
	( 25.04, 83.85) --
	( 25.45, 84.08) --
	( 25.87, 84.29) --
	( 26.28, 84.49) --
	( 26.69, 84.71) --
	( 27.11, 84.90) --
	( 27.52, 85.07) --
	( 27.94, 85.25) --
	( 28.35, 85.42) --
	( 28.76, 85.58) --
	( 29.18, 85.74) --
	( 29.59, 85.91) --
	( 30.01, 86.06) --
	( 30.42, 86.21) --
	( 30.83, 86.36) --
	( 31.25, 86.52) --
	( 31.66, 86.67) --
	( 32.08, 86.80) --
	( 32.49, 86.94) --
	( 32.90, 87.05) --
	( 33.32, 87.20) --
	( 33.73, 87.32) --
	( 34.14, 87.44) --
	( 34.56, 87.55) --
	( 34.97, 87.64) --
	( 35.39, 87.74) --
	( 35.80, 87.82) --
	( 36.21, 87.90) --
	( 36.63, 87.99) --
	( 37.04, 88.04) --
	( 37.46, 88.11) --
	( 37.87, 88.17) --
	( 38.28, 88.23) --
	( 38.70, 88.28) --
	( 39.11, 88.35) --
	( 39.53, 88.40) --
	( 39.94, 88.46) --
	( 40.35, 88.53) --
	( 40.77, 88.58) --
	( 41.18, 88.66) --
	( 41.60, 88.76) --
	( 42.01, 88.86) --
	( 42.42, 88.97) --
	( 42.84, 89.07) --
	( 43.25, 89.20) --
	( 43.67, 89.34) --
	( 44.08, 89.49) --
	( 44.49, 89.64) --
	( 44.91, 89.77) --
	( 45.32, 89.93) --
	( 45.74, 90.09) --
	( 46.15, 90.24) --
	( 46.56, 90.37) --
	( 46.98, 90.52) --
	( 47.39, 90.66) --
	( 47.81, 90.76) --
	( 48.22, 90.88) --
	( 48.63, 90.96) --
	( 49.05, 91.03) --
	( 49.46, 91.09) --
	( 49.87, 91.16) --
	( 50.29, 91.19) --
	( 50.70, 91.23) --
	( 51.12, 91.26) --
	( 51.53, 91.29) --
	( 51.94, 91.31) --
	( 52.36, 91.31) --
	( 52.77, 91.28) --
	( 53.19, 91.28) --
	( 53.60, 91.26) --
	( 54.01, 91.24) --
	( 54.43, 91.23) --
	( 54.84, 91.19) --
	( 55.26, 91.15) --
	( 55.67, 91.12) --
	( 56.08, 91.06) --
	( 56.50, 91.00) --
	( 56.91, 90.94) --
	( 57.33, 90.88) --
	( 57.74, 90.81) --
	( 58.15, 90.73) --
	( 58.57, 90.65) --
	( 58.98, 90.54) --
	( 59.40, 90.43) --
	( 59.81, 90.31) --
	( 60.22, 90.20) --
	( 60.64, 90.06) --
	( 61.05, 89.89) --
	( 61.47, 89.76) --
	( 61.88, 89.61) --
	( 62.29, 89.46) --
	( 62.71, 89.28) --
	( 63.12, 89.09) --
	( 63.53, 88.90) --
	( 63.95, 88.69) --
	( 64.36, 88.47) --
	( 64.78, 88.25) --
	( 65.19, 87.97) --
	( 65.60, 87.70) --
	( 66.02, 87.41) --
	( 66.43, 87.13) --
	( 66.85, 86.83) --
	( 67.26, 86.52) --
	( 67.67, 86.21) --
	( 68.09, 85.89) --
	( 68.50, 85.59) --
	( 68.92, 85.31) --
	( 69.33, 84.98) --
	( 69.74, 84.61) --
	( 70.16, 84.26) --
	( 70.57, 83.92) --
	( 70.99, 83.59) --
	( 71.40, 83.27) --
	( 71.81, 82.91) --
	( 72.23, 82.55) --
	( 72.64, 82.13) --
	( 73.06, 81.74) --
	( 73.47, 81.35) --
	( 73.88, 80.94) --
	( 74.30, 80.58) --
	( 74.71, 80.21) --
	( 75.13, 79.84) --
	( 75.54, 79.46) --
	( 75.95, 79.08) --
	( 76.37, 78.67) --
	( 76.78, 78.22) --
	( 77.20, 77.72) --
	( 77.61, 77.31) --
	( 78.02, 76.91) --
	( 78.44, 76.37) --
	( 78.85, 75.88) --
	( 79.26, 75.38) --
	( 79.68, 74.90) --
	( 80.09, 74.31) --
	( 80.51, 73.77) --
	( 80.92, 73.23) --
	( 81.33, 72.74) --
	( 81.75, 72.14) --
	( 82.16, 71.51) --
	( 82.58, 70.95) --
	( 82.99, 70.30) --
	( 83.40, 69.74) --
	( 83.82, 69.12) --
	( 84.23, 68.50) --
	( 84.65, 67.88) --
	( 85.06, 67.32) --
	( 85.47, 66.70) --
	( 85.89, 65.97) --
	( 86.30, 65.25) --
	( 86.72, 64.55) --
	( 87.13, 63.85) --
	( 87.54, 63.14) --
	( 87.96, 62.45) --
	( 88.37, 61.75) --
	( 88.79, 60.92) --
	( 89.20, 60.23) --
	( 89.61, 59.57) --
	( 90.03, 58.66) --
	( 90.44, 58.05) --
	( 90.86, 57.19) --
	( 91.27, 56.47) --
	( 91.68, 55.70) --
	( 92.10, 54.96) --
	( 92.51, 54.27) --
	( 92.93, 53.52) --
	( 93.34, 52.70) --
	( 93.75, 51.82) --
	( 94.17, 50.98) --
	( 94.58, 50.08) --
	( 94.99, 49.12) --
	( 95.41, 48.09) --
	( 95.82, 47.05) --
	( 96.24, 46.26) --
	( 96.65, 45.53) --
	( 97.06, 44.68) --
	( 97.48, 43.86) --
	( 97.89, 42.83) --
	( 98.31, 41.79) --
	( 98.72, 41.08) --
	( 99.13, 40.25) --
	( 99.55, 39.45) --
	( 99.96, 38.50) --
	(100.38, 37.58) --
	(100.79, 36.60);

\draw[color=drawColor,line width= 0.8pt,dash pattern=on 4pt off 4pt ,line cap=round,line join=round,fill opacity=0.00,] ( 18.42, 82.88) --
	( 18.83, 83.02) --
	( 19.24, 83.15) --
	( 19.66, 83.30) --
	( 20.07, 83.43) --
	( 20.48, 83.60) --
	( 20.90, 83.77) --
	( 21.31, 83.93) --
	( 21.73, 84.12) --
	( 22.14, 84.33) --
	( 22.55, 84.56) --
	( 22.97, 84.81) --
	( 23.38, 85.09) --
	( 23.80, 85.39) --
	( 24.21, 85.72) --
	( 24.62, 86.08) --
	( 25.04, 86.44) --
	( 25.45, 86.81) --
	( 25.87, 87.13) --
	( 26.28, 87.49) --
	( 26.69, 87.81) --
	( 27.11, 88.13) --
	( 27.52, 88.44) --
	( 27.94, 88.74) --
	( 28.35, 89.05) --
	( 28.76, 89.32) --
	( 29.18, 89.56) --
	( 29.59, 89.80) --
	( 30.01, 90.03) --
	( 30.42, 90.26) --
	( 30.83, 90.45) --
	( 31.25, 90.61) --
	( 31.66, 90.78) --
	( 32.08, 90.93) --
	( 32.49, 91.04) --
	( 32.90, 91.18) --
	( 33.32, 91.30) --
	( 33.73, 91.45) --
	( 34.14, 91.59) --
	( 34.56, 91.70) --
	( 34.97, 91.84) --
	( 35.39, 91.97) --
	( 35.80, 92.08) --
	( 36.21, 92.22) --
	( 36.63, 92.35) --
	( 37.04, 92.50) --
	( 37.46, 92.65) --
	( 37.87, 92.78) --
	( 38.28, 92.91) --
	( 38.70, 93.05) --
	( 39.11, 93.17) --
	( 39.53, 93.31) --
	( 39.94, 93.43) --
	( 40.35, 93.57) --
	( 40.77, 93.74) --
	( 41.18, 93.85) --
	( 41.60, 94.00) --
	( 42.01, 94.13) --
	( 42.42, 94.26) --
	( 42.84, 94.40) --
	( 43.25, 94.53) --
	( 43.67, 94.61) --
	( 44.08, 94.72) --
	( 44.49, 94.87) --
	( 44.91, 94.99) --
	( 45.32, 95.10) --
	( 45.74, 95.22) --
	( 46.15, 95.35) --
	( 46.56, 95.48) --
	( 46.98, 95.65) --
	( 47.39, 95.77) --
	( 47.81, 95.88) --
	( 48.22, 96.01) --
	( 48.63, 96.18) --
	( 49.05, 96.35) --
	( 49.46, 96.55) --
	( 49.87, 96.75) --
	( 50.29, 96.95) --
	( 50.70, 97.11) --
	( 51.12, 97.29) --
	( 51.53, 97.47) --
	( 51.94, 97.65) --
	( 52.36, 97.85) --
	( 52.77, 98.01) --
	( 53.19, 98.16) --
	( 53.60, 98.31) --
	( 54.01, 98.46) --
	( 54.43, 98.58) --
	( 54.84, 98.67) --
	( 55.26, 98.72) --
	( 55.67, 98.73) --
	( 56.08, 98.74) --
	( 56.50, 98.76) --
	( 56.91, 98.73) --
	( 57.33, 98.68) --
	( 57.74, 98.62) --
	( 58.15, 98.55) --
	( 58.57, 98.43) --
	( 58.98, 98.31) --
	( 59.40, 98.19) --
	( 59.81, 98.05) --
	( 60.22, 97.90) --
	( 60.64, 97.72) --
	( 61.05, 97.60) --
	( 61.47, 97.44) --
	( 61.88, 97.28) --
	( 62.29, 97.10) --
	( 62.71, 96.93) --
	( 63.12, 96.79) --
	( 63.53, 96.61) --
	( 63.95, 96.46) --
	( 64.36, 96.32) --
	( 64.78, 96.18) --
	( 65.19, 96.02) --
	( 65.60, 95.85) --
	( 66.02, 95.72) --
	( 66.43, 95.59) --
	( 66.85, 95.47) --
	( 67.26, 95.37) --
	( 67.67, 95.26) --
	( 68.09, 95.15) --
	( 68.50, 95.08) --
	( 68.92, 94.97) --
	( 69.33, 94.88) --
	( 69.74, 94.78) --
	( 70.16, 94.70) --
	( 70.57, 94.63) --
	( 70.99, 94.54) --
	( 71.40, 94.46) --
	( 71.81, 94.42) --
	( 72.23, 94.35) --
	( 72.64, 94.28) --
	( 73.06, 94.19) --
	( 73.47, 94.07) --
	( 73.88, 94.01) --
	( 74.30, 93.96) --
	( 74.71, 93.91) --
	( 75.13, 93.83) --
	( 75.54, 93.76) --
	( 75.95, 93.69) --
	( 76.37, 93.65) --
	( 76.78, 93.63) --
	( 77.20, 93.58) --
	( 77.61, 93.53) --
	( 78.02, 93.52) --
	( 78.44, 93.54) --
	( 78.85, 93.52) --
	( 79.26, 93.49) --
	( 79.68, 93.47) --
	( 80.09, 93.48) --
	( 80.51, 93.48) --
	( 80.92, 93.51) --
	( 81.33, 93.52) --
	( 81.75, 93.45) --
	( 82.16, 93.42) --
	( 82.58, 93.45) --
	( 82.99, 93.48) --
	( 83.40, 93.50) --
	( 83.82, 93.54) --
	( 84.23, 93.58) --
	( 84.65, 93.55) --
	( 85.06, 93.58) --
	( 85.47, 93.60) --
	( 85.89, 93.66) --
	( 86.30, 93.68) --
	( 86.72, 93.76) --
	( 87.13, 93.81) --
	( 87.54, 93.89) --
	( 87.96, 93.97) --
	( 88.37, 94.04) --
	( 88.79, 94.13) --
	( 89.20, 94.27) --
	( 89.61, 94.38) --
	( 90.03, 94.42) --
	( 90.44, 94.52) --
	( 90.86, 94.64) --
	( 91.27, 94.77) --
	( 91.68, 94.89) --
	( 92.10, 95.04) --
	( 92.51, 95.25) --
	( 92.93, 95.43) --
	( 93.34, 95.58) --
	( 93.75, 95.72) --
	( 94.17, 95.83) --
	( 94.58, 95.94) --
	( 94.99, 96.10) --
	( 95.41, 96.26) --
	( 95.82, 96.48) --
	( 96.24, 96.70) --
	( 96.65, 96.91) --
	( 97.06, 97.13) --
	( 97.48, 97.36) --
	( 97.89, 97.61) --
	( 98.31, 97.87) --
	( 98.72, 98.12) --
	( 99.13, 98.31) --
	( 99.55, 98.54) --
	( 99.96, 98.76) --
	(100.38, 99.01) --
	(100.79, 99.23);

\draw[color=drawColor,line width= 0.8pt,dash pattern=on 1pt off 3pt ,line cap=round,line join=round,fill opacity=0.00,] ( 18.42, 75.07) --
	( 18.83, 75.86) --
	( 19.24, 76.61) --
	( 19.66, 77.34) --
	( 20.07, 78.06) --
	( 20.48, 78.76) --
	( 20.90, 79.44) --
	( 21.31, 80.11) --
	( 21.73, 80.63) --
	( 22.14, 81.07) --
	( 22.55, 81.48) --
	( 22.97, 81.86) --
	( 23.38, 82.20) --
	( 23.80, 82.55) --
	( 24.21, 82.89) --
	( 24.62, 83.14) --
	( 25.04, 83.38) --
	( 25.45, 83.60) --
	( 25.87, 83.78) --
	( 26.28, 83.95) --
	( 26.69, 84.12) --
	( 27.11, 84.30) --
	( 27.52, 84.46) --
	( 27.94, 84.61) --
	( 28.35, 84.77) --
	( 28.76, 84.94) --
	( 29.18, 85.10) --
	( 29.59, 85.27) --
	( 30.01, 85.44) --
	( 30.42, 85.60) --
	( 30.83, 85.73) --
	( 31.25, 85.90) --
	( 31.66, 86.03) --
	( 32.08, 86.15) --
	( 32.49, 86.28) --
	( 32.90, 86.40) --
	( 33.32, 86.47) --
	( 33.73, 86.55) --
	( 34.14, 86.62) --
	( 34.56, 86.69) --
	( 34.97, 86.77) --
	( 35.39, 86.89) --
	( 35.80, 86.94) --
	( 36.21, 86.97) --
	( 36.63, 87.00) --
	( 37.04, 87.05) --
	( 37.46, 87.05) --
	( 37.87, 87.09) --
	( 38.28, 87.05) --
	( 38.70, 87.04) --
	( 39.11, 87.08) --
	( 39.53, 87.09) --
	( 39.94, 87.15) --
	( 40.35, 87.18) --
	( 40.77, 87.23) --
	( 41.18, 87.27) --
	( 41.60, 87.37) --
	( 42.01, 87.52) --
	( 42.42, 87.71) --
	( 42.84, 87.91) --
	( 43.25, 87.99) --
	( 43.67, 88.20) --
	( 44.08, 88.33) --
	( 44.49, 88.48) --
	( 44.91, 88.71) --
	( 45.32, 88.84) --
	( 45.74, 88.99) --
	( 46.15, 89.21) --
	( 46.56, 89.38) --
	( 46.98, 89.52) --
	( 47.39, 89.64) --
	( 47.81, 89.75) --
	( 48.22, 89.78) --
	( 48.63, 89.84) --
	( 49.05, 89.91) --
	( 49.46, 89.95) --
	( 49.87, 90.01) --
	( 50.29, 90.02) --
	( 50.70, 90.04) --
	( 51.12, 90.06) --
	( 51.53, 90.06) --
	( 51.94, 90.03) --
	( 52.36, 90.03) --
	( 52.77, 90.03) --
	( 53.19, 89.98) --
	( 53.60, 89.96) --
	( 54.01, 89.93) --
	( 54.43, 89.88) --
	( 54.84, 89.86) --
	( 55.26, 89.79) --
	( 55.67, 89.72) --
	( 56.08, 89.64) --
	( 56.50, 89.57) --
	( 56.91, 89.54) --
	( 57.33, 89.50) --
	( 57.74, 89.39) --
	( 58.15, 89.31) --
	( 58.57, 89.21) --
	( 58.98, 89.11) --
	( 59.40, 89.05) --
	( 59.81, 88.94) --
	( 60.22, 88.83) --
	( 60.64, 88.70) --
	( 61.05, 88.57) --
	( 61.47, 88.41) --
	( 61.88, 88.19) --
	( 62.29, 88.02) --
	( 62.71, 87.77) --
	( 63.12, 87.55) --
	( 63.53, 87.31) --
	( 63.95, 87.11) --
	( 64.36, 86.87) --
	( 64.78, 86.64) --
	( 65.19, 86.32) --
	( 65.60, 85.98) --
	( 66.02, 85.64) --
	( 66.43, 85.34) --
	( 66.85, 84.97) --
	( 67.26, 84.70) --
	( 67.67, 84.39) --
	( 68.09, 84.11) --
	( 68.50, 83.83) --
	( 68.92, 83.45) --
	( 69.33, 83.21) --
	( 69.74, 82.82) --
	( 70.16, 82.40) --
	( 70.57, 81.85) --
	( 70.99, 81.38) --
	( 71.40, 80.79) --
	( 71.81, 80.45) --
	( 72.23, 80.01) --
	( 72.64, 79.51) --
	( 73.06, 79.05) --
	( 73.47, 78.66) --
	( 73.88, 78.24) --
	( 74.30, 77.78) --
	( 74.71, 77.07) --
	( 75.13, 76.35) --
	( 75.54, 75.89) --
	( 75.95, 74.87) --
	( 76.37, 74.12) --
	( 76.78, 73.35) --
	( 77.20, 72.73) --
	( 77.61, 72.15) --
	( 78.02, 71.85) --
	( 78.44, 70.98) --
	( 78.85, 70.28) --
	( 79.26, 69.42) --
	( 79.68, 68.78) --
	( 80.09, 68.30) --
	( 80.51, 67.73) --
	( 80.92, 67.06) --
	( 81.33, 66.38) --
	( 81.75, 65.43) --
	( 82.16, 64.73) --
	( 82.58, 64.05) --
	( 82.99, 63.37) --
	( 83.40, 62.68) --
	( 83.82, 61.93) --
	( 84.23, 61.05) --
	( 84.65, 60.06) --
	( 85.06, 59.02) --
	( 85.47, 58.35) --
	( 85.89, 57.44) --
	( 86.30, 56.51) --
	( 86.72, 55.58) --
	( 87.13, 54.64) --
	( 87.54, 53.62) --
	( 87.96, 52.59) --
	( 88.37, 51.55) --
	( 88.79, 50.47) --
	( 89.20, 49.02) --
	( 89.61, 47.94) --
	( 90.03, 47.18) --
	( 90.44, 46.24) --
	( 90.86, 45.16) --
	( 91.27, 44.07) --
	( 91.68, 42.98) --
	( 92.10, 41.87) --
	( 92.51, 40.77) --
	( 92.93, 39.81) --
	( 93.34, 38.95) --
	( 93.75, 38.10) --
	( 94.17, 37.25) --
	( 94.58, 36.30) --
	( 94.99, 35.21) --
	( 95.41, 34.12) --
	( 95.82, 33.02) --
	( 96.24, 31.91) --
	( 96.65, 30.80) --
	( 97.06, 29.68) --
	( 97.48, 28.56) --
	( 97.89, 27.41) --
	( 98.31, 26.11) --
	( 98.72, 24.79) --
	( 99.13, 23.47) --
	( 99.55, 22.14) --
	( 99.96, 20.80) --
	(100.38, 19.45) --
	(100.79, 18.10);

\draw[color=drawColor,line width= 0.8pt,dash pattern=on 1pt off 3pt ,line cap=round,line join=round,fill opacity=0.00,] ( 18.42, 84.02) --
	( 18.83, 84.09) --
	( 19.24, 84.17) --
	( 19.66, 84.21) --
	( 20.07, 84.29) --
	( 20.48, 84.40) --
	( 20.90, 84.49) --
	( 21.31, 84.60) --
	( 21.73, 84.75) --
	( 22.14, 84.86) --
	( 22.55, 85.14) --
	( 22.97, 85.36) --
	( 23.38, 85.66) --
	( 23.80, 85.95) --
	( 24.21, 86.34) --
	( 24.62, 86.75) --
	( 25.04, 87.17) --
	( 25.45, 87.57) --
	( 25.87, 87.98) --
	( 26.28, 88.40) --
	( 26.69, 88.77) --
	( 27.11, 89.14) --
	( 27.52, 89.44) --
	( 27.94, 89.76) --
	( 28.35, 90.09) --
	( 28.76, 90.34) --
	( 29.18, 90.55) --
	( 29.59, 90.74) --
	( 30.01, 90.95) --
	( 30.42, 91.11) --
	( 30.83, 91.35) --
	( 31.25, 91.55) --
	( 31.66, 91.73) --
	( 32.08, 91.89) --
	( 32.49, 92.01) --
	( 32.90, 92.18) --
	( 33.32, 92.27) --
	( 33.73, 92.36) --
	( 34.14, 92.53) --
	( 34.56, 92.68) --
	( 34.97, 92.84) --
	( 35.39, 92.98) --
	( 35.80, 93.14) --
	( 36.21, 93.23) --
	( 36.63, 93.36) --
	( 37.04, 93.49) --
	( 37.46, 93.67) --
	( 37.87, 93.83) --
	( 38.28, 93.91) --
	( 38.70, 94.06) --
	( 39.11, 94.20) --
	( 39.53, 94.34) --
	( 39.94, 94.48) --
	( 40.35, 94.67) --
	( 40.77, 94.81) --
	( 41.18, 94.92) --
	( 41.60, 95.04) --
	( 42.01, 95.23) --
	( 42.42, 95.37) --
	( 42.84, 95.55) --
	( 43.25, 95.67) --
	( 43.67, 95.78) --
	( 44.08, 95.93) --
	( 44.49, 96.04) --
	( 44.91, 96.16) --
	( 45.32, 96.33) --
	( 45.74, 96.50) --
	( 46.15, 96.64) --
	( 46.56, 96.76) --
	( 46.98, 96.88) --
	( 47.39, 97.03) --
	( 47.81, 97.14) --
	( 48.22, 97.35) --
	( 48.63, 97.54) --
	( 49.05, 97.79) --
	( 49.46, 97.97) --
	( 49.87, 98.12) --
	( 50.29, 98.35) --
	( 50.70, 98.61) --
	( 51.12, 98.80) --
	( 51.53, 99.01) --
	( 51.94, 99.17) --
	( 52.36, 99.39) --
	( 52.77, 99.60) --
	( 53.19, 99.79) --
	( 53.60,100.12) --
	( 54.01,100.35) --
	( 54.43,100.57) --
	( 54.84,100.76) --
	( 55.26,100.90) --
	( 55.67,100.98) --
	( 56.08,101.01) --
	( 56.50,101.00) --
	( 56.91,100.96) --
	( 57.33,100.98) --
	( 57.74,100.91) --
	( 58.15,100.73) --
	( 58.57,100.45) --
	( 58.98,100.36) --
	( 59.40,100.24) --
	( 59.81,100.01) --
	( 60.22, 99.73) --
	( 60.64, 99.74) --
	( 61.05, 99.69) --
	( 61.47, 99.41) --
	( 61.88, 99.19) --
	( 62.29, 99.03) --
	( 62.71, 98.74) --
	( 63.12, 98.63) --
	( 63.53, 98.43) --
	( 63.95, 98.33) --
	( 64.36, 98.15) --
	( 64.78, 98.15) --
	( 65.19, 98.15) --
	( 65.60, 98.10) --
	( 66.02, 97.97) --
	( 66.43, 97.84) --
	( 66.85, 97.70) --
	( 67.26, 97.53) --
	( 67.67, 97.51) --
	( 68.09, 97.42) --
	( 68.50, 97.42) --
	( 68.92, 97.33) --
	( 69.33, 97.16) --
	( 69.74, 96.95) --
	( 70.16, 96.81) --
	( 70.57, 96.78) --
	( 70.99, 96.83) --
	( 71.40, 96.95) --
	( 71.81, 96.82) --
	( 72.23, 96.88) --
	( 72.64, 96.95) --
	( 73.06, 96.86) --
	( 73.47, 96.76) --
	( 73.88, 96.66) --
	( 74.30, 96.54) --
	( 74.71, 96.47) --
	( 75.13, 96.51) --
	( 75.54, 96.44) --
	( 75.95, 96.44) --
	( 76.37, 96.42) --
	( 76.78, 96.51) --
	( 77.20, 96.52) --
	( 77.61, 96.61) --
	( 78.02, 96.67) --
	( 78.44, 96.70) --
	( 78.85, 96.69) --
	( 79.26, 96.74) --
	( 79.68, 96.79) --
	( 80.09, 96.84) --
	( 80.51, 96.84) --
	( 80.92, 96.94) --
	( 81.33, 96.98) --
	( 81.75, 97.04) --
	( 82.16, 97.09) --
	( 82.58, 97.25) --
	( 82.99, 97.38) --
	( 83.40, 97.35) --
	( 83.82, 97.41) --
	( 84.23, 97.57) --
	( 84.65, 97.63) --
	( 85.06, 97.81) --
	( 85.47, 97.90) --
	( 85.89, 97.98) --
	( 86.30, 98.16) --
	( 86.72, 98.25) --
	( 87.13, 98.36) --
	( 87.54, 98.59) --
	( 87.96, 98.78) --
	( 88.37, 98.99) --
	( 88.79, 99.11) --
	( 89.20, 99.37) --
	( 89.61, 99.58) --
	( 90.03, 99.70) --
	( 90.44,100.05) --
	( 90.86,100.21) --
	( 91.27,100.48) --
	( 91.68,100.82) --
	( 92.10,101.11) --
	( 92.51,101.29) --
	( 92.93,101.52) --
	( 93.34,101.98) --
	( 93.75,102.35) --
	( 94.17,102.73) --
	( 94.58,103.13) --
	( 94.99,103.58) --
	( 95.41,103.82) --
	( 95.82,104.09) --
	( 96.24,104.34) --
	( 96.65,104.60) --
	( 97.06,104.96) --
	( 97.48,105.30) --
	( 97.89,105.71) --
	( 98.31,106.08) --
	( 98.72,106.46) --
	( 99.13,106.91) --
	( 99.55,107.40) --
	( 99.96,107.88) --
	(100.38,108.25) --
	(100.79,108.48);
\end{scope}
\end{tikzpicture}

%% file: appendix.tex

\appendix{}
\part*{\appendixname}
\pdfbookmark[0]{\appendixname}{\appendixname}

Section~\ref{sec:impl-deta:marg-lik-comp} gives details on the computation of
the marginal likelihood~\eqref{eq:normal-models:marginal-likelihood} for normal
additive models. Section~\ref{sec:impl-deta:proposal-probs} derives the proposal
probabilities for the stochastic search described in
Section~\ref{sec:model-prior-and-search}.

\section{Marginal likelihood computation}
\label{sec:impl-deta:marg-lik-comp}

Under the hyper-$g$ prior, which assumes a uniform prior on the shrinkage
coefficient $g/(g + 1)$, the marginal likelihood of the transformed response
vector is \citep{LiangPauloMolinaClydeBerger2008}
\begin{equation}
  f(\tilde{\boldsymbol{y}} \given \boldsymbol{d})
  \propto  
  \norm{\boldsymbol{V}_{\boldsymbol{d}}^{-T/2}(\boldsymbol{y} -
    \boldsymbol{1}_{n}\bar{y})}^{-(n-1)} 
  (I + 2)^{-1} 
  {}_{2}\mathrm{F}_{1}
  \left(
    \frac{n-1}{2}; 1; \frac{I + 4}{2};
    \tilde{R}_{\boldsymbol{d}}^{2} 
  \right)
  \label{eq:normal-models:hyper-g-marginal-likelihood}  
\end{equation}
where $\bar{y} = n^{-1}\sum_{i=1}^{n}y_{i}$, ${}_{2}\mathrm{F}_{1}$ is the
Gaussian hypergeometric function \citep[p.~558]{Abramowitz:Stegun:1964} and
$\tilde{R}_{\boldsymbol{d}}^{2}$ is the classical coefficient of determination
in model~\eqref{eq:normal-models:marginal}. Under the hyper-$g/n$ prior, which
assumes a uniform prior on the term $(g/n) / \{(g/n) + 1\}$, the marginal
likelihood in the standard linear model is \citep[p.~155]{forte2011}
\begin{align}
  f(\tilde{\boldsymbol{y}} \given \boldsymbol{d})
  &\propto 
  n^{-I/2}(1 - \tilde{R}_{\boldsymbol{d}}^{2})^{-(n-1)/2}
  \frac{2}{I + 2}
  \notag \\
  &\quad \times
  \mathrm{AF}_{1}
  \left(
    \frac{I}{2} + 1;
    \frac{I + 1 - n}{2};
    \frac{n - 1}{2};
    \frac{I}{2} + 2;
    \frac{n - 1}{n},
    \frac{n - (1 - \tilde{R}_{\boldsymbol{d}}^{2})^{-1}}{n} 
  \right),
  \label{eq:normal-models:hyper-g-over-n-marg-lik}  
\end{align}
where $\mathrm{AF}_{1}$ is the Appell hypergeometric function of
the first kind \citep{appell1925}. \citet{colavecchia.gasaneo2004} provide
Fortran code for computing this special function, which is accessible in
\texttt{R} via the package ``\texttt{appell}'' \citep{sabanesbove2012}. 
For large sample sizes $n > 100$ or when the numerical computations of the
special functions in \eqref{eq:normal-models:hyper-g-marginal-likelihood} or
\eqref{eq:normal-models:hyper-g-over-n-marg-lik} fail, we instead use Laplace
approximations as described by
\citet[appendix~A]{LiangPauloMolinaClydeBerger2008}.

For the coefficient of determination $\tilde{R}^{2}_{\boldsymbol{d}} =
SSM_{\boldsymbol{d}} / SST_{\boldsymbol{d}}$ required in
\eqref{eq:normal-models:hyper-g-marginal-likelihood} or
\eqref{eq:normal-models:hyper-g-over-n-marg-lik}, we need to compute the sum of
squares in total ($SST_{\boldsymbol{d}}$) and the sum of squares explained by
the model ($SSM_{\boldsymbol{d}}$). For $SST_{\boldsymbol{d}}$, we have
\begin{align*}
  SST_{\boldsymbol{d}} &= 
  (\boldsymbol{y} - \boldsymbol{1}_{n}\bar{y})^{T}
  \boldsymbol{V}_{\boldsymbol{d}}^{-1}
  (\boldsymbol{y} - \boldsymbol{1}_{n}\bar{y})
  \\
  &= 
  \normsmall{\boldsymbol{y} - \boldsymbol{1}_{n}\bar{y}}^{2}
  -
  \normsmall{\boldsymbol{W}_{\boldsymbol{d}}^{T}(\boldsymbol{y} -
    \boldsymbol{1}_{n}\bar{y})}^{2}.
\end{align*}
Note that the first term in~\eqref{eq:normal-models:marginal-likelihood} can be
written as $\norm{\boldsymbol{V}_{\boldsymbol{d}}^{-T/2}(\boldsymbol{y} -
  \boldsymbol{1}_{n}\bar{y})}^{-(n-1)} = SST_{\boldsymbol{d}}^{-(n-1)/2}$. For
$SSM_{\boldsymbol{d}}$, note that the fit of the general linear model is
$\hat{\boldsymbol{y}}_{\boldsymbol{d}} = \boldsymbol{1}_{n}\bar{y} +
\boldsymbol{X}_{\boldsymbol{d}}\hat{\boldsymbol{\beta}}_{\boldsymbol{d}}$, where
\[
\hat{\boldsymbol{\beta}}_{\boldsymbol{d}} = 
(\boldsymbol{X}_{\boldsymbol{d}}^{T} \boldsymbol{V}_{\boldsymbol{d}}^{-1}
\boldsymbol{X}_{\boldsymbol{d}})^{-1}
\boldsymbol{X}_{\boldsymbol{d}}^{T}\boldsymbol{V}_{\boldsymbol{d}}^{-1}\boldsymbol{y} 
\]
is the weighted least squares estimate of $\boldsymbol{\beta}_{\boldsymbol{d}}$.
Therefore
\begin{align*}
  SSM_{\boldsymbol{d}} 
  &=
  (\hat{\boldsymbol{y}}_{\boldsymbol{d}} - \boldsymbol{1}_{n}\bar{y})^{T}
  \boldsymbol{V}_{\boldsymbol{d}}^{-1}
  (\hat{\boldsymbol{y}}_{\boldsymbol{d}} - \boldsymbol{1}_{n}\bar{y})
  \\
  &=
  \hat{\boldsymbol{\beta}}_{\boldsymbol{d}}^{T}
  \boldsymbol{X}_{\boldsymbol{d}}^{T}
  \boldsymbol{V}_{\boldsymbol{d}}^{-1}
  \boldsymbol{X}_{\boldsymbol{d}}
  \hat{\boldsymbol{\beta}}_{\boldsymbol{d}}
\end{align*}
can be computed by Cholesky factorising $\boldsymbol{X}_{\boldsymbol{d}}^{T}
\boldsymbol{V}_{\boldsymbol{d}}^{-1} \boldsymbol{X}_{\boldsymbol{d}} =
\boldsymbol{C}_{\boldsymbol{d}}^{T}\boldsymbol{C}_{\boldsymbol{d}}$, solving the
triangular system $\boldsymbol{C}_{\boldsymbol{d}}^{T}
\boldsymbol{v}_{\boldsymbol{d}} =
\boldsymbol{X}_{\boldsymbol{d}}^{T}\boldsymbol{V}_{\boldsymbol{d}}^{-1}\boldsymbol{y}$
and setting $SSM_{\boldsymbol{d}} =
\norm{\boldsymbol{v}_{\boldsymbol{d}}}^{2}$.

For the computations above, we need the inverse of the covariance matrix
$\boldsymbol{V}_{\boldsymbol{d}} \in \R^{n\times n}$. While usually a Cholesky
factorisation would be done, here it is advisable to avoid it because it has
complexity $\mathcal{O}(n^{3})$ and is therefore computationally expensive.
Therefore, we instead work with the formula
\[
\boldsymbol{V}_{\boldsymbol{d}}^{-1} = 
\boldsymbol{I}_{n} -
\boldsymbol{Z}_{\boldsymbol{d}}
\boldsymbol{M}_{\boldsymbol{d}}^{-1}
\boldsymbol{Z}_{\boldsymbol{d}}^{T} 
\]
for the precision matrix, where $\boldsymbol{M}_{\boldsymbol{d}} =
\boldsymbol{Z}_{\boldsymbol{d}}^{T}\boldsymbol{Z}_{\boldsymbol{d}} +
\boldsymbol{D}_{\boldsymbol{d}}^{-1}$. The latter matrix has dimension $JK$,
which is usually smaller than $n$, provided the spline basis dimension $K$ is
small. Thus, the Cholesky factorisation $\boldsymbol{M}_{\boldsymbol{d}} =
\boldsymbol{M}_{\boldsymbol{d}}^{T/2}\boldsymbol{M}_{\boldsymbol{d}}^{1/2}$ is
relatively fast, and we compute $\boldsymbol{W}_{\boldsymbol{d}} =
\boldsymbol{Z}_{\boldsymbol{d}}\boldsymbol{M}_{\boldsymbol{d}}^{-1/2}$ such that
$\boldsymbol{V}_{\boldsymbol{d}}^{-1} = \boldsymbol{I}_{n} -
\boldsymbol{W}_{\boldsymbol{d}}\boldsymbol{W}_{\boldsymbol{d}}^{T}$.

Finally, to compute the determinant term
in~\eqref{eq:normal-models:marginal-likelihood}, we can again avoid factorising
$\boldsymbol{V}_{\boldsymbol{d}}$, because we have
\[
\abs{\boldsymbol{V}_{\boldsymbol{d}}^{1/2}}^{-1} 
= \abs{\boldsymbol{V}_{\boldsymbol{d}}^{-1}}^{1/2}
= \abs{\boldsymbol{I}_{n} -
  \boldsymbol{W}_{\boldsymbol{d}}\boldsymbol{W}_{\boldsymbol{d}}^{T}}^{1/2} 
= \abs{\boldsymbol{I}_{JK} -
  \boldsymbol{W}_{\boldsymbol{d}}^{T}\boldsymbol{W}_{\boldsymbol{d}}}^{1/2}, 
\]
see \citet[p.~416]{Harville1997} for the last equality. So again only a matrix
of dimension $JK$, namely $\boldsymbol{I}_{JK} -
\boldsymbol{W}_{\boldsymbol{d}}^{T}\boldsymbol{W}_{\boldsymbol{d}}$, needs to be
factorised. Here, a LU factorisation can be used.

\section{Proposal probabilities}
\label{sec:impl-deta:proposal-probs}

First note that the two proposal types `Move' and `Swap' do not overlap,
because a `Move' always changes exactly one $d_{j}$, while a `Swap' either
changes none or two $d_{j}$'s. Denote with $p_{m}$ the probability to choose a
`Move'. 

Suppose a `Move' was proposed for covariate $j \in \{0, 1, \dotsc, p\}$. We then
have
\[
q(\boldsymbol{d}' \given \boldsymbol{d}) = p_{m} \cdot \frac{1}{p} \cdot
\begin{cases}
  1, & d_{j} \in \{0, K\},\\
  \frac{1}{2}, & \text{else}
\end{cases}
\]
and analogously
\[
q(\boldsymbol{d} \given \boldsymbol{d}') = p_{m} \cdot \frac{1}{p} \cdot
\begin{cases}
  1, & d_{j}' \in \{0, K\},\\
  \frac{1}{2}, & \text{else}
\end{cases}
\]
with proposal ratio
\[
\frac
{q(\boldsymbol{d}' \given \boldsymbol{d})}
{q(\boldsymbol{d} \given \boldsymbol{d}')} 
=
\begin{cases}
  \frac{1}{2}, & d_{j}' \in \{0, K\},\\
  2, & d_{j} \in \{0, K\},\\
  1, & \text{else}.
\end{cases}
\]

For the `Swap' proposal, suppose covariates $i$ and $j$ are proposed to
interchange their model parameters $d_{i}$ and $d_{j}$. Of course, if $d_{i} =
d_{j}$, then the proposal ratio equals unity because $\boldsymbol{d}' =
\boldsymbol{d}$. In the other case, both model parameters are changed, and
\[
q(\boldsymbol{d}' \given \boldsymbol{d}) =
q(\boldsymbol{d} \given \boldsymbol{d}') =
(1 - p_{m}) \cdot \binom{p}{2}^{-1},
\]
so that for a `Swap' we always have $q(\boldsymbol{d}' \given \boldsymbol{d})
/ q(\boldsymbol{d} \given \boldsymbol{d}') = 1$.


%% file: literatur.tex
\bibliographystyle{abbrvnat}
\bibliography{literatur}
